%% file: manuscript.tex
\documentclass[preprint]{elsarticle}

\usepackage{graphicx}
\usepackage{xcolor}
\usepackage{amsmath,amsthm}
\usepackage{filecontents}
\usepackage{lipsum}
\usepackage{pgfplots}
\pgfplotsset{compat=1.5.1}
\usepackage{tikz}
\usepackage{mathdots}
\usepackage{yhmath}
\usepackage{cancel}
\usepackage{xcolor}
\usepackage{siunitx}
\usepackage{array}
\usepackage{multirow}
\usepackage{textcomp}
\usepackage{gensymb}
\usepackage{tabularx}
\usepackage{booktabs}
\usepackage[hidelinks,breaklinks]{hyperref}
\usepackage{xargs}                      
\usepackage{todonotes}
\usepackage{subfigure}
\usepackage{cleveref}

\usepackage{amssymb}
\usepackage{amsthm}

\usepackage{lineno}

\usetikzlibrary{spy}

\biboptions{sort&compress}

\journal{Computers and Mathematics with Applications}

\newcommand{\db}{\boldsymbol{d}}
\newcommand{\bE}{\mathbf{E}}
\newcommand{\bF}{\mathbf{F}}
\newcommand{\bS}{\mathbf{S}}
\newcommand{\bI}{\mathbf{I}}
\newcommand{\bT}{\mathbf{T}}

\newcommand{\bvec}{\boldsymbol{b}}
\newcommand{\dvec}{\boldsymbol{d}}
\newcommand{\dvech}{\dvec^h}
\newcommand{\gvec}{\boldsymbol{g}}
\newcommand{\hvec}{\boldsymbol{h}}
\newcommand{\uvec}{\boldsymbol{u}}
\newcommand{\xvec}{\boldsymbol{x}}
\newcommand{\ovec}{\boldsymbol{0}}
\newcommand{\nvec}{\boldsymbol{n}}
\newcommand{\Ivec}{\boldsymbol{I}}

\newcommand{\tvec}{\boldsymbol{t}}
\newcommand{\thetavec}{\boldsymbol{\theta}}
\newcommand{\uvech}{\uvec^h}
\newcommand{\wvec}{\boldsymbol{w}}
\newcommand{\wvech}{\wvec^h}
\newcommand{\fvec}{\boldsymbol{f}}
\newcommand{\fvech}{\fvec^h}
\newcommand{\ph}{p^h}
\newcommand{\qh}{q^h}
\newcommand{\Pb}{\boldsymbol{P}}
\newcommand{\grad}{\boldsymbol{\nabla}}
\newcommand{\Div}{\boldsymbol{\nabla}\cdot}
\newcommand{\DivO}{\boldsymbol{\nabla}_0\cdot}

\newcommand{\defGrad}{\boldsymbol{F}}

\usepackage{stackengine,scalerel}
\stackMath
\newcommand\qfrac[3][1pt]{\frac{%
        \ThisStyle{\addstackgap[#1]{\SavedStyle#2}}}{%
        \ThisStyle{\addstackgap[#1]{\SavedStyle#3}}%
}}

\newtheorem{remark}{Remark}

\tikzset {_6b01hkhlk/.code = {\pgfsetadditionalshadetransform{ \pgftransformshift{\pgfpoint{0 bp } { 0 bp }  }  \pgftransformscale{1 }  }}}
\pgfdeclareradialshading{_3nha1xokn}{\pgfpoint{0bp}{0bp}}{rgb(0bp)=(1,1,1);
rgb(0.08928571428571429bp)=(1,1,1);
rgb(25bp)=(0.43,0.43,0.43);
rgb(400bp)=(0.43,0.43,0.43)}
\tikzset{_tdutiy9ph/.code = {\pgfsetadditionalshadetransform{\pgftransformshift{\pgfpoint{0 bp } { 0 bp }  }  \pgftransformscale{1 } }}}
\pgfdeclareradialshading{_h9ys7hqx8} { \pgfpoint{0bp} {0bp}} {color(0bp)=(transparent!28);
color(0.08928571428571429bp)=(transparent!28);
color(25bp)=(transparent!30);
color(400bp)=(transparent!30)}
\pgfdeclarefading{_jdoi03xod}{\tikz \fill[shading=_h9ys7hqx8,_tdutiy9ph] (0,0) rectangle (50bp,50bp); }
\tikzset{every picture/.style={line width=0.75pt}} 

\tikzset {_crarpyw4h/.code = {\pgfsetadditionalshadetransform{ \pgftransformshift{\pgfpoint{0 bp } { 0 bp }  }  \pgftransformrotate{0 }  \pgftransformscale{2 }  }}}
\pgfdeclarehorizontalshading{_cdyt0duuv}{150bp}{rgb(0bp)=(0,0,0);
rgb(37.5bp)=(0,0,0);
rgb(37.5bp)=(0,0,0);
rgb(57.5bp)=(1,1,1);
rgb(100bp)=(1,1,1)}
\tikzset{_bl5llyq4a/.code = {\pgfsetadditionalshadetransform{\pgftransformshift{\pgfpoint{0 bp } { 0 bp }  }  \pgftransformrotate{0 }  \pgftransformscale{2 } }}}
\pgfdeclarehorizontalshading{_fpxgbj5sz} {150bp} {color(0bp)=(transparent!73);
color(37.5bp)=(transparent!73);
color(37.5bp)=(transparent!73);
color(57.5bp)=(transparent!73);
color(100bp)=(transparent!73) }
\pgfdeclarefading{_16ok8mlny}{\tikz \fill[shading=_fpxgbj5sz,_bl5llyq4a] (0,0) rectangle (50bp,50bp); }
\tikzset{every picture/.style={line width=0.75pt}} 

\begin{document}

\begin{frontmatter}




\title{Spline-Based Space-Time Finite Element Approach for Fluid-Structure Interaction Problems With a Focus on Fully Enclosed Domains}


\author[rwth]{Michel Make\corref{cor1}}
\ead{make@cats.rwth-aachen.de}
\cortext[cor1]{Corresponing Author}
\author[rwth]{Thomas Spenke}
\ead{spenke@cats.rwth-aachen.de}
\author[rwth]{Norbert Hosters}
\ead{hosters@cats.rwth-aachen.de}
\author[rwth]{Marek Behr}
\ead{behr@cats.rwth-aachen.de}

\address[rwth]{Chair for Computational Analysis of Technical Systems (CATS), \\ JARA Center for Simulation and Data Science (JARA-CSD), \\ RWTH Aachen University, Schinkelstrasse 2, 52062 Aachen, Germany}

\begin{abstract}
    Non-Uniform Rational B-Spline (NURBS) surfaces are commonly used within Computer-Aided Design (CAD) tools to represent geometric objects. When using isogeometric analysis (IGA), it is possible to use such NURBS geometries for numerical analysis directly. Analyzing fluid flows, however, requires complex three-dimensional geometries to represent flow domains. Defining a parametrization of such volumetric domains using NURBS can be challenging and is still an ongoing topic in the IGA community.

    With the recently developed NURBS-enhanced finite element method (NEFEM), the favorable geometric characteristics of NURBS are used within a standard finite element method. This is achieved by enhancing the elements touching the boundary by using the NURBS geometry itself. In the current work, a new variation of NEFEM is introduced, which is suitable for three-dimensional space-time finite element formulations. The proposed method makes use of a new mapping which results in a non-Cartesian formulation suitable for fluid-structure interaction (FSI).

    This is demonstrated by combining the method with an IGA formulation in a strongly-coupled partitioned framework for solving FSI problems. The framework yields a fully spline-based representation of the fluid-structure interface through a single  NURBS.

    The coupling conditions at the fluid-structure interface are enforced through a Robin-Neumann type coupling scheme. This scheme is particularly useful when considering incompressible fluids in fully Dirichlet-bounded and curved problems, as it satisfies the incompressibility constraint on the fluid for each step within the coupling procedure.

    The accuracy and performance of the introduced spline-based space-time finite element approach and its use within the proposed coupled FSI framework are demonstrated using a series of two- and three-dimensional benchmark problems.

\end{abstract}

\begin{keyword}
Non-Cartesian NURBS-Enhanced Finite Elements \sep
Spline-Based Methods \sep
Fluid-Structure Interaction \sep
Exact Geometry Representation \sep
Fully Dirichlet-Bounded Problems \sep
Fully Enclosed Domains \sep
Curved Domains
\end{keyword}
\end{frontmatter}

\section{Introduction}
\label{sec: instroduction}

Engineering tools such as CAD software are the modern-day standard in many engineering design processes. These software tools often use Non-Uniform Rational B-Splines, or NURBS, to represent geometric shapes and objects. With the introduction of IGA \cite{hughes2005}, it became possible to directly perform numerical analysis on such objects without the need for meshing the geometric model under consideration.

However, most of the available CAD tools do not represent three-dimensional objects through volume splines, since generating such splines for complex shapes is challenging \cite{zhang2012solid, schillinger2012isogeometric}. Instead, objects are represented with surface splines only. A surface-based alternative to isogeometric analysis for volume problems can therefore be advantageous.

With the recently developed NEFEM \cite{sevilla2011, sevilla2008NEFEM, sevilla20113dNEFEM, stavrev2016STNEFEM}, the difficulties that arise when using IGA for complex volumetric domains are avoided. NEFEM allows favorable geometric characteristics of NURBS to be utilized within a standard finite element method (FEM)\footnote{In the current work,  \textit{standard finite element method}, is referring to classic isoparametric finite element methods in conjunction with linear Lagrangian finite elements.}. This is achieved by \textit{enhancing} boundary elements using the NURBS geometry, which is assumed to be exact.

Where IGA uses a NURBS basis to represent both geometry and numerical solution \cite{cottrell2009isogeometric},  NEFEM uses a NURBS representation of the domain boundary only. Consequently, for problems in $\mathbb{R}^{n_{sd}}$ (with $n_{sd}$ being the number of spatial dimensions), at most an $\mathbb{R}^{n_{sd}-1}$ NURBS domain boundary surface is needed.

Furthermore, the elements which do not have a common interface with the NURBS boundary are treated as \textit{standard} finite elements.  As a result, only a small number of elements are of NEFEM type, keeping the potential increase in computational cost to a minimum. While the approach avoids the need for specialized mesh generation tools since the method requires only a standard finite element mesh as a starting point, some care during the meshing step is needed to assure the proposed NEFEM mapping is one-to-one (see, e.g., \cite{Sevilla2016}).

A space-time variant of NEFEM for three-dimensional problems is presented in the current work, which extends the formulation in \cite{hosters2018} to three-dimensional problems. For this, a new mapping from local to global coordinates is proposed. An extension of the mapping to allow for space-time finite elements is also presented.

The proposed spline-based formulation differs from the original NEFEM method by defining the numerical solution as well as the variational form on the reference element $I$ using local coordinates $(\hat{\xi}, \hat{\eta}, \hat{\zeta})$. Consequently, when evaluating boundary integrals, negative shape function contributions are avoided, and the partition of unity property is maintained. This is favorable when considering fluid-structure interaction (FSI) problems where surface quantities such as fluid tractions need to be evaluated accurately.

As will become apparent in the remainder of this work, strictly speaking, the proposed method can be classified as a blending function method (see, e.g., \cite{szabo2004pfem}). However, to emphasize the similarities with NEFEM \cite{sevilla2011, sevilla2008NEFEM, sevilla20113dNEFEM,stavrev2016STNEFEM}, the proposed formulation is termed \textit{non-Cartesian NEFEM}.

To demonstrate its favorable properties in relation to FSI problems, the method is employed within a strongly-coupled partitioned FSI solver framework. The framework combines non-Cartesian NEFEM and IGA to obtain an exact smooth geometric representation of the fluid-structure interface using a single NURBS definition.

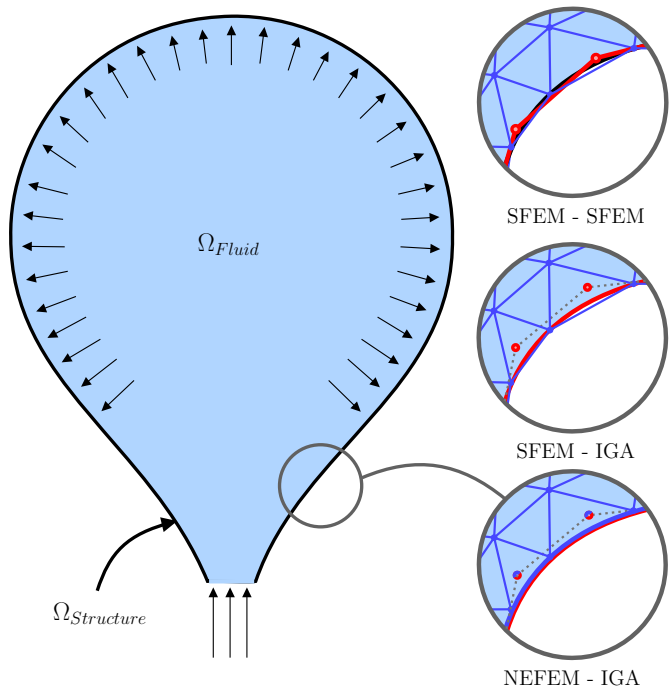
\begin{figure}[ht]
    \centering
    \resizebox{0.75\textwidth}{!}{
        \centering{\input{figures/spline-based-methods.tikz}}
    }
    \caption{Various combinations of discretizations for an inflating balloon. Here standard linear finite elements (SFEM), non-Cartesian NEFEM, and IGA are combined in various ways. Note that here the thin-walled structure of the balloon is represented by an IGA shell. The presented spline-based framework also allows for structures represented by volume splines. In that case, not the complete volume spline, but rather its surfaces are used for the non-Cartesian NEFEM formulation. In Section \ref{sec: numerical-examples}, this concept is used for a two-dimensional test case.
        \label{fig:spline-based-methods}}
\end{figure}

Having a common spline representation for the fluid and structural problem allows for a direct transfer of coupling quantities while still permitting different refinement levels for the individual domain discretizations. An example in which conventional and spline-based methods are combined in various ways is depicted in Figure \ref{fig:spline-based-methods}.

Within a strongly-coupled partitioned procedure, the fluid and structural problems are solved separately in an iterative fashion. The individual problems are then coupled by imposing a set of coupling conditions at the fluid-structure interface.

A popular choice to enforce these conditions is the Dirichlet-Neumann (DN) scheme. With a DN scheme, the computed structural deformation and velocity are imposed as a Dirichlet boundary condition to the fluid problem, while the fluid tractions at the coupling interface are imposed as a Neumann boundary condition to the structural problem. Despite its straightforward implementation, as shown in \cite{causin2005}, DN schemes typically require a significant number of iterations and relaxation to obtain converged solutions for high density ratio problems \cite{causin2005}. Additional measures, such as the artificial compressibility method or interface quasi-Newton methods could further improve stability \cite{degroote2010, spenke2020}. Additionally, for fully Dirichlet-bounded problems, DN procedures fail altogether \cite{kuttler2006}.

A problem is fully Dirichlet-bounded if Dirichlet boundary conditions are applied to all boundaries of the computational domain. This occurs, e.g., for balloon type problems where a structure is filled with a fluid or for flexible tubular structures with prescribed velocity profiles at the inflow and outflow boundaries.

Such problems can only be solved if the prescribed velocities along the domain boundary satisfy the mass balance and when the fluid pressure level is fixed by an additional constraint \cite{kuttler2006}. DN-type procedures cannot solve enclosed, fully Dirichlet-bounded problems as they do not meet these requirements \cite{kuttler2006}.

A solution to the mass balance dilemma is to use Robin boundary conditions, as they allow for artificial fluxes over the FSI interface \cite{badia2008,gerardo-giorda2010,nobile2008, hosters2018thesis}. In this work, a Robin-Neumann (RN) procedure is used to introduce such an artificial flux. By minimizing this artificial flux during the coupling, a solution that satisfies the incompressibility constraint can be obtained.

The spline-based coupling strategy introduced in this work is used to solve the incompressible Navier-Stokes equations on the fluid domain using the Deforming Spatial Domain/Stabilized Space-Time (DSD/SST) formulation \cite{TEZDUYAR1992339} complemented with non-Cartesian NURBS-enhanced finite elements. On the structural domain, a thin-walled elastic structure is solved using isogeometric analysis. Both solvers are coupled in a strong partitioned manner in combination with an RN type coupling. The framework's superior accuracy is demonstrated by a set of benchmark problems involving enclosed, fully Dirichlet-bounded, and curved domains.

The rest of this paper is organized as follows: In Section \ref{sec: spline-based-methods}, a basic review of NURBS theory is presented followed by the concept of Cartesian and non-Cartesian NEFEM. In section \ref{sec: gov-eq}, the governing equations used to describe both the fluid and structural problems in the context of FSI are given. The computational framework used to solve FSI problems is given in Section \ref{sec: num-meth}. The proposed spline-based framework is demonstrated by means of a set of numerical examples in Section \ref{sec: numerical-examples}. Finally, concluding remarks are given in Section \ref{sec: conclusions}.

\section{Spline-Based Methods for Interface Coupled Problems}
\label{sec: spline-based-methods}

This section will discuss the general concept of NEFEM and the newly proposed three-dimensional non-Cartesian NEFEM formulation, followed by a summary of the IGA approach.
However, before continuing on the topic of NEFEM, a brief introduction to NURBS will be presented first.

\subsection{Review of Non-Uniform Rational B-Splines}
\label{sec: nurbs}

Geometries can be described by means of NURBS \cite{piegl1997nurbs}, or occasionally T-splines \cite{BAZILEVS2010229}. Due to their favorable mathematical properties, such descriptions are commonly used in CAD tools to represent geometric objects.

A NURBS curve $\boldsymbol{C}(\theta)$ is a function of parametric coordinate $\theta$ and describes a geometric curve in $\mathbb{R}^{n_{sd}}$.  Such a parametrization is constructed using a NURBS basis $R_{i}^p$, and a set of control points $\Pb_i(\xvec)$ in $\mathbb{R}^{n_{sd}}$:
\begin{align}
\boldsymbol{C}(\theta) = \sum_{i=1}^{n_{cp}} R_{i}^p(\theta)\Pb_i(\xvec).
\label{eq:nurbs}
\end{align}
Here, $n_{cp}$ is the number of control points, and superscript $p$ the order of the NURBS basis. $R_{i}^p(\theta)$ are rational functions which are constructed using $p^{th}$-degree B-spline basis functions $N_i^p(\theta)$, and a set of control weights $w_i$:
\begin{align}
R_i^p(\theta) = \frac{N_i^p(\theta) w_i}{ \sum_{j=1}^{n_{cp}} N_j^p(\theta) w_j}.
\label{eq:nurbs-basis}
\end{align}

The B-spline basis functions $N_i^p(\theta)$ are generated using the Cox-de Boor recursion formula. An extensive discussion on the construction of a NURBS curve and the necessary bases can be found in the work of Piegl and Tiller \cite{piegl1997nurbs}.

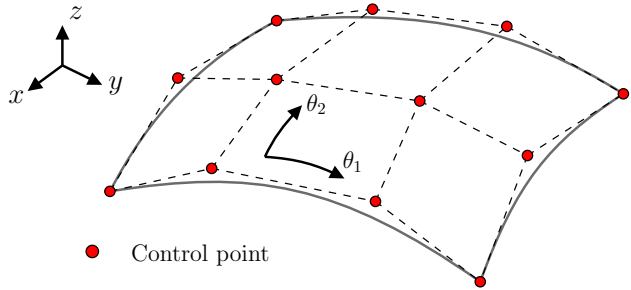
\begin{figure}
    \centering
    \resizebox{0.7\textwidth}{!}{%
        \input{figures/nurbs-surf.tikz}
    }
    \caption{NURBS surface in $\mathbb{R}^3$, and its corresponding parametric coordinate system $\thetavec=(\theta_1,\theta_2)^T$.\label{fig:nurbs-surf}}
\end{figure}

Analogous to Equations \eqref{eq:nurbs} and \eqref{eq:nurbs-basis}, a NURBS surface $\boldsymbol{S}(\thetavec)$ can be obtained by taking the tensor product of two NURBS curves \cite{cottrell2009isogeometric}. The resulting description for $\boldsymbol{S}(\thetavec)$ is then defined using the basis $R_{i,j}^{p,q}(\,\thetavec\,)$ and a net of control points $\Pb_{ij}(\xvec)$:
\begin{align}
\mathbf{S}(\,\thetavec\,) = \sum_{i=1}^{n_{cp}} \sum_{j=1}^{m_{cp}}R_{i,j}^{p,q}(\,\thetavec\,)\Pb_{ij}(\xvec).
\label{eq:nurbs2d}
\end{align}
Note that in this case two parametric coordinates are used along the principal directions of the NURBS surface, i.e., $\thetavec=(\theta_1,\theta_2)^T$ (see Figure \ref{fig:nurbs-surf}). Since the NURBS surface is constructed by means of a tensor product, the order of the basis $R_{i,j}^{p,q}(\,\thetavec\,)$ can be chosen independently in each parametric coordinate direction given by $p$ and $q$. As for the NURBS curve, an extensive discussion on NURBS surfaces is given in \cite{piegl1997nurbs}.

\subsection{General Concept of NEFEM}

\begin{figure}
    \centering
    \resizebox{0.7\textwidth}{!}{%
        \input{figures/element-types.tikz}
    }
    \caption{Possible configuration of element faces or edges with a NURBS boundary. Note that the nodal coordinates of the element nodes touching the NURBS can be expressed using the parametric coordinates and the NURBS itself. \label{fig:element-types}}
\end{figure}
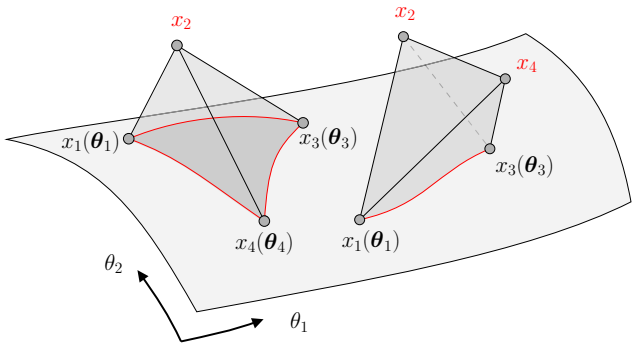

As a starting point for NEFEM, a standard finite element mesh is used. Such a mesh is typically generated using a NURBS geometry coming, e.g., from a CAD model. With NEFEM, the NURBS itself is used to modify the elements touching the boundary. Note that this leads to two groups of elements. The first group consists of the interior elements treated as standard finite elements with no additional modifications. The second group consists of the enhanced elements touching the NURBS boundary  (see Figure \ref{fig:element-types}). It is this second group of elements through which the NURBS geometry is incorporated into the finite element formulation.

\begin{remark}
    The occurrence of trimmed NURBS as well as elements with multiple faces coinciding with one or more NURBS surfaces would require special treatment. To reduce casuistics, it is assumed in this work that NURBS are untrimmed and the bounding edges of the NURBS surface are not intersecting with the edges of the NEFEM elements. In other words, all common interfaces of the NEFEM elements with the NURBS boundary are within the bounds of the NURBS surface itself. Although not applied here, a face-splitting procedure as proposed in \cite{sevilla20113dNEFEM} can be applied to avoid multiple element faces coinciding with a NURBS surface. To allow for NEFEM using trimmed NURBS, additional measures are needed as discussed in detail for both NEFEM and p-FEM in \cite{sevilla2009thesis}.
\end{remark}

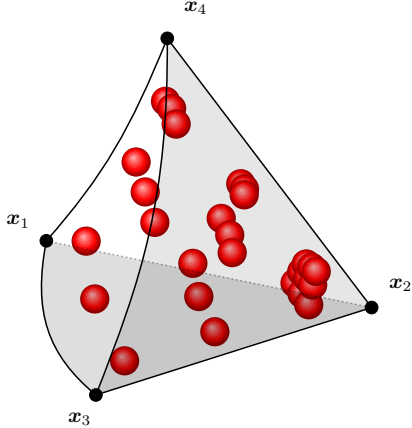
\begin{figure}
    \centering
    \resizebox{0.5\textwidth}{!}{%
        \input{figures/quad-points-3d.tikz}
    }
    \caption{Integration point positioning within an NEFEM element along a curved NURBS surface. Here, nodes $\boldsymbol{x}_{1}$, $\boldsymbol{x}_{3}$ and $\boldsymbol{x}_{4}$ are the nodes on the NURBS surface.     \label{fig:gauss-points}}
\end{figure}

The NURBS geometry itself is made available during element integration using a suitable mapping. By using such a mapping, the position of integration points is no longer determined by an approximate geometry but by the exact NURBS geometry instead. This leads to a shift of the quadrature points (see Figure \ref{fig:gauss-points}).

Furthermore, the NURBS geometry is exploited for the evaluation of boundary integrals. In this case, integration points and surface normals are defined along the NURBS geometry rather than the approximate geometry. Doing so increases the accuracy of evaluating these integrals. This can be particularly beneficial for the mapping of fluid forces along a coupling interface when considering FSI problems, as demonstrated in \cite{hosters2018}.

\subsection{Cartesian NEFEM}

In the original formulation of NEFEM, the test and interpolation functions are Lagrange polynomials defined in physical space using coordinates $\xvec$ in $\mathbb{R}^{n_{sd}}$. This so-called \textit{Cartesian} approach ensures the reproducibility of polynomials in both the reference and physical space. This is independent of the order of the polynomials itself \cite{sevilla2011Comparison}.

\begin{figure}[ht]
    \centering
    \subfigure[Cartesian NEFEM.]{
        \label{fig:shape-function-cartesian}\input{figures/shape-func-car.tikz}
    }
    \subfigure[Non-Cartesian NEFEM.]{
        \label{fig:shape-function-noncartesian}\input{figures/shape-func-noncar.tikz}
    }
    \caption{Shape function $\phi(\xvec)$ corresponding to the interior node of a 2D Cartesian and non-Cartesian NEFEM element. As can be seen here, the non-zero shape-function contributions along the domain boundary are avoided when using the proposed non-Cartesian NEFEM formulation. \label{fig:shape-function}}
\end{figure}
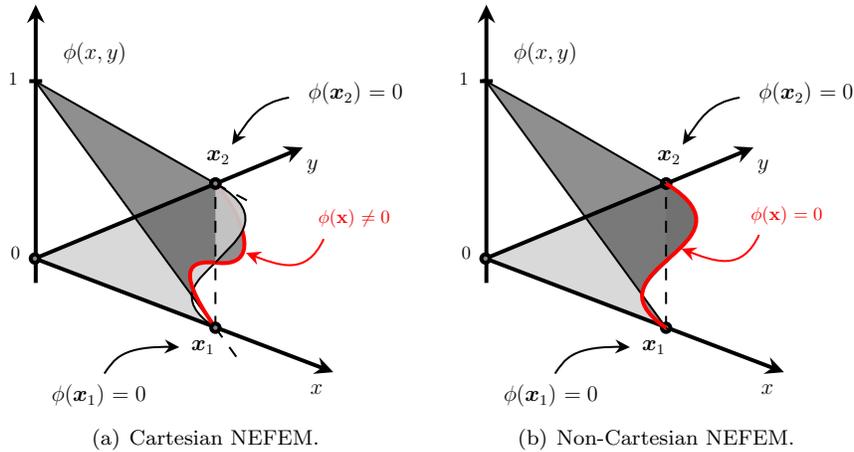

An anomaly that occurs specifically for linear finite elements and the Cartesian approach, as was first discussed in \cite{hosters2018}, is that of non-zero interior shape function contributions along the NURBS boundary. While higher-order elements are able to represent curved boundaries, linear finite elements are preferred in this work due to computational efficiency and straightforward implementation in the context of fluid flow problems. This may cause the partition of unity property to be no longer fulfilled (see Figure \ref{fig:shape-function-cartesian}). Such behavior introduces unwanted errors when evaluating elements along the boundary. This is of particular importance when evaluating boundary quantities, e.g., for FSI problems.

\subsection{Non-Cartesian NEFEM}

The non-zero boundary integral contributions of the interior shape functions can be avoided by using a \textit{non-Cartesian} approach instead. Within such an approach, e.g., the p-FEM formulation \cite{sevilla2011Comparison}, or the formulation introduced for 2D problems in \cite{hosters2018}, shape functions are no longer defined in the global space.

Instead, just as the integration points, the shape functions are defined on the reference element. As a result, the partition of unity property is fulfilled, and shape functions corresponding to the interior element nodes are zero along the NURBS boundary (see Figure \ref{fig:shape-function-noncartesian}).

Due to the mapping used, this approach might lead to distorted shape functions in the physical space when using higher order polynomials \cite{sevilla2011Comparison}. In the current work, this is avoided by limiting ourselves to the use of first-order Lagrange polynomials only.

The NURBS geometry is made available for integration of the shape functions using a mapping from the physical space to the reference domain, which will be discussed next.

\subsubsection{Tetrahedron-Hexahedron-Tetrahedron Geometric Mapping}
\label{sec: tht-mapping}

\begin{figure}
    \centering
    \resizebox{0.8\textwidth}{!}{%
        \input{figures/mapping-face.tikz}
    }
    \caption{THT mapping from a reference tetrahedron to a tetrahedral element with a spline face in the physical space. Here, the element face $\hat{\xi}+\hat{\eta}+\hat{\zeta}=1$ corresponds to the NURBS surface of the physical element spanned by vertices $\xvec_1$, $\xvec_3$ and $\xvec_4$. Vertex $\xvec_2$ corresponds to the interior element node. \label{fig:mapping-face}}
\end{figure}
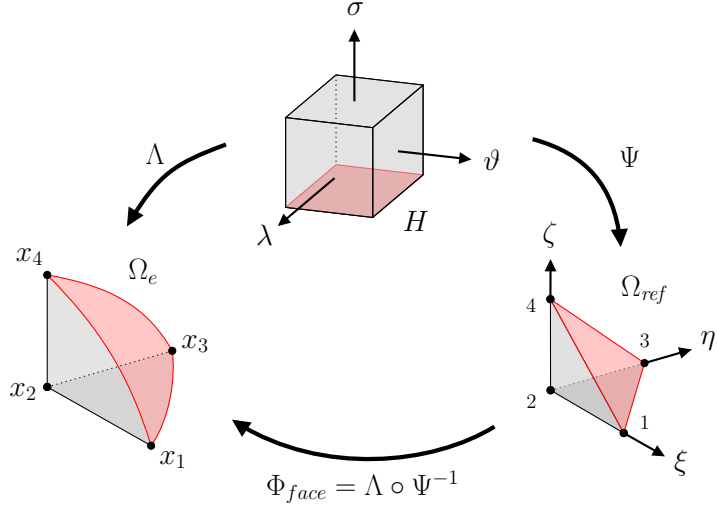

To include the NURBS geometry into the 3D non-Cartesian NEFEM formulation, a geometric mapping needs to be defined. Using the concept of degeneration, a standard tri-linear hexahedral element $H$ is mapped to a tetrahedral element by coalescing certain nodes. The inverse of this mapping is used first to map from a reference tetrahedron $\Omega_{ref}$ to the hexahedron $H$. In the next step, the mapping is employed to map from the hexahedron $H$ to a tetrahedral element $\Omega^e$ in the physical space (see Figure \ref{fig:mapping-face}). This mapping is analogous to the mapping presented for two-dimensional problems in \cite{hosters2018} and is termed \textit{Tetrahedron-Hexahedron-Tetrahedron (THT) mapping}. Note that, by incorporating the hexahedron into the mapping, it can be ensured that the NURBS direction and the interior direction are clearly separated. This leads to straight interior element surfaces, even though the boundary surface is curved.

To obtain the final THT mapping, the mapping from a reference hexahedron $H$ to a tetrahedron $\Omega_{ref}$ is defined first.
This mapping $\Psi$ is obtained by  coalescing nodes and is defined as:
\begin{equation}
\begin{aligned}
\Psi :\quad  & H \mapsto \Omega_{ref}, \\
\quad & \hat{\xi} = \frac{1}{8}[1+\lambda][1-\vartheta][1-\sigma],\\
&\hat{\eta} =\frac{1}{8}[1-\lambda][1-\vartheta][1-\sigma],\\
&\hat{\zeta}=\frac{1}{4}[1+\vartheta][1-\sigma].
\end{aligned}\label{eq:mapping-to-ref-elm}
\end{equation}
Here, $\hat{\xi}$, $\hat{\eta}$ and $\hat{\zeta}$ are the coordinates defined on the reference tetrahedron $I$, where $\hat{\xi},\hat{\eta},\hat{\zeta} \in [0,1]$. Moreover, $\lambda$, $\vartheta$ and $\sigma$ are the coordinates defined on the hexahedral reference element $H$, where $\lambda, \vartheta, \sigma \in [-1,1]$. Similar to \eqref{eq:mapping-to-ref-elm}, a mapping from hexahedral $H$ to the elements in global space $\Omega^e$ can be defined:
\begin{equation}
\begin{aligned}
\Lambda :\quad  & H \mapsto \Omega^e,\\
\quad & \Lambda(\lambda,\vartheta,\sigma) = \frac{1}{2}(1-\sigma)\,\mathbf{S}\,(\thetavec(\lambda,\vartheta))+\frac{1}{2}(1+\sigma)\,\xvec_2.
\end{aligned}\label{eq:mapping-to-global-element}
\end{equation}
Note that this expression already includes the spline geometry. The parametric coordinates $\thetavec$ of the spline are, in this case, aligned with the coordinates $\lambda$ and $\vartheta$ of $H$. To obtain $\thetavec(\lambda,\vartheta)$ for the particular element in the global space, a linear interpolation is used:
\begin{align}
\thetavec(\lambda,\vartheta) = \frac{(1-\lambda)(1-\vartheta)}{4}\thetavec_1 + \frac{(1+\lambda)(1-\vartheta)}{4}\thetavec_2 + \frac{(1+\vartheta)}{2}\thetavec_4,\label{eq:param-coord}
\end{align}
where $\thetavec_i$ are the parametric coordinates of the element nodes touching the spline.

Combining \Cref{eq:mapping-to-ref-elm,eq:mapping-to-global-element,eq:param-coord} results in the mapping $\Phi_{face} = \Lambda \,\circ \,\Psi^{-1}$, for elements that have a common face with the NURBS boundary surface:
\begin{align}
\Phi_{face}(\hat{\xi}, \hat{\eta}, \hat{\zeta}) = (1-\hat{\xi}- \hat{\eta}- \hat{\zeta})\,\xvec_2 + (\hat{\xi}+ \hat{\eta}+ \hat{\zeta})
\,\mathbf{S}\,\left(\frac{\thetavec_1 \hat{\xi} +
    \thetavec_3 \hat{\eta} +
    \thetavec_4 \hat{\zeta} }
{\hat{\xi}+ \hat{\eta}+ \hat{\zeta}}\right).
\label{eq:face-mapping}
\end{align}
Here, $\Phi_{face}$ is a function of the local coordinates $\hat{\xi}$, $\hat{\eta}$ and $\hat{\zeta}$ on the reference element $\Omega_{ref}$. $\thetavec_i$ represent the values of the parametric coordinates at the element nodes along the NURBS surface (see Figure \ref{fig:element-types}).

As already mentioned, the mapping in Equation \eqref{eq:face-mapping} is specific for elements with a face on the NURBS geometry (e.g., the left-hand side element in Figure \ref{fig:element-types}).

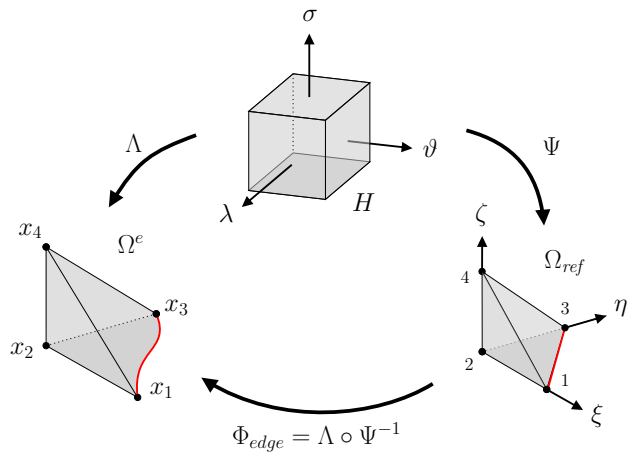
\begin{figure}
    \centering
    \resizebox{0.7\textwidth}{!}{%
        \input{figures/mapping-edge.tikz}
    }
    \caption{THT mapping from a reference tetrahedron to a global tetrahedral element with a spline-based edge. Here, the element edge $\hat{\xi}+\hat{\eta}=1$ corresponds to the edge on the NURBS surface of the global element spanned by vertices $\xvec_1$, and $\xvec_3$. Vertices $\xvec_2$ and $\xvec_4$ correspond to the interior element nodes. \label{fig:mapping-edge}}
\end{figure}

For elements with only an edge on the NURBS (see right-hand side element in Figure \ref{fig:element-types}), a slightly modified mapping is needed, as shown in Figure \ref{fig:mapping-edge}. In this particular case, only one edge of the NEFEM element is touching the NURBS geometry. Following a similar derivation as for $\Phi_{face}$, the THT mapping for edge elements is:
\begin{align}
\Phi_{edge}(\hat{\xi}, \hat{\eta}, \hat{\zeta}) = (1-\hat{\xi}- \hat{\eta}- \hat{\zeta})\,\xvec_2 + (\hat{\xi}+ \hat{\eta})
\,\mathbf{S}\,\left(\frac{\thetavec_1 \hat{\xi} +
    \thetavec_3 \hat{\eta}}
{\hat{\xi}+ \hat{\eta}}\right) + \hat{\zeta}\,\xvec_4.
\label{eq:edge-mapping}
\end{align}
Note that this mapping has only two parametric coordinates ($\thetavec_1$ and $\thetavec_3$) since only two element nodes touch the NURBS surface.

Using Equations \eqref{eq:face-mapping} and \eqref{eq:edge-mapping}, the three-dimensional non-Cartesian NEFEM approach can be incorporated into a standard finite element framework. The THT mappings are then used to reposition the integration points. Additionally, the mappings are used when defining the element Jacobians $\mathbf{J}_\Phi$. These are needed for coordinate transformations and evaluating derivatives within the weak form of the model problem. Note that the Jacobian inverse is a non-polynomial expression for face and edge mapping.

\begin{remark}
    The mappings given by Equations \eqref{eq:face-mapping} and \eqref{eq:edge-mapping} can be used directly when computing element Jacobians in space. When applying the mappings to space-time finite elements, a linear combination of the mappings at multiple time levels is constructed. These additional steps are discussed in Section \ref{sec:space-time-nefem} and are analogous to those presented in  \cite{stavrev2016STNEFEM} for Cartesian NEFEM.
\end{remark}

Due to the convenient representation of domain boundaries by NURBS, geometric properties such as tangents, normals, and curvature along these boundaries can be evaluated in an exact manner. Note, however, since strictly linear basis functions are used, properties such as higher-order spatial derivatives along the boundary are not directly available. For this, additional post-processing of the numerical solution is required.

\begin{remark}
    Apart from the curved boundary faces and edges, the NEFEM elements are also allowed to have curved faces which do not coincide with the NURBS boundary (see domain $\Omega_{ref}$ in Figure \ref{fig:mapping-edge}).  These curved faces follow naturally from the applied mapping in \Cref{eq:edge-mapping} due to the incorporated NURBS geometry. This ensures that no overlaps occur between neighboring elements, i.e., $\Omega_i \cap \Omega_j = \emptyset$ for $i \neq j$, where $\Omega_i$ is the $i$-th element in the regular domain partition $\Omega = \textstyle\bigcup_e \Omega_e$.
    Mappings \eqref{eq:face-mapping} and \eqref{eq:edge-mapping} naturally allow for such curved internal element faces.
\end{remark}

\subsection{On Numerical Integration}
\label{sec:numerical-integration}

The inverse and determinant of the Jacobian $\mathbf{J}_{\Phi}$, which contain the mappings $\Phi_{face}$ and $\Phi_{edge}$, are non-polynomial functions. Consequently, exact integration using a standard Gauss quadrature rule is not possible. This issue is discussed in more detail for Cartesian NEFEM and p-FEM in \cite{sevilla2011Comparison}.

Furthermore, it is also shown that such non-polynomial mappings combined with non-Cartesian formulations can result in a loss of consistency for higher-order elements ($p>1$) \cite{sevilla2011Comparison}.  This is since such formulations do not fulfill the higher-order patch test \cite{zienkiewicz2005finite}.

This, however, is of no concern for the formulation presented in this work since only linear ($p=1$) finite elements are considered. Nevertheless, choosing a suitable integration rule remains of great importance. Therefore, in Section \ref{sec: numerical-examples}, the performance of various quadrature rules, such as the symmetric quadrature rule proposed in \cite{williams2014symmetric}, is studied in the context of non-Cartesian NEFEM. The approach in \cite{williams2014symmetric} uses symmetrically distributed quadrature points for which the weights and coordinates are obtained using an optimization algorithm. As pointed out in \cite{shunn2012}, the tensor product Gaussian rule applied to triangular and tetrahedral elements results in an asymmetric distribution of quadrature points with dense clustering near one of the vertices (see \Cref{fig:gauss-points}). Due to the symmetric nature of the integration used in this work,  the orientation of NEFEM elements (and with that the applied quadrature rule) does not influence the accuracy of integration.

\subsection{General Concept of Isogeometric Analysis}
\label{sec: iga}

To solve the elastodynamic problem, to be discussed in Section \ref{sec: gov-eq-struct}, a standard isogeometric analysis (IGA) method is used (see, e.g, \cite{hughes2005,cottrell2009isogeometric}). The idea of IGA is to apply the NURBS basis $R_{i,j}^{p,q}(\thetavec)$ directly to the weak form resulting from an isoparametric finite element formulation.

This means that not only the geometry (e.g. entities in \eqref{eq:nurbs} or \eqref{eq:nurbs2d}), but also the solution $\db(\,\thetavec\,)$ of the finite element problem is interpolated using $R_{i,j}^{p,q}$:
\begin{align}
\db(\,\thetavec\,) = \sum_{i=1}^{n_{cp}} \sum_{j=1}^{m_{cp}}R_{i,j}^{p,q}(\,\thetavec\,)\db_{ij}(\xvec).
\label{eq:nurbssol2d}
\end{align}
Here, the control points $\Pb_{ij}$ are replaced with the discrete solutions $\db_{ij}$.

Depending on the order of a given NURBS, quantities such as derivatives can be computed accurately and in a straightforward fashion. For a full and detailed discussion on IGA, the authors refer to the original publications \cite{hughes2005,cottrell2009isogeometric}.

\begin{remark}
    One of the numerical examples in Section \ref{sec: numerical-examples} uses a shell formulation for the structural problem. As a result, only NURBS surfaces and their corresponding NURBS bases $R_{i,j}^{p,q}(\thetavec)$ are needed. Note that for structural problems involving 3D volume domains, a volume NURBS basis would be needed. Similar to Equation \eqref{eq:nurbs2d}, a tensor product can be defined for volume splines.
\end{remark}

\section{Governing Equations for Fluid-Structure Interaction}
\label{sec: gov-eq}

Generally speaking, FSI problems involve a fluid and a structure, each defined on their own domain, $\Omega^f_t \subset \mathbb{R}^{n_{sd}}$ and $\Omega^s_t \subset \mathbb{R}^{n_{sd}}$ respectively. The subscript $t$ refers to time and $n_{sd}$ to the number of spatial dimensions.  $\Omega^f_t $ and $\Omega^s_t $ are connected via a common fluid-structure interface $\Gamma^{fs}_t=\Gamma^f_t \cap \Gamma^s_t \subset \mathbb{R}^{n_{sd}}$, where $\Gamma = \partial\Omega$ is the domain boundary. The fluid and structural domains are equipped with outward normal unit vectors $\nvec^f$ and $\nvec^s$ on boundaries $\Gamma^f_t$ and $\Gamma^s_t$, respectively.

In the remainder of this section the equations that govern the individual problems on their domains $\Omega^f_t $ and $\Omega^s_t$, as well as the coupling conditions at the FSI interface will be presented.

\subsection{Equations of Fluids in Motion}
\label{sec: gov-eq-fluid}

Within the fluid domain, we assume an incompressible Newtonian fluid, described by the Navier-Stokes equations:
\begin{subequations}\label{eq:ns-set}
    \begin{align}
    \!\!\!\!\!\!\!\!\!\!\!\!\!\!\!\rho^f \left(\uvec^f_t + (\uvec^f \cdot \grad)\,\uvec^f  - \fvec^f \:   \right) - \Div  (\:\nu^f \:\grad\uvec^f\:) +\grad p^f &= \ovec&&\textrm{in} \; \Omega^f_t,\;t > 0, \label{eq:ns-equations-mom}\\
    \Div\uvec^f &= 0 &&\textrm{in} \; \Omega^f_t,\;t >0,\label{eq:ns-equations-cont}\\
    \uvec^f &= \gvec^f &&\textrm{on} \; \Gamma^f_{D,t},\;t >0,\label{eq:ns-equations-diri}\\
    -p^f\nvec^f + \nu^f\, (\nvec^f \cdot \grad)\,\uvec^f &= \mathbf{h}^f &&\textrm{on} \; \Gamma^f_{N,t},\;t >0,\label{eq:ns-equations-neum}\\
    \uvec^f(\xvec,0) &= \uvec^f_0 &&\textrm{in}\;\Omega^f_0.\label{eq:ns-equations-init}
    \end{align}
    \label{eq:ns-equations}
\end{subequations}
Here, $\rho^f$, $\nu^f$, and $\fvec^f$ represent the fluid density, kinematic viscosity, and the body force vector. Moreover, the subscript $t$ indicates the derivative with respect to time and $\Gamma^f_{D,t}$ and $\Gamma^f_{N,t}$ represent the Dirichlet and Neumann portion of the domain boundary $\Gamma^f_{t}$. Equation system \eqref{eq:ns-equations} is solved for the fluid velocity $\uvec^f(\xvec,t)$ and the pressure $p^f(\xvec,t)$.

The system is discretized using using the DSD/SST-formulation \cite{TEZDUYAR1992339} as shown in Section \ref{sec: num-meth}. This method naturally allows for deformable spatial domains and is therefore a suitable choice for FSI problems (see, e.g., \cite{hosters2018}).

\subsection{Equations of Structure Elastodynamics}
\label{sec: gov-eq-struct}

On the structural domain $\Omega^s$, the deformations $\dvec(\xvec,t)$ are governed by Newton's second law of motion. Using a Lagrangian framework, the deformation $\dvec(\xvec,t)$ relative to the reference state $\Omega^s_0$ with boundary $\Gamma^s_0$ is given by:
\begin{subequations}
    \begin{align}
    \rho^s  \frac{\text{d}^2\dvec}{\text{d}t^2} &=\DivO \left(\bS\bF^T\right) + \bvec^s  &&\textrm{in} \; \Omega^s_0,\;t > 0, \label{eq:init-stuct-equation}\\
    \dvec^s &= \gvec^s &&\textrm{on} \; \Gamma^s_{D,0},\;t >0,\label{eq:init-struct-equations-diri}\\
    \bF \bS \nvec^s_0 &= \hvec^s &&\textrm{on} \; \Gamma^s_{N,0},\;t >0,\label{eq:init-struct-equations-neum}\\
    \dvec(\xvec,0) &= \dvec^s_0 &&\textrm{in}\;\Omega^s_0.\label{eq:init-struct-equations-init}
    \end{align}
    \label{eq:init-struct-equations}
\end{subequations}
Here, $\rho^s$ denotes the material density, $\bS$ the second Piola-Kirchhoff stress tensor, $\bF$ the deformation gradient, and $\bvec^s$ the body forces acting on the structure. $\gvec^s$ and $\hvec^s$ represent the prescribed displacement and tractions on the Dirichlet and Neumann part of the boundary $\Gamma^s_0$; $\nvec^s_0$ is the outward unit normal in the reference state.

The Saint Venant-Kirchhoff material model relates the second Piola-Kirchhoff stresses $\bS$ to the Green-Lagrange strains $\bE$:
\begin{align}
\bS = \lambda^s \,tr(\bE) + 2\mu^s \bE.
\end{align}
Here, $\lambda^s$ and $\mu^s$ are the Lam\'{e} parameters. The Green-Lagrange strains are given by
\begin{align}
\bE &= \frac{1}{2} \left[(\bF)^T\,\defGrad -\bI\right].
\end{align}

Since the Green-Lagrange strain definition is a nonlinear kinematic relation, \eqref{eq:init-struct-equations} is geometrically nonlinear; it allows for large displacements and rotations with only small strains \cite{Fu2001NonlinearE, bathe1996finite}.

The elastodynamic equation \eqref{eq:init-struct-equations} is discretized using an IGA formulation as shown in Section \ref{sec: num-meth-struct}.

\subsection{Coupling Conditions at the Fluid-Structure Interface}
\label{sec: gov-eq-coupling}

When considering FSI, the fluid and structural domains are coupled at their common interface. Hence, a set of coupling conditions needs to be considered:

\begin{itemize}
    \item Kinematic coupling conditions ensure the continuity of displacement and velocity across the interface:
    \begin{subequations}
        \begin{align}
        \dvec^f &= \dvec^s && \textrm{on} \; \Gamma^{fs}_t,\;t > 0,\label{eq:coupl-cond-kin-disp}\\
        \uvec^f &= \uvec^s && \textrm{on} \; \Gamma^{fs}_t,\;t > 0.\label{eq:coupl-cond-kin-vel}
        \end{align}
        \label{eq:coupl-cond-kin}
    \end{subequations}
    \item Following Newton's third law of motion, the dynamic coupling condition ensures that the tractions are continuous across the interface:
        \begin{align}
        \bT^f \nvec^f &= \bT^s \nvec^s && \textrm{on} \; \Gamma^{fs}_t,\;t > 0,\label{eq:coupl-cond-dyn-disp}
        \end{align}
    where $\bT^f$ and $\bT^s$ represent the Cauchy stress tensors of the fluid and structure, respectively.
\end{itemize}
Satisfying the above conditions for continuous elastodynamic FSI problems ensures the conservation of mass, momentum, and energy across $\Gamma_t^{fs}$ \cite{hosters2018thesis}.

\section{Computational Framework}
\label{sec: num-meth}

In this section, the discretization strategy for the fluid and structural problems and the coupling of the two are presented.

\subsection{Space-Time NURBS-Enhanced Finite Element Formulation for the Incompressible Navier-Stokes Equations}
\label{sec: governing-equation}

The presented non-Cartesian NEFEM approach combined with the DSD/SST-formulation \cite{TEZDUYAR1992339} is applied to solve the Navier-Stokes equations \eqref{eq:ns-equations} numerically. For this, the governing equations need to be cast into a variational form.

For this, the time interval $[0,T]$ is divided into subintervals $I_n=(t_n,t_{n+1})$, with $0<\cdots<t_n < t_{n+1}<\cdots<T$. The spatial domain boundary, as it traverses from $\Gamma_n$ to $\Gamma_{n+1}$, is represented by $P_n$.

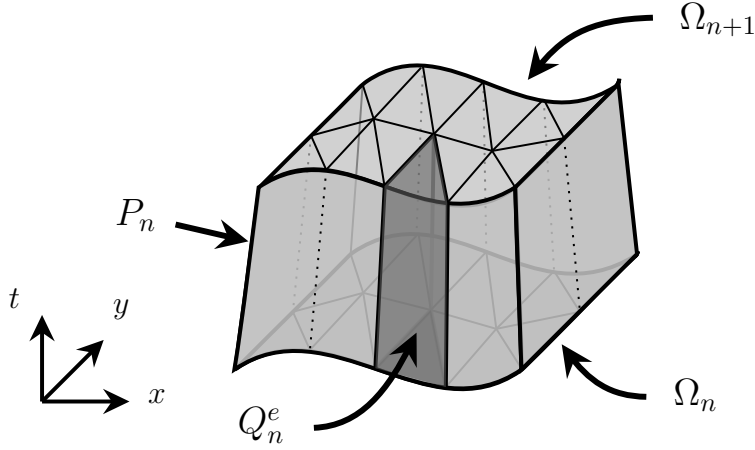
\begin{figure}
    \centerline{\input{figures/space-time-slab.tikz}}
    \caption{Space-time slab $Q_n$ enclosed by $\Omega_n$, $\Omega_{n+1}$, and $P_n$ for a two-dimensional spatial domain.\label{fig:space-time-slab}}

\end{figure}

As shown in Figure \ref{fig:space-time-slab}, the domain enclosed by $P_n$ and the spatial domains $\Omega_n$ and $\Omega_{n+1}$  at $t_n$ and $t_{n+1}$, defines a \textit{space-time slab} $Q_n$ with its corresponding elements $Q_n^e$.

For a flat space-time domain, as used here, the finite elements are simply spatial elements extruded in time (see, e.g., \cite{karyofylli2019simplex-space-time}). Note that for two-dimensional problems, a three-dimensional space-time slab, as shown in Figure \ref{fig:space-time-slab}, is constructed. Correspondingly, for three-dimensional problems, a four-dimensional space-time slab is needed.

The function spaces required for the variational form consist of linear $C^0$-continuous functions in space and linear but discontinuous functions in time. This yields the following definition for the interpolation and test function spaces for velocity $\uvec$ and pressure $p$:
\begin{subequations}
\begin{align}
(\mathcal{S}_{\uvec}^h)_n &= \{\; \uvech |\; \uvech\in [H^{1h}(Q_n)]^{n_{sd}}, \uvech = \uvec_D \; on \; (P_n)_D\},
\label{eq:interpolation-space-v}\\
(\mathcal{V}_{\uvec}^h)_n &= \{\; \wvech |\; \wvech\in [H^{1h}(Q_n)]^{n_{sd}}, \wvech = \mathbf{0} \; on \; (P_n)_D\},
\label{eq:test-space-v}\\
(\mathcal{S}_p^h)_n &= \{\; \ph |\; \ph\in H^{1h}(Q_n)\},\label{eq:interpolation-test-space-P}\\
(\mathcal{V}_q^h)_n &= \{\; \qh |\; \qh\in H^{1h}(Q_n)\}.
\label{eq:interpolation-test-space-q}
\end{align}
\label{eq:functions-spaces}
\end{subequations}
Here, $H^{1h}$ represents a finite-dimensional Sobolev space. Using the function spaces \eqref{eq:functions-spaces} and the strong form \eqref{eq:ns-set}, the stabilized space-time formulation of the incompressible Navier-Stokes equations can be stated as:

Given $(\uvech)_n^-$, find $\uvech\in(\mathcal{S}_{\uvec}^h)_n$ and $\ph\in(\mathcal{S}_p^h)_n$, such that $\forall \wvech\in(\mathcal{V}_{\uvec}^h)_n$ and $\forall \qh \in (\mathcal{V}_q^h)_n$:
\begin{equation}
\begin{aligned}
&\int_{Q_n} \wvec^h \cdot \rho \left(\uvec^h_{t} +
\uvec^h\cdot \grad\uvec^h -\fvec^h \right) \text{d}Q\\
&+\int_{Q_n} \grad\wvech : \bT^h (\ph,\uvech)   \;\text{d}Q \\
&+\int_{Q_n} \qh \Div\uvech \text{d}Q
+\int_{\Omega_n} (\wvech)_n^+ \cdot \rho \left( (\uvech)_n^+ - (\uvech)_n^-\right)   \text{d}\Omega \; \\
&+\sum_{e=1}^{(n_{el})_n}\int_{Q_n^e} \tau_{MOM} \frac{1}{\rho}\left[ \rho \left( \wvech_t +\uvech \cdot\grad\wvech \right) - \Div\bT^h(q^h,\wvech)  \right] \\
&\cdot \left[\rho \left( \uvech_t +\uvech \cdot\grad\uvech -\fvech \right) - \Div\bT^h(p^h,\uvech )  \right]\text{d}Q  \\
&+\sum_{e=1}^{(n_{el})_n}\int_{Q_n^e} \tau_{CONT} \Div \wvech \rho \Div \uvech \text{d}Q  = \int_{(P_n)_N} \wvech \cdot \mathbf{h}\,\text{d}P.
\end{aligned}\label{eq:ns-weakforms}
\end{equation}
Here, the stress tensor is given by $\bT (\ph,\uvech) = -\ph \Ivec +\mu \left( \grad \uvech + (\grad \uvech) ^T\right)$. The first three integrals on the left-hand side, combined with the right-hand side integral, represent the Galerkin weak form of the Navier-Stokes equations. The fourth integral is known as the \textit{jump term}, which is added to enforce continuity between consecutive time slabs weakly. The fifth integral is added to stabilize the formulation using the consistent Galerkin/Least Squares method. A detailed discussion on this specific formulation and the corresponding stabilization parameters is given in\cite{pauli2017}.

In \eqref{eq:ns-weakforms}, the following notation is used:
\begin{align}
(\uvech)_n^{\pm} &= \lim\limits_{\epsilon \to 0} \uvec(t_n \pm \epsilon), \\
\int_{Q_n} \cdots dQ &= \int_{I_n} \int_{\Omega_t^h}\cdots d\Omega dt,\\
\int_{P_n} \cdots dP &= \int_{I_n} \int_{\Gamma_t^h}\cdots d\Gamma dt.
\label{eq:integral-definition}
\end{align}

In case of linear finite elements, the higher-order spatial derivatives present in Equation \eqref{eq:ns-weakforms} are recovered using a least-squares technique as proposed in \cite{JANSEN1999153}.

\begin{remark}
    Note that in this section the superscript $f$, used to denote fluid-specific properties, is dropped for brevity.
\end{remark}

\subsubsection{On NEFEM and Space-Time Formulation}
\label{sec:space-time-nefem}

So far, the presented work on three-dimensional non-Cartesian NEFEM only considers spatial finite elements. For space-time finite elements, additional measures are to be taken analogously to the work presented for two-dimensional Cartesian-NEFEM in \cite{stavrev2016STNEFEM}.

\begin{figure}
    \centering
    \resizebox{0.5\textwidth}{!}{%
        \input{figures/space-time-nefem-element.tikz}
    }
    \caption{Space-time slab for a two-dimensional non-Cartesian NEFEM element. The NURBS boundary is now defined by a linear interpolation of the NURBS geometry between the upper and lower time level $n$ and $n+1$.}
    \label{fig:space-time-nefem-elm}
\end{figure}
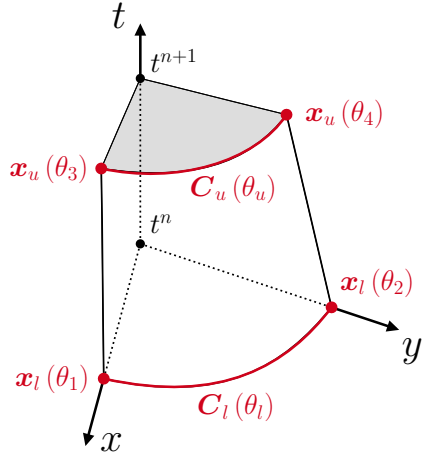

A linear interpolation between the mappings at the lower and upper time level, $n$ and $n+1$, is constructed for space-time finite elements. Using an additional reference coordinate for the time dimension, $\hat{\tau} \in [-1,1]$, this yields the following for expression \eqref{eq:face-mapping}:
\begin{equation}
\begin{aligned}
\Phi^{st}_{face}(\hat{\xi}, \hat{\eta}, \hat{\zeta}, \hat{\tau})  &=\frac{1-\hat{\tau}}{2}
\begin{pmatrix}
(1-\hat{\xi}- \hat{\eta}- \hat{\zeta})\xvec_2 +
(\hat{\xi}+ \hat{\eta}+ \hat{\zeta})
\;\mathbf{S}_{l}\left(\qfrac{\thetavec_1 \hat{\xi} +
    \thetavec_3 \hat{\eta} +
    \thetavec_4 \hat{\zeta} }
{\hat{\xi}+ \hat{\eta}+ \hat{\zeta}}\right) \\
t_l
\end{pmatrix}  \\
&+\frac{1+\hat{\tau}}{2}
\begin{pmatrix}
(1+\hat{\xi}- \hat{\eta}- \hat{\zeta})\xvec_6 +
(\hat{\xi}+ \hat{\eta}+ \hat{\zeta})
\;\mathbf{S}_{u}\left(\qfrac{\thetavec_5 \hat{\xi} +
    \thetavec_7 \hat{\eta} +
    \thetavec_8 \hat{\zeta} }
{\hat{\xi}+ \hat{\eta}+ \hat{\zeta}}\right) \\
t_u
\end{pmatrix}.
\end{aligned}\label{eq: space-time-face-mapping}
\end{equation}
Here, subscripts $l$ and $u$ refer to the upper and lower time levels $n$ and $n+1$, respectively. Furthermore, as is the case for standard space-time finite elements, the total number of element nodes is doubled. This is due to the fact that the spatial finite element is \textit{extruded} in the time dimension. A graphical explanation is shown in Figure \ref{fig:space-time-nefem-elm} for the two-dimensional case.

\begin{remark}
    Note that, when applying the presented space-time NEFEM formulation to deforming domains, in principle, a separate NURBS description of the upper and lower level is needed.
\end{remark}

For the edge-only NEFEM element, the situation is similar, resulting in the following expression for the mapping:
\begin{equation}
\begin{aligned}
\Phi^{st}_{edge}(\hat{\xi}, \hat{\eta}, \hat{\zeta}, \hat{\tau})
&=\frac{1-\hat{\tau}}{2}
\begin{pmatrix}
(1-\hat{\xi}- \hat{\eta}- \hat{\zeta})\xvec_2 +
(\hat{\xi}+ \hat{\eta})
\;\mathbf{S}_{l}\left(\qfrac{\thetavec_1 \hat{\xi} +
    \thetavec_3 \hat{\eta} }
{\hat{\xi}+ \hat{\eta}}\right) + \hat{\zeta}\xvec_4\\
t_l
\end{pmatrix}  \\
&+\frac{1+\hat{\tau}}{2}
\begin{pmatrix}
(1-\hat{\xi}- \hat{\eta}- \hat{\zeta})\xvec_6 +
(\hat{\xi}+ \hat{\eta})
\;\mathbf{S}_{u}\left(\qfrac{\thetavec_5 \hat{\xi} +
    \thetavec_7 \hat{\eta} }
{\hat{\xi}+ \hat{\eta}}\right) + \hat{\zeta}\xvec_8\\
t_l
\end{pmatrix}.
\end{aligned}\label{eq: space-time-edge-mapping}
\end{equation}

As for the semi-discrete case, the THT mappings \eqref{eq: space-time-face-mapping} and \eqref{eq: space-time-edge-mapping} can directly be used for the positioning of integration points and the evaluation of element Jacobians.

\begin{remark}
    As stated before, in the presented non-Cartesian NEFEM formulation, the shape functions are defined on the reference element. Hence the definition of these shape functions is identical to those of standard finite elements. Only the positioning of the integration points and the evaluation of the element Jacobians are modified by means of the presented mappings.
\end{remark}

\subsection{Isogeometric Analysis for Non-Linear Structural Mechanics}
\label{sec: num-meth-struct}

The spatial discretization of \eqref{eq:init-struct-equations} is done using isogeometric analysis \cite{cottrell2009isogeometric}, whereas the time integration is performed using a \textit{generalized-$\alpha$} scheme \cite{chung1993, kuhl1999}.

For the isogeometric analysis formulation used in this work, the test and interpolation functions are defined by a NURBS basis. The discrete function spaces are defined as follows:
\begin{subequations}
\begin{align}
\mathcal{S}_{\dvec}^h &= \{\; \dvech |\; \dvech\in [H^{1h}(\Omega)]^{n_{sd}}, \dvech = \gvec \; on \; \Gamma_D\},\\
\mathcal{V}_{\dvec}^h &= \{\; \wvech |\; \wvech\in [H^{1h}(\Omega)]^{n_{sd}}, \wvech = \mathbf{0} \; on \; \Gamma_D\}.\label{eq:test-space-d}
\end{align}
\end{subequations}

The weak form of the structural problem is obtained in the usual way by multiplying Equation \eqref{eq:init-struct-equations} with a test function $\wvec$, integration over the domain, and integrating the stress term by parts:

Find $\dvech \in \mathcal{S}_{\dvec}^h$ such that $\forall \,\wvech \in \mathcal{S}_{\dvec}^h$ the following holds:
\begin{align}
\int_{\Omega} \wvec^h \cdot \frac{\partial^2 \dvech}{\partial t^2}
\;\text{d}\Omega + \int_{\Omega} \grad_0\wvec^h : \bS\bF^T\, \text{d}\Omega =  \int_{\Gamma_N} \wvec^h \cdot \hvec^h \text{d}\Gamma.
\label{eq:struct-weakform}
\end{align}

\begin{remark}
    Note that in this section the superscript $s$, used to denote structure specific properties, is dropped for brevity.
\end{remark}

\subsubsection{Application to Thin-Walled Structures}

When a thin-walled structure is considered (e.g., the example in Figure \ref{fig:spline-based-methods}), the structure can often be represented by a shell model. Shell models provide a cost-effective approach to many engineering applications by exploiting the specific geometric characteristics of thin-walled structures \cite{ramm2004}.

Shell models typically use the limited thickness of a structure to reduce a volumetric description to a mid-surface description. When combining shell theory with IGA, a volumetric spline is no longer necessary, and a surface description of the structure can be used instead. In the numerical example given in Section \ref{sec: numerical-examples-3d}, isogeometric Reissner-Mindlin shell elements are used to represent the structure. For more details regarding isogeometric Reissner-Mindlin shell elements, see e.g., \cite{benson2010}.

\subsection{Coupling Approach}
\label{sec: num-meth-coupling}

To solve the FSI problem, a partitioned coupling approach is used, meaning that the fluid and structural problems are solved each by separate solvers. The two single-field solvers are coupled through a coupling module, which handles the exchange of interface data in accordance with the coupling conditions presented in Section \ref{sec: gov-eq-coupling}.

This staggered approach provides flexibility and modularity. However, as will be discussed next, this approach comes at the price of taking additional measures for the temporal and spatial data exchange between the sub-problems.

\subsubsection{Spatial Coupling}
\label{sec:spatial-coupling}

The partitioned FSI approach naturally allows for non-matching meshes. The nodes of the respective fluid and structural meshes do not have to coincide at the coupling interface. This, however, requires additional measures to correctly transfer the interface data between the single-field domains in a conservative manner.

In the current work, a NURBS-based variant of the \textit{finite interpolation elements} method is employed. This method uses the NURBS basis on the coupling interface to transfer nodal data between the single-field solvers. A detailed description of this approach is given in \cite{hosters2018}, where it was applied to two-dimensional FSI problems.

Furthermore, a fully spline-based coupling allows for the direct transfer of coupling data between the fluid and structural problem, as has been demonstrated in \cite{hosters2018, hosters2018thesis}.

\subsubsection{Temporal Coupling}

To consider the potentially strong interdependence between the single field problems, a \textit{strong coupling} procedure is employed where an identical and constant time step size for both problems is assumed. This procedure ensures that the coupling conditions are fulfilled after each time step by employing an energy-conserving fixed-point iteration scheme. A schematic of the partitioned procedure is given in Figure \ref{fig:strong-coupling}, where two coupling approaches are provided in which slightly different data sets are transferred between the individual solvers. Both approaches are discussed next:

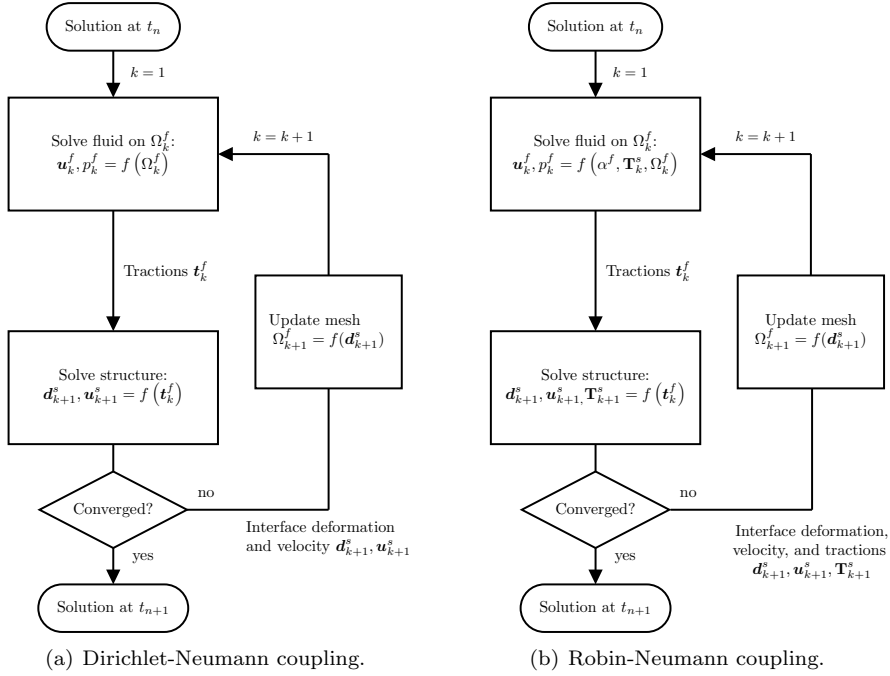
\begin{figure}[ht]
    \centering
    \subfigure[Dirichlet-Neumann coupling.]{
        \label{fig:strong-coupling-DN}\input{figures/strong-coupling-scheme-d-n.tikz}}
    \hfill
    \subfigure[Robin-Neumann coupling.]{
        \label{fig:strong-coupling-RN}\input{figures/strong-coupling-scheme-r-n.tikz}}
    \caption{Schematic of the strong temporal coupling procedures. Here, $k$ represents the current coupling iteration.}
    \label{fig:strong-coupling}
\end{figure}

\paragraph{Dirichlet-Neumann Coupling}
\label{sec:dirichlet-neumann}

There exist various coupling schemes to ensure that the interface conditions are met at each time step. The most common approach for FSI problems is the \textit{Dirichlet-Neumann} (DN) scheme. An example of such a scheme is given in Figure \ref{fig:strong-coupling-DN}.

For each fixed-point iteration $k$, the fluid tractions $\tvec^f_k(u^f_k,p^f_k)$ are used to impose a Neumann boundary condition onto the structural problem. Vice versa, the deformations $\dvec^s_{k+1}$ and velocities $\uvec^s_{k+1}$ obtained from the structural problem are applied to the fluid mesh and the fluid problem as Dirichlet boundary conditions. The Dirichlet and Neumann boundary conditions naturally enforce the kinematic and dynamic coupling conditions presented in Section \ref{sec: gov-eq-coupling} once the coupling loop has converged.

As already discussed, a DN coupling scheme is suitable for most common FSI problems, but leads to numerical instabilities when considering enclosed, fully Dirichlet-bounded fluid domains. In fact, without taking extra measures, such problems are not solvable using DN coupling \cite{hosters2018,kuttler2006}.

\paragraph{Robin-Neumann Coupling}
\label{sec:robin-neumann}

To successfully compute enclosed, fully Dirichlet-bounded fluid domain problems, a \textit{Robin-Neumann} (RN) type coupling can be used as depicted in Figure \ref{fig:strong-coupling-RN}. The Robin boundary condition can be interpreted as a linear combination of the kinematic and dynamic coupling conditions and has been proposed as an alternative to DN schemes in \cite{badia2008, gerardo-giorda2010,nobile2008,hosters2018thesis}. In this particular case, both the tractions and local velocity are combined for the fluid and structural problem as follows:
\begin{subequations}
    \begin{align}
        \alpha^f \uvec^f + \bT^f\nvec^f &= \alpha^f\frac{\partial\dvec^s}{\partial t} - \bT^s  \nvec^f, & \textrm{on} \; \Gamma^{fs,f},\label{eq:robin-fluid}\\
        \alpha^s \frac{\partial\dvec^s}{\partial t} + \bT^s \nvec^s &= \alpha^s \uvec^f - \bT^f\nvec^s,  & \textrm{on} \; \Gamma^{fs,s},  \label{eq:robin-struct}
    \end{align}
\end{subequations}
where for scalar coefficients $\alpha$, $\alpha^f \neq \alpha^s$ and $\alpha^f \alpha^s\geq 0$ must hold. It can be observed that the DN coupling is a special case of \eqref{eq:robin-fluid} and \eqref{eq:robin-struct}, where $\alpha^s=0$ and $\alpha^f \rightarrow \infty$.

The RN coupling is obtained when $\alpha^s=0$ and $\alpha^f>0$:
\begin{subequations}
    \begin{align}
        \alpha^f \uvec^f + \bT^f\nvec^f &= \alpha^f\frac{\partial\dvec^s}{\partial t} - \bT^s \nvec^f, & \textrm{on} \; \Gamma^{fs,f},\label{eq:robin-neumann-fluid}\\
        \bT^s \nvec^s &= -\bT^f \nvec^s,  & \textrm{on} \; \Gamma^{fs,s}.  \label{eq:robin-neumann-struct}
    \end{align}
\end{subequations}
Equation \eqref{eq:robin-neumann-fluid} results in the following boundary condition for the fluid problem:
\begin{align}
\bT^f  \nvec^f &= \alpha^f\left(\frac{\partial\dvec^s}{\partial t} - \uvec^f\right) - \bT^s \nvec^f.
\end{align}
Inserting this expression into the boundary integral of \eqref{eq:ns-weakforms} results in a weakly enforced condition on the fluid velocity at the coupling interface. The benefit of using the Robin condition is that the violation of mass conservation caused by errors in the structural solution is counterbalanced by allowing an artificial flux over the FSI interface. During the coupling iterations within a single time step, this flux is minimized until a converged and mass-conserving solution is obtained. A detailed discussion on the RN coupling approach, used here, is given in \cite{hosters2018thesis}.

\section{Numerical Examples}
\label{sec: numerical-examples}

Next, the capabilities of the non-Cartesian NEFEM formulation and its use within the proposed strongly-coupled solver framework are demonstrated. This is done by means of a series of benchmark problems. The obtained results are compared against the results from a standard finite element formulation using linear Lagrangian finite elements.

\subsection{Application of non-Cartesian NEFEM to Fluid Flow Problems}
\label{sec: numerical-examples-flow}

In this section, the performance of the non-Cartesian NEFEM formulation is compared against a standard space-time finite element approach using two rigid benchmark problems. Hence, FSI phenomena are not considered in this part of the numerical study. The considered benchmark problems involve a flow over a cylindrical bump and a flow around a circular cylinder.

\subsubsection{Square Channel with Cylindrical Bump}
\label{sec:cylinder-segment}

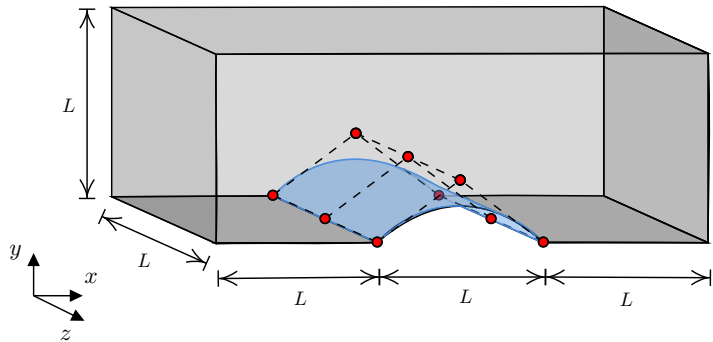
\begin{figure}
    \centering
    \resizebox{0.8\textwidth}{!}{%
        \input{figures/3d-bump-domain.tikz}
    }
    \caption{Computational domain for the square channel with a cylindrical bump.\label{fig:3d-bump-domain}}

\end{figure}

The first test case considers the volumetric domain $\Omega$ depicted in Figure \ref{fig:3d-bump-domain}. This domain consists of a cuboid with characteristic length $L$, from which a 120-degree cylindrical segment is subtracted.

The cylindrical boundary can be represented exactly with a NURBS surface, making this problem very suitable for non-Cartesian NEFEM. On the other hand, using linear finite elements would require a large number of elements along the boundary to capture the computational geometry with reasonable accuracy.

This test case is used to study solely the numerical volume of the presented domain. For this, the volume is calculated using non-Cartesian NEFEM and standard finite elements (SFEM) for a series of six mesh refinements, as presented in Table \ref{tab:bump-mesh-refinements}.
\begin{center}
    \begin{table}[t]
        \centering
        \caption{Grids used for the volume computation comparison of the square channel with a cylindrical bump domain. Parameters $n_e$ and $n_{e,face}$ represent the total number of elements and the number of non-Cartesian NEFEM face elements along the cylindrical segment of the domain.\label{tab:bump-mesh-refinements}}
        \begin{tabular}{c r r}
            \toprule
            Mesh & $n_e$  & $n_{e,face}$ \\
            \midrule
            1 & 47 & 2 \\
            2 & 376 & 8 \\
            3 & 3,008 & 32 \\
            4 & 24,064 & 128 \\
            5 & 192,512 & 512 \\
            6 & 1,540,096 & 2048 \\
            \bottomrule
        \end{tabular}
    \end{table}
\end{center}
The wall of the cylindrical segment is represented by a NURBS that is spanned by a 3 by 3 control net, as shown in Figure \ref{fig:3d-bump-domain}. This NURBS surface is of second degree along both parametric coordinate directions, i.e., $p=q=2$ (see Equation \eqref{eq:nurbs2d}).

The computed domain volumes are compared against the exact volume, for which the analytic expression based on the characteristic length $L$ is given by
\begin{align}
V_{\Omega} = \left( 3 - \frac{\pi}{9} + \frac{\sqrt{3}}{12} \right) L^3.
\label{eq: exact-bump-volume}
\end{align}

\begin{figure}
    \centerline{
        \begin{tikzpicture}[scale=1.0]
        \begin{loglogaxis}[
        xlabel={$n_{e,face}$},
        ylabel={$\varepsilon_{rel} = \frac{\left|\tilde{V}_{\Omega} - V_{\Omega} \right|}{\left|V_{\Omega}\right|}$ },
        legend cell align={left},
        xmin=1,
        xmax=10000,
        ymax=1
        ]
        \addplot [mark=square*,blue]table [x=ne, y=errorSFEM, col sep=comma] {./data/vol-bump-case.csv};
        \addplot [mark=*,red]table [x=ne, y=errorQuad, col sep=comma] {./data/vol-bump-case-quad.csv};
        \addplot [mark=triangle* ]table [x=ne, y=errorNEFEM, col sep=comma] {./data/vol-bump-case.csv};
        \legend{\scriptsize SFEM$_{tensor}$,\scriptsize non-Cart. NEFEM$_{tensor}$,\scriptsize non-Cart. NEFEM$_{sym}$}
        \end{loglogaxis}
        \end{tikzpicture}
    }
    \caption{The relative error $\varepsilon_{rel}$ between the numerical and exact volume $\tilde{V}_{\Omega}$ and $V_{\Omega}$.\label{fig:bump-volume-error}}
\end{figure}
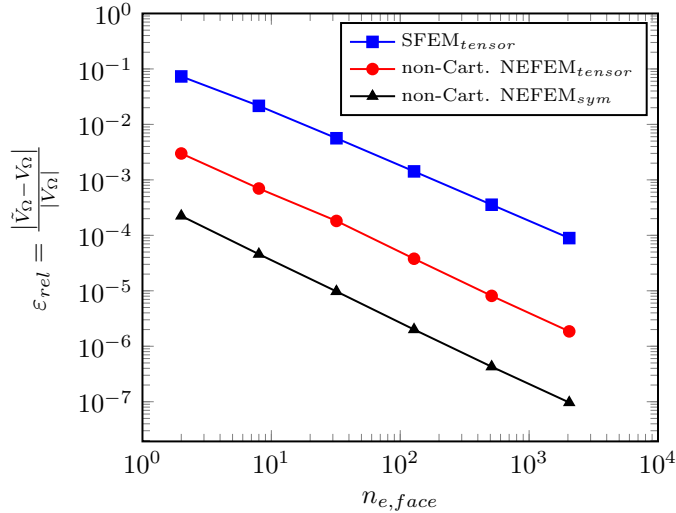

The results of the refinement study are presented in Figure \ref{fig:bump-volume-error}. Here, the relative error $\varepsilon_{rel}$,  between the numerical volume $\tilde{V}_{\Omega}$ and exact volume $V_{\Omega}$ is presented.

The non-Cartesian NEFEM results are obtained with two different integration rules: (1) The non-Cartesian  NEFEM$_{tensor}$ results were obtained using a tensor product of a standard 1D Gauss integration rule projected onto the reference tetrahedron. This rule results in clustered integration points within the element, as was shown in Figure \ref{fig:gauss-points}. (2) The non-Cartesian NEFEM$_{sym}$ results are obtained using the symmetric quadrature rule proposed in \cite{williams2014symmetric}, where the integration points are distributed symmetrically within the reference element. The presented SFEM results are also obtained using the symmetric integration rule.

It is evident that, when comparing non-Cartesian NEFEM and SFEM, the NURBS-based method requires significantly fewer elements along the spline to reach a certain level of accuracy. Furthermore, the presented results show that choosing a suitable integration rule can further improve the accuracy of the method. In this particular case, the symmetric integration rule results in improved accuracy compared to the tensor product 1D Gauss integration rule.

\subsubsection{Flow Around a 3D Cylinder}

The next test case considers the flow around a cylinder, as first proposed in \cite{Schaefer1996}, and further studied in \cite{BRAACK2006372, john2002}. Using a NURBS surface, it is possible to describe the geometry of a cylinder exactly. Similarly to the previous test case, this problem is interesting as it is not possible to describe the geometry exactly using linear finite elements.

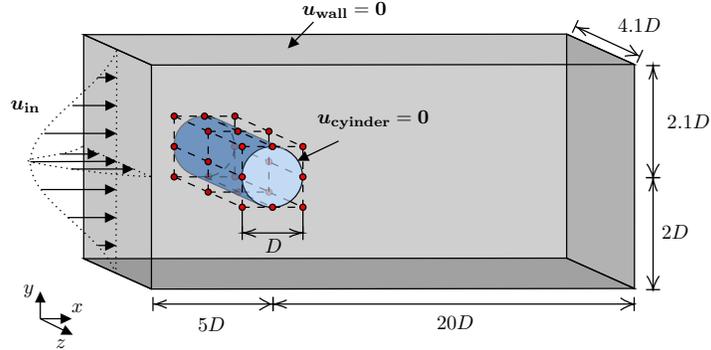
\begin{figure}
    \centering
    \resizebox{0.8\textwidth}{!}{%
        \input{figures/3d-cylinder-domain.tikz}
    }
    \caption{Computational domain for the flow around a 3D cylinder problem.\label{fig:3d-cylinder-domain}}
\end{figure}

The computational domain is shown in Figure \ref{fig:3d-cylinder-domain} and is parametrized with the cylinder diameter $D$. A similar case was studied in 2D using non-Cartesian NEFEM in \cite{hosters2018}, where it was also extended in the context of interface-coupled FSI simulations.

\paragraph{Steady Flow}
\label{sec:steady-case}

First, a steady flow around the cylinder is compared for a series of grids with increasing refinement levels, as shown in Table \ref{tab:cylinder-mesh-refinements}.

\begin{center}
    \begin{table}[ht]
        \centering
        \caption{Grids used  for the flow around a 3D cylinder problem. Parameters $n_e$ and $n_{e,face}$  represent the total number of elements and the number of non-Cartesian NEFEM face elements along the cylinder wall of the domain.\label{tab:cylinder-mesh-refinements}}
        \begin{tabular}{crr}
            \toprule
            Mesh & $n_e$  & $n_{e,face}$ \\
            \midrule
            1 & 7,437 & 496 \\
            2 & 59,496 & 1,984 \\
            3 & 475,968 & 7,936 \\
            4 & 3,807,744& 31,744 \\
            \bottomrule
        \end{tabular}
    \end{table}
\end{center}

The cylinder wall is represented by a NURBS spaned by a 9 by 3 control net, as shown in Figure \ref{fig:3d-cylinder-domain}. Note that this NURBS surface has two edges that coincide to obtain the cylindrical shape. For this, one row of control points along the $z$-axis is therefore doubled. The NURBS cylinder is of second degree along both parametric coordinate directions, i.e., $p=q=2$ (see Equation \eqref{eq:nurbs2d}).

Using the finite element formulation described in Section \ref{sec: governing-equation}, the Navier-Stokes equations are solved for a Reynolds number $Re=20$. The Reynolds number itself is based on the cylinder diameter and the mean inflow velocity $\bar{U}$:
\begin{align}
Re=\frac{\rho D \bar{U} }{\mu}.
\label{eq: Reynolds-number}
\end{align}

The inflow velocity profile itself is given by $\uvec=(U,0,0)^T$ with
\begin{align}
U = 16\, U_{m}\, y\,z(H-y)(H-z)/H^4
\label{eq: inflow-profile-cylinder-steady}.
\end{align}
Here, $H$ is set to $4.1D$, and $U_{m}$ is the maximum inflow velocity. The flow velocity is set to zero on the sidewalls and the cylinder itself, resulting in a no-slip boundary condition.

The problem parameters used for the presented benchmark problem are shown in Table \ref{tab:cylinder-parameters}. These parameters yield a steady flow that excludes any type of vortex shedding.

\begin{center}
    \begin{table}[ht]
        \centering
        \caption{Parameters used for the unsteady flow around a 3D cylinder benchmark case.\label{tab:cylinder-parameters}}
        \begin{tabular}{lccc}
            \toprule
            Parameter               & Variable  & Magnitude         & Dimension\\
            \midrule
            Reynolds number         & $Re$      & $20$              & $[-]$\\
            Cylinder diameter       & $D$       & $0.1$             & $[\text{m}]$\\
            Fluid density           & $\rho$    & $1.0$             & $[\text{kg/m}^3]$ \\
            Dynamic viscosity       & $\mu$     & $1.0\times10^{-3}$& $[\text{kg/m/s}]$ \\
            Maximum inflow velocity & $U_m$     & $0.45$            & $[\text{m/s}]$ \\
            \bottomrule
        \end{tabular}
    \end{table}
\end{center}

A quantitative comparison can be made by looking at the drag coefficient given by
\begin{align}
C_D=\frac{F_D}{\frac{1}{2}\,\rho \,||\uvec||^2 \,S}.
\label{eq: drag-coeff}
\end{align}
Here, $F_D$ is the resulting fluid force acting on the cylinder in $x$-direction. $S$ is the frontal surface area of the cylinder given by $4.1D^2$ (see Figure \ref{fig:3d-cylinder-domain}).

The reference value that serves as a basis for comparing non-Cartesian NEFEM and SFEM results is the drag coefficient $C_D = 6.18533$ \cite{BRAACK2006372}.  In both cases -- non-Cartesian NEFEM and SFEM -- the drag coefficient computed on the most refined grid is in good agreement with the reference value (non-Cartesian NEFEM: $C_D = 6.17527$ and SFEM: $C_D = 6.17279$).

\begin{figure}
    \centerline{
        \begin{tikzpicture}[scale=1.0]
        \begin{loglogaxis}[
        xlabel={$n_{e,face}$},
        ylabel={$\varepsilon_{rel} = \frac{\left|C_{D_{i}} - C_{D_{ref}} \right|}{\left|C_{D_{ref}}\right|}$ },
        legend cell align={left},
        ]
        \addplot [mark=square*,blue]table [x=ne, y=errorSFEM, col sep=comma] {./data/cylinder-steady.csv};
        \addplot[mark=*,red] table [x=ne, y=errorNEFEM, col sep=comma] {./data/cylinder-steady.csv};
        \legend{SFEM,non-Cart. NEFEM}
        \end{loglogaxis}
        \end{tikzpicture}
    }
    \caption{The error $\varepsilon_{rel}$ of the steady drag coefficient $C_D$ relative to the reference solution \cite{BRAACK2006372}\label{fig:cylinder-drag-error}.}
\end{figure}
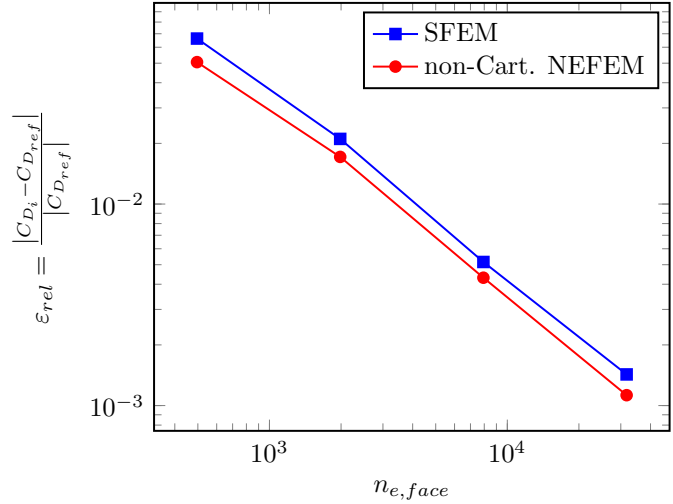

For the complete set of computed grids, the error in $C_D$ relative to the reference value in \cite{BRAACK2006372} is presented in Figure \ref{fig:cylinder-drag-error}. Although less significant than for the previous benchmark problem, it can be seen that for the complete range of grids, the non-Cartesian NEFEM solution shows a reduced relative error. Similar behavior was previously observed for 2D in \cite{hosters2018}.

\paragraph{Unsteady Flow}

Next, the unsteady variant of the cylinder benchmark case is presented. The setup is analogous to the steady case. In this case, however, the inflow velocity is varied over time according to the following expression:
\begin{align}
U = 16\, U_{m} \sin(\pi t/8)\,y\,z\,(H-y)(H-z)/H^4,
\label{eq: inflow-profile-cylinder-unsteady}
\end{align}
where again $H=4.1D$. Furthermore, the maximum inflow velocity is $U_m=2.25$. The resulting mean velocity $\bar{U}(t)=\sin(\pi t/8)$, which is now time-dependent, yields a maximum Reynolds number $Re=100$.

\begin{figure}
    \centerline{
        \begin{tikzpicture}[scale=1.0,spy using outlines=
        {rectangle, magnification=4, connect spies}]
        \begin{axis}[
        xlabel={Time $[t]$},
        ylabel={$C_D$ },
        ymax=5,
        xmax=8,
        xmin =0,
        no markers,
        legend style={at={(0.99,0.99)},anchor=north east,font=\scriptsize},
        legend cell align={left},
        ]
        \addplot [mark=none, blue] table [x=itt, y=Cd, col sep=comma] {./data/unsteady-sfem.csv};
        \addplot [mark=none, dotted, red] table [x=itt, y=Cd, col sep=comma] {./data/unsteady-nefem.csv};
        \addplot [mark=none, , black] table [x=itt, y=Cd, col sep=comma] {./data/unsteady-reference.csv};
        \legend{\scriptsize SFEM,\scriptsize non-Cart. NEFEM, \scriptsize John \cite{john2006efficiency}}
        \coordinate (spypoint) at (axis cs:4,3.28);
        \coordinate (magnifyglass) at (axis cs:4,0.7);
        \end{axis}
        \spy [black, size=2.5cm] on (spypoint)
        in node[fill=white] at (magnifyglass);
        \end{tikzpicture}
    }
    \caption{Time history of the drag coefficient $C_D$ of the cylinder.\label{fig:cylinder-transient-drag-coeff}}
\end{figure}
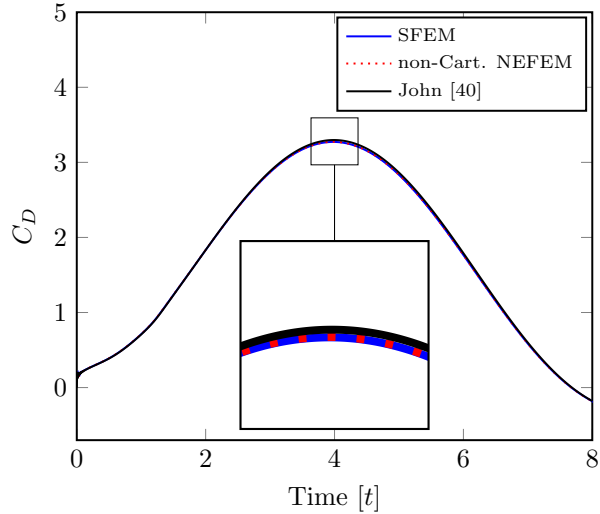

The parameters used for the time-dependent problem are given in Table \ref{tab:cylinder-parameters-unsteady}.
\begin{center}
    \begin{table}[ht]
        \centering
        \caption{Parameters used for the unsteady flow around a 3D cylinder benchmark case.\label{tab:cylinder-parameters-unsteady}}
        \begin{tabular}{lccc}
            \toprule
            Parameter               & Variable  & Magnitude         & Dimension\\
            \midrule
            Reynolds number         & $Re$      & $100$             & $[-]$\\
            Cylinder diameter       & $D$       & $0.1$             & $[\text{m}]$\\
            Fluid density           & $\rho$    & $1.0$             & $[\text{kg/m}^3]$ \\
            Dynamic viscosity       & $\mu$     &$1.0\times 10^{-3}$& $[\text{kg/m/s}]$ \\
            Maximum inflow velocity & $U_m$     & $2.25$            & $[\text{m/s}]$ \\
            Time increment          & $\Delta t$& $0.01$            & $[\text{s}]$ \\
            \bottomrule
        \end{tabular}
    \end{table}
\end{center}
The flow problem is computed for a total time of $T=8.0$ seconds, resulting in the flow changing in a sinusoidal fashion over a half period. The time history of the drag coefficient is presented in Figure \ref{fig:cylinder-transient-drag-coeff} for both non-Cartesian NEFEM and SFEM. For these simulations grid 4 from Table \ref{tab:cylinder-mesh-refinements} is used.

\begin{figure}
    \centering
    \resizebox{1.0\textwidth}{!}{%
        \begin{tikzpicture}[scale=1.0]
        \begin{loglogaxis}[
        xlabel={$n_{e,face}$ },
        ylabel={$\varepsilon_{rel} = \frac{\left|C_{D,max_{i}} - C_{D,max_{ref}} \right|}{\left|C_{D,max_{ref}}\right|}$ },
        legend cell align={left},
        legend style={at={(0.05,0.05)},anchor=south west,font=\scriptsize}]
        \addplot[mark=square*,blue] table [x=ne, y=CD-error, col sep=comma] {./data/turek-error-sfem.csv};
        \addplot [mark=*,red]table [x=ne, y=CD-error, col sep=comma] {./data/turek-error-nefem.csv};
        \legend{SFEM,non-Cart. NEFEM}
        \end{loglogaxis}
        \end{tikzpicture}
    }
    \caption{The error $\varepsilon_{rel}$ of the maximum unsteady drag coefficient $C_{D,max}$ relative to the reference solution \cite{john2006efficiency}.}
    \label{fig:cylinder-transient-error}
\end{figure}
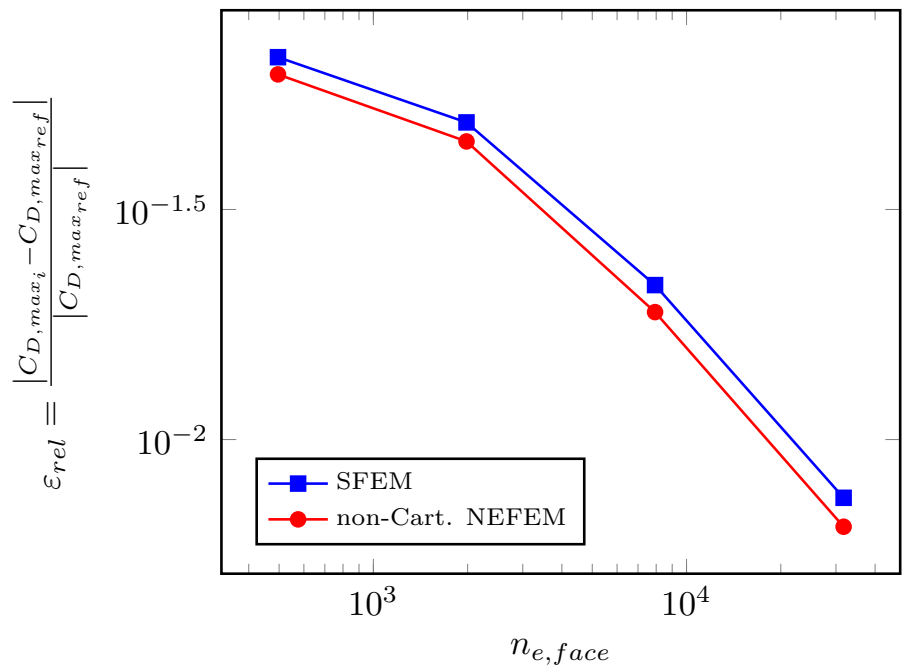

The results in Figure \ref{fig:cylinder-transient-error} show a good agreement of both non-Cartesian NEFEM and SFEM simulations with the reference solution. Furthermore, only minimal differences between the non-Cartesian NEFEM and SFEM solution are observed.

A quantitative comparison between the methods is obtained by means of the maximum drag coefficient.  The maximum occurs at time $t=4.0$ seconds. For the non-Cartesian NEFEM and SFEM simulations, the respective maximum drag coefficient $C_{D,max}=3.272$ and $C_{D,max}=3.275$, are in good agreement with the reference solution $C_{D,max}=3,2968$ as provided by \cite{john2006efficiency}.

\subsection{Application to Fluid-Structure Interaction Problems}
\label{sec: numerical-examples-fsi}

In Section \ref{sec: numerical-examples-flow}, the strengths of the proposed non-Cartesian NEFEM formulation are demonstrated. In this section, these benefits are further studied in the context of FSI problems. For this, a test case for fully enclosed FSI problems in both 2D and 3D is analyzed. The studied problem considers an inflating circular domain enclosed by a thin-walled elastic structure. As already stated, incompressible flow problems involving enclosed, fully Dirichlet-bounded domains are not solvable with  Dirichlet-Neumann coupling schemes. Hence, a Robin-Neumann coupling as presented in \ref{sec: num-meth-coupling} is used.

\subsubsection{Inflation of a Closed 2D Circular Domain}
\label{sec: numerical-examples-2d}

\begin{figure}[ht]
    \centering
    \subfigure[Problem setup.]{
        \label{fig:2d-balloon-problem-domain}\input{figures/2d-inflating-balloon-domain.tikz}
    }
    \hfill
    \subfigure[Analytic time-dependent solution of the disk radius $r_{exact}$.]{
        \label{fig:2d-balloon-problem-domain-analytic}
        \begin{tikzpicture}[scale=0.65, every node/.style={scale=1.2}]
        \begin{axis}[/pgf/number format/.cd,fixed,precision=3,
        xlabel={Time $t$},
        ylabel={Disk radius $r_{exact}$}]
        \addplot[black] table [x=t, y=R, col sep=comma] {./data/balloon-exact.csv};
        \end{axis}
        \end{tikzpicture}
    }
    \caption{Two-dimensional inflatable circular domain enclosed by thin-walled structure.}
\label{fig:2d-balloon-problem}
\end{figure}
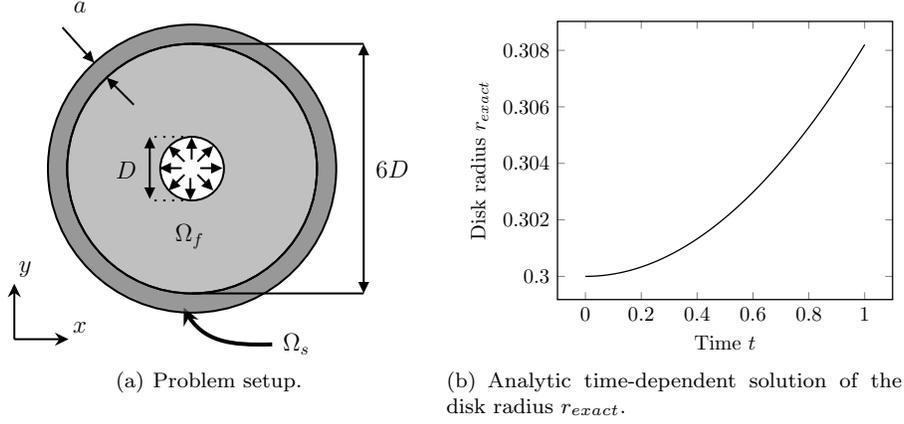

As shown in Figure \ref{fig:2d-balloon-problem-domain}, the two-dimensional example involves an incompressible fluid entering a circular domain enclosed by a thin-walled structure. A uniform radial inflow condition is applied at the inner boundary. The dimensions and material properties of the problem are presented in Table \ref{tab:2d-balloon-parameters-unsteady}.

\begin{table}[ht]
    \centering
    \caption{Parameters of the two-dimensional inflatable circular domain benchmark case.}
    \begin{tabular}{l c  c c}
        \toprule
        Parameter               & Variable  & Magnitude                 & Dimension\\
        \midrule
        Characteristic length   & $D$       & $0.1$                     & $[\text{m}]$\\
        Inflow velocity         & $\uvec(t)$    & $0.1\cdot t \cdot \nvec$  & $[\text{m/s}]$\\
        Dynamic fluid viscosity &$\mu^f$    & $1.0$                     & $[\text{kg/m/s}]$\\
        Fluid density           &$\rho^f$   & $1000$                    & $[\text{kg/m}^3]$ \\
        \midrule
        Young's modulus         &$E$        & $1.4\times 10^{6}$        & $[\text{Pa}]$ \\
        Poisson ratio           &$\nu$      & $0.3$                     & $[-]$ \\
        Structural density      &$\rho^s$   & $10000$                   & $[\text{kg/m}^3]$ \\
        Cylinder wall thickness &$a$        & $0.02$                    & $[\text{m}]$ \\
        \bottomrule
    \end{tabular}
    \label{tab:2d-balloon-parameters-unsteady}
\end{table}

The given inflow boundary condition results in an increase of the domain radius $r(t)$. As the structure fully encloses the domain and the inflow condition is known, $r(t)$ can be computed exactly as a function of time:
\begin{align}
r_{exact}(t) &= \sqrt{\frac{1}{\pi}A_{in}\int_0^t u(t) dt + R_{0}^2  }.\label{eq:balloon-exact}
\end{align}
Here, $A_{in}$ is the area (circumference in 2D) of the inflow boundary, and $R_0$ represents the initial radius at $t=0$. The exact solution is depicted in Figure \ref{fig:2d-balloon-problem-domain-analytic}.

When discretizing the fluid domain using linear Lagrangian finite elements, a geometric error is introduced. This error depends on the number of linear elements $n$ along the circumferential direction of the outer domain boundary. The resulting piecewise linear approximation of the domain yields a slightly different radius, which is given by
\begin{align}
r_{polygon}(t)&= \sqrt{\frac{1} {\frac{n}{2}\sin\frac{2\pi}{n}}A_{in}\int_0^t u(t) dt + R^2_{0}}.\label{eq:balloon-polygon}
\end{align}
Note that the geometric error, and hence the error in $r(t)$, will vanish only when $n\rightarrow\infty$.

The circular structure of our problem is discretized in a geometrically exact manner using 225 quadratic NURBS elements. Similarly, the fluid domain is discretized by $2400$ structured triangular elements. This results in a polygonal-type geometry along the circumferential direction with $n=60$ linear segments. An exact geometric representation is also obtained for the fluid problem by enhancing the elements along the fluid-structure interface via the non-Cartesian NEFEM formulation. To allow for a proper comparison, the inflow boundary is discretized using standard finite elements for both the non-Cartesian NEFEM and SFEM case.

\begin{remark}
    For the two-dimensional inflating circular domain problem, instead of the THT mapping from Section \ref{sec: tht-mapping}, its two-dimensional equivalent introduced in \cite{hosters2018} is used. Apart from the mapping, the non-Cartesian NEFEM formulation for two-dimensional problems remains unchanged compared to the three-dimensional formulation in Section \ref{sec: tht-mapping}.
\end{remark}

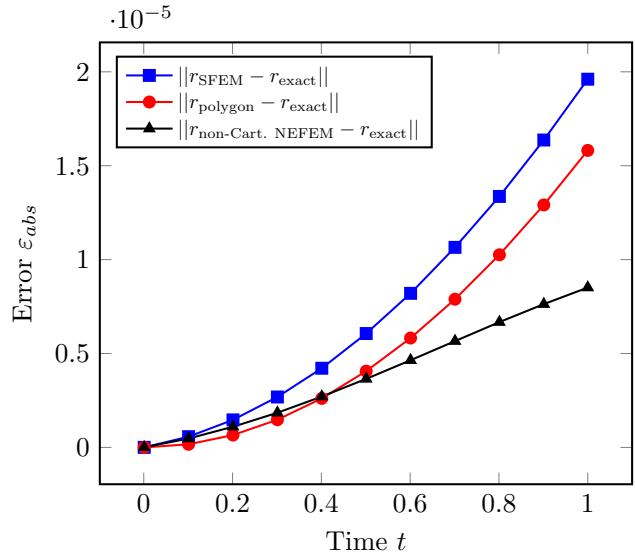
\begin{figure}[ht]
    \centering
    \resizebox{.7\textwidth}{!}{%
        \begin{tikzpicture}[scale=1.0]
        \begin{axis}[
        xlabel={Time $t$ },
        ylabel={Error $\varepsilon_{abs}$},
        legend pos={north west},
        legend style={nodes={scale=0.75}},
        legend cell align={left},
        ]
        \addplot[mark=square* ,blue ,mark options={solid,draw=blue}] table [x=dt, y=SFEM-exact, col sep=comma] {./data/2d-balloon-error-t=1.0-dt0.001.csv};
        \addplot[mark=*,red ,mark options={solid,draw=red}] table [x=dt, y=Polygon-Exact, col sep=comma] {./data/2d-balloon-error-t=1.0-dt0.001.csv};
         \addplot [mark=triangle* ,mark options={solid,draw=black},black]  table [x=dt, y=NEFEM-exact, col sep=comma] {./data/2d-balloon-error-t=1.0-dt0.001.csv};
        \legend{$||r_{\text{SFEM}}-r_{\text{exact}}||$,$||r_{\text{polygon}}-r_{\text{exact}}||$,$||r_{\text{non-Cart. NEFEM}}-r_{\text{exact}}||$}
        \end{axis}
        \end{tikzpicture}
    }
    \caption{The absolute error $\varepsilon_{abs}$ of the two-dimensional circular domain radius with respect to the exact analytic radius.}
    \label{fig:2d-cylinder-transient-error}
\end{figure}

For the non-Cartesian NEFEM solution, the exact NURBS geometry is included in the numerical integration. Hence, the domain volume can be computed exactly, limited only by the accuracy of numerical integration (see Section \ref{sec:numerical-integration} and Figure \ref{fig:bump-volume-error})). When using standard linear finite elements, the discretized domain is slightly smaller than the exact one. This leads to an increase in deformations for a given inflow condition. This error is shown in Figure \ref{fig:2d-cylinder-transient-error}, where a comparison between Non-Cartesian NEFEM and SFEM simulations is presented over time.

The fact that the non-Cartesian NEFEM solution also shows an error with respect to the exact solution can be attributed to time integration.

\begin{figure}[ht]
    \centering
    \resizebox{0.7\textwidth}{!}{%
        \begin{tikzpicture}[scale=1.0]
        \begin{loglogaxis}[
        xlabel={Time-step size $\Delta t$ },
        ylabel={Error $\varepsilon_{abs}$},
        legend pos={south west},
        legend style={nodes={scale=0.75}},
        legend cell align={left},
        grid=major,
        xmax=0.021,
        xmin =0.0004,
        x dir = reverse,
        ]
        \addplot[mark=square*,red ,mark options={solid,draw=red}] table [x=Dt, y=R_SFEM-R_exact, col sep=comma] {./data/2d-balloon-time-ref.csv};
        \addplot[mark=none] coordinates {(0.000001,0.0000568808) (1,0.0000568808)};
        \addplot[mark=*,blue ,mark options={solid,draw=blue}] table [x=Dt, y=R_NEFEM-R_exact, col sep=comma] {./data/2d-balloon-time-ref.csv};
        \addplot[mark=triangle* ,mark options={solid,draw=black},black] table [x=Dt, y=R_SFEM-R_Polygon, col sep=comma] {./data/2d-balloon-time-ref.csv};
        \legend{$||r_{\text{SFEM}}-r_{\text{exact}}||$,$||r_{\text{polygon}}-r_{\text{exact}}||$,$||r_{\text{non-Cart. NEFEM}}-r_{\text{exact}}||$,$||r_{\text{SFEM}}-r_{\text{Polygon}}||$}
        \end{loglogaxis}
        \end{tikzpicture}
    }
    \caption{The absolute error $\varepsilon_{abs}$ of the two-dimensional circular domain radius time $t=1.0$ for a range of time-step sizes. Note here that the SFEM solution approaches the polygon solution, while the non-Cartesian NEFEM solution converges towards the geometrically exact solution when $\Delta t \to 0$.}
    \label{fig:2d-final-error}
\end{figure}
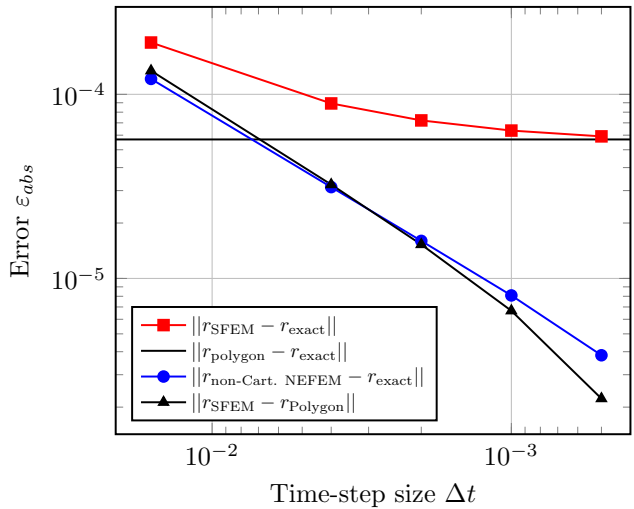

This is corroborated by looking at the cylinder radius at the final simulation time ($t=1.0\:s$) for a range of time-step sizes $\Delta t$. From Figure \ref{fig:2d-final-error}, it can be seen that for $\Delta t \rightarrow 0$ the SFEM solution approaches its best possible approximation, the polygonal solution given by \eqref{eq:balloon-polygon}. On the other hand, the non-Cartesian NEFEM results approach the exact solution given by \eqref{eq:balloon-exact}.

The presented results show the importance of considering the correct geometry within a numerical formulation. The accuracy of standard linear finite elements heavily depends on their ability to represent the exact computational domain. In this particular case, this is only possible when the number of elements along the domain boundaries is increased to $n\to \infty$. The accuracy of the NURBS-enhanced finite element formulation, on the other hand, is mainly affected by the error introduced by the time discretization.

\subsubsection{Inflation of a Closed 3D Cylindrical Domain}
\label{sec: numerical-examples-3d}

\begin{figure}[ht]
    \centering
    \subfigure[Problem setup.]{
        \label{fig:balloon-problem-domain}\input{figures/inflating-balloon-domain.tikz}
    }
    \subfigure[Analytic time-dependant solution of the cylinder radius $r_{exact}$.]{
        \label{fig:balloon-problem-analytic}
        \begin{tikzpicture}[scale=0.55, every node/.style={scale=1.2}]
        \begin{axis}[/pgf/number format/.cd,fixed,precision=3,
        xlabel={Time $t$},
        ylabel={Cylinder radius $r_{exact}$}]
        \addplot[black] table [x=t, y=R, col sep=comma] {./data/balloon-exact.csv};
        \end{axis}
        \end{tikzpicture}
    }
    \caption{Three-dimensional inflating circular domain enclosed by thin-walled structure.}
    \label{fig:balloon-problem}
\end{figure}
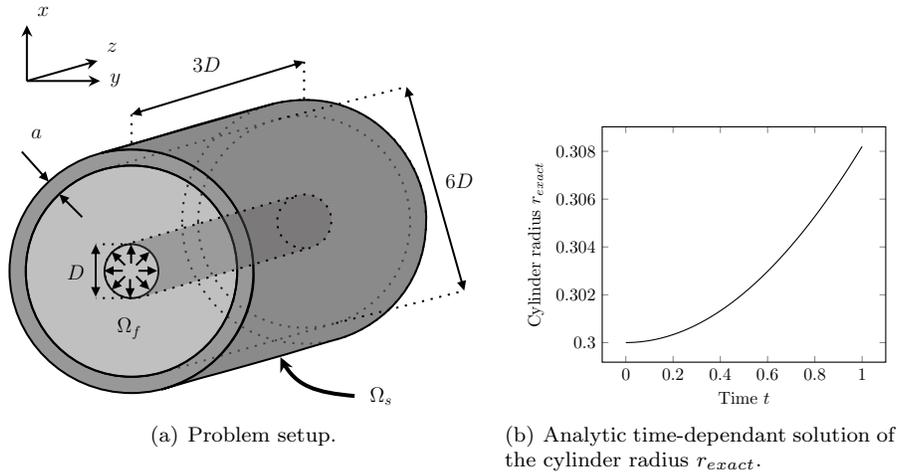

The three-dimensional equivalent of the previous test case is studied next. By extruding the 2D circular domain in the third spatial dimension, a  three-dimensional cylindrical domain is obtained (see Figure \ref{fig:balloon-problem-domain}). The cylindrical domain is enclosed by a thin-walled shell structure and two rigid walls at the cylinder's ends. At these ends, the no-slip boundary condition is enforced on the fluid problem while the shell structure can freely expand in the radial direction. As depicted in Figure \ref{fig:balloon-problem-domain}, an inflow condition is enforced at the remaining inner boundary. The exact dimensions and material properties of the problem are given in Table \ref{tab:balloon-parameters-unsteady}.

\begin{table}[ht]
    \centering
    \caption{Parameters used for the 3D inflatable cylindrical domain problem.}
    \begin{tabular}{l c  c c}
        \toprule
        Parameter               & Variable  & Magnitude             & Dimension\\
        \midrule
        Characteristic length   & $D$       & $0.1$                 & $[\text{m}]$\\
        Inflow velocity         & $\uvec(t)$    & $0.1\cdot t\cdot\nvec$& $[\text{m/s}]$\\
        Dynamic fluid viscosity & $\mu^f$   & $1.0$                 & $[\text{kg/m/s}]$\\
        Fluid density           & $\rho^f$  &$1000$                 & $[\text{kg/m}^3]$ \\
        \midrule
        Young's modulus         & $E$       & $1.4\times 10^{6}$    & $[\text{Pa}]$ \\
        Poisson ratio           & $\nu$     & $0.3$                 & $[-]$ \\
        Structural density      & $\rho^s$  &$10000$                & $[\text{kg/m}^3]$ \\
        Cylinder wall thickness & $a$       &$0.02$                 & $[\text{m}]$ \\
        \bottomrule
    \end{tabular}
    \label{tab:balloon-parameters-unsteady}
\end{table}

As observed for the two-dimensional problem in Section \ref{sec: numerical-examples-2d}, the given inflow boundary condition causes the cylinder to inflate over time. By using the correct three-dimensional inflow area $A_{in}$, the cylinder radius can be computed using Equation \eqref{eq:balloon-exact} and \eqref{eq:balloon-polygon} from Section \ref{sec: numerical-examples-2d}. The exact radius $r(t)$ corresponding to the parameters presented in Table \ref{tab:balloon-parameters-unsteady} is plotted as a function of time in Figure \ref{fig:balloon-problem-analytic}.

\begin{figure}[ht]
    \centering
    \def\svgwidth{0.75\textwidth}
    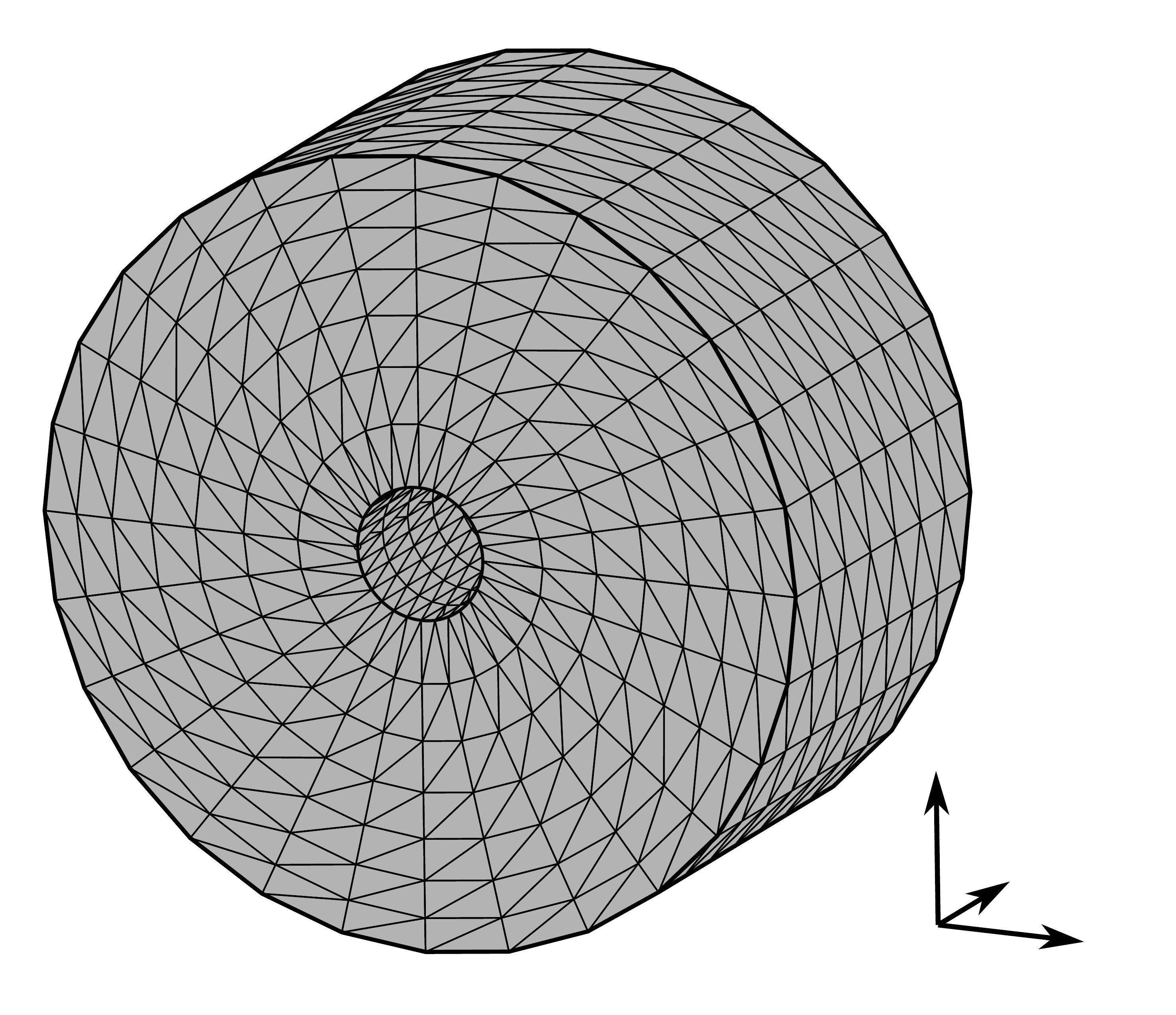
    \caption{Computational grid with $n=28$ linear tetrahedral elements in the circumferential direction.}
    \label{fig:balloon-grid}
\end{figure}

The cylindrical structure is represented in a geometrically exact manner using a second-order NURBS surface of $136 \times 10$ Reissner-Mindlin shell elements \cite{benson2010} in the circumferential and axial direction, respectively. The discretized fluid domain consists of $11,540$ structured tetrahedral elements resulting in a polygonal-type geometry along the circumferential direction with $n=28$. A render of the corresponding mesh is depicted in Figure \ref{fig:balloon-grid}. For sake of comparison, the inflow boundary is discretized using standard finite elements for both the non-Cartesian NEFEM and SFEM cases.

Analogous to the previous test case, the discretized domain is slightly smaller than the exact domain when using linear finite elements. The resulting discrepancy in structural deformations with respect to the exact solution is shown in Figure \ref{fig:error-with-exact-sol}. In this figure, a comparison between the time-dependent non-Cartesian NEFEM and SFEM simulations is presented. Similar to the two-dimensional example, the non-Cartesian NEFEM solution shows an error with respect to the exact solution, which can be attributed to the time integration.

\begin{figure}[ht]
    \centering
    \resizebox{.7\textwidth}{!}{%
        \begin{tikzpicture}[scale=1.0]
        \begin{axis}[
        xlabel={Time $t$ },
        ylabel={Error $\varepsilon_{abs}$},
        legend pos={north west},
        legend style={nodes={scale=0.75}},
        legend cell align={left},
        ]
        \addplot[mark=square* ,mark options={solid,draw=blue},blue] table [x=t, y=SFEM, col sep=comma] {./data/balloon-error-t=1.0-dt0.001.csv};
        \addplot[mark=* ,mark options={solid,draw=red},red] table [x=t, y=Polygon, col sep=comma] {./data/balloon-error-t=1.0-dt0.001.csv};
        \addplot [mark=triangle* ,mark options={solid,draw=black},black]  table [x=t, y=NEFEM, col sep=comma] {./data/balloon-error-t=1.0-dt0.001.csv};
        \legend{$||r_{\text{SFEM}}-r_{\text{exact}}||$,$||r_{\text{polygon}}-r_{\text{exact}}||$,$||r_{\text{non-Cart. NEFEM}}-r_{\text{exact}}||$}
        \end{axis}
        \end{tikzpicture}
    }
    \caption{The absolute error $\varepsilon_{abs}$ of the three-dimensional cylindrical domain radius withe respect to the exact analytic radius.}
    \label{fig:error-with-exact-sol}
\end{figure}

\begin{figure}[ht]
    \centering
    \resizebox{0.7\textwidth}{!}{%
        \begin{tikzpicture}[scale=1.0]
        \begin{loglogaxis}[
        xlabel={Time-step size $\Delta t$ },
        ylabel={Error $\varepsilon_{abs}$},
        legend pos={south west},
        legend style={nodes={scale=0.75}},
        legend cell align={left},
        grid=major,
        xmax=2e-1,
        xmin =5e-5,
        x dir = reverse,
        ]
        \addplot[mark=square* ,mark options={solid,draw=red},red] table [x=Dt, y=R_SFEM-R_exact, col sep=comma] {./data/balloon-time-ref.csv};
        \addplot[mark=none] coordinates {(0.000001,0.000068326994556) (1,0.000068326994556)};
        \addplot[mark=* ,mark options={solid,draw=blue},blue] table [x=Dt, y=R_NEFEM-R_exact, col sep=comma] {./data/balloon-time-ref.csv};
        \addplot[mark=triangle* ,mark options={solid,draw=black},black] table [x=Dt, y=R_SFEM-R_Polygon, col sep=comma] {./data/balloon-time-ref.csv};
        \legend{$||r_{\text{SFEM}}-r_{\text{exact}}||$,$||r_{\text{polygon}}-r_{\text{exact}}||$,$||r_{\text{non-Cart. NEFEM}}-r_{\text{exact}}||$,$||r_{\text{SFEM}}-r_{\text{Polygon}}||$}
        \end{loglogaxis}
        \end{tikzpicture}
     }
    \caption{The error $\varepsilon_{abs}$ of the three-dimensional cylindrical domain radius at time $t=1.0$.}
    \label{fig:final-error}
\end{figure}
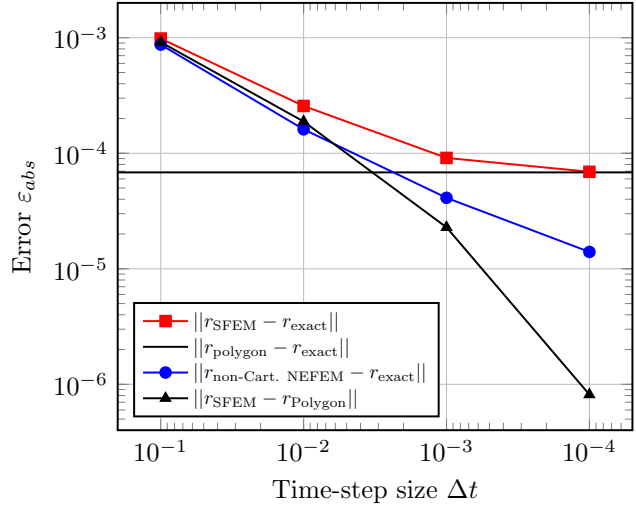

This becomes evident when reducing the time step size, which causes the time integration error to approach zero. This is shown in Figure \ref{fig:final-error}, where it can be seen that for $\Delta t \to 0$ the non-Cartesian NEFEM solution at $t=1.0$ approaches the exact solution given by \eqref{eq:balloon-exact}. The SFEM solution, on the other hand, approaches its best possible approximation, the polygonal solution \eqref{eq:balloon-polygon}.

Similar to the observations in Section \ref{sec: numerical-examples-2d}, the presented results show that the employed RN coupling allows us to simulate enclosed, fully Dirichlet-bounded FSI problems. Furthermore, the study shows the importance of using an accurate geometric representation within a numerical formulation. The accuracy of a standard linear finite element method heavily depends on the ability to represent the exact computational domain by the element type used. For curved domains used here, a geometric error remains even for highly refined grids. The use of non-Cartesian NEFEM elements can significantly reduce the geometric error, as the accuracy of the NURBS-enhanced finite element formulation is mainly affected by the time-discretization error.

\section{Conclusions}
\label{sec: conclusions}

In the current work, a non-Cartesian NURBS-Enhanced Finite Element Method is presented. The method allows for exact geometric representation, while maintaining the proven computational efficiency of standard finite elements in the interior domain. This is achieved by using the NURBS-based geometry to enhance only those elements that have a common interface with the NURBS. By doing so, the dependence on volumetric splines is avoided. Furthermore, since the non-Cartesian NEFEM approach builds upon conventional FE meshes, standard grid generation tools can be used.

The non-Cartesian nature of the presented formulation ensures that the interior shape function contributions along the NURBS boundary are zero. Consequently, the partition of unity property is maintained. This is of special importance when considering interface-coupled problems, where an accurate evaluation of boundary quantities is crucial.

The proposed method is used within a three-dimensional spline-based, strongly-coupled partitioned procedure to solve fluid-structure interaction problems. By combining space-time non-Cartesian NEFEM and IGA for the fluid and structural sub-problems, respectively, the proposed procedure relies on an exact smooth geometric representation of the fluid-structure interface. The individual sub-problems use a shared spline definition, which allows for an accurate and direct transfer of coupling data.

The performance benefits of non-Cartesian NEFEM and the presented coupled framework are demonstrated by comparing it with a standard linear space-time finite element approach. For this, the numerical results of a series of benchmark problems are presented.

From the results obtained in this study we draw the following conclusions:
\begin{itemize}
    \item The proposed THT mapping is a suitable choice for solving three dimensional problems using the non-Cartesian Space-Time NEFEM formulation.
    \item The performance of the non-Cartesian NEFEM approach is in close agreement with both the SFEM and reference data available in the literature for the presented fluid flow problems. While the accuracy increase in fluid flow quantities is small, more significant improvement gains of the non-Cartesian NEFEM approach are observed for domain volume computations and the evaluation of surface quantities. For the volume computation, an improvement of three orders of magnitude between standard FE and NEFEM is observed.
    \item The benefit of the exact geometry representation of the non-Cartesian NEFEM formulation becomes more apparent when used within fluid-structure interaction. For enclosed, fully Dirichlet-bounded FSI problems involving curved domain boundaries, a significant reduction in the error relative to the analytic solution is observed.
    \item The use of the RN scheme within the presented spline-based solver framework allows for solving enclosed, fully Dirichlet-bounded problems involving incompressible fluids. This holds for both standard linear finite methods and for non-Cartesian NEFEM.
    \item Following an investigation of the time-integration error, the improved accuracy of the spline-based solver framework can be attributed to a reduction in spatial discretization errors.
    \item The choice of integration rules can significantly affect the accuracy of non-Cartesian NEFEM for volume problems. For example, it was shown in this work that a symmetric quadrature rule results in an accuracy improvement when compared to a 1D Gauss quadrature rule projection on the reference element. However, to get a deeper understanding of the influence of quadrature rules on the performance of non-Cartesian NEFEM, an additional in-depth study regarding this matter is needed.
    \item The influence of the number of elements along curved boundaries is less important when considering the geometric error. For finite element methods in conjunction with standard linear finite elements, this does not hold.
\end{itemize}

The presented spline-based FSI solver framework, in combination with an RN coupling scheme, is a promising candidate for accurately solving FSI problems involving enclosed, fully Dirichlet-bounded and curved domains. Hence, based on the current work, further research will focus on applying the NURBS-based method to interface-coupled problems in various fields of application. Future research involving more complex NURBS geometries, including trimmed NURBS, is needed. This will allow usage of the method in increasingly complex problems and applications.

\section*{Acknowledgments}

This work is funded by the Federal Ministry of Education and Research (BMBF) and the state of North Rhine-Westphalia as part of the NHR Program. The authors also gratefully acknowledge support from the German Research Foundation (DFG) grant BE 3689/10 "Geometrically Exact Methods for Fluid-Structure Interaction.", and the computing time granted through JARA-HPC on the supercomputers CLAIX at the RWTH Aachen University IT Center and JURECA at Forschungszentrum J\"{u}lich.

\bibliographystyle{model1-num-names}
\bibliography{sample.bib}



\end{document}

%% file: figures/spline-based-methods.tikz
\tikzset{every picture/.style={line width=0.75pt}} 

\begin{tikzpicture}[x=0.75pt,y=0.75pt,yscale=-1,xscale=1]

\draw  [draw opacity=0][fill={rgb, 255:red, 176; green, 213; blue, 255 }  ,fill opacity=1 ] (524.67,222.63) .. controls (567.59,226.03) and (573.47,229.62) .. (593.9,254.96) .. controls (528.57,277.62) and (496.73,288.99) .. (466.74,363.68) .. controls (439.22,332.34) and (443.95,324.67) .. (444.1,298.07) .. controls (457.07,223.22) and (513.25,224.27) .. (524.67,222.63) -- cycle ;
\draw [color={rgb, 255:red, 255; green, 0; blue, 0 }  ,draw opacity=1 ][line width=3]    (466.68,364.4) .. controls (485.69,304.37) and (533.52,270.13) .. (593.9,254.96) ;
\draw [color={rgb, 255:red, 70; green, 70; blue, 255 }  ,draw opacity=1 ][line width=1.5]    (506.67,238.7) -- (582.99,257.51) ;
\draw [color={rgb, 255:red, 70; green, 70; blue, 255 }  ,draw opacity=1 ][line width=1.5]    (584.94,244.32) -- (582.99,257.51) ;
\draw [color={rgb, 255:red, 70; green, 70; blue, 255 }  ,draw opacity=1 ][line width=1.5]    (472.18,346.92) -- (453.08,346.48) ;
\draw [color={rgb, 255:red, 70; green, 70; blue, 255 }  ,draw opacity=1 ][line width=1.5]    (456.92,281.97) -- (472.18,346.92) ;
\draw [color={rgb, 255:red, 70; green, 70; blue, 255 }  ,draw opacity=1 ][line width=1.5]    (506.67,238.7) -- (456.92,281.97) ;
\draw [color={rgb, 255:red, 70; green, 70; blue, 255 }  ,draw opacity=1 ][line width=1.5]    (541.15,222.8) -- (506.67,238.7) ;
\draw [color={rgb, 255:red, 70; green, 70; blue, 255 }  ,draw opacity=1 ][line width=1.5]    (456.92,281.97) -- (443.29,301.78) ;
\draw [color={rgb, 255:red, 70; green, 70; blue, 255 }  ,draw opacity=1 ][line width=1.5]    (446.53,280.71) -- (456.92,281.97) ;
\draw [color={rgb, 255:red, 70; green, 70; blue, 255 }  ,draw opacity=1 ][line width=1.5]    (458.28,256.96) -- (456.92,281.97) ;
\draw [color={rgb, 255:red, 70; green, 70; blue, 255 }  ,draw opacity=1 ][line width=1.5]    (506.67,238.7) -- (483.2,234.63) ;
\draw [color={rgb, 255:red, 70; green, 70; blue, 255 }  ,draw opacity=1 ][line width=1.5]    (507.43,224.8) -- (506.67,238.7) ;
\draw [color={rgb, 255:red, 70; green, 70; blue, 255 }  ,draw opacity=1 ][line width=1.5]    (472.18,346.92) -- (507.53,299.37) ;
\draw [color={rgb, 255:red, 70; green, 70; blue, 255 }  ,draw opacity=1 ][line width=1.5]    (469.14,367.71) -- (472.18,346.92) ;
\draw [color={rgb, 255:red, 70; green, 70; blue, 255 }  ,draw opacity=1 ][line width=1.5]    (582.99,257.51) -- (597.62,257.61) ;
\draw  [color={rgb, 255:red, 97; green, 97; blue, 97 }  ,draw opacity=1 ][line width=3]  (443.29,301.78) .. controls (445.84,255.16) and (485.7,219.44) .. (532.32,221.98) .. controls (578.94,224.53) and (614.67,264.39) .. (612.12,311.02) .. controls (609.57,357.64) and (569.71,393.36) .. (523.09,390.82) .. controls (476.47,388.27) and (440.74,348.41) .. (443.29,301.78) -- cycle ;
\draw  [color={rgb, 255:red, 70; green, 70; blue, 255 }  ,draw opacity=1 ][fill={rgb, 255:red, 134; green, 134; blue, 255 }  ,fill opacity=1 ][line width=2.25]  (470.71,346.43) .. controls (470.98,345.62) and (471.86,345.18) .. (472.67,345.45) .. controls (473.48,345.72) and (473.92,346.59) .. (473.65,347.41) .. controls (473.38,348.22) and (472.5,348.66) .. (471.69,348.39) .. controls (470.88,348.12) and (470.44,347.24) .. (470.71,346.43) -- cycle ;
\draw  [color={rgb, 255:red, 70; green, 70; blue, 255 }  ,draw opacity=1 ][fill={rgb, 255:red, 134; green, 134; blue, 255 }  ,fill opacity=1 ][line width=2.25]  (455.45,281.48) .. controls (455.72,280.67) and (456.59,280.23) .. (457.41,280.5) .. controls (458.22,280.77) and (458.66,281.65) .. (458.39,282.46) .. controls (458.12,283.27) and (457.24,283.71) .. (456.43,283.44) .. controls (455.62,283.17) and (455.18,282.29) .. (455.45,281.48) -- cycle ;
\draw  [color={rgb, 255:red, 70; green, 70; blue, 255 }  ,draw opacity=1 ][fill={rgb, 255:red, 134; green, 134; blue, 255 }  ,fill opacity=1 ][line width=2.25]  (505.2,238.21) .. controls (505.47,237.4) and (506.35,236.96) .. (507.16,237.23) .. controls (507.97,237.5) and (508.41,238.38) .. (508.14,239.19) .. controls (507.87,240) and (506.99,240.44) .. (506.18,240.17) .. controls (505.37,239.9) and (504.93,239.02) .. (505.2,238.21) -- cycle ;
\draw  [color={rgb, 255:red, 70; green, 70; blue, 255 }  ,draw opacity=1 ][fill={rgb, 255:red, 134; green, 134; blue, 255 }  ,fill opacity=1 ][line width=2.25]  (581.52,257.02) .. controls (581.79,256.21) and (582.67,255.77) .. (583.48,256.04) .. controls (584.29,256.31) and (584.73,257.18) .. (584.46,258) .. controls (584.19,258.81) and (583.31,259.25) .. (582.5,258.98) .. controls (581.69,258.71) and (581.25,257.83) .. (581.52,257.02) -- cycle ;
\draw [color={rgb, 255:red, 117; green, 117; blue, 117 }  ,draw opacity=1 ][line width=1.5]  [dash pattern={on 1.69pt off 2.76pt}]  (593.9,254.96) -- (541.4,260.79) ;
\draw [color={rgb, 255:red, 117; green, 117; blue, 117 }  ,draw opacity=1 ][line width=1.5]  [dash pattern={on 1.69pt off 2.76pt}]  (541.4,260.79) -- (476.5,315.25) ;
\draw [color={rgb, 255:red, 117; green, 117; blue, 117 }  ,draw opacity=1 ][line width=1.5]  [dash pattern={on 1.69pt off 2.76pt}]  (466.74,363.68) -- (476.5,315.25) ;
\draw  [color={rgb, 255:red, 255; green, 0; blue, 0 }  ,draw opacity=1 ][fill={rgb, 255:red, 255; green, 161; blue, 161 }  ,fill opacity=1 ][line width=2.25]  (474.03,314.43) .. controls (474.48,313.06) and (475.95,312.33) .. (477.32,312.78) .. controls (478.68,313.23) and (479.42,314.7) .. (478.96,316.07) .. controls (478.51,317.43) and (477.04,318.17) .. (475.68,317.71) .. controls (474.31,317.26) and (473.58,315.79) .. (474.03,314.43) -- cycle ;
\draw  [color={rgb, 255:red, 255; green, 0; blue, 0 }  ,draw opacity=1 ][fill={rgb, 255:red, 255; green, 161; blue, 161 }  ,fill opacity=1 ][line width=2.25]  (538.93,259.97) .. controls (539.38,258.6) and (540.86,257.87) .. (542.22,258.32) .. controls (543.58,258.77) and (544.32,260.24) .. (543.87,261.61) .. controls (543.41,262.97) and (541.94,263.71) .. (540.58,263.25) .. controls (539.22,262.8) and (538.48,261.33) .. (538.93,259.97) -- cycle ;
\draw [color={rgb, 255:red, 70; green, 70; blue, 255 }  ,draw opacity=1 ][line width=1.5]    (507.53,299.37) -- (582.99,257.51) ;
\draw [color={rgb, 255:red, 70; green, 70; blue, 255 }  ,draw opacity=1 ][line width=1.5]    (507.53,299.37) -- (506.67,238.7) ;
\draw [color={rgb, 255:red, 70; green, 70; blue, 255 }  ,draw opacity=1 ][line width=1.5]    (456.92,281.97) -- (507.53,299.37) ;
\draw  [color={rgb, 255:red, 70; green, 70; blue, 255 }  ,draw opacity=1 ][fill={rgb, 255:red, 134; green, 134; blue, 255 }  ,fill opacity=1 ][line width=2.25]  (506.06,298.88) .. controls (506.33,298.07) and (507.21,297.63) .. (508.02,297.9) .. controls (508.83,298.17) and (509.27,299.05) .. (509,299.86) .. controls (508.73,300.67) and (507.86,301.11) .. (507.04,300.84) .. controls (506.23,300.57) and (505.79,299.7) .. (506.06,298.88) -- cycle ;

\draw  [draw opacity=0][fill={rgb, 255:red, 176; green, 213; blue, 255 }  ,fill opacity=1 ] (524.99,9.79) .. controls (563.19,7.84) and (576.04,19.95) .. (594.23,42.13) .. controls (534.12,57.49) and (487.6,86.76) .. (467.07,150.84) .. controls (448.97,132.64) and (443.87,111.19) .. (444.02,84.59) .. controls (458.66,11.15) and (513.57,11.43) .. (524.99,9.79) -- cycle ;
\draw [color={rgb, 255:red, 70; green, 70; blue, 255 }  ,draw opacity=1 ][line width=1.5]    (506.99,25.86) -- (583.32,44.67) ;
\draw [color={rgb, 255:red, 70; green, 70; blue, 255 }  ,draw opacity=1 ][line width=1.5]    (457.24,69.14) -- (507.86,86.54) ;
\draw [color={rgb, 255:red, 70; green, 70; blue, 255 }  ,draw opacity=1 ][line width=1.5]    (585.26,31.48) -- (583.32,44.67) ;
\draw [color={rgb, 255:red, 70; green, 70; blue, 255 }  ,draw opacity=1 ][line width=1.5]    (507.86,86.54) -- (506.99,25.86) ;
\draw [color={rgb, 255:red, 70; green, 70; blue, 255 }  ,draw opacity=1 ][line width=1.5]    (506.99,25.86) -- (457.24,69.14) ;
\draw [color={rgb, 255:red, 70; green, 70; blue, 255 }  ,draw opacity=1 ][line width=1.5]    (541.48,9.97) -- (506.99,25.86) ;
\draw [color={rgb, 255:red, 70; green, 70; blue, 255 }  ,draw opacity=1 ][line width=1.5]    (457.24,69.14) -- (443.62,88.95) ;
\draw [color={rgb, 255:red, 70; green, 70; blue, 255 }  ,draw opacity=1 ][line width=1.5]    (446.86,67.87) -- (457.24,69.14) ;
\draw [color={rgb, 255:red, 70; green, 70; blue, 255 }  ,draw opacity=1 ][line width=1.5]    (458.61,44.12) -- (457.24,69.14) ;
\draw [color={rgb, 255:red, 70; green, 70; blue, 255 }  ,draw opacity=1 ][line width=1.5]    (506.99,25.86) -- (483.52,21.79) ;
\draw [color={rgb, 255:red, 70; green, 70; blue, 255 }  ,draw opacity=1 ][line width=1.5]    (507.75,11.97) -- (506.99,25.86) ;
\draw [color={rgb, 255:red, 70; green, 70; blue, 255 }  ,draw opacity=1 ][line width=1.5]    (469.47,154.87) -- (472.51,134.08) ;
\draw [color={rgb, 255:red, 0; green, 0; blue, 0 }  ,draw opacity=1 ][line width=3]    (467.07,150.84) .. controls (486.08,90.81) and (533.85,57.3) .. (594.23,42.13) ;
\draw  [color={rgb, 255:red, 70; green, 70; blue, 255 }  ,draw opacity=1 ][fill={rgb, 255:red, 134; green, 134; blue, 255 }  ,fill opacity=1 ][line width=2.25]  (506.39,86.05) .. controls (506.66,85.24) and (507.54,84.8) .. (508.35,85.07) .. controls (509.16,85.34) and (509.6,86.21) .. (509.33,87.03) .. controls (509.06,87.84) and (508.18,88.28) .. (507.37,88.01) .. controls (506.56,87.74) and (506.12,86.86) .. (506.39,86.05) -- cycle ;
\draw  [color={rgb, 255:red, 70; green, 70; blue, 255 }  ,draw opacity=1 ][fill={rgb, 255:red, 134; green, 134; blue, 255 }  ,fill opacity=1 ][line width=2.25]  (455.77,68.65) .. controls (456.04,67.84) and (456.92,67.4) .. (457.73,67.67) .. controls (458.54,67.94) and (458.98,68.81) .. (458.71,69.62) .. controls (458.44,70.44) and (457.57,70.87) .. (456.75,70.6) .. controls (455.94,70.33) and (455.5,69.46) .. (455.77,68.65) -- cycle ;
\draw  [color={rgb, 255:red, 70; green, 70; blue, 255 }  ,draw opacity=1 ][fill={rgb, 255:red, 134; green, 134; blue, 255 }  ,fill opacity=1 ][line width=2.25]  (505.53,25.38) .. controls (505.8,24.56) and (506.67,24.13) .. (507.48,24.39) .. controls (508.29,24.66) and (508.73,25.54) .. (508.46,26.35) .. controls (508.19,27.16) and (507.32,27.6) .. (506.51,27.33) .. controls (505.7,27.06) and (505.26,26.19) .. (505.53,25.38) -- cycle ;
\draw [color={rgb, 255:red, 255; green, 0; blue, 0 }  ,draw opacity=1 ][fill={rgb, 255:red, 255; green, 161; blue, 161 }  ,fill opacity=1 ][line width=3]    (476.5,117.92) -- (549,53.59) ;
\draw [color={rgb, 255:red, 255; green, 0; blue, 0 }  ,draw opacity=1 ][fill={rgb, 255:red, 255; green, 161; blue, 161 }  ,fill opacity=1 ][line width=3]    (467,151.57) -- (476.5,117.92) ;
\draw [color={rgb, 255:red, 255; green, 0; blue, 0 }  ,draw opacity=1 ][fill={rgb, 255:red, 255; green, 161; blue, 161 }  ,fill opacity=1 ][line width=3]    (594.23,42.13) -- (549,53.59) ;
\draw  [color={rgb, 255:red, 255; green, 0; blue, 0 }  ,draw opacity=1 ][fill={rgb, 255:red, 255; green, 161; blue, 161 }  ,fill opacity=1 ][line width=2.25]  (473.34,116.87) .. controls (473.92,115.13) and (475.8,114.18) .. (477.55,114.76) .. controls (479.29,115.34) and (480.23,117.22) .. (479.65,118.97) .. controls (479.07,120.71) and (477.19,121.65) .. (475.45,121.07) .. controls (473.71,120.49) and (472.76,118.61) .. (473.34,116.87) -- cycle ;
\draw  [color={rgb, 255:red, 255; green, 0; blue, 0 }  ,draw opacity=1 ][fill={rgb, 255:red, 255; green, 161; blue, 161 }  ,fill opacity=1 ][line width=2.25]  (545.85,52.54) .. controls (546.43,50.8) and (548.31,49.86) .. (550.05,50.44) .. controls (551.79,51.02) and (552.74,52.9) .. (552.16,54.64) .. controls (551.58,56.38) and (549.7,57.32) .. (547.95,56.74) .. controls (546.21,56.16) and (545.27,54.28) .. (545.85,52.54) -- cycle ;
\draw [color={rgb, 255:red, 70; green, 70; blue, 255 }  ,draw opacity=1 ][line width=1.5]    (472.51,134.08) -- (453.83,133.62) ;
\draw [color={rgb, 255:red, 70; green, 70; blue, 255 }  ,draw opacity=1 ][line width=1.5]    (457.24,69.14) -- (472.51,134.08) ;
\draw  [color={rgb, 255:red, 70; green, 70; blue, 255 }  ,draw opacity=1 ][fill={rgb, 255:red, 134; green, 134; blue, 255 }  ,fill opacity=1 ][line width=2.25]  (471.04,133.59) .. controls (471.31,132.78) and (472.18,132.34) .. (472.99,132.61) .. controls (473.81,132.88) and (474.25,133.76) .. (473.98,134.57) .. controls (473.71,135.38) and (472.83,135.82) .. (472.02,135.55) .. controls (471.21,135.28) and (470.77,134.4) .. (471.04,133.59) -- cycle ;
\draw [color={rgb, 255:red, 70; green, 70; blue, 255 }  ,draw opacity=1 ][line width=1.5]    (472.51,134.08) -- (507.86,86.54) ;
\draw [color={rgb, 255:red, 70; green, 70; blue, 255 }  ,draw opacity=1 ][line width=1.5]    (583.32,44.67) -- (597.95,44.77) ;
\draw [color={rgb, 255:red, 70; green, 70; blue, 255 }  ,draw opacity=1 ][line width=1.5]    (507.86,86.54) -- (583.32,44.67) ;
\draw  [color={rgb, 255:red, 70; green, 70; blue, 255 }  ,draw opacity=1 ][fill={rgb, 255:red, 134; green, 134; blue, 255 }  ,fill opacity=1 ][line width=2.25]  (581.85,44.18) .. controls (582.12,43.37) and (582.99,42.93) .. (583.81,43.2) .. controls (584.62,43.47) and (585.06,44.35) .. (584.79,45.16) .. controls (584.52,45.97) and (583.64,46.41) .. (582.83,46.14) .. controls (582.02,45.87) and (581.58,44.99) .. (581.85,44.18) -- cycle ;
\draw  [color={rgb, 255:red, 97; green, 97; blue, 97 }  ,draw opacity=1 ][line width=3]  (443.62,88.95) .. controls (446.17,42.33) and (486.03,6.6) .. (532.65,9.15) .. controls (579.27,11.7) and (615,51.56) .. (612.45,98.18) .. controls (609.9,144.8) and (570.04,180.53) .. (523.42,177.98) .. controls (476.79,175.43) and (441.07,135.57) .. (443.62,88.95) -- cycle ;

\draw  [draw opacity=0][fill={rgb, 255:red, 176; green, 213; blue, 255 }  ,fill opacity=1 ] (524.99,428.22) .. controls (561.88,430.19) and (573.65,435.19) .. (594.04,460.47) .. controls (528.88,483.06) and (498.34,495.37) .. (467.22,568.88) .. controls (439.78,537.64) and (444.49,529.99) .. (444.64,503.46) .. controls (456.87,429.4) and (513.6,429.86) .. (524.99,428.22) -- cycle ;
\draw [color={rgb, 255:red, 117; green, 117; blue, 117 }  ,draw opacity=1 ][line width=1.5]  [dash pattern={on 1.69pt off 2.76pt}]  (542.73,466.73) -- (478,521.04) ;
\draw [color={rgb, 255:red, 255; green, 0; blue, 0 }  ,draw opacity=1 ][line width=3]    (468.87,570.86) .. controls (487.84,510.99) and (535.54,476.84) .. (595.75,461.71) ;
\draw [color={rgb, 255:red, 117; green, 117; blue, 117 }  ,draw opacity=1 ][line width=1.5]  [dash pattern={on 1.69pt off 2.76pt}]  (595.09,460.92) -- (542.73,466.73) ;
\draw  [color={rgb, 255:red, 255; green, 0; blue, 0 }  ,draw opacity=1 ][fill={rgb, 255:red, 255; green, 161; blue, 161 }  ,fill opacity=1 ][line width=2.25]  (475.54,520.23) .. controls (476,518.87) and (477.46,518.13) .. (478.82,518.58) .. controls (480.18,519.04) and (480.92,520.5) .. (480.46,521.86) .. controls (480.01,523.22) and (478.54,523.96) .. (477.19,523.5) .. controls (475.83,523.05) and (475.09,521.58) .. (475.54,520.23) -- cycle ;
\draw  [color={rgb, 255:red, 255; green, 0; blue, 0 }  ,draw opacity=1 ][fill={rgb, 255:red, 255; green, 161; blue, 161 }  ,fill opacity=1 ][line width=2.25]  (540.27,465.91) .. controls (540.72,464.56) and (542.19,463.82) .. (543.55,464.27) .. controls (544.91,464.72) and (545.64,466.19) .. (545.19,467.55) .. controls (544.74,468.91) and (543.27,469.65) .. (541.91,469.19) .. controls (540.55,468.74) and (539.82,467.27) .. (540.27,465.91) -- cycle ;
\draw  [draw opacity=0][fill={rgb, 255:red, 129; green, 129; blue, 248 }  ,fill opacity=1 ][line width=2.25]  (540.71,468.43) .. controls (540.71,468.43) and (540.71,468.43) .. (540.71,468.43) .. controls (540.71,468.43) and (540.71,468.43) .. (540.71,468.43) .. controls (539.78,467.31) and (539.92,465.65) .. (541.04,464.71) .. controls (542.15,463.78) and (543.81,463.92) .. (544.75,465.04) -- (542.73,466.73) -- cycle ; \draw  [color={rgb, 255:red, 70; green, 70; blue, 255 }  ,draw opacity=1 ][line width=2.25]  (540.71,468.43) .. controls (540.71,468.43) and (540.71,468.43) .. (540.71,468.43) .. controls (540.71,468.43) and (540.71,468.43) .. (540.71,468.43) .. controls (539.78,467.31) and (539.92,465.65) .. (541.04,464.71) .. controls (542.15,463.78) and (543.81,463.92) .. (544.75,465.04) ;
\draw  [draw opacity=0][fill={rgb, 255:red, 129; green, 129; blue, 248 }  ,fill opacity=1 ][line width=2.25]  (475.99,522.74) .. controls (475.99,522.74) and (475.99,522.74) .. (475.99,522.74) .. controls (475.05,521.62) and (475.2,519.96) .. (476.31,519.03) .. controls (477.43,518.09) and (479.09,518.24) .. (480.02,519.35) -- (478,521.04) -- cycle ; \draw  [color={rgb, 255:red, 70; green, 70; blue, 255 }  ,draw opacity=1 ][line width=2.25]  (475.99,522.74) .. controls (475.99,522.74) and (475.99,522.74) .. (475.99,522.74) .. controls (475.05,521.62) and (475.2,519.96) .. (476.31,519.03) .. controls (477.43,518.09) and (479.09,518.24) .. (480.02,519.35) ;
\draw [color={rgb, 255:red, 70; green, 70; blue, 255 }  ,draw opacity=1 ][line width=3]    (467.16,569.61) .. controls (486.12,509.74) and (533.82,475.6) .. (594.04,460.47) ;
\draw [color={rgb, 255:red, 70; green, 70; blue, 255 }  ,draw opacity=1 ][line width=1.5]    (506.71,443.85) -- (582.82,462.61) ;
\draw [color={rgb, 255:red, 70; green, 70; blue, 255 }  ,draw opacity=1 ][line width=1.5]    (457.09,487.01) -- (507.57,504.36) ;
\draw [color={rgb, 255:red, 70; green, 70; blue, 255 }  ,draw opacity=1 ][line width=1.5]    (584.77,449.46) -- (582.82,462.61) ;
\draw [color={rgb, 255:red, 70; green, 70; blue, 255 }  ,draw opacity=1 ][line width=1.5]    (472.31,551.78) -- (453.46,551.84) ;
\draw [color={rgb, 255:red, 70; green, 70; blue, 255 }  ,draw opacity=1 ][line width=1.5]    (507.57,504.36) -- (506.71,443.85) ;
\draw [color={rgb, 255:red, 70; green, 70; blue, 255 }  ,draw opacity=1 ][line width=1.5]    (457.09,487.01) -- (472.31,551.78) ;
\draw [color={rgb, 255:red, 70; green, 70; blue, 255 }  ,draw opacity=1 ][line width=1.5]    (506.71,443.85) -- (457.09,487.01) ;
\draw [color={rgb, 255:red, 70; green, 70; blue, 255 }  ,draw opacity=1 ][line width=1.5]    (541.1,428) -- (506.71,443.85) ;
\draw [color={rgb, 255:red, 70; green, 70; blue, 255 }  ,draw opacity=1 ][line width=1.5]    (457.09,487.01) -- (443.5,506.77) ;
\draw [color={rgb, 255:red, 70; green, 70; blue, 255 }  ,draw opacity=1 ][line width=1.5]    (446.74,485.75) -- (457.09,487.01) ;
\draw [color={rgb, 255:red, 70; green, 70; blue, 255 }  ,draw opacity=1 ][line width=1.5]    (458.46,462.06) -- (457.09,487.01) ;
\draw [color={rgb, 255:red, 70; green, 70; blue, 255 }  ,draw opacity=1 ][line width=1.5]    (506.71,443.85) -- (483.3,439.79) ;
\draw [color={rgb, 255:red, 70; green, 70; blue, 255 }  ,draw opacity=1 ][line width=1.5]    (507.47,430) -- (506.71,443.85) ;
\draw  [color={rgb, 255:red, 70; green, 70; blue, 255 }  ,draw opacity=1 ][fill={rgb, 255:red, 134; green, 134; blue, 255 }  ,fill opacity=1 ][line width=2.25]  (506.1,503.88) .. controls (506.37,503.07) and (507.25,502.63) .. (508.06,502.9) .. controls (508.87,503.17) and (509.3,504.04) .. (509.03,504.85) .. controls (508.77,505.66) and (507.89,506.1) .. (507.08,505.83) .. controls (506.27,505.56) and (505.84,504.68) .. (506.1,503.88) -- cycle ;
\draw  [color={rgb, 255:red, 70; green, 70; blue, 255 }  ,draw opacity=1 ][fill={rgb, 255:red, 134; green, 134; blue, 255 }  ,fill opacity=1 ][line width=2.25]  (470.85,551.29) .. controls (471.12,550.48) and (471.99,550.04) .. (472.8,550.31) .. controls (473.61,550.58) and (474.05,551.46) .. (473.78,552.26) .. controls (473.51,553.07) and (472.64,553.51) .. (471.83,553.24) .. controls (471.02,552.97) and (470.58,552.1) .. (470.85,551.29) -- cycle ;
\draw  [color={rgb, 255:red, 70; green, 70; blue, 255 }  ,draw opacity=1 ][fill={rgb, 255:red, 134; green, 134; blue, 255 }  ,fill opacity=1 ][line width=2.25]  (455.63,486.52) .. controls (455.9,485.71) and (456.77,485.27) .. (457.58,485.54) .. controls (458.39,485.81) and (458.83,486.69) .. (458.56,487.5) .. controls (458.29,488.31) and (457.41,488.74) .. (456.6,488.47) .. controls (455.8,488.2) and (455.36,487.33) .. (455.63,486.52) -- cycle ;
\draw  [color={rgb, 255:red, 70; green, 70; blue, 255 }  ,draw opacity=1 ][fill={rgb, 255:red, 134; green, 134; blue, 255 }  ,fill opacity=1 ][line width=2.25]  (505.24,443.37) .. controls (505.51,442.56) and (506.39,442.12) .. (507.2,442.39) .. controls (508.01,442.66) and (508.44,443.53) .. (508.17,444.34) .. controls (507.9,445.15) and (507.03,445.59) .. (506.22,445.32) .. controls (505.41,445.05) and (504.97,444.18) .. (505.24,443.37) -- cycle ;
\draw  [color={rgb, 255:red, 70; green, 70; blue, 255 }  ,draw opacity=1 ][fill={rgb, 255:red, 134; green, 134; blue, 255 }  ,fill opacity=1 ][line width=2.25]  (581.36,462.12) .. controls (581.63,461.31) and (582.5,460.88) .. (583.31,461.15) .. controls (584.12,461.42) and (584.56,462.29) .. (584.29,463.1) .. controls (584.02,463.91) and (583.15,464.35) .. (582.34,464.08) .. controls (581.53,463.81) and (581.09,462.93) .. (581.36,462.12) -- cycle ;
\draw  [color={rgb, 255:red, 97; green, 97; blue, 97 }  ,draw opacity=1 ][line width=3]  (443.5,506.77) .. controls (446.04,460.27) and (485.8,424.64) .. (532.29,427.18) .. controls (578.79,429.73) and (614.42,469.48) .. (611.87,515.97) .. controls (609.33,562.47) and (569.58,598.1) .. (523.08,595.56) .. controls (476.59,593.01) and (440.96,553.26) .. (443.5,506.77) -- cycle ;
\draw [color={rgb, 255:red, 117; green, 117; blue, 117 }  ,draw opacity=1 ][line width=1.5]  [dash pattern={on 1.69pt off 2.76pt}]  (468.28,569.34) -- (478,521.04) ;

\draw  [fill={rgb, 255:red, 176; green, 213; blue, 255 }  ,fill opacity=1 ][line width=2.25]  (419.47,215.19) .. controls (421.54,353.92) and (283.5,404.99) .. (241.74,526.46) .. controls (236.57,527.84) and (202.75,525.77) .. (198.26,526.46) .. controls (145.46,399.47) and (19.85,348.4) .. (20.54,215.19) .. controls (18.47,130.3) and (84.73,18.49) .. (220,15.73) .. controls (330.43,14.35) and (421.54,105.45) .. (419.47,215.19) -- cycle ;
\draw [line width=0.75]    (218.62,595.48) -- (218.62,534.98) ;
\draw [shift={(218.62,531.98)}, rotate = 450] [fill={rgb, 255:red, 0; green, 0; blue, 0 }  ][line width=0.08]  [draw opacity=0] (8.93,-4.29) -- (0,0) -- (8.93,4.29) -- cycle    ;
\draw [line width=0.75]    (233.81,595.48) -- (233.81,534.98) ;
\draw [shift={(233.81,531.98)}, rotate = 450] [fill={rgb, 255:red, 0; green, 0; blue, 0 }  ][line width=0.08]  [draw opacity=0] (8.93,-4.29) -- (0,0) -- (8.93,4.29) -- cycle    ;
\draw [line width=0.75]    (203.44,595.48) -- (203.44,534.98) ;
\draw [shift={(203.44,531.98)}, rotate = 450] [fill={rgb, 255:red, 0; green, 0; blue, 0 }  ][line width=0.08]  [draw opacity=0] (8.93,-4.29) -- (0,0) -- (8.93,4.29) -- cycle    ;

\draw [color={rgb, 255:red, 176; green, 213; blue, 255 }  ,draw opacity=1 ][line width=3]    (240.36,526.46) -- (199.64,526.46) ;
\draw [line width=0.75]    (219.8,59.9) -- (219.8,27.01) ;
\draw [shift={(219.8,24.01)}, rotate = 450] [fill={rgb, 255:red, 0; green, 0; blue, 0 }  ][line width=0.08]  [draw opacity=0] (8.93,-4.29) -- (0,0) -- (8.93,4.29) -- cycle    ;
\draw [line width=0.75]    (245.54,58.52) -- (249.31,28.37) ;
\draw [shift={(249.68,25.39)}, rotate = 457.13] [fill={rgb, 255:red, 0; green, 0; blue, 0 }  ][line width=0.08]  [draw opacity=0] (8.93,-4.29) -- (0,0) -- (8.93,4.29) -- cycle    ;
\draw [line width=0.75]    (271.08,64.73) -- (279.83,37.22) ;
\draw [shift={(280.74,34.36)}, rotate = 467.65] [fill={rgb, 255:red, 0; green, 0; blue, 0 }  ][line width=0.08]  [draw opacity=0] (8.93,-4.29) -- (0,0) -- (8.93,4.29) -- cycle    ;
\draw [line width=0.75]    (293.85,73.7) -- (306.26,50.13) ;
\draw [shift={(307.66,47.48)}, rotate = 477.76] [fill={rgb, 255:red, 0; green, 0; blue, 0 }  ][line width=0.08]  [draw opacity=0] (8.93,-4.29) -- (0,0) -- (8.93,4.29) -- cycle    ;
\draw [line width=0.75]    (316.63,89.58) -- (332.91,65.16) ;
\draw [shift={(334.57,62.66)}, rotate = 483.69] [fill={rgb, 255:red, 0; green, 0; blue, 0 }  ][line width=0.08]  [draw opacity=0] (8.93,-4.29) -- (0,0) -- (8.93,4.29) -- cycle    ;
\draw [line width=0.75]    (335.95,108.21) -- (357.08,87.53) ;
\draw [shift={(359.22,85.43)}, rotate = 495.61] [fill={rgb, 255:red, 0; green, 0; blue, 0 }  ][line width=0.08]  [draw opacity=0] (8.93,-4.29) -- (0,0) -- (8.93,4.29) -- cycle    ;
\draw [line width=0.75]    (351.14,129.61) -- (375.59,112.68) ;
\draw [shift={(378.05,110.97)}, rotate = 505.3] [fill={rgb, 255:red, 0; green, 0; blue, 0 }  ][line width=0.08]  [draw opacity=0] (8.93,-4.29) -- (0,0) -- (8.93,4.29) -- cycle    ;
\draw [line width=0.75]    (363.56,152.38) -- (391.15,141.1) ;
\draw [shift={(393.93,139.96)}, rotate = 517.75] [fill={rgb, 255:red, 0; green, 0; blue, 0 }  ][line width=0.08]  [draw opacity=0] (8.93,-4.29) -- (0,0) -- (8.93,4.29) -- cycle    ;
\draw [line width=0.75]    (370.46,176.54) -- (400.67,169.62) ;
\draw [shift={(403.59,168.95)}, rotate = 527.0899999999999] [fill={rgb, 255:red, 0; green, 0; blue, 0 }  ][line width=0.08]  [draw opacity=0] (8.93,-4.29) -- (0,0) -- (8.93,4.29) -- cycle    ;
\draw [line width=0.75]    (373.22,199.32) -- (404.74,197.43) ;
\draw [shift={(407.73,197.25)}, rotate = 536.5699999999999] [fill={rgb, 255:red, 0; green, 0; blue, 0 }  ][line width=0.08]  [draw opacity=0] (8.93,-4.29) -- (0,0) -- (8.93,4.29) -- cycle    ;
\draw [line width=0.75]    (372.53,224.16) -- (405.43,226.06) ;
\draw [shift={(408.42,226.23)}, rotate = 183.3] [fill={rgb, 255:red, 0; green, 0; blue, 0 }  ][line width=0.08]  [draw opacity=0] (8.93,-4.29) -- (0,0) -- (8.93,4.29) -- cycle    ;
\draw [line width=0.75]    (369.77,246.25) -- (401.39,255.1) ;
\draw [shift={(404.28,255.91)}, rotate = 195.64] [fill={rgb, 255:red, 0; green, 0; blue, 0 }  ][line width=0.08]  [draw opacity=0] (8.93,-4.29) -- (0,0) -- (8.93,4.29) -- cycle    ;
\draw [line width=0.75]    (362.87,269.02) -- (393.19,280.39) ;
\draw [shift={(396,281.45)}, rotate = 200.56] [fill={rgb, 255:red, 0; green, 0; blue, 0 }  ][line width=0.08]  [draw opacity=0] (8.93,-4.29) -- (0,0) -- (8.93,4.29) -- cycle    ;
\draw [line width=0.75]    (351.83,287.66) -- (381.01,305.43) ;
\draw [shift={(383.58,306.98)}, rotate = 211.32999999999998] [fill={rgb, 255:red, 0; green, 0; blue, 0 }  ][line width=0.08]  [draw opacity=0] (8.93,-4.29) -- (0,0) -- (8.93,4.29) -- cycle    ;
\draw [line width=0.75]    (337.33,307.68) -- (368.66,328.78) ;
\draw [shift={(371.15,330.45)}, rotate = 213.96] [fill={rgb, 255:red, 0; green, 0; blue, 0 }  ][line width=0.08]  [draw opacity=0] (8.93,-4.29) -- (0,0) -- (8.93,4.29) -- cycle    ;
\draw [line width=0.75]    (324.91,325.62) -- (351.65,349.17) ;
\draw [shift={(353.9,351.16)}, rotate = 221.38] [fill={rgb, 255:red, 0; green, 0; blue, 0 }  ][line width=0.08]  [draw opacity=0] (8.93,-4.29) -- (0,0) -- (8.93,4.29) -- cycle    ;
\draw [line width=0.75]    (307.66,344.94) -- (334.5,371.15) ;
\draw [shift={(336.64,373.24)}, rotate = 224.31] [fill={rgb, 255:red, 0; green, 0; blue, 0 }  ][line width=0.08]  [draw opacity=0] (8.93,-4.29) -- (0,0) -- (8.93,4.29) -- cycle    ;
\draw [line width=0.75]    (173.26,64.73) -- (166.58,36.12) ;
\draw [shift={(165.89,33.2)}, rotate = 436.86] [fill={rgb, 255:red, 0; green, 0; blue, 0 }  ][line width=0.08]  [draw opacity=0] (8.93,-4.29) -- (0,0) -- (8.93,4.29) -- cycle    ;
\draw [line width=0.75]    (151.17,73.7) -- (136.01,45.06) ;
\draw [shift={(134.61,42.4)}, rotate = 422.11] [fill={rgb, 255:red, 0; green, 0; blue, 0 }  ][line width=0.08]  [draw opacity=0] (8.93,-4.29) -- (0,0) -- (8.93,4.29) -- cycle    ;
\draw [line width=0.75]    (126.32,89.58) -- (109.43,60.64) ;
\draw [shift={(107.92,58.05)}, rotate = 419.73] [fill={rgb, 255:red, 0; green, 0; blue, 0 }  ][line width=0.08]  [draw opacity=0] (8.93,-4.29) -- (0,0) -- (8.93,4.29) -- cycle    ;
\draw [line width=0.75]    (197.18,58.52) -- (193.86,30.66) ;
\draw [shift={(193.5,27.68)}, rotate = 443.19] [fill={rgb, 255:red, 0; green, 0; blue, 0 }  ][line width=0.08]  [draw opacity=0] (8.93,-4.29) -- (0,0) -- (8.93,4.29) -- cycle    ;
\draw [line width=0.75]    (106.08,108.21) -- (88.45,82.61) ;
\draw [shift={(86.75,80.13)}, rotate = 415.46000000000004] [fill={rgb, 255:red, 0; green, 0; blue, 0 }  ][line width=0.08]  [draw opacity=0] (8.93,-4.29) -- (0,0) -- (8.93,4.29) -- cycle    ;
\draw [line width=0.75]    (90.43,129.61) -- (66.88,107.93) ;
\draw [shift={(64.67,105.9)}, rotate = 402.62] [fill={rgb, 255:red, 0; green, 0; blue, 0 }  ][line width=0.08]  [draw opacity=0] (8.93,-4.29) -- (0,0) -- (8.93,4.29) -- cycle    ;
\draw [line width=0.75]    (79.39,152.38) -- (51.5,133.36) ;
\draw [shift={(49.02,131.67)}, rotate = 394.3] [fill={rgb, 255:red, 0; green, 0; blue, 0 }  ][line width=0.08]  [draw opacity=0] (8.93,-4.29) -- (0,0) -- (8.93,4.29) -- cycle    ;
\draw [line width=0.75]    (72.03,176.54) -- (39.05,168.29) ;
\draw [shift={(36.14,167.56)}, rotate = 374.05] [fill={rgb, 255:red, 0; green, 0; blue, 0 }  ][line width=0.08]  [draw opacity=0] (8.93,-4.29) -- (0,0) -- (8.93,4.29) -- cycle    ;
\draw [line width=0.75]    (67.79,199.32) -- (34.52,195.51) ;
\draw [shift={(31.54,195.16)}, rotate = 366.53] [fill={rgb, 255:red, 0; green, 0; blue, 0 }  ][line width=0.08]  [draw opacity=0] (8.93,-4.29) -- (0,0) -- (8.93,4.29) -- cycle    ;
\draw [line width=0.75]    (69.27,224.16) -- (33.62,223.73) ;
\draw [shift={(30.62,223.69)}, rotate = 360.7] [fill={rgb, 255:red, 0; green, 0; blue, 0 }  ][line width=0.08]  [draw opacity=0] (8.93,-4.29) -- (0,0) -- (8.93,4.29) -- cycle    ;
\draw [line width=0.75]    (72.03,246.25) -- (37.24,253.45) ;
\draw [shift={(34.3,254.06)}, rotate = 348.3] [fill={rgb, 255:red, 0; green, 0; blue, 0 }  ][line width=0.08]  [draw opacity=0] (8.93,-4.29) -- (0,0) -- (8.93,4.29) -- cycle    ;
\draw [line width=0.75]    (79.39,269.02) -- (46.31,281.53) ;
\draw [shift={(43.5,282.59)}, rotate = 339.3] [fill={rgb, 255:red, 0; green, 0; blue, 0 }  ][line width=0.08]  [draw opacity=0] (8.93,-4.29) -- (0,0) -- (8.93,4.29) -- cycle    ;
\draw [line width=0.75]    (88.59,287.66) -- (56.27,305.09) ;
\draw [shift={(53.63,306.51)}, rotate = 331.66999999999996] [fill={rgb, 255:red, 0; green, 0; blue, 0 }  ][line width=0.08]  [draw opacity=0] (8.93,-4.29) -- (0,0) -- (8.93,4.29) -- cycle    ;
\draw [line width=0.75]    (101.48,307.68) -- (69.08,327.06) ;
\draw [shift={(66.51,328.6)}, rotate = 329.1] [fill={rgb, 255:red, 0; green, 0; blue, 0 }  ][line width=0.08]  [draw opacity=0] (8.93,-4.29) -- (0,0) -- (8.93,4.29) -- cycle    ;
\draw [line width=0.75]    (116.2,325.62) -- (86.36,348.84) ;
\draw [shift={(83.99,350.69)}, rotate = 322.11] [fill={rgb, 255:red, 0; green, 0; blue, 0 }  ][line width=0.08]  [draw opacity=0] (8.93,-4.29) -- (0,0) -- (8.93,4.29) -- cycle    ;
\draw [line width=0.75]    (131.85,344.94) -- (104.58,370.71) ;
\draw [shift={(102.4,372.77)}, rotate = 316.62] [fill={rgb, 255:red, 0; green, 0; blue, 0 }  ][line width=0.08]  [draw opacity=0] (8.93,-4.29) -- (0,0) -- (8.93,4.29) -- cycle    ;

\draw [line width=2.25]    (165.75,473.3) .. controls (125.32,483.85) and (109.83,500.28) .. (98.63,534.99) ;
\draw [shift={(170.94,472.01)}, rotate = 166.68] [fill={rgb, 255:red, 0; green, 0; blue, 0 }  ][line width=0.08]  [draw opacity=0] (14.29,-6.86) -- (0,0) -- (14.29,6.86) -- cycle    ;
\draw  [color={rgb, 255:red, 97; green, 97; blue, 97 }  ,draw opacity=1 ][line width=2.25]  (263.27,440.15) .. controls (263.27,419.66) and (279.89,403.05) .. (300.38,403.05) .. controls (320.87,403.05) and (337.49,419.66) .. (337.49,440.15) .. controls (337.49,460.65) and (320.87,477.26) .. (300.38,477.26) .. controls (279.89,477.26) and (263.27,460.65) .. (263.27,440.15) -- cycle ;
\draw [color={rgb, 255:red, 97; green, 97; blue, 97 }  ,draw opacity=1 ][line width=2.25]    (337.5,433) .. controls (387.5,412) and (426.5,419) .. (466.74,451.75) ;

\draw (101.32,555.38) node  [font=\LARGE]  {$\Omega _{Structure}$};
\draw (217.31,221.72) node  [font=\LARGE]  {$\Omega _{Fluid}$};
\draw (531.07,195.75) node  [font=\Large] [align=left] {SFEM - SFEM};
\draw (530.56,613.24) node  [font=\Large] [align=left] {NEFEM - IGA };
\draw (530.82,408.42) node  [font=\Large] [align=left] {SFEM - IGA};

\end{tikzpicture}

%% file: figures/nurbs-surf.tikz
\begin{tikzpicture}[x=0.75pt,y=0.75pt,yscale=-1,xscale=1]

\path  [shading=_3nha1xokn,_6b01hkhlk,path fading= _jdoi03xod ,fading transform={xshift=2}] (87.15,160.68) .. controls (119.16,99.53) and (158.29,56.84) .. (221.13,23.38) .. controls (343.26,13) and (419.14,31.46) .. (499.76,82.22) .. controls (439.3,121.45) and (413.21,154.91) .. (384.75,233.36) .. controls (270.93,153.75) and (202.16,142.22) .. (87.15,160.68) -- cycle ; 
 \draw  [color={rgb, 255:red, 107; green, 107; blue, 107 }  ,draw opacity=1 ][line width=1.5]  (87.15,160.68) .. controls (119.16,99.53) and (158.29,56.84) .. (221.13,23.38) .. controls (343.26,13) and (419.14,31.46) .. (499.76,82.22) .. controls (439.3,121.45) and (413.21,154.91) .. (384.75,233.36) .. controls (270.93,153.75) and (202.16,142.22) .. (87.15,160.68) -- cycle ; 

\draw [color={rgb, 255:red, 0; green, 0; blue, 0 }  ,draw opacity=1 ] [dash pattern={on 4.5pt off 4.5pt}]  (87.15,160.68) -- (168.96,142.22) ;
\draw [shift={(168.96,142.22)}, rotate = 347.28] [color={rgb, 255:red, 0; green, 0; blue, 0 }  ,draw opacity=1 ][fill={rgb, 255:red, 0; green, 0; blue, 0 }  ,fill opacity=1 ][line width=0.75]      (0, 0) circle [x radius= 3.35, y radius= 3.35]   ;
\draw [shift={(87.15,160.68)}, rotate = 347.28] [color={rgb, 255:red, 0; green, 0; blue, 0 }  ,draw opacity=1 ][fill={rgb, 255:red, 0; green, 0; blue, 0 }  ,fill opacity=1 ][line width=0.75]      (0, 0) circle [x radius= 3.35, y radius= 3.35]   ;
\draw  [dash pattern={on 4.5pt off 4.5pt}]  (300.57,168.75) -- (384.75,233.36) ;
\draw [shift={(384.75,233.36)}, rotate = 37.51] [color={rgb, 255:red, 0; green, 0; blue, 0 }  ][fill={rgb, 255:red, 0; green, 0; blue, 0 }  ][line width=0.75]      (0, 0) circle [x radius= 3.35, y radius= 3.35]   ;
\draw [shift={(300.57,168.75)}, rotate = 37.51] [color={rgb, 255:red, 0; green, 0; blue, 0 }  ][fill={rgb, 255:red, 0; green, 0; blue, 0 }  ][line width=0.75]      (0, 0) circle [x radius= 3.35, y radius= 3.35]   ;
\draw  [dash pattern={on 4.5pt off 4.5pt}]  (168.96,142.22) -- (300.57,168.75) ;
\draw [shift={(300.57,168.75)}, rotate = 11.4] [color={rgb, 255:red, 0; green, 0; blue, 0 }  ][fill={rgb, 255:red, 0; green, 0; blue, 0 }  ][line width=0.75]      (0, 0) circle [x radius= 3.35, y radius= 3.35]   ;
\draw [shift={(168.96,142.22)}, rotate = 11.4] [color={rgb, 255:red, 0; green, 0; blue, 0 }  ][fill={rgb, 255:red, 0; green, 0; blue, 0 }  ][line width=0.75]      (0, 0) circle [x radius= 3.35, y radius= 3.35]   ;
\draw [color={rgb, 255:red, 0; green, 0; blue, 0 }  ,draw opacity=1 ] [dash pattern={on 4.5pt off 4.5pt}]  (221.13,23.38) -- (298.2,14.15) ;
\draw [shift={(298.2,14.15)}, rotate = 353.17] [color={rgb, 255:red, 0; green, 0; blue, 0 }  ,draw opacity=1 ][fill={rgb, 255:red, 0; green, 0; blue, 0 }  ,fill opacity=1 ][line width=0.75]      (0, 0) circle [x radius= 3.35, y radius= 3.35]   ;
\draw [shift={(221.13,23.38)}, rotate = 353.17] [color={rgb, 255:red, 0; green, 0; blue, 0 }  ,draw opacity=1 ][fill={rgb, 255:red, 0; green, 0; blue, 0 }  ,fill opacity=1 ][line width=0.75]      (0, 0) circle [x radius= 3.35, y radius= 3.35]   ;
\draw  [dash pattern={on 4.5pt off 4.5pt}]  (406.1,28) -- (499.76,82.22) ;
\draw [shift={(499.76,82.22)}, rotate = 30.07] [color={rgb, 255:red, 0; green, 0; blue, 0 }  ][fill={rgb, 255:red, 0; green, 0; blue, 0 }  ][line width=0.75]      (0, 0) circle [x radius= 3.35, y radius= 3.35]   ;
\draw [shift={(406.1,28)}, rotate = 30.07] [color={rgb, 255:red, 0; green, 0; blue, 0 }  ][fill={rgb, 255:red, 0; green, 0; blue, 0 }  ][line width=0.75]      (0, 0) circle [x radius= 3.35, y radius= 3.35]   ;
\draw [color={rgb, 255:red, 0; green, 0; blue, 0 }  ,draw opacity=1 ] [dash pattern={on 4.5pt off 4.5pt}]  (298.2,14.15) -- (406.1,28) ;
\draw [shift={(406.1,28)}, rotate = 7.31] [color={rgb, 255:red, 0; green, 0; blue, 0 }  ,draw opacity=1 ][fill={rgb, 255:red, 0; green, 0; blue, 0 }  ,fill opacity=1 ][line width=0.75]      (0, 0) circle [x radius= 3.35, y radius= 3.35]   ;
\draw [shift={(298.2,14.15)}, rotate = 7.31] [color={rgb, 255:red, 0; green, 0; blue, 0 }  ,draw opacity=1 ][fill={rgb, 255:red, 0; green, 0; blue, 0 }  ,fill opacity=1 ][line width=0.75]      (0, 0) circle [x radius= 3.35, y radius= 3.35]   ;
\draw [color={rgb, 255:red, 0; green, 0; blue, 0 }  ,draw opacity=1 ] [dash pattern={on 4.5pt off 4.5pt}]  (141.69,69.53) -- (221.13,70.69) ;
\draw [shift={(221.13,70.69)}, rotate = 0.83] [color={rgb, 255:red, 0; green, 0; blue, 0 }  ,draw opacity=1 ][fill={rgb, 255:red, 0; green, 0; blue, 0 }  ,fill opacity=1 ][line width=0.75]      (0, 0) circle [x radius= 3.35, y radius= 3.35]   ;
\draw [shift={(141.69,69.53)}, rotate = 0.83] [color={rgb, 255:red, 0; green, 0; blue, 0 }  ,draw opacity=1 ][fill={rgb, 255:red, 0; green, 0; blue, 0 }  ,fill opacity=1 ][line width=0.75]      (0, 0) circle [x radius= 3.35, y radius= 3.35]   ;
\draw  [dash pattern={on 4.5pt off 4.5pt}]  (336.14,87.99) -- (422.7,131.83) ;
\draw [shift={(422.7,131.83)}, rotate = 26.86] [color={rgb, 255:red, 0; green, 0; blue, 0 }  ][fill={rgb, 255:red, 0; green, 0; blue, 0 }  ][line width=0.75]      (0, 0) circle [x radius= 3.35, y radius= 3.35]   ;
\draw [shift={(336.14,87.99)}, rotate = 26.86] [color={rgb, 255:red, 0; green, 0; blue, 0 }  ][fill={rgb, 255:red, 0; green, 0; blue, 0 }  ][line width=0.75]      (0, 0) circle [x radius= 3.35, y radius= 3.35]   ;
\draw [color={rgb, 255:red, 0; green, 0; blue, 0 }  ,draw opacity=1 ] [dash pattern={on 4.5pt off 4.5pt}]  (221.13,70.69) -- (336.14,87.99) ;
\draw [shift={(336.14,87.99)}, rotate = 8.56] [color={rgb, 255:red, 0; green, 0; blue, 0 }  ,draw opacity=1 ][fill={rgb, 255:red, 0; green, 0; blue, 0 }  ,fill opacity=1 ][line width=0.75]      (0, 0) circle [x radius= 3.35, y radius= 3.35]   ;
\draw [shift={(221.13,70.69)}, rotate = 8.56] [color={rgb, 255:red, 0; green, 0; blue, 0 }  ,draw opacity=1 ][fill={rgb, 255:red, 0; green, 0; blue, 0 }  ,fill opacity=1 ][line width=0.75]      (0, 0) circle [x radius= 3.35, y radius= 3.35]   ;
\draw [color={rgb, 255:red, 0; green, 0; blue, 0 }  ,draw opacity=1 ] [dash pattern={on 4.5pt off 4.5pt}]  (221.13,70.69) -- (298.2,14.15) ;
\draw [shift={(298.2,14.15)}, rotate = 323.74] [color={rgb, 255:red, 0; green, 0; blue, 0 }  ,draw opacity=1 ][fill={rgb, 255:red, 0; green, 0; blue, 0 }  ,fill opacity=1 ][line width=0.75]      (0, 0) circle [x radius= 3.35, y radius= 3.35]   ;
\draw [shift={(221.13,70.69)}, rotate = 323.74] [color={rgb, 255:red, 0; green, 0; blue, 0 }  ,draw opacity=1 ][fill={rgb, 255:red, 0; green, 0; blue, 0 }  ,fill opacity=1 ][line width=0.75]      (0, 0) circle [x radius= 3.35, y radius= 3.35]   ;
\draw  [dash pattern={on 4.5pt off 4.5pt}]  (336.14,87.99) -- (406.1,28) ;
\draw [shift={(406.1,28)}, rotate = 319.38] [color={rgb, 255:red, 0; green, 0; blue, 0 }  ][fill={rgb, 255:red, 0; green, 0; blue, 0 }  ][line width=0.75]      (0, 0) circle [x radius= 3.35, y radius= 3.35]   ;
\draw [shift={(336.14,87.99)}, rotate = 319.38] [color={rgb, 255:red, 0; green, 0; blue, 0 }  ][fill={rgb, 255:red, 0; green, 0; blue, 0 }  ][line width=0.75]      (0, 0) circle [x radius= 3.35, y radius= 3.35]   ;
\draw [color={rgb, 255:red, 0; green, 0; blue, 0 }  ,draw opacity=1 ] [dash pattern={on 4.5pt off 4.5pt}]  (168.96,142.22) -- (221.13,70.69) ;
\draw [shift={(221.13,70.69)}, rotate = 306.1] [color={rgb, 255:red, 0; green, 0; blue, 0 }  ,draw opacity=1 ][fill={rgb, 255:red, 0; green, 0; blue, 0 }  ,fill opacity=1 ][line width=0.75]      (0, 0) circle [x radius= 3.35, y radius= 3.35]   ;
\draw [shift={(168.96,142.22)}, rotate = 306.1] [color={rgb, 255:red, 0; green, 0; blue, 0 }  ,draw opacity=1 ][fill={rgb, 255:red, 0; green, 0; blue, 0 }  ,fill opacity=1 ][line width=0.75]      (0, 0) circle [x radius= 3.35, y radius= 3.35]   ;
\draw  [dash pattern={on 4.5pt off 4.5pt}]  (300.57,168.75) -- (336.14,87.99) ;
\draw [shift={(336.14,87.99)}, rotate = 293.77] [color={rgb, 255:red, 0; green, 0; blue, 0 }  ][fill={rgb, 255:red, 0; green, 0; blue, 0 }  ][line width=0.75]      (0, 0) circle [x radius= 3.35, y radius= 3.35]   ;
\draw [shift={(300.57,168.75)}, rotate = 293.77] [color={rgb, 255:red, 0; green, 0; blue, 0 }  ][fill={rgb, 255:red, 0; green, 0; blue, 0 }  ][line width=0.75]      (0, 0) circle [x radius= 3.35, y radius= 3.35]   ;
\draw [color={rgb, 255:red, 0; green, 0; blue, 0 }  ,draw opacity=1 ] [dash pattern={on 4.5pt off 4.5pt}]  (422.7,131.83) -- (499.76,82.22) ;
\draw [shift={(499.76,82.22)}, rotate = 327.23] [color={rgb, 255:red, 0; green, 0; blue, 0 }  ,draw opacity=1 ][fill={rgb, 255:red, 0; green, 0; blue, 0 }  ,fill opacity=1 ][line width=0.75]      (0, 0) circle [x radius= 3.35, y radius= 3.35]   ;
\draw [shift={(422.7,131.83)}, rotate = 327.23] [color={rgb, 255:red, 0; green, 0; blue, 0 }  ,draw opacity=1 ][fill={rgb, 255:red, 0; green, 0; blue, 0 }  ,fill opacity=1 ][line width=0.75]      (0, 0) circle [x radius= 3.35, y radius= 3.35]   ;
\draw [color={rgb, 255:red, 0; green, 0; blue, 0 }  ,draw opacity=1 ] [dash pattern={on 4.5pt off 4.5pt}]  (141.69,69.53) -- (221.13,23.38) ;
\draw [shift={(221.13,23.38)}, rotate = 329.85] [color={rgb, 255:red, 0; green, 0; blue, 0 }  ,draw opacity=1 ][fill={rgb, 255:red, 0; green, 0; blue, 0 }  ,fill opacity=1 ][line width=0.75]      (0, 0) circle [x radius= 3.35, y radius= 3.35]   ;
\draw [shift={(141.69,69.53)}, rotate = 329.85] [color={rgb, 255:red, 0; green, 0; blue, 0 }  ,draw opacity=1 ][fill={rgb, 255:red, 0; green, 0; blue, 0 }  ,fill opacity=1 ][line width=0.75]      (0, 0) circle [x radius= 3.35, y radius= 3.35]   ;
\draw [color={rgb, 255:red, 0; green, 0; blue, 0 }  ,draw opacity=1 ] [dash pattern={on 4.5pt off 4.5pt}]  (87.15,160.68) -- (141.69,69.53) ;
\draw [shift={(141.69,69.53)}, rotate = 300.9] [color={rgb, 255:red, 0; green, 0; blue, 0 }  ,draw opacity=1 ][fill={rgb, 255:red, 0; green, 0; blue, 0 }  ,fill opacity=1 ][line width=0.75]      (0, 0) circle [x radius= 3.35, y radius= 3.35]   ;
\draw [shift={(87.15,160.68)}, rotate = 300.9] [color={rgb, 255:red, 0; green, 0; blue, 0 }  ,draw opacity=1 ][fill={rgb, 255:red, 0; green, 0; blue, 0 }  ,fill opacity=1 ][line width=0.75]      (0, 0) circle [x radius= 3.35, y radius= 3.35]   ;
\draw [color={rgb, 255:red, 0; green, 0; blue, 0 }  ,draw opacity=1 ] [dash pattern={on 4.5pt off 4.5pt}]  (422.7,131.83) -- (384.75,233.36) ;
\draw [shift={(384.75,233.36)}, rotate = 110.49] [color={rgb, 255:red, 0; green, 0; blue, 0 }  ,draw opacity=1 ][fill={rgb, 255:red, 0; green, 0; blue, 0 }  ,fill opacity=1 ][line width=0.75]      (0, 0) circle [x radius= 3.35, y radius= 3.35]   ;
\draw [shift={(422.7,131.83)}, rotate = 110.49] [color={rgb, 255:red, 0; green, 0; blue, 0 }  ,draw opacity=1 ][fill={rgb, 255:red, 0; green, 0; blue, 0 }  ,fill opacity=1 ][line width=0.75]      (0, 0) circle [x radius= 3.35, y radius= 3.35]   ;
\draw [line width=1.5]    (211.7,132.89) .. controls (237.21,134.49) and (253.54,139.14) .. (273.17,149.06) ;
\draw [shift={(275.64,150.33)}, rotate = 207.47] [fill={rgb, 255:red, 0; green, 0; blue, 0 }  ][line width=1.5]  [draw opacity=0] (11.61,-5.58) -- (0,0) -- (11.61,5.58) -- cycle    ;

\draw [line width=1.5]    (211.7,132.89) .. controls (218.84,117.82) and (225.98,108.06) .. (239.62,93.47) ;
\draw [shift={(241.6,91.37)}, rotate = 493.45] [fill={rgb, 255:red, 0; green, 0; blue, 0 }  ][line width=1.5]  [draw opacity=0] (11.61,-5.58) -- (0,0) -- (11.61,5.58) -- cycle    ;

\draw  [color={rgb, 255:red, 0; green, 0; blue, 0 }  ,draw opacity=1 ][fill={rgb, 255:red, 255; green, 0; blue, 0 }  ,fill opacity=1 ] (83,160.68) .. controls (83,158.38) and (84.86,156.53) .. (87.15,156.53) .. controls (89.44,156.53) and (91.3,158.38) .. (91.3,160.68) .. controls (91.3,162.97) and (89.44,164.83) .. (87.15,164.83) .. controls (84.86,164.83) and (83,162.97) .. (83,160.68) -- cycle ;
\draw  [color={rgb, 255:red, 0; green, 0; blue, 0 }  ,draw opacity=1 ][fill={rgb, 255:red, 255; green, 0; blue, 0 }  ,fill opacity=1 ] (164.81,142.22) .. controls (164.81,139.92) and (166.67,138.07) .. (168.96,138.07) .. controls (171.26,138.07) and (173.11,139.92) .. (173.11,142.22) .. controls (173.11,144.51) and (171.26,146.37) .. (168.96,146.37) .. controls (166.67,146.37) and (164.81,144.51) .. (164.81,142.22) -- cycle ;
\draw  [color={rgb, 255:red, 0; green, 0; blue, 0 }  ,draw opacity=1 ][fill={rgb, 255:red, 255; green, 0; blue, 0 }  ,fill opacity=1 ] (296.42,168.75) .. controls (296.42,166.46) and (298.28,164.6) .. (300.57,164.6) .. controls (302.86,164.6) and (304.72,166.46) .. (304.72,168.75) .. controls (304.72,171.05) and (302.86,172.91) .. (300.57,172.91) .. controls (298.28,172.91) and (296.42,171.05) .. (296.42,168.75) -- cycle ;
\draw  [color={rgb, 255:red, 0; green, 0; blue, 0 }  ,draw opacity=1 ][fill={rgb, 255:red, 255; green, 0; blue, 0 }  ,fill opacity=1 ] (418.54,131.83) .. controls (418.54,129.54) and (420.4,127.68) .. (422.7,127.68) .. controls (424.99,127.68) and (426.85,129.54) .. (426.85,131.83) .. controls (426.85,134.13) and (424.99,135.99) .. (422.7,135.99) .. controls (420.4,135.99) and (418.54,134.13) .. (418.54,131.83) -- cycle ;
\draw  [color={rgb, 255:red, 0; green, 0; blue, 0 }  ,draw opacity=1 ][fill={rgb, 255:red, 255; green, 0; blue, 0 }  ,fill opacity=1 ] (380.6,233.36) .. controls (380.6,231.07) and (382.46,229.21) .. (384.75,229.21) .. controls (387.05,229.21) and (388.91,231.07) .. (388.91,233.36) .. controls (388.91,235.66) and (387.05,237.51) .. (384.75,237.51) .. controls (382.46,237.51) and (380.6,235.66) .. (380.6,233.36) -- cycle ;
\draw  [color={rgb, 255:red, 0; green, 0; blue, 0 }  ,draw opacity=1 ][fill={rgb, 255:red, 255; green, 0; blue, 0 }  ,fill opacity=1 ] (495.61,82.22) .. controls (495.61,79.93) and (497.47,78.07) .. (499.76,78.07) .. controls (502.06,78.07) and (503.92,79.93) .. (503.92,82.22) .. controls (503.92,84.52) and (502.06,86.37) .. (499.76,86.37) .. controls (497.47,86.37) and (495.61,84.52) .. (495.61,82.22) -- cycle ;
\draw  [color={rgb, 255:red, 0; green, 0; blue, 0 }  ,draw opacity=1 ][fill={rgb, 255:red, 255; green, 0; blue, 0 }  ,fill opacity=1 ] (331.99,87.99) .. controls (331.99,85.7) and (333.85,83.84) .. (336.14,83.84) .. controls (338.43,83.84) and (340.29,85.7) .. (340.29,87.99) .. controls (340.29,90.28) and (338.43,92.14) .. (336.14,92.14) .. controls (333.85,92.14) and (331.99,90.28) .. (331.99,87.99) -- cycle ;
\draw  [color={rgb, 255:red, 0; green, 0; blue, 0 }  ,draw opacity=1 ][fill={rgb, 255:red, 255; green, 0; blue, 0 }  ,fill opacity=1 ] (401.94,28) .. controls (401.94,25.7) and (403.8,23.84) .. (406.1,23.84) .. controls (408.39,23.84) and (410.25,25.7) .. (410.25,28) .. controls (410.25,30.29) and (408.39,32.15) .. (406.1,32.15) .. controls (403.8,32.15) and (401.94,30.29) .. (401.94,28) -- cycle ;
\draw  [color={rgb, 255:red, 0; green, 0; blue, 0 }  ,draw opacity=1 ][fill={rgb, 255:red, 255; green, 0; blue, 0 }  ,fill opacity=1 ] (216.98,70.69) .. controls (216.98,68.39) and (218.84,66.53) .. (221.13,66.53) .. controls (223.43,66.53) and (225.28,68.39) .. (225.28,70.69) .. controls (225.28,72.98) and (223.43,74.84) .. (221.13,74.84) .. controls (218.84,74.84) and (216.98,72.98) .. (216.98,70.69) -- cycle ;
\draw  [color={rgb, 255:red, 0; green, 0; blue, 0 }  ,draw opacity=1 ][fill={rgb, 255:red, 255; green, 0; blue, 0 }  ,fill opacity=1 ] (294.05,14.15) .. controls (294.05,11.86) and (295.91,10) .. (298.2,10) .. controls (300.49,10) and (302.35,11.86) .. (302.35,14.15) .. controls (302.35,16.44) and (300.49,18.3) .. (298.2,18.3) .. controls (295.91,18.3) and (294.05,16.44) .. (294.05,14.15) -- cycle ;
\draw  [color={rgb, 255:red, 0; green, 0; blue, 0 }  ,draw opacity=1 ][fill={rgb, 255:red, 255; green, 0; blue, 0 }  ,fill opacity=1 ] (216.98,23.38) .. controls (216.98,21.09) and (218.84,19.23) .. (221.13,19.23) .. controls (223.43,19.23) and (225.28,21.09) .. (225.28,23.38) .. controls (225.28,25.67) and (223.43,27.53) .. (221.13,27.53) .. controls (218.84,27.53) and (216.98,25.67) .. (216.98,23.38) -- cycle ;
\draw  [color={rgb, 255:red, 0; green, 0; blue, 0 }  ,draw opacity=1 ][fill={rgb, 255:red, 255; green, 0; blue, 0 }  ,fill opacity=1 ] (137.54,69.53) .. controls (137.54,67.24) and (139.4,65.38) .. (141.69,65.38) .. controls (143.99,65.38) and (145.84,67.24) .. (145.84,69.53) .. controls (145.84,71.82) and (143.99,73.68) .. (141.69,73.68) .. controls (139.4,73.68) and (137.54,71.82) .. (137.54,69.53) -- cycle ;

\draw [line width=1.5]    (48.97,59.39) -- (78.3,73.69) ;
\draw [shift={(81,75)}, rotate = 205.99] [fill={rgb, 255:red, 0; green, 0; blue, 0 }  ][line width=1.5]  [draw opacity=0] (11.61,-5.58) -- (0,0) -- (11.61,5.58) -- cycle    ;

\draw [line width=1.5]    (23.42,78.22) -- (48.97,59.39) ;

\draw [shift={(21,80)}, rotate = 323.62] [fill={rgb, 255:red, 0; green, 0; blue, 0 }  ][line width=1.5]  [draw opacity=0] (11.61,-5.58) -- (0,0) -- (11.61,5.58) -- cycle    ;
\draw [line width=1.5]    (48.97,59.39) -- (48.97,30) ;
\draw [shift={(48.97,27)}, rotate = 450] [fill={rgb, 255:red, 0; green, 0; blue, 0 }  ][line width=1.5]  [draw opacity=0] (11.61,-5.58) -- (0,0) -- (11.61,5.58) -- cycle    ;

\draw  [color={rgb, 255:red, 0; green, 0; blue, 0 }  ,draw opacity=1 ][fill={rgb, 255:red, 255; green, 0; blue, 0 }  ,fill opacity=1 ] (68,209) .. controls (68,206.24) and (70.24,204) .. (73,204) .. controls (75.76,204) and (78,206.24) .. (78,209) .. controls (78,211.76) and (75.76,214) .. (73,214) .. controls (70.24,214) and (68,211.76) .. (68,209) -- cycle ;

\draw (12,85) node [scale=1.2]  {\Large$x$};
\draw (92,76) node [scale=1.2]  {\Large$y$};
\draw (60,17) node [scale=1.2]  {\Large$z$};
\draw (282.28,137.04) node   {\Large$\theta _{1}$};
\draw (253.22,90.54) node   {\Large$\theta _{2}$};
\draw (163,212) node  [align=left] {\Large Control point };

\end{tikzpicture}

%% file: figures/element-types.tikz
\tikzset{every picture/.style={line width=0.75pt}} 

\begin{tikzpicture}[x=0.75pt,y=0.75pt,yscale=-1,xscale=1]

\draw  [fill={rgb, 255:red, 0; green, 0; blue, 0 }  ,fill opacity=0.05 ][line width=0.75]  (7,149.28) .. controls (92.34,130.38) and (406.84,96.42) .. (535.18,41.68) .. controls (595.99,78.85) and (627.53,132.9) .. (642.17,212.86) .. controls (537.43,265.79) and (367.38,293.95) .. (200.42,324.8) .. controls (143.38,224) and (102.62,168.18) .. (7,149.28) -- cycle ;
\draw [color={rgb, 255:red, 139; green, 139; blue, 139 }  ,draw opacity=1 ]   (265.86,107.14) .. controls (213.19,116.4) and (183.13,120.13) .. (142.43,126.86) ;
\draw  [draw opacity=0][fill={rgb, 255:red, 170; green, 170; blue, 170 }  ,fill opacity=0.27 ][line width=0.75]  (131.2,150.36) .. controls (180.05,55.13) and (131.52,150.69) .. (180.26,55.36) .. controls (268.86,234.05) and (180.05,55.39) .. (269.17,234.2) .. controls (198.28,183.78) and (178.31,166.61) .. (131.2,150.36) -- cycle ;
\draw  [draw opacity=0][fill={rgb, 255:red, 170; green, 170; blue, 170 }  ,fill opacity=0.27 ][line width=0.75]  (269.17,232.44) .. controls (180.57,53.64) and (269.12,232.56) .. (180.26,53.61) .. controls (208.96,71.35) and (247.5,95.31) .. (308.57,132.45) .. controls (279.18,158.25) and (272.73,180.85) .. (269.17,232.44) -- cycle ;
\draw  [fill={rgb, 255:red, 0; green, 0; blue, 0 }  ,fill opacity=1 ] (127.63,148.6) .. controls (127.63,146.64) and (129.23,145.04) .. (131.2,145.04) .. controls (133.16,145.04) and (134.76,146.64) .. (134.76,148.6) .. controls (134.76,150.57) and (133.16,152.17) .. (131.2,152.17) .. controls (129.23,152.17) and (127.63,150.57) .. (127.63,148.6) -- cycle ;
\draw  [fill={rgb, 255:red, 0; green, 0; blue, 0 }  ,fill opacity=1 ] (176.7,53.61) .. controls (176.7,51.64) and (178.29,50.04) .. (180.26,50.04) .. controls (182.23,50.04) and (183.83,51.64) .. (183.83,53.61) .. controls (183.83,55.57) and (182.23,57.17) .. (180.26,57.17) .. controls (178.29,57.17) and (176.7,55.57) .. (176.7,53.61) -- cycle ;
\draw  [fill={rgb, 255:red, 0; green, 0; blue, 0 }  ,fill opacity=1 ] (305,132.45) .. controls (305,130.48) and (306.6,128.89) .. (308.57,128.89) .. controls (310.53,128.89) and (312.13,130.48) .. (312.13,132.45) .. controls (312.13,134.42) and (310.53,136.01) .. (308.57,136.01) .. controls (306.6,136.01) and (305,134.42) .. (305,132.45) -- cycle ;
\draw  [fill={rgb, 255:red, 0; green, 0; blue, 0 }  ,fill opacity=1 ] (265.61,230.63) .. controls (265.61,228.67) and (267.2,227.07) .. (269.17,227.07) .. controls (271.14,227.07) and (272.74,228.67) .. (272.74,230.63) .. controls (272.74,232.6) and (271.14,234.2) .. (269.17,234.2) .. controls (267.2,234.2) and (265.61,232.6) .. (265.61,230.63) -- cycle ;
\draw [color={rgb, 255:red, 139; green, 139; blue, 139 }  ,draw opacity=1 ]   (462.47,66.13) .. controls (437.56,72.6) and (430.33,74.05) .. (401.97,80.7) ;
\draw  [draw opacity=0][fill={rgb, 255:red, 170; green, 170; blue, 170 }  ,fill opacity=0.27 ] (514.31,87.6) -- (366.12,231.08) -- (410.63,44.66) -- cycle ;
\draw [color={rgb, 255:red, 0; green, 0; blue, 0 }  ,draw opacity=0.27 ] [dash pattern={on 4.5pt off 4.5pt}]  (410.63,44.66) -- (498.6,158.82) ;
\draw  [draw opacity=0][fill={rgb, 255:red, 170; green, 170; blue, 170 }  ,fill opacity=0.27 ] (366.12,231.08) .. controls (514.67,87.61) and (367.16,228.99) .. (514.31,87.6) .. controls (498.08,158.82) and (513.99,87.61) .. (498.6,158.82) .. controls (445.71,179.77) and (440.36,212.85) .. (366.12,231.08) -- cycle ;
\draw [color={rgb, 255:red, 255; green, 0; blue, 0 }  ,draw opacity=1 ][line width=0.75]    (366.12,231.08) .. controls (433.14,216.42) and (443.62,181.86) .. (498.6,158.82) ;
\draw  [color={rgb, 255:red, 255; green, 0; blue, 0 }  ,draw opacity=1 ][fill={rgb, 255:red, 170; green, 170; blue, 170 }  ,fill opacity=0.34 ][line width=0.75]  (131.2,148.6) .. controls (186.56,125.52) and (253.39,120.12) .. (308.57,132.45) .. controls (277.45,161.46) and (273.28,178.13) .. (269.17,232.44) .. controls (206.71,188.86) and (183,167.71) .. (131.2,148.6) -- cycle ;
\draw [line width=1.5]    (184.41,355) .. controls (171.73,327.82) and (159.91,310.21) .. (142.29,285.47) ;
\draw [shift={(140.06,282.34)}, rotate = 414.46000000000004] [fill={rgb, 255:red, 0; green, 0; blue, 0 }  ][line width=0.08]  [draw opacity=0] (11.61,-5.58) -- (0,0) -- (11.61,5.58) -- cycle    ;
\draw [line width=1.5]    (184.41,355) .. controls (212.64,348.63) and (234.72,344.89) .. (265.06,333.61) ;
\draw [shift={(268.4,332.35)}, rotate = 519.0799999999999] [fill={rgb, 255:red, 0; green, 0; blue, 0 }  ][line width=0.08]  [draw opacity=0] (11.61,-5.58) -- (0,0) -- (11.61,5.58) -- cycle    ;
\draw    (180.26,53.61) -- (308.57,132.45) ;
\draw    (131.2,148.6) -- (180.26,53.61) ;
\draw    (269.17,232.44) -- (180.26,53.61) ;
\draw  [fill={rgb, 255:red, 177; green, 177; blue, 177 }  ,fill opacity=1 ] (174.89,53.61) .. controls (174.89,50.64) and (177.29,48.23) .. (180.26,48.23) .. controls (183.23,48.23) and (185.64,50.64) .. (185.64,53.61) .. controls (185.64,56.57) and (183.23,58.98) .. (180.26,58.98) .. controls (177.29,58.98) and (174.89,56.57) .. (174.89,53.61) -- cycle ;
\draw  [fill={rgb, 255:red, 177; green, 177; blue, 177 }  ,fill opacity=1 ] (263.8,232.44) .. controls (263.8,229.48) and (266.2,227.07) .. (269.17,227.07) .. controls (272.14,227.07) and (274.55,229.48) .. (274.55,232.44) .. controls (274.55,235.41) and (272.14,237.82) .. (269.17,237.82) .. controls (266.2,237.82) and (263.8,235.41) .. (263.8,232.44) -- cycle ;
\draw  [fill={rgb, 255:red, 177; green, 177; blue, 177 }  ,fill opacity=1 ] (303.19,132.45) .. controls (303.19,129.48) and (305.6,127.08) .. (308.57,127.08) .. controls (311.53,127.08) and (313.94,129.48) .. (313.94,132.45) .. controls (313.94,135.42) and (311.53,137.82) .. (308.57,137.82) .. controls (305.6,137.82) and (303.19,135.42) .. (303.19,132.45) -- cycle ;
\draw  [fill={rgb, 255:red, 177; green, 177; blue, 177 }  ,fill opacity=1 ] (125.82,148.6) .. controls (125.82,145.64) and (128.23,143.23) .. (131.2,143.23) .. controls (134.16,143.23) and (136.57,145.64) .. (136.57,148.6) .. controls (136.57,151.57) and (134.16,153.98) .. (131.2,153.98) .. controls (128.23,153.98) and (125.82,151.57) .. (125.82,148.6) -- cycle ;
\draw    (410.63,44.66) -- (514.31,87.6) ;
\draw    (514.31,87.6) -- (498.6,158.82) ;
\draw    (410.63,44.66) -- (366.12,231.08) ;
\draw  [fill={rgb, 255:red, 177; green, 177; blue, 177 }  ,fill opacity=1 ] (405.56,44.66) .. controls (405.56,41.86) and (407.83,39.6) .. (410.63,39.6) .. controls (413.43,39.6) and (415.7,41.86) .. (415.7,44.66) .. controls (415.7,47.46) and (413.43,49.73) .. (410.63,49.73) .. controls (407.83,49.73) and (405.56,47.46) .. (405.56,44.66) -- cycle ;
\draw  [fill={rgb, 255:red, 177; green, 177; blue, 177 }  ,fill opacity=1 ] (493.53,158.82) .. controls (493.53,156.02) and (495.8,153.75) .. (498.6,153.75) .. controls (501.4,153.75) and (503.67,156.02) .. (503.67,158.82) .. controls (503.67,161.62) and (501.4,163.89) .. (498.6,163.89) .. controls (495.8,163.89) and (493.53,161.62) .. (493.53,158.82) -- cycle ;
\draw    (514.31,87.6) -- (366.12,231.08) ;
\draw  [fill={rgb, 255:red, 177; green, 177; blue, 177 }  ,fill opacity=1 ] (509.24,87.6) .. controls (509.24,84.8) and (511.51,82.53) .. (514.31,82.53) .. controls (517.11,82.53) and (519.38,84.8) .. (519.38,87.6) .. controls (519.38,90.4) and (517.11,92.67) .. (514.31,92.67) .. controls (511.51,92.67) and (509.24,90.4) .. (509.24,87.6) -- cycle ;
\draw  [fill={rgb, 255:red, 177; green, 177; blue, 177 }  ,fill opacity=1 ] (361.05,231.08) .. controls (361.05,228.28) and (363.32,226.02) .. (366.12,226.02) .. controls (368.92,226.02) and (371.19,228.28) .. (371.19,231.08) .. controls (371.19,233.88) and (368.92,236.15) .. (366.12,236.15) .. controls (363.32,236.15) and (361.05,233.88) .. (361.05,231.08) -- cycle ;

\draw (92.77,152.22) node  [font=\normalsize]  {\LARGE$x_{1}(\thetavec_1)$};
\draw (334.57,148.36) node  [font=\normalsize]  {\LARGE$x_{3}(\thetavec_3)$};
\draw (269.4,255.11) node  [font=\normalsize,color={rgb, 255:red, 0; green, 0; blue, 0 }  ,opacity=1 ]  {\LARGE$x_{4}(\thetavec_4)$};
\draw (184.89,30.11) node  [font=\normalsize]  {\LARGE$\textcolor[rgb]{1,0,0}{x}\textcolor[rgb]{1,0,0}{_{2}}$};
\draw (377.74,253.37) node  [font=\normalsize]  {\LARGE$x_{1}(\thetavec_1)$};
\draw (534.29,176.69) node  [font=\normalsize]  {\LARGE$\textcolor[rgb]{0,0,0}{x_{3}}(\thetavec_3)$};
\draw (536.29,74.44) node  [font=\normalsize]  {\LARGE$\textcolor[rgb]{1,0,0}{x_{4}}$};
\draw (412.65,21.21) node  [font=\normalsize]  {\LARGE$\textcolor[rgb]{1,0,0}{x}\textcolor[rgb]{1,0,0}{_{2}}$};
\draw (304.79,335.26) node    {\LARGE$\theta _{1}$};
\draw (115.94,275.81) node    {\LARGE$\theta _{2}$};

\end{tikzpicture}

%% file: figures/quad-points-3d.tikz

\tikzset {_4pjsqdaw7/.code = {\pgfsetadditionalshadetransform{ \pgftransformshift{\pgfpoint{89.1 bp } { -128.7 bp }  }  \pgftransformscale{1.32 }  }}}
\pgfdeclareradialshading{_0leox0kss}{\pgfpoint{-72bp}{104bp}}{rgb(0bp)=(1,0.65,0.65);
rgb(0bp)=(1,0.65,0.65);
rgb(12.053571428571429bp)=(1,0,0);
rgb(17.142857142857142bp)=(0.63,0,0);
rgb(400bp)=(0.63,0,0)}


\tikzset {_rayp314gs/.code = {\pgfsetadditionalshadetransform{ \pgftransformshift{\pgfpoint{89.1 bp } { -128.7 bp }  }  \pgftransformscale{1.32 }  }}}
\pgfdeclareradialshading{_aubufyszb}{\pgfpoint{-72bp}{104bp}}{rgb(0bp)=(1,0.65,0.65);
rgb(0bp)=(1,0.65,0.65);
rgb(12.053571428571429bp)=(1,0,0);
rgb(17.142857142857142bp)=(0.63,0,0);
rgb(400bp)=(0.63,0,0)}


\tikzset {_y4crx1svf/.code = {\pgfsetadditionalshadetransform{ \pgftransformshift{\pgfpoint{89.1 bp } { -128.7 bp }  }  \pgftransformscale{1.32 }  }}}
\pgfdeclareradialshading{_4elnd7gtm}{\pgfpoint{-72bp}{104bp}}{rgb(0bp)=(1,0.65,0.65);
rgb(0bp)=(1,0.65,0.65);
rgb(12.053571428571429bp)=(1,0,0);
rgb(17.142857142857142bp)=(0.63,0,0);
rgb(400bp)=(0.63,0,0)}


\tikzset {_5rofg47fb/.code = {\pgfsetadditionalshadetransform{ \pgftransformshift{\pgfpoint{89.1 bp } { -128.7 bp }  }  \pgftransformscale{1.32 }  }}}
\pgfdeclareradialshading{_cwslkbl90}{\pgfpoint{-72bp}{104bp}}{rgb(0bp)=(1,0.65,0.65);
rgb(0bp)=(1,0.65,0.65);
rgb(12.053571428571429bp)=(1,0,0);
rgb(17.142857142857142bp)=(0.63,0,0);
rgb(400bp)=(0.63,0,0)}


\tikzset {_sxfdyg33n/.code = {\pgfsetadditionalshadetransform{ \pgftransformshift{\pgfpoint{89.1 bp } { -128.7 bp }  }  \pgftransformscale{1.32 }  }}}
\pgfdeclareradialshading{_187nczpdu}{\pgfpoint{-72bp}{104bp}}{rgb(0bp)=(1,0.65,0.65);
rgb(0bp)=(1,0.65,0.65);
rgb(12.053571428571429bp)=(1,0,0);
rgb(17.142857142857142bp)=(0.63,0,0);
rgb(400bp)=(0.63,0,0)}


\tikzset {_vna8046j8/.code = {\pgfsetadditionalshadetransform{ \pgftransformshift{\pgfpoint{89.1 bp } { -128.7 bp }  }  \pgftransformscale{1.32 }  }}}
\pgfdeclareradialshading{_o5w44rhyi}{\pgfpoint{-72bp}{104bp}}{rgb(0bp)=(1,0.65,0.65);
rgb(0bp)=(1,0.65,0.65);
rgb(12.053571428571429bp)=(1,0,0);
rgb(17.142857142857142bp)=(0.63,0,0);
rgb(400bp)=(0.63,0,0)}


\tikzset {_btx3tjry5/.code = {\pgfsetadditionalshadetransform{ \pgftransformshift{\pgfpoint{89.1 bp } { -128.7 bp }  }  \pgftransformscale{1.32 }  }}}
\pgfdeclareradialshading{_ljzw8fn0a}{\pgfpoint{-72bp}{104bp}}{rgb(0bp)=(1,0.65,0.65);
rgb(0bp)=(1,0.65,0.65);
rgb(12.053571428571429bp)=(1,0,0);
rgb(17.142857142857142bp)=(0.63,0,0);
rgb(400bp)=(0.63,0,0)}


\tikzset {_menf7kmgu/.code = {\pgfsetadditionalshadetransform{ \pgftransformshift{\pgfpoint{89.1 bp } { -128.7 bp }  }  \pgftransformscale{1.32 }  }}}
\pgfdeclareradialshading{_u8bj4mzdb}{\pgfpoint{-72bp}{104bp}}{rgb(0bp)=(1,0.65,0.65);
rgb(0bp)=(1,0.65,0.65);
rgb(12.053571428571429bp)=(1,0,0);
rgb(17.142857142857142bp)=(0.63,0,0);
rgb(400bp)=(0.63,0,0)}


\tikzset {_g2yuulour/.code = {\pgfsetadditionalshadetransform{ \pgftransformshift{\pgfpoint{89.1 bp } { -128.7 bp }  }  \pgftransformscale{1.32 }  }}}
\pgfdeclareradialshading{_3td5tqxli}{\pgfpoint{-72bp}{104bp}}{rgb(0bp)=(1,0.65,0.65);
rgb(0bp)=(1,0.65,0.65);
rgb(12.053571428571429bp)=(1,0,0);
rgb(17.142857142857142bp)=(0.63,0,0);
rgb(400bp)=(0.63,0,0)}


\tikzset {_tu7f9ybf7/.code = {\pgfsetadditionalshadetransform{ \pgftransformshift{\pgfpoint{89.1 bp } { -128.7 bp }  }  \pgftransformscale{1.32 }  }}}
\pgfdeclareradialshading{_0e6dfoijv}{\pgfpoint{-72bp}{104bp}}{rgb(0bp)=(1,0.65,0.65);
rgb(0bp)=(1,0.65,0.65);
rgb(12.053571428571429bp)=(1,0,0);
rgb(17.142857142857142bp)=(0.63,0,0);
rgb(400bp)=(0.63,0,0)}


\tikzset {_kugv0aa0g/.code = {\pgfsetadditionalshadetransform{ \pgftransformshift{\pgfpoint{89.1 bp } { -128.7 bp }  }  \pgftransformscale{1.32 }  }}}
\pgfdeclareradialshading{_9joh1s4qg}{\pgfpoint{-72bp}{104bp}}{rgb(0bp)=(1,0.65,0.65);
rgb(0bp)=(1,0.65,0.65);
rgb(12.053571428571429bp)=(1,0,0);
rgb(17.142857142857142bp)=(0.63,0,0);
rgb(400bp)=(0.63,0,0)}


\tikzset {_gxavjewue/.code = {\pgfsetadditionalshadetransform{ \pgftransformshift{\pgfpoint{89.1 bp } { -128.7 bp }  }  \pgftransformscale{1.32 }  }}}
\pgfdeclareradialshading{_q74e9t26p}{\pgfpoint{-72bp}{104bp}}{rgb(0bp)=(1,0.65,0.65);
rgb(0bp)=(1,0.65,0.65);
rgb(12.053571428571429bp)=(1,0,0);
rgb(17.142857142857142bp)=(0.63,0,0);
rgb(400bp)=(0.63,0,0)}


\tikzset {_v6ejr9cpd/.code = {\pgfsetadditionalshadetransform{ \pgftransformshift{\pgfpoint{89.1 bp } { -128.7 bp }  }  \pgftransformscale{1.32 }  }}}
\pgfdeclareradialshading{_d85mdg4lb}{\pgfpoint{-72bp}{104bp}}{rgb(0bp)=(1,0.65,0.65);
rgb(0bp)=(1,0.65,0.65);
rgb(12.053571428571429bp)=(1,0,0);
rgb(17.142857142857142bp)=(0.63,0,0);
rgb(400bp)=(0.63,0,0)}


\tikzset {_z633frs80/.code = {\pgfsetadditionalshadetransform{ \pgftransformshift{\pgfpoint{89.1 bp } { -128.7 bp }  }  \pgftransformscale{1.32 }  }}}
\pgfdeclareradialshading{_mp5l6gvea}{\pgfpoint{-72bp}{104bp}}{rgb(0bp)=(1,0.65,0.65);
rgb(0bp)=(1,0.65,0.65);
rgb(12.053571428571429bp)=(1,0,0);
rgb(17.142857142857142bp)=(0.63,0,0);
rgb(400bp)=(0.63,0,0)}


\tikzset {_8mbzpzyzz/.code = {\pgfsetadditionalshadetransform{ \pgftransformshift{\pgfpoint{89.1 bp } { -128.7 bp }  }  \pgftransformscale{1.32 }  }}}
\pgfdeclareradialshading{_9dmvyes44}{\pgfpoint{-72bp}{104bp}}{rgb(0bp)=(1,0.65,0.65);
rgb(0bp)=(1,0.65,0.65);
rgb(12.053571428571429bp)=(1,0,0);
rgb(17.142857142857142bp)=(0.63,0,0);
rgb(400bp)=(0.63,0,0)}


\tikzset {_inrstl3w2/.code = {\pgfsetadditionalshadetransform{ \pgftransformshift{\pgfpoint{89.1 bp } { -128.7 bp }  }  \pgftransformscale{1.32 }  }}}
\pgfdeclareradialshading{_s4cb77aa0}{\pgfpoint{-72bp}{104bp}}{rgb(0bp)=(1,0.65,0.65);
rgb(0bp)=(1,0.65,0.65);
rgb(12.053571428571429bp)=(1,0,0);
rgb(17.142857142857142bp)=(0.63,0,0);
rgb(400bp)=(0.63,0,0)}


\tikzset {_i0ivqcdiq/.code = {\pgfsetadditionalshadetransform{ \pgftransformshift{\pgfpoint{89.1 bp } { -128.7 bp }  }  \pgftransformscale{1.32 }  }}}
\pgfdeclareradialshading{_vmqg54jdr}{\pgfpoint{-72bp}{104bp}}{rgb(0bp)=(1,0.65,0.65);
rgb(0bp)=(1,0.65,0.65);
rgb(12.053571428571429bp)=(1,0,0);
rgb(17.142857142857142bp)=(0.63,0,0);
rgb(400bp)=(0.63,0,0)}


\tikzset {_p2ob5oo9h/.code = {\pgfsetadditionalshadetransform{ \pgftransformshift{\pgfpoint{89.1 bp } { -128.7 bp }  }  \pgftransformscale{1.32 }  }}}
\pgfdeclareradialshading{_a4k5yp2uq}{\pgfpoint{-72bp}{104bp}}{rgb(0bp)=(1,0.65,0.65);
rgb(0bp)=(1,0.65,0.65);
rgb(12.053571428571429bp)=(1,0,0);
rgb(17.142857142857142bp)=(0.63,0,0);
rgb(400bp)=(0.63,0,0)}


\tikzset {_c1q5dsyxy/.code = {\pgfsetadditionalshadetransform{ \pgftransformshift{\pgfpoint{89.1 bp } { -128.7 bp }  }  \pgftransformscale{1.32 }  }}}
\pgfdeclareradialshading{_fqn9gxfq6}{\pgfpoint{-72bp}{104bp}}{rgb(0bp)=(1,0.65,0.65);
rgb(0bp)=(1,0.65,0.65);
rgb(12.053571428571429bp)=(1,0,0);
rgb(17.142857142857142bp)=(0.63,0,0);
rgb(400bp)=(0.63,0,0)}


\tikzset {_e40hmjl4h/.code = {\pgfsetadditionalshadetransform{ \pgftransformshift{\pgfpoint{89.1 bp } { -128.7 bp }  }  \pgftransformscale{1.32 }  }}}
\pgfdeclareradialshading{_hnk9rm7t5}{\pgfpoint{-72bp}{104bp}}{rgb(0bp)=(1,0.65,0.65);
rgb(0bp)=(1,0.65,0.65);
rgb(12.053571428571429bp)=(1,0,0);
rgb(17.142857142857142bp)=(0.63,0,0);
rgb(400bp)=(0.63,0,0)}


\tikzset {_vcxjud5pw/.code = {\pgfsetadditionalshadetransform{ \pgftransformshift{\pgfpoint{89.1 bp } { -128.7 bp }  }  \pgftransformscale{1.32 }  }}}
\pgfdeclareradialshading{_wcpsa44is}{\pgfpoint{-72bp}{104bp}}{rgb(0bp)=(1,0.65,0.65);
rgb(0bp)=(1,0.65,0.65);
rgb(12.053571428571429bp)=(1,0,0);
rgb(17.142857142857142bp)=(0.63,0,0);
rgb(400bp)=(0.63,0,0)}


\tikzset {_zrm5mwu45/.code = {\pgfsetadditionalshadetransform{ \pgftransformshift{\pgfpoint{89.1 bp } { -128.7 bp }  }  \pgftransformscale{1.32 }  }}}
\pgfdeclareradialshading{_7d7nul3bl}{\pgfpoint{-72bp}{104bp}}{rgb(0bp)=(1,0.65,0.65);
rgb(0bp)=(1,0.65,0.65);
rgb(12.053571428571429bp)=(1,0,0);
rgb(17.142857142857142bp)=(0.63,0,0);
rgb(400bp)=(0.63,0,0)}


\tikzset {_oa3okideq/.code = {\pgfsetadditionalshadetransform{ \pgftransformshift{\pgfpoint{89.1 bp } { -128.7 bp }  }  \pgftransformscale{1.32 }  }}}
\pgfdeclareradialshading{_fmuae1ial}{\pgfpoint{-72bp}{104bp}}{rgb(0bp)=(1,0.65,0.65);
rgb(0bp)=(1,0.65,0.65);
rgb(12.053571428571429bp)=(1,0,0);
rgb(17.142857142857142bp)=(0.63,0,0);
rgb(400bp)=(0.63,0,0)}


\tikzset {_5kcqy7x39/.code = {\pgfsetadditionalshadetransform{ \pgftransformshift{\pgfpoint{89.1 bp } { -128.7 bp }  }  \pgftransformscale{1.32 }  }}}
\pgfdeclareradialshading{_2ctvkyf0g}{\pgfpoint{-72bp}{104bp}}{rgb(0bp)=(1,0.65,0.65);
rgb(0bp)=(1,0.65,0.65);
rgb(12.053571428571429bp)=(1,0,0);
rgb(17.142857142857142bp)=(0.63,0,0);
rgb(400bp)=(0.63,0,0)}


\tikzset {_71tb7groi/.code = {\pgfsetadditionalshadetransform{ \pgftransformshift{\pgfpoint{89.1 bp } { -128.7 bp }  }  \pgftransformscale{1.32 }  }}}
\pgfdeclareradialshading{_tv6j6pw8v}{\pgfpoint{-72bp}{104bp}}{rgb(0bp)=(1,0.65,0.65);
rgb(0bp)=(1,0.65,0.65);
rgb(12.053571428571429bp)=(1,0,0);
rgb(17.142857142857142bp)=(0.63,0,0);
rgb(400bp)=(0.63,0,0)}


\tikzset {_6fg74akqd/.code = {\pgfsetadditionalshadetransform{ \pgftransformshift{\pgfpoint{89.1 bp } { -128.7 bp }  }  \pgftransformscale{1.32 }  }}}
\pgfdeclareradialshading{_s6o1714va}{\pgfpoint{-72bp}{104bp}}{rgb(0bp)=(1,0.65,0.65);
rgb(0bp)=(1,0.65,0.65);
rgb(12.053571428571429bp)=(1,0,0);
rgb(17.142857142857142bp)=(0.63,0,0);
rgb(400bp)=(0.63,0,0)}


\tikzset {_no01xt5eh/.code = {\pgfsetadditionalshadetransform{ \pgftransformshift{\pgfpoint{89.1 bp } { -128.7 bp }  }  \pgftransformscale{1.32 }  }}}
\pgfdeclareradialshading{_jt00dtbet}{\pgfpoint{-72bp}{104bp}}{rgb(0bp)=(1,0.65,0.65);
rgb(0bp)=(1,0.65,0.65);
rgb(12.053571428571429bp)=(1,0,0);
rgb(17.142857142857142bp)=(0.63,0,0);
rgb(400bp)=(0.63,0,0)}


\tikzset {_msueephh0/.code = {\pgfsetadditionalshadetransform{ \pgftransformshift{\pgfpoint{4.5 bp } { 0 bp }  }  \pgftransformrotate{-46 }  \pgftransformscale{2 }  }}}
\pgfdeclarehorizontalshading{_guh1gfw47}{150bp}{rgb(0bp)=(1,1,1);
rgb(37.589285714285715bp)=(1,1,1);
rgb(62.5bp)=(0,0,0);
rgb(100bp)=(0,0,0)}
\tikzset{_cdu28vbi8/.code = {\pgfsetadditionalshadetransform{\pgftransformshift{\pgfpoint{4.5 bp } { 0 bp }  }  \pgftransformrotate{-46 }  \pgftransformscale{2 } }}}
\pgfdeclarehorizontalshading{_m9spj3q8p} {150bp} {color(0bp)=(transparent!43.00000000000001);
color(37.589285714285715bp)=(transparent!43.00000000000001);
color(62.5bp)=(transparent!63);
color(100bp)=(transparent!63) }
\pgfdeclarefading{_4y2o5vc27}{\tikz \fill[shading=_m9spj3q8p,_cdu28vbi8] (0,0) rectangle (50bp,50bp); }
\tikzset{every picture/.style={line width=0.75pt}} 
    \hspace*{-6cm}%
\begin{tikzpicture}[x=0.75pt,y=0.75pt,yscale=-1,xscale=1]


\draw  [draw opacity=0][shading=_0leox0kss,_4pjsqdaw7] (303.54,251.58) .. controls (303.54,240.93) and (312.17,232.3) .. (322.82,232.3) .. controls (333.46,232.3) and (342.09,240.93) .. (342.09,251.58) .. controls (342.09,262.23) and (333.46,270.86) .. (322.82,270.86) .. controls (312.17,270.86) and (303.54,262.23) .. (303.54,251.58) -- cycle ;
\draw  [draw opacity=0][shading=_aubufyszb,_rayp314gs] (308.45,258.51) .. controls (308.45,247.86) and (317.08,239.23) .. (327.72,239.23) .. controls (338.37,239.23) and (347,247.86) .. (347,258.51) .. controls (347,269.15) and (338.37,277.79) .. (327.72,277.79) .. controls (317.08,277.79) and (308.45,269.15) .. (308.45,258.51) -- cycle ;
\draw  [draw opacity=0][shading=_4elnd7gtm,_y4crx1svf] (308.04,266.25) .. controls (308.04,255.6) and (316.67,246.97) .. (327.32,246.97) .. controls (337.96,246.97) and (346.59,255.6) .. (346.59,266.25) .. controls (346.59,276.89) and (337.96,285.52) .. (327.32,285.52) .. controls (316.67,285.52) and (308.04,276.89) .. (308.04,266.25) -- cycle ;
\draw  [draw opacity=0][shading=_cwslkbl90,_5rofg47fb] (277.75,297.36) .. controls (277.75,286.71) and (286.38,278.08) .. (297.02,278.08) .. controls (307.67,278.08) and (316.3,286.71) .. (316.3,297.36) .. controls (316.3,308.01) and (307.67,316.64) .. (297.02,316.64) .. controls (286.38,316.64) and (277.75,308.01) .. (277.75,297.36) -- cycle ;
\draw  [draw opacity=0][shading=_187nczpdu,_sxfdyg33n] (288.98,319.26) .. controls (288.98,308.61) and (297.61,299.98) .. (308.26,299.98) .. controls (318.91,299.98) and (327.54,308.61) .. (327.54,319.26) .. controls (327.54,329.91) and (318.91,338.54) .. (308.26,338.54) .. controls (297.61,338.54) and (288.98,329.91) .. (288.98,319.26) -- cycle ;
\draw  [draw opacity=0][shading=_o5w44rhyi,_vna8046j8] (292.23,342.49) .. controls (292.23,331.84) and (300.86,323.21) .. (311.51,323.21) .. controls (322.15,323.21) and (330.78,331.84) .. (330.78,342.49) .. controls (330.78,353.14) and (322.15,361.77) .. (311.51,361.77) .. controls (300.86,361.77) and (292.23,353.14) .. (292.23,342.49) -- cycle ;
\draw  [draw opacity=0][shading=_ljzw8fn0a,_btx3tjry5] (240.39,356.63) .. controls (240.39,345.99) and (249.02,337.35) .. (259.67,337.35) .. controls (270.31,337.35) and (278.94,345.99) .. (278.94,356.63) .. controls (278.94,367.28) and (270.31,375.91) .. (259.67,375.91) .. controls (249.02,375.91) and (240.39,367.28) .. (240.39,356.63) -- cycle ;
\draw  [draw opacity=0][shading=_u8bj4mzdb,_menf7kmgu] (248.21,400.43) .. controls (248.21,389.78) and (256.84,381.15) .. (267.49,381.15) .. controls (278.13,381.15) and (286.76,389.78) .. (286.76,400.43) .. controls (286.76,411.08) and (278.13,419.71) .. (267.49,419.71) .. controls (256.84,419.71) and (248.21,411.08) .. (248.21,400.43) -- cycle ;
\draw  [draw opacity=0][shading=_3td5tqxli,_g2yuulour] (269.35,446.89) .. controls (269.35,436.24) and (277.98,427.61) .. (288.63,427.61) .. controls (299.27,427.61) and (307.91,436.24) .. (307.91,446.89) .. controls (307.91,457.53) and (299.27,466.17) .. (288.63,466.17) .. controls (277.98,466.17) and (269.35,457.53) .. (269.35,446.89) -- cycle ;
\draw  [draw opacity=0][shading=_0e6dfoijv,_tu7f9ybf7] (204.31,142.82) .. controls (204.31,132.17) and (212.94,123.54) .. (223.59,123.54) .. controls (234.23,123.54) and (242.86,132.17) .. (242.86,142.82) .. controls (242.86,153.47) and (234.23,162.1) .. (223.59,162.1) .. controls (212.94,162.1) and (204.31,153.47) .. (204.31,142.82) -- cycle ;
\draw  [draw opacity=0][shading=_9joh1s4qg,_kugv0aa0g] (212.34,152.3) .. controls (212.34,141.65) and (220.97,133.02) .. (231.62,133.02) .. controls (242.27,133.02) and (250.9,141.65) .. (250.9,152.3) .. controls (250.9,162.95) and (242.27,171.58) .. (231.62,171.58) .. controls (220.97,171.58) and (212.34,162.95) .. (212.34,152.3) -- cycle ;
\draw  [draw opacity=0][shading=_q74e9t26p,_gxavjewue] (218.34,172.91) .. controls (218.34,162.26) and (226.97,153.63) .. (237.62,153.63) .. controls (248.26,153.63) and (256.89,162.26) .. (256.89,172.91) .. controls (256.89,183.56) and (248.26,192.19) .. (237.62,192.19) .. controls (226.97,192.19) and (218.34,183.56) .. (218.34,172.91) -- cycle ;
\draw  [draw opacity=0][shading=_d85mdg4lb,_v6ejr9cpd] (165.56,222.94) .. controls (165.56,212.3) and (174.19,203.66) .. (184.83,203.66) .. controls (195.48,203.66) and (204.11,212.3) .. (204.11,222.94) .. controls (204.11,233.59) and (195.48,242.22) .. (184.83,242.22) .. controls (174.19,242.22) and (165.56,233.59) .. (165.56,222.94) -- cycle ;
\draw  [draw opacity=0][shading=_mp5l6gvea,_z633frs80] (177.56,262.43) .. controls (177.56,251.78) and (186.19,243.15) .. (196.84,243.15) .. controls (207.49,243.15) and (216.12,251.78) .. (216.12,262.43) .. controls (216.12,273.08) and (207.49,281.71) .. (196.84,281.71) .. controls (186.19,281.71) and (177.56,273.08) .. (177.56,262.43) -- cycle ;
\draw  [draw opacity=0][shading=_9dmvyes44,_8mbzpzyzz] (190.23,302.58) .. controls (190.23,291.93) and (198.86,283.3) .. (209.51,283.3) .. controls (220.16,283.3) and (228.79,291.93) .. (228.79,302.58) .. controls (228.79,313.23) and (220.16,321.86) .. (209.51,321.86) .. controls (198.86,321.86) and (190.23,313.23) .. (190.23,302.58) -- cycle ;
\draw  [draw opacity=0][shading=_s4cb77aa0,_inrstl3w2] (99.68,327.33) .. controls (99.68,316.69) and (108.31,308.06) .. (118.96,308.06) .. controls (129.6,308.06) and (138.24,316.69) .. (138.24,327.33) .. controls (138.24,337.98) and (129.6,346.61) .. (118.96,346.61) .. controls (108.31,346.61) and (99.68,337.98) .. (99.68,327.33) -- cycle ;
\draw  [draw opacity=0][shading=_vmqg54jdr,_i0ivqcdiq] (111.03,403.64) .. controls (111.03,393) and (119.66,384.36) .. (130.31,384.36) .. controls (140.96,384.36) and (149.59,393) .. (149.59,403.64) .. controls (149.59,414.29) and (140.96,422.92) .. (130.31,422.92) .. controls (119.66,422.92) and (111.03,414.29) .. (111.03,403.64) -- cycle ;
\draw  [draw opacity=0][shading=_a4k5yp2uq,_p2ob5oo9h] (150.36,485.28) .. controls (150.36,474.63) and (158.99,466) .. (169.64,466) .. controls (180.29,466) and (188.92,474.63) .. (188.92,485.28) .. controls (188.92,495.93) and (180.29,504.56) .. (169.64,504.56) .. controls (158.99,504.56) and (150.36,495.93) .. (150.36,485.28) -- cycle ;
\draw  [draw opacity=0][shading=_fqn9gxfq6,_c1q5dsyxy] (375.03,382.58) .. controls (375.03,371.93) and (383.66,363.3) .. (394.31,363.3) .. controls (404.96,363.3) and (413.59,371.93) .. (413.59,382.58) .. controls (413.59,393.23) and (404.96,401.86) .. (394.31,401.86) .. controls (383.66,401.86) and (375.03,393.23) .. (375.03,382.58) -- cycle ;
\draw  [draw opacity=0][shading=_hnk9rm7t5,_e40hmjl4h] (383.87,367.76) .. controls (383.87,357.11) and (392.5,348.48) .. (403.15,348.48) .. controls (413.8,348.48) and (422.43,357.11) .. (422.43,367.76) .. controls (422.43,378.41) and (413.8,387.04) .. (403.15,387.04) .. controls (392.5,387.04) and (383.87,378.41) .. (383.87,367.76) -- cycle ;
\draw  [draw opacity=0][shading=_wcpsa44is,_vcxjud5pw] (389.89,357.44) .. controls (389.89,346.8) and (398.52,338.17) .. (409.17,338.17) .. controls (419.82,338.17) and (428.45,346.8) .. (428.45,357.44) .. controls (428.45,368.09) and (419.82,376.72) .. (409.17,376.72) .. controls (398.52,376.72) and (389.89,368.09) .. (389.89,357.44) -- cycle ;
\draw  [draw opacity=0][shading=_7d7nul3bl,_zrm5mwu45] (384.3,396.74) .. controls (384.3,386.09) and (392.94,377.46) .. (403.58,377.46) .. controls (414.23,377.46) and (422.86,386.09) .. (422.86,396.74) .. controls (422.86,407.39) and (414.23,416.02) .. (403.58,416.02) .. controls (392.94,416.02) and (384.3,407.39) .. (384.3,396.74) -- cycle ;
\draw  [draw opacity=0][shading=_fmuae1ial,_oa3okideq] (391.81,376.33) .. controls (391.81,365.69) and (400.44,357.06) .. (411.09,357.06) .. controls (421.74,357.06) and (430.37,365.69) .. (430.37,376.33) .. controls (430.37,386.98) and (421.74,395.61) .. (411.09,395.61) .. controls (400.44,395.61) and (391.81,386.98) .. (391.81,376.33) -- cycle ;
\draw  [draw opacity=0][shading=_2ctvkyf0g,_5kcqy7x39] (396.94,361.97) .. controls (396.94,351.33) and (405.57,342.69) .. (416.22,342.69) .. controls (426.87,342.69) and (435.5,351.33) .. (435.5,361.97) .. controls (435.5,372.62) and (426.87,381.25) .. (416.22,381.25) .. controls (405.57,381.25) and (396.94,372.62) .. (396.94,361.97) -- cycle ;
\draw  [draw opacity=0][shading=_tv6j6pw8v,_71tb7groi] (393.11,411.33) .. controls (393.11,400.69) and (401.74,392.05) .. (412.38,392.05) .. controls (423.03,392.05) and (431.66,400.69) .. (431.66,411.33) .. controls (431.66,421.98) and (423.03,430.61) .. (412.38,430.61) .. controls (401.74,430.61) and (393.11,421.98) .. (393.11,411.33) -- cycle ;
\draw  [draw opacity=0][shading=_s6o1714va,_6fg74akqd] (398.33,385.23) .. controls (398.33,374.59) and (406.96,365.95) .. (417.6,365.95) .. controls (428.25,365.95) and (436.88,374.59) .. (436.88,385.23) .. controls (436.88,395.88) and (428.25,404.51) .. (417.6,404.51) .. controls (406.96,404.51) and (398.33,395.88) .. (398.33,385.23) -- cycle ;
\draw  [draw opacity=0][shading=_jt00dtbet,_no01xt5eh] (403.02,367.36) .. controls (403.02,356.71) and (411.65,348.08) .. (422.29,348.08) .. controls (432.94,348.08) and (441.57,356.71) .. (441.57,367.36) .. controls (441.57,378.01) and (432.94,386.64) .. (422.29,386.64) .. controls (411.65,386.64) and (403.02,378.01) .. (403.02,367.36) -- cycle ;
\draw [line width=1.5]    (225.9,59.74) -- (495.35,414.96) ;

\draw [line width=1.5]    (66.51,326.91) .. controls (142.41,240.39) and (177.33,179.67) .. (225.9,59.74) ;

\draw [line width=1.5]    (131.79,530.32) .. controls (192.51,375.49) and (227.42,272.26) .. (225.9,59.74) ;

\draw [line width=1.5]    (66.51,326.91) .. controls (48.3,421.03) and (68.03,477.19) .. (131.79,530.32) ;

\draw [color={rgb, 255:red, 0; green, 0; blue, 0 }  ,draw opacity=0.34 ][line width=1.5]  [dash pattern={on 1.69pt off 2.76pt}]  (66.51,326.91) -- (495.35,414.96) ;

\draw [line width=1.5]    (131.79,530.32) -- (495.35,414.96) ;

\draw  [draw opacity=0][fill={rgb, 255:red, 0; green, 0; blue, 0 }  ,fill opacity=0.1 ] (225.9,59.74) .. controls (307.87,169.04) and (442.98,345.13) .. (495.35,414.96) .. controls (357.97,455.94) and (230.46,498.45) .. (131.79,530.32) .. controls (186.43,407.37) and (230.46,238.87) .. (225.9,59.74) -- cycle ;
\draw  [draw opacity=0][fill={rgb, 255:red, 0; green, 0; blue, 0 }  ,fill opacity=0.13 ] (66.51,326.91) .. controls (175.81,349.68) and (323.05,378.52) .. (495.35,414.96) .. controls (357.97,455.94) and (230.46,498.45) .. (131.79,530.32) .. controls (69.55,475.68) and (48.3,431.65) .. (66.51,326.91) -- cycle ;
\draw  [draw opacity=0][shading=_guh1gfw47,_msueephh0,path fading= _4y2o5vc27 ,fading transform={xshift=2}] (66.51,326.91) .. controls (116.61,282.89) and (177.33,182.7) .. (225.9,59.74) .. controls (233.49,235.83) and (190.99,390.67) .. (131.79,530.32) .. controls (69.55,475.68) and (48.3,431.65) .. (66.51,326.91) -- cycle ;
\draw  [fill={rgb, 255:red, 0; green, 0; blue, 0 }  ,fill opacity=1 ] (217.56,59.74) .. controls (217.56,55.14) and (221.29,51.4) .. (225.9,51.4) .. controls (230.51,51.4) and (234.25,55.14) .. (234.25,59.74) .. controls (234.25,64.35) and (230.51,68.09) .. (225.9,68.09) .. controls (221.29,68.09) and (217.56,64.35) .. (217.56,59.74) -- cycle ;
\draw  [fill={rgb, 255:red, 0; green, 0; blue, 0 }  ,fill opacity=1 ] (58.17,326.91) .. controls (58.17,322.3) and (61.9,318.57) .. (66.51,318.57) .. controls (71.12,318.57) and (74.86,322.3) .. (74.86,326.91) .. controls (74.86,331.52) and (71.12,335.26) .. (66.51,335.26) .. controls (61.9,335.26) and (58.17,331.52) .. (58.17,326.91) -- cycle ;
\draw  [fill={rgb, 255:red, 0; green, 0; blue, 0 }  ,fill opacity=1 ] (123.44,530.32) .. controls (123.44,525.72) and (127.18,521.98) .. (131.79,521.98) .. controls (136.39,521.98) and (140.13,525.72) .. (140.13,530.32) .. controls (140.13,534.93) and (136.39,538.67) .. (131.79,538.67) .. controls (127.18,538.67) and (123.44,534.93) .. (123.44,530.32) -- cycle ;
\draw  [fill={rgb, 255:red, 0; green, 0; blue, 0 }  ,fill opacity=1 ] (487,414.96) .. controls (487,410.35) and (490.74,406.61) .. (495.35,406.61) .. controls (499.96,406.61) and (503.69,410.35) .. (503.69,414.96) .. controls (503.69,419.56) and (499.96,423.3) .. (495.35,423.3) .. controls (490.74,423.3) and (487,419.56) .. (487,414.96) -- cycle ;

\draw (29.38,301.27) node [scale=2.074]  {$\boldsymbol{x}_{1}$};
\draw (110.42,561.26) node [scale=2.074]  {$\boldsymbol{x}_{3}$};
\draw (263.49,19.89) node [scale=2.074]  {$\boldsymbol{x}_{4}$};
\draw (533.62,384.55) node [scale=2.074]  {$\boldsymbol{x}_{2}$};

\end{tikzpicture}

%% file: figures/shape-func-car.tikz
\tikzset{every picture/.style={line width=0.75pt}} 

\begin{tikzpicture}[x=0.75pt,y=0.75pt,yscale=-0.7,xscale=0.7, every node/.style={scale=0.7}]

\draw  [draw opacity=0][fill={rgb, 255:red, 202; green, 202; blue, 202 }  ,fill opacity=0.7019607843137254 ] (167,140) .. controls (238,185) and (106,205) .. (167,244) .. controls (106,221) and (97,215) .. (36.92,194.07) .. controls (114,162) and (121,159) .. (167,140) -- cycle ;
\draw  [draw opacity=0][fill={rgb, 255:red, 0; green, 0; blue, 0 }  ,fill opacity=0.44 ] (37.18,66.17) .. controls (95.8,98.5) and (114.2,108.1) .. (167,140) .. controls (237,246.5) and (103.8,150.1) .. (167,244) .. controls (106,160) and (80,125) .. (37.18,66.17) -- cycle ;
\draw [line width=1.5]    (36.92,194.07) -- (226.31,115.53) ;
\draw [shift={(230,114)}, rotate = 517.48] [fill={rgb, 255:red, 0; green, 0; blue, 0 }  ][line width=0.08]  [draw opacity=0] (13.4,-6.43) -- (0,0) -- (13.4,6.44) -- (8.9,0) -- cycle    ;
\draw [line width=1.5]    (36.92,194.07) -- (248.26,274.58) ;
\draw [shift={(252,276)}, rotate = 200.85] [fill={rgb, 255:red, 0; green, 0; blue, 0 }  ][line width=0.08]  [draw opacity=0] (13.4,-6.43) -- (0,0) -- (13.4,6.44) -- (8.9,0) -- cycle    ;
\draw [line width=1.5]    (32.91,66.17) -- (42.18,66.17) ;
\draw [line width=1.5]    (37.55,211.66) -- (37.55,15) ;
\draw [shift={(37.55,11)}, rotate = 450] [fill={rgb, 255:red, 0; green, 0; blue, 0 }  ][line width=0.08]  [draw opacity=0] (13.4,-6.43) -- (0,0) -- (13.4,6.44) -- (8.9,0) -- cycle    ;
\draw [color={rgb, 255:red, 255; green, 0; blue, 0 }  ,draw opacity=1 ][line width=1.5]    (167,197) .. controls (206,202) and (181,155) .. (167,140) ;
\draw  [draw opacity=0][fill={rgb, 255:red, 202; green, 202; blue, 202 }  ,fill opacity=0.91 ] (167,140) .. controls (167,161) and (167,178) .. (167,197) .. controls (184,179) and (206,161) .. (167,140) -- cycle ;
\draw [line width=0.75]    (167,197) .. controls (179,183) and (209,164) .. (167,140) ;
\draw [color={rgb, 255:red, 0; green, 0; blue, 0 }  ,draw opacity=1 ][line width=0.75]    (167,244) .. controls (135,223) and (156,208) .. (167,197) ;
\draw  [fill={rgb, 255:red, 136; green, 136; blue, 136 }  ,fill opacity=1 ][line width=1.5]  (164,140) .. controls (164,138.34) and (165.34,137) .. (167,137) .. controls (168.66,137) and (170,138.34) .. (170,140) .. controls (170,141.66) and (168.66,143) .. (167,143) .. controls (165.34,143) and (164,141.66) .. (164,140) -- cycle ;
\draw [color={rgb, 255:red, 0; green, 0; blue, 0 }  ,draw opacity=1 ][line width=0.75]    (87,277) .. controls (93.86,267.2) and (98.8,256.44) .. (138.52,257.9) ;
\draw [shift={(141,258)}, rotate = 182.73] [fill={rgb, 255:red, 0; green, 0; blue, 0 }  ,fill opacity=1 ][line width=0.08]  [draw opacity=0] (10.72,-5.15) -- (0,0) -- (10.72,5.15) -- (7.12,0) -- cycle    ;
\draw [color={rgb, 255:red, 0; green, 0; blue, 0 }  ,draw opacity=1 ][line width=0.75]    (220,78) .. controls (198.03,81.82) and (191.57,92.94) .. (180.59,109.61) ;
\draw [shift={(179,112)}, rotate = 303.69] [fill={rgb, 255:red, 0; green, 0; blue, 0 }  ,fill opacity=1 ][line width=0.08]  [draw opacity=0] (10.72,-5.15) -- (0,0) -- (10.72,5.15) -- (7.12,0) -- cycle    ;
\draw    (37.18,66.17) -- (167,244) ;
\draw    (37.18,66.17) -- (167,140) ;
\draw  [dash pattern={on 4.5pt off 4.5pt}]  (167,140) -- (195,155) ;
\draw  [dash pattern={on 4.5pt off 4.5pt}]  (167,244) -- (185,269) ;
\draw [color={rgb, 255:red, 255; green, 0; blue, 0 }  ,draw opacity=1 ][line width=0.75]    (192.09,191.21) .. controls (213.79,199.74) and (232.59,206.74) .. (245,180) ;
\draw [shift={(189,190)}, rotate = 21.37] [fill={rgb, 255:red, 255; green, 0; blue, 0 }  ,fill opacity=1 ][line width=0.08]  [draw opacity=0] (10.72,-5.15) -- (0,0) -- (10.72,5.15) -- (7.12,0) -- cycle    ;
\draw [color={rgb, 255:red, 255; green, 0; blue, 0 }  ,draw opacity=1 ][line width=1.5]    (167,244) .. controls (134,197) and (153,196) .. (167,197) ;
\draw [color={rgb, 255:red, 0; green, 0; blue, 0 }  ,draw opacity=1 ][line width=0.75]  [dash pattern={on 4.5pt off 4.5pt}]  (167,244) -- (167,140) ;
\draw  [fill={rgb, 255:red, 136; green, 136; blue, 136 }  ,fill opacity=1 ][line width=1.5]  (164,244) .. controls (164,242.34) and (165.34,241) .. (167,241) .. controls (168.66,241) and (170,242.34) .. (170,244) .. controls (170,245.66) and (168.66,247) .. (167,247) .. controls (165.34,247) and (164,245.66) .. (164,244) -- cycle ;
\draw  [fill={rgb, 255:red, 136; green, 136; blue, 136 }  ,fill opacity=1 ][line width=1.5]  (33.92,194.07) .. controls (33.92,192.42) and (35.26,191.07) .. (36.92,191.07) .. controls (38.58,191.07) and (39.92,192.42) .. (39.92,194.07) .. controls (39.92,195.73) and (38.58,197.07) .. (36.92,197.07) .. controls (35.26,197.07) and (33.92,195.73) .. (33.92,194.07) -- cycle ;
\draw  [fill={rgb, 255:red, 136; green, 136; blue, 136 }  ,fill opacity=1 ][line width=1.5]  (164,140) .. controls (164,138.34) and (165.34,137) .. (167,137) .. controls (168.66,137) and (170,138.34) .. (170,140) .. controls (170,141.66) and (168.66,143) .. (167,143) .. controls (165.34,143) and (164,141.66) .. (164,140) -- cycle ;

\draw (80.12,46.09) node  [font=\large]  {$\phi ( x,y)$};
\draw (21.29,64.17) node  [font=\normalsize]  {$1$};
\draw (23.05,190.06) node  [font=\normalsize]  {$0$};
\draw (267,165) node  [font=\normalsize]  {$\textcolor[rgb]{0.97,0,0}{\phi }\textcolor[rgb]{0.97,0,0}{(}\mathbf{\textcolor[rgb]{0.97,0,0}{x}}\textcolor[rgb]{0.97,0,0}{)}\textcolor[rgb]{0.97,0,0}{\ \neq 0}$};
\draw (169.12,121.09) node  [font=\large]  {$\boldsymbol{x}_{2}$};
\draw (158.12,258.09) node  [font=\large]  {$\boldsymbol{x}_{1}$};
\draw (83,293) node  [font=\large,color={rgb, 255:red, 0; green, 0; blue, 0 }  ,opacity=1 ]  {$\textcolor[rgb]{0,0,0}{\phi }\textcolor[rgb]{0,0,0}{(}\textcolor[rgb]{0,0,0}{\boldsymbol{x}}\textcolor[rgb]{0,0,0}{_{1}}\textcolor[rgb]{0,0,0}{)}\textcolor[rgb]{0,0,0}{\ =0}$};
\draw (240,288) node  [font=\large,color={rgb, 255:red, 0; green, 0; blue, 0 }  ,opacity=1 ]  {$x$};
\draw (237,128) node  [font=\large,color={rgb, 255:red, 0; green, 0; blue, 0 }  ,opacity=1 ]  {$y$};
\draw (268,75) node  [font=\large,color={rgb, 255:red, 0; green, 0; blue, 0 }  ,opacity=1 ]  {$\textcolor[rgb]{0,0,0}{\phi }\textcolor[rgb]{0,0,0}{(}\textcolor[rgb]{0,0,0}{\boldsymbol{x}}\textcolor[rgb]{0,0,0}{_{2}}\textcolor[rgb]{0,0,0}{)}\textcolor[rgb]{0,0,0}{\ =0}$};

\end{tikzpicture}

%% file: figures/shape-func-noncar.tikz
\tikzset{every picture/.style={line width=0.75pt}} 

\begin{tikzpicture}[x=0.75pt,y=0.75pt,yscale=-0.7,xscale=0.7, every node/.style={scale=0.7}]

\draw  [draw opacity=0][fill={rgb, 255:red, 202; green, 202; blue, 202 }  ,fill opacity=0.7019607843137254 ] (169,138) .. controls (240,183) and (108,203) .. (169,242) .. controls (108,219) and (99,213) .. (38.92,192.07) .. controls (116,160) and (123,157) .. (169,138) -- cycle ;
\draw [line width=1.5]    (38.92,192.07) -- (228.31,113.53) ;
\draw [shift={(232,112)}, rotate = 517.48] [fill={rgb, 255:red, 0; green, 0; blue, 0 }  ][line width=0.08]  [draw opacity=0] (13.4,-6.43) -- (0,0) -- (13.4,6.44) -- (8.9,0) -- cycle    ;
\draw [line width=1.5]    (38.92,192.07) -- (250.26,272.58) ;
\draw [shift={(254,274)}, rotate = 200.85] [fill={rgb, 255:red, 0; green, 0; blue, 0 }  ][line width=0.08]  [draw opacity=0] (13.4,-6.43) -- (0,0) -- (13.4,6.44) -- (8.9,0) -- cycle    ;
\draw [line width=1.5]    (34.91,64.17) -- (44.18,64.17) ;
\draw [line width=1.5]    (39.55,209.66) -- (39.55,13) ;
\draw [shift={(39.55,9)}, rotate = 450] [fill={rgb, 255:red, 0; green, 0; blue, 0 }  ][line width=0.08]  [draw opacity=0] (13.4,-6.43) -- (0,0) -- (13.4,6.44) -- (8.9,0) -- cycle    ;
\draw  [draw opacity=0][fill={rgb, 255:red, 202; green, 202; blue, 202 }  ,fill opacity=0.91 ] (169,138) .. controls (169,159) and (169,176) .. (169,195) .. controls (186,177) and (208,159) .. (169,138) -- cycle ;
\draw [line width=0.75]    (169,195) .. controls (181,181) and (211,162) .. (169,138) ;
\draw [color={rgb, 255:red, 0; green, 0; blue, 0 }  ,draw opacity=0.37 ][line width=0.75]    (169,242) .. controls (137,221) and (158,206) .. (169,195) ;
\draw  [fill={rgb, 255:red, 136; green, 136; blue, 136 }  ,fill opacity=1 ][line width=1.5]  (166,138) .. controls (166,136.34) and (167.34,135) .. (169,135) .. controls (170.66,135) and (172,136.34) .. (172,138) .. controls (172,139.66) and (170.66,141) .. (169,141) .. controls (167.34,141) and (166,139.66) .. (166,138) -- cycle ;
\draw [color={rgb, 255:red, 0; green, 0; blue, 0 }  ,draw opacity=1 ][line width=0.75]    (89,275) .. controls (95.86,265.2) and (100.8,254.44) .. (140.52,255.9) ;
\draw [shift={(143,256)}, rotate = 182.73] [fill={rgb, 255:red, 0; green, 0; blue, 0 }  ,fill opacity=1 ][line width=0.08]  [draw opacity=0] (10.72,-5.15) -- (0,0) -- (10.72,5.15) -- (7.12,0) -- cycle    ;
\draw [color={rgb, 255:red, 0; green, 0; blue, 0 }  ,draw opacity=1 ][line width=0.75]    (222,76) .. controls (200.03,79.82) and (193.57,90.94) .. (182.59,107.61) ;
\draw [shift={(181,110)}, rotate = 303.69] [fill={rgb, 255:red, 0; green, 0; blue, 0 }  ,fill opacity=1 ][line width=0.08]  [draw opacity=0] (10.72,-5.15) -- (0,0) -- (10.72,5.15) -- (7.12,0) -- cycle    ;
\draw [color={rgb, 255:red, 255; green, 0; blue, 0 }  ,draw opacity=1 ][line width=0.75]    (183.09,186.21) .. controls (204.79,194.74) and (223.59,201.74) .. (236,175) ;
\draw [shift={(180,185)}, rotate = 21.37] [fill={rgb, 255:red, 255; green, 0; blue, 0 }  ,fill opacity=1 ][line width=0.08]  [draw opacity=0] (10.72,-5.15) -- (0,0) -- (10.72,5.15) -- (7.12,0) -- cycle    ;
\draw [color={rgb, 255:red, 0; green, 0; blue, 0 }  ,draw opacity=1 ][line width=0.75]  [dash pattern={on 4.5pt off 4.5pt}]  (169,242) -- (169,138) ;
\draw  [fill={rgb, 255:red, 136; green, 136; blue, 136 }  ,fill opacity=1 ][line width=1.5]  (166,242) .. controls (166,240.34) and (167.34,239) .. (169,239) .. controls (170.66,239) and (172,240.34) .. (172,242) .. controls (172,243.66) and (170.66,245) .. (169,245) .. controls (167.34,245) and (166,243.66) .. (166,242) -- cycle ;
\draw  [fill={rgb, 255:red, 136; green, 136; blue, 136 }  ,fill opacity=1 ][line width=1.5]  (166,138) .. controls (166,136.34) and (167.34,135) .. (169,135) .. controls (170.66,135) and (172,136.34) .. (172,138) .. controls (172,139.66) and (170.66,141) .. (169,141) .. controls (167.34,141) and (166,139.66) .. (166,138) -- cycle ;
\draw  [fill={rgb, 255:red, 136; green, 136; blue, 136 }  ,fill opacity=1 ][line width=1.5]  (35.92,192.07) .. controls (35.92,190.42) and (37.26,189.07) .. (38.92,189.07) .. controls (40.58,189.07) and (41.92,190.42) .. (41.92,192.07) .. controls (41.92,193.73) and (40.58,195.07) .. (38.92,195.07) .. controls (37.26,195.07) and (35.92,193.73) .. (35.92,192.07) -- cycle ;
\draw  [fill={rgb, 255:red, 0; green, 0; blue, 0 }  ,fill opacity=0.44 ] (39.18,64.17) .. controls (98,96) and (115,107) .. (169,138) .. controls (240,183) and (107,200) .. (169,242) .. controls (108,158) and (82,123) .. (39.18,64.17) -- cycle ;
\draw [color={rgb, 255:red, 255; green, 0; blue, 0 }  ,draw opacity=1 ][line width=1.5]    (169,242) .. controls (106,199) and (241,183) .. (169,138) ;

\draw (82.12,44.09) node  [font=\large]  {$\phi ( x,y)$};
\draw (23.29,62.17) node  [font=\normalsize]  {$1$};
\draw (25.05,188.06) node  [font=\normalsize]  {$0$};
\draw (256,161) node  [font=\normalsize]  {$\textcolor[rgb]{0.97,0,0}{\phi }\textcolor[rgb]{0.97,0,0}{(}\mathbf{\textcolor[rgb]{0.97,0,0}{x}}\textcolor[rgb]{0.97,0,0}{)}\textcolor[rgb]{0.97,0,0}{\ =0}$};
\draw (171.12,119.09) node  [font=\large]  {$\boldsymbol{x}_{2}$};
\draw (160.12,256.09) node  [font=\large]  {$\boldsymbol{x}_{1}$};
\draw (86,291) node  [font=\large,color={rgb, 255:red, 0; green, 0; blue, 0 }  ,opacity=1 ]  {$\textcolor[rgb]{0,0,0}{\phi }\textcolor[rgb]{0,0,0}{(}\textcolor[rgb]{0,0,0}{\boldsymbol{x}}\textcolor[rgb]{0,0,0}{_{1}}\textcolor[rgb]{0,0,0}{)}\textcolor[rgb]{0,0,0}{\ =0}$};
\draw (242,286) node  [font=\large,color={rgb, 255:red, 0; green, 0; blue, 0 }  ,opacity=1 ]  {$x$};
\draw (239,126) node  [font=\large,color={rgb, 255:red, 0; green, 0; blue, 0 }  ,opacity=1 ]  {$y$};
\draw (270,73) node  [font=\large,color={rgb, 255:red, 0; green, 0; blue, 0 }  ,opacity=1 ]  {$\textcolor[rgb]{0,0,0}{\phi }\textcolor[rgb]{0,0,0}{(}\textcolor[rgb]{0,0,0}{\boldsymbol{x}}\textcolor[rgb]{0,0,0}{_{2}}\textcolor[rgb]{0,0,0}{)}\textcolor[rgb]{0,0,0}{\ =0}$};

\end{tikzpicture}

%% file: figures/mapping-face.tikz
\tikzset{every picture/.style={line width=0.75pt}} 

\begin{tikzpicture}[x=0.75pt,y=0.75pt,yscale=-1,xscale=1]

    \draw  [color={rgb, 255:red, 254; green, 0; blue, 0 }  ,draw opacity=1 ][fill={rgb, 255:red, 255; green, 147; blue, 147 }  ,fill opacity=1 ] (324.31,168.47) -- (406.04,177.93) -- (358.74,217.97) -- (277.02,208.51) -- cycle ;
    \draw  [draw opacity=0][fill={rgb, 255:red, 204; green, 203; blue, 203 }  ,fill opacity=1 ][line width=1.5]  (613.87,354.82) -- (594.37,420.53) -- (525.68,380.32) -- cycle ;
    \draw  [draw opacity=0][fill={rgb, 255:red, 204; green, 203; blue, 203 }  ,fill opacity=0.59 ][line width=0.75]  (53.23,377.24) .. controls (105,362.4) and (126.6,355.6) .. (170.56,343.31) .. controls (173.84,368.26) and (168.71,406.75) .. (150.74,432.41) .. controls (107.8,407.6) and (84.2,394.8) .. (53.23,377.24) -- cycle ;
    \draw  [draw opacity=0][fill={rgb, 255:red, 204; green, 203; blue, 203 }  ,fill opacity=0.52 ] (53.23,272.02) .. controls (102.92,321.1) and (129.87,366.65) .. (150.74,432.41) .. controls (126.02,418.3) and (61.86,382.37) .. (53.23,377.24) .. controls (53.55,368.9) and (52.26,280.36) .. (53.23,272.02) -- cycle ;
    \draw  [color={rgb, 255:red, 255; green, 0; blue, 0 }  ,draw opacity=1 ][fill={rgb, 255:red, 255; green, 147; blue, 147 }  ,fill opacity=0.52 ][line width=0.75]  (53.23,272.02) .. controls (112.25,282.29) and (146.89,301.53) .. (170.56,343.31) .. controls (173.84,368.26) and (168.71,406.75) .. (150.74,432.41) .. controls (130.21,365.69) and (101.98,320.78) .. (53.23,272.02) -- cycle ;
    \draw [line width=0.75]    (53.23,377.24) -- (150.74,432.41) ;
    \draw [shift={(150.74,432.41)}, rotate = 29.5] [color={rgb, 255:red, 0; green, 0; blue, 0 }  ][fill={rgb, 255:red, 0; green, 0; blue, 0 }  ][line width=0.75]      (0, 0) circle [x radius= 3.35, y radius= 3.35]   ;
    \draw [color={rgb, 255:red, 0; green, 0; blue, 0 }  ,draw opacity=1 ][fill={rgb, 255:red, 0; green, 0; blue, 0 }  ,fill opacity=1 ][line width=0.75]  [dash pattern={on 0.84pt off 2.51pt}]  (525.68,380.32) -- (613.87,354.82) ;
    \draw [line width=0.75]    (525.54,294.8) -- (594.37,420.53) ;
    \draw [shift={(594.37,420.53)}, rotate = 61.3] [color={rgb, 255:red, 0; green, 0; blue, 0 }  ][fill={rgb, 255:red, 0; green, 0; blue, 0 }  ][line width=0.75]      (0, 0) circle [x radius= 3.35, y radius= 3.35]   ;
    \draw [shift={(525.54,294.8)}, rotate = 61.3] [color={rgb, 255:red, 0; green, 0; blue, 0 }  ][fill={rgb, 255:red, 0; green, 0; blue, 0 }  ][line width=0.75]      (0, 0) circle [x radius= 3.35, y radius= 3.35]   ;
    \draw [line width=1.5]    (525.66,261.05) -- (525.54,294.8) ;
    \draw [shift={(525.68,257.05)}, rotate = 90.21] [fill={rgb, 255:red, 0; green, 0; blue, 0 }  ][line width=0.08]  [draw opacity=0] (11.61,-5.58) -- (0,0) -- (11.61,5.58) -- cycle    ;
    \draw [line width=1.5]    (594.37,420.53) -- (629.6,440.5) ;
    \draw [shift={(633.08,442.48)}, rotate = 209.55] [fill={rgb, 255:red, 0; green, 0; blue, 0 }  ][line width=0.08]  [draw opacity=0] (11.61,-5.58) -- (0,0) -- (11.61,5.58) -- cycle    ;
    \draw  [draw opacity=0][fill={rgb, 255:red, 204; green, 203; blue, 203 }  ,fill opacity=0.59 ][line width=1.5]  (525.54,294.8) -- (594.37,420.53) -- (525.68,380.32) -- cycle ;
    \draw  [color={rgb, 255:red, 0; green, 0; blue, 0 }  ,draw opacity=1 ][fill={rgb, 255:red, 204; green, 203; blue, 203 }  ,fill opacity=0.53 ] (406.04,93.63) -- (406.04,177.93) -- (358.74,217.97) -- (358.74,133.66) -- cycle ;
    \draw    (324.31,84.16) -- (406.04,93.63) ;
    \draw    (324.31,84.16) -- (277.02,124.2) ;
    \draw    (277.02,124.2) -- (277.02,208.51) ;
    \draw    (406.04,93.63) -- (406.04,177.93) ;
    \draw    (406.04,93.63) -- (358.74,133.66) ;
    \draw    (277.02,124.2) -- (358.74,133.66) ;
    \draw    (406.04,177.93) -- (358.74,217.97) ;
    \draw    (277.02,208.51) -- (358.74,217.97) ;
    \draw    (358.74,133.66) -- (358.74,217.97) ;
    \draw  [color={rgb, 255:red, 0; green, 0; blue, 0 }  ,draw opacity=1 ][fill={rgb, 255:red, 204; green, 203; blue, 203 }  ,fill opacity=0.53 ] (324.31,84.16) -- (406.04,93.63) -- (358.74,133.66) -- (277.02,124.2) -- cycle ;
    \draw  [color={rgb, 255:red, 0; green, 0; blue, 0 }  ,draw opacity=1 ][fill={rgb, 255:red, 204; green, 203; blue, 203 }  ,fill opacity=0.53 ] (277.02,124.2) -- (358.74,133.66) -- (358.74,217.97) -- (277.02,208.51) -- cycle ;
    \draw [line width=1.5]    (341.53,45.21) -- (341.53,108.91) ;
    \draw [shift={(341.53,41.21)}, rotate = 90] [fill={rgb, 255:red, 0; green, 0; blue, 0 }  ][line width=0.08]  [draw opacity=0] (11.61,-5.58) -- (0,0) -- (11.61,5.58) -- cycle    ;
    \draw [line width=1.5]    (272.73,223.54) -- (322.88,181.09) ;
    \draw [shift={(269.68,226.13)}, rotate = 319.75] [fill={rgb, 255:red, 0; green, 0; blue, 0 }  ][line width=0.08]  [draw opacity=0] (11.61,-5.58) -- (0,0) -- (11.61,5.58) -- cycle    ;
    \draw [line width=1.5]    (382.39,155.8) -- (447.51,163.07) ;
    \draw [shift={(451.48,163.51)}, rotate = 186.37] [fill={rgb, 255:red, 0; green, 0; blue, 0 }  ][line width=0.08]  [draw opacity=0] (11.61,-5.58) -- (0,0) -- (11.61,5.58) -- cycle    ;
    \draw [line width=3]    (221,149.5) .. controls (173.71,166.87) and (158.09,176.79) .. (131,221.5) ;
    \draw [shift={(128,226.5)}, rotate = 300.62] [fill={rgb, 255:red, 0; green, 0; blue, 0 }  ][line width=0.08]  [draw opacity=0] (16.97,-8.15) -- (0,0) -- (16.97,8.15) -- cycle    ;
    \draw [line width=3]    (590.92,239.44) .. controls (585.12,188.6) and (559.54,158.21) .. (508.84,144.42) ;
    \draw [shift={(591.54,245.9)}, rotate = 265.47] [fill={rgb, 255:red, 0; green, 0; blue, 0 }  ][line width=0.08]  [draw opacity=0] (16.97,-8.15) -- (0,0) -- (16.97,8.15) -- cycle    ;
    \draw [line width=3]    (232.52,411.77) .. controls (311.01,460.61) and (415.03,450.41) .. (473,414.5) ;
    \draw [shift={(226.53,407.9)}, rotate = 33.92] [fill={rgb, 255:red, 0; green, 0; blue, 0 }  ][line width=0.08]  [draw opacity=0] (16.97,-8.15) -- (0,0) -- (16.97,8.15) -- cycle    ;
    \draw [color={rgb, 255:red, 0; green, 0; blue, 0 }  ,draw opacity=1 ][line width=0.75]  [dash pattern={on 0.84pt off 2.51pt}]  (53.23,377.24) -- (170.56,343.31) ;
    \draw [shift={(170.56,343.31)}, rotate = 343.87] [color={rgb, 255:red, 0; green, 0; blue, 0 }  ,draw opacity=1 ][fill={rgb, 255:red, 0; green, 0; blue, 0 }  ,fill opacity=1 ][line width=0.75]      (0, 0) circle [x radius= 3.35, y radius= 3.35]   ;
    \draw [shift={(53.23,377.24)}, rotate = 343.87] [color={rgb, 255:red, 0; green, 0; blue, 0 }  ,draw opacity=1 ][fill={rgb, 255:red, 0; green, 0; blue, 0 }  ,fill opacity=1 ][line width=0.75]      (0, 0) circle [x radius= 3.35, y radius= 3.35]   ;
    \draw    (53.23,272.02) -- (53.23,377.24) ;
    \draw [shift={(53.23,377.24)}, rotate = 90] [color={rgb, 255:red, 0; green, 0; blue, 0 }  ][fill={rgb, 255:red, 0; green, 0; blue, 0 }  ][line width=0.75]      (0, 0) circle [x radius= 3.35, y radius= 3.35]   ;
    \draw [shift={(53.23,272.02)}, rotate = 90] [color={rgb, 255:red, 0; green, 0; blue, 0 }  ][fill={rgb, 255:red, 0; green, 0; blue, 0 }  ][line width=0.75]      (0, 0) circle [x radius= 3.35, y radius= 3.35]   ;
    \draw  [color={rgb, 255:red, 255; green, 0; blue, 0 }  ,draw opacity=1 ][fill={rgb, 255:red, 255; green, 147; blue, 147 }  ,fill opacity=0.52 ][line width=0.75]  (525.54,294.8) -- (613.87,354.82) -- (594.37,420.53) -- cycle ;
    \draw [line width=0.75]    (594.37,420.53) -- (525.68,380.32) ;
    \draw [shift={(525.68,380.32)}, rotate = 210.34] [color={rgb, 255:red, 0; green, 0; blue, 0 }  ][fill={rgb, 255:red, 0; green, 0; blue, 0 }  ][line width=0.75]      (0, 0) circle [x radius= 3.35, y radius= 3.35]   ;
    \draw [shift={(594.37,420.53)}, rotate = 210.34] [color={rgb, 255:red, 0; green, 0; blue, 0 }  ][fill={rgb, 255:red, 0; green, 0; blue, 0 }  ][line width=0.75]      (0, 0) circle [x radius= 3.35, y radius= 3.35]   ;
    \draw [color={rgb, 255:red, 0; green, 0; blue, 0 }  ,draw opacity=1 ][line width=0.75]    (525.54,294.8) -- (525.68,380.32) ;
    \draw [shift={(525.68,380.32)}, rotate = 89.91] [color={rgb, 255:red, 0; green, 0; blue, 0 }  ,draw opacity=1 ][fill={rgb, 255:red, 0; green, 0; blue, 0 }  ,fill opacity=1 ][line width=0.75]      (0, 0) circle [x radius= 3.35, y radius= 3.35]   ;
    \draw [shift={(525.54,294.8)}, rotate = 89.91] [color={rgb, 255:red, 0; green, 0; blue, 0 }  ,draw opacity=1 ][fill={rgb, 255:red, 0; green, 0; blue, 0 }  ,fill opacity=1 ][line width=0.75]      (0, 0) circle [x radius= 3.35, y radius= 3.35]   ;
    \draw [line width=1.5]    (654.12,343.18) -- (613.87,354.82) ;
    \draw [shift={(657.97,342.06)}, rotate = 163.87] [fill={rgb, 255:red, 0; green, 0; blue, 0 }  ][line width=0.08]  [draw opacity=0] (11.61,-5.58) -- (0,0) -- (11.61,5.58) -- cycle    ;
    \draw [line width=0.75]    (657.97,342.06) -- (613.87,354.82) ;
    \draw [shift={(613.87,354.82)}, rotate = 163.87] [color={rgb, 255:red, 0; green, 0; blue, 0 }  ][fill={rgb, 255:red, 0; green, 0; blue, 0 }  ][line width=0.75]      (0, 0) circle [x radius= 3.35, y radius= 3.35]   ;
    \draw  [dash pattern={on 0.84pt off 2.51pt}]  (324.31,84.16) -- (324.31,168.47) ;

    \draw (154.65,158.12) node  [font=\huge]  {$\Lambda $};
    \draw (600,161.38) node  [font=\huge]  {$\Psi $};
    \draw (349.34,471.12) node  [font=\huge]  {$\Phi_{face} =\Lambda \circ  \Psi^{-1} $};
    \draw (143.52,268.83) node  [font=\huge]  {$\Omega _{e}$};
    \draw (171.31,445.36) node  [color={rgb, 255:red, 0; green, 0; blue, 0}  ,opacity=1 ,font=\huge]  {$x_{1}$};
    \draw (192.13,336.44) node  [color={rgb, 255:red, 0; green, 0; blue, 0 }  ,opacity=1  ,font=\huge]  {$x_{3}$};
    \draw (36.58,254.93) node  [color={rgb, 255:red, 0; green, 0; blue, 0 }  ,opacity=1  ,font=\huge]  {$x_{4}$};
    \draw (30.64,381.9) node  [color={rgb, 255:red, 0; green, 0; blue, 0 }  ,opacity=1  ,font=\huge]  {$x_{2}$};
    \draw (648.24,447.95) node  [font=\huge]  {$\xi $};
    \draw (525.2,233.97) node  [font=\huge]  {$\zeta $};
    \draw (673.47,333.99) node  [font=\huge]  {$\eta $};
    \draw (614.28,285.95) node  [font=\huge]  {$\Omega_{\mathit{ref}}$};
    \draw (614.19,410.72) node  [color={rgb, 255:red, 0; green, 0; blue, 0 }  ,opacity=1 ,font=\Large ]  {$1$};
    \draw (507.75,390.72) node  [color={rgb, 255:red, 0; green, 0; blue, 0 }  ,opacity=1  ,font=\Large]  {$2$};
    \draw (614.19,333.44) node  [color={rgb, 255:red, 0; green, 0; blue, 0 }  ,opacity=1 ,font=\Large ]  {$3$};
    \draw (507.46,300.57) node  [color={rgb, 255:red, 0; green, 0; blue, 0 }  ,opacity=1 ,font=\Large ]  {$4$};
    \draw (257.15,234.42) node  [font=\huge]  {$\lambda $};
    \draw (342.55,22.16) node  [font=\huge]  {$\sigma $};
    \draw (470.57,163.7) node  [font=\huge]  {$\vartheta $};
    \draw (399.44,221.37) node  [font=\huge]  {$H$};

\end{tikzpicture}

%% file: figures/mapping-edge.tikz
\tikzset{every picture/.style={line width=0.75pt}} 

\begin{tikzpicture}[x=0.75pt,y=0.75pt,yscale=-1,xscale=1]

    \draw  [draw opacity=0][fill={rgb, 255:red, 204; green, 203; blue, 203 }  ,fill opacity=0.6 ][line width=0.75]  (61.23,268.67) .. controls (179.02,339.86) and (62.02,269.19) .. (178.56,339.95) .. controls (195.69,380.19) and (152.69,374.19) .. (158.74,429.06) .. controls (61.02,269.19) and (158.33,428.5) .. (61.23,268.67) -- cycle ;
    \draw  [draw opacity=0][fill={rgb, 255:red, 204; green, 203; blue, 203 }  ,fill opacity=0.58 ][line width=0.75]  (61.23,373.88) .. controls (136.2,351.6) and (62,374) .. (178.56,339.95) .. controls (195.69,380.19) and (152.69,374.19) .. (158.74,429.06) .. controls (61,374) and (101,395.6) .. (61.23,373.88) -- cycle ;
    \draw [color={rgb, 255:red, 255; green, 0; blue, 0 }  ,draw opacity=1 ][line width=1.5]    (158.74,429.06) .. controls (151.67,376.5) and (195.33,380.17) .. (178.56,339.95) ;
    \draw  [draw opacity=0][fill={rgb, 255:red, 204; green, 203; blue, 203 }  ,fill opacity=0.58 ][line width=0.75]  (613.87,354.82) -- (594.37,420.53) -- (525.68,380.32) -- cycle ;
    \draw [color={rgb, 255:red, 0; green, 0; blue, 0 }  ,draw opacity=1 ][line width=0.75]  [dash pattern={on 0.84pt off 2.51pt}]  (525.68,380.32) -- (613.87,354.82) ;
    \draw [line width=0.75]    (525.54,294.8) -- (594.37,420.53) ;
    \draw [shift={(594.37,420.53)}, rotate = 61.3] [color={rgb, 255:red, 0; green, 0; blue, 0 }  ][fill={rgb, 255:red, 0; green, 0; blue, 0 }  ][line width=0.75]      (0, 0) circle [x radius= 3.35, y radius= 3.35]   ;
    \draw [shift={(525.54,294.8)}, rotate = 61.3] [color={rgb, 255:red, 0; green, 0; blue, 0 }  ][fill={rgb, 255:red, 0; green, 0; blue, 0 }  ][line width=0.75]      (0, 0) circle [x radius= 3.35, y radius= 3.35]   ;
    \draw [line width=1.5]    (525.66,261.05) -- (525.54,294.8) ;
    \draw [shift={(525.68,257.05)}, rotate = 90.21] [fill={rgb, 255:red, 0; green, 0; blue, 0 }  ][line width=0.08]  [draw opacity=0] (11.61,-5.58) -- (0,0) -- (11.61,5.58) -- cycle    ;
    \draw [line width=1.5]    (654.12,343.18) -- (613.87,354.82) ;
    \draw [shift={(657.97,342.06)}, rotate = 163.87] [fill={rgb, 255:red, 0; green, 0; blue, 0 }  ][line width=0.08]  [draw opacity=0] (11.61,-5.58) -- (0,0) -- (11.61,5.58) -- cycle    ;
    \draw  [color={rgb, 255:red, 0; green, 0; blue, 0 }  ,draw opacity=1 ][fill={rgb, 255:red, 204; green, 203; blue, 203 }  ,fill opacity=0.52 ][line width=0.75]  (525.54,294.8) -- (613.87,354.82) -- (594.37,420.53) -- cycle ;
    \draw [line width=1.5]    (594.37,420.53) -- (629.6,440.5) ;
    \draw [shift={(633.08,442.48)}, rotate = 209.55] [fill={rgb, 255:red, 0; green, 0; blue, 0 }  ][line width=0.08]  [draw opacity=0] (11.61,-5.58) -- (0,0) -- (11.61,5.58) -- cycle    ;
    \draw  [draw opacity=0][fill={rgb, 255:red, 204; green, 203; blue, 203 }  ,fill opacity=0.59 ][line width=0.75]  (525.54,294.8) -- (594.37,420.53) -- (525.68,380.32) -- cycle ;
    \draw [color={rgb, 255:red, 0; green, 0; blue, 0 }  ,draw opacity=1 ][line width=0.75]    (525.54,294.8) -- (525.68,380.32) ;
    \draw [shift={(525.68,380.32)}, rotate = 89.91] [color={rgb, 255:red, 0; green, 0; blue, 0 }  ,draw opacity=1 ][fill={rgb, 255:red, 0; green, 0; blue, 0 }  ,fill opacity=1 ][line width=0.75]      (0, 0) circle [x radius= 3.35, y radius= 3.35]   ;
    \draw    (324.31,84.16) -- (406.04,93.63) ;
    \draw    (324.31,84.16) -- (277.02,124.2) ;
    \draw    (406.04,93.63) -- (358.74,133.66) ;
    \draw    (277.02,124.2) -- (358.74,133.66) ;
    \draw  [color={rgb, 255:red, 0; green, 0; blue, 0 }  ,draw opacity=1 ][fill={rgb, 255:red, 204; green, 203; blue, 203 }  ,fill opacity=0.53 ] (324.31,84.16) -- (406.04,93.63) -- (358.74,133.66) -- (277.02,124.2) -- cycle ;
    \draw [line width=1.5]    (341.53,45.21) -- (341.53,108.91) ;
    \draw [shift={(341.53,41.21)}, rotate = 90] [fill={rgb, 255:red, 0; green, 0; blue, 0 }  ][line width=0.08]  [draw opacity=0] (11.61,-5.58) -- (0,0) -- (11.61,5.58) -- cycle    ;
    \draw [line width=1.5]    (382.39,155.8) -- (447.51,163.07) ;
    \draw [shift={(451.48,163.51)}, rotate = 186.37] [fill={rgb, 255:red, 0; green, 0; blue, 0 }  ][line width=0.08]  [draw opacity=0] (11.61,-5.58) -- (0,0) -- (11.61,5.58) -- cycle    ;
    \draw [line width=3]    (221,149.5) .. controls (173.71,166.87) and (158.09,176.79) .. (131,221.5) ;
    \draw [shift={(128,226.5)}, rotate = 300.62] [fill={rgb, 255:red, 0; green, 0; blue, 0 }  ][line width=0.08]  [draw opacity=0] (16.97,-8.15) -- (0,0) -- (16.97,8.15) -- cycle    ;
    \draw [line width=3]    (590.92,239.44) .. controls (585.12,188.6) and (559.54,158.21) .. (508.84,144.42) ;
    \draw [shift={(591.54,245.9)}, rotate = 265.47] [fill={rgb, 255:red, 0; green, 0; blue, 0 }  ][line width=0.08]  [draw opacity=0] (16.97,-8.15) -- (0,0) -- (16.97,8.15) -- cycle    ;
    \draw [line width=3]    (232.52,411.77) .. controls (311.01,460.61) and (415.03,450.41) .. (473,414.5) ;
    \draw [shift={(226.53,407.9)}, rotate = 33.92] [fill={rgb, 255:red, 0; green, 0; blue, 0 }  ][line width=0.08]  [draw opacity=0] (16.97,-8.15) -- (0,0) -- (16.97,8.15) -- cycle    ;
    \draw  [draw opacity=0][fill={rgb, 255:red, 204; green, 203; blue, 203 }  ,fill opacity=0.59 ] (61.23,268.67) .. controls (157.69,427.86) and (118.36,363.19) .. (158.74,429.06) .. controls (134.02,414.94) and (69.86,379.01) .. (61.23,373.88) .. controls (61.55,365.54) and (60.26,277.01) .. (61.23,268.67) -- cycle ;
    \draw    (61.23,268.67) -- (61.23,373.88) ;
    \draw [shift={(61.23,373.88)}, rotate = 90] [color={rgb, 255:red, 0; green, 0; blue, 0 }  ][fill={rgb, 255:red, 0; green, 0; blue, 0 }  ][line width=0.75]      (0, 0) circle [x radius= 3.35, y radius= 3.35]   ;
    \draw [shift={(61.23,268.67)}, rotate = 90] [color={rgb, 255:red, 0; green, 0; blue, 0 }  ][fill={rgb, 255:red, 0; green, 0; blue, 0 }  ][line width=0.75]      (0, 0) circle [x radius= 3.35, y radius= 3.35]   ;
    \draw [color={rgb, 255:red, 0; green, 0; blue, 0 }  ,draw opacity=1 ][line width=0.75]  [dash pattern={on 0.84pt off 2.51pt}]  (61.23,373.88) -- (178.56,339.95) ;
    \draw [shift={(178.56,339.95)}, rotate = 343.87] [color={rgb, 255:red, 0; green, 0; blue, 0 }  ,draw opacity=1 ][fill={rgb, 255:red, 0; green, 0; blue, 0 }  ,fill opacity=1 ][line width=0.75]      (0, 0) circle [x radius= 3.35, y radius= 3.35]   ;
    \draw  [color={rgb, 255:red, 0; green, 0; blue, 0 }  ,draw opacity=1 ][fill={rgb, 255:red, 204; green, 203; blue, 203 }  ,fill opacity=0.58 ] (324.31,168.47) -- (406.04,177.93) -- (358.74,217.97) -- (277.02,208.51) -- cycle ;
    \draw  [color={rgb, 255:red, 0; green, 0; blue, 0 }  ,draw opacity=1 ][fill={rgb, 255:red, 204; green, 203; blue, 203 }  ,fill opacity=0.53 ] (406.04,93.63) -- (406.04,177.93) -- (358.74,217.97) -- (358.74,133.66) -- cycle ;
    \draw  [color={rgb, 255:red, 0; green, 0; blue, 0 }  ,draw opacity=1 ][fill={rgb, 255:red, 204; green, 203; blue, 203 }  ,fill opacity=0.53 ] (277.02,124.2) -- (358.74,133.66) -- (358.74,217.97) -- (277.02,208.51) -- cycle ;
    \draw  [dash pattern={on 0.84pt off 2.51pt}]  (324.31,84.16) -- (324.31,168.47) ;
    \draw [line width=1.5]    (272.73,223.54) -- (322.88,181.09) ;
    \draw [shift={(269.68,226.13)}, rotate = 319.75] [fill={rgb, 255:red, 0; green, 0; blue, 0 }  ][line width=0.08]  [draw opacity=0] (11.61,-5.58) -- (0,0) -- (11.61,5.58) -- cycle    ;
    \draw    (61.23,268.67) -- (158.74,429.06) ;
    \draw    (61.23,268.67) -- (178.56,339.95) ;
    \draw [shift={(178.56,339.95)}, rotate = 31.28] [color={rgb, 255:red, 0; green, 0; blue, 0 }  ][fill={rgb, 255:red, 0; green, 0; blue, 0 }  ][line width=0.75]      (0, 0) circle [x radius= 3.35, y radius= 3.35]   ;
    \draw [shift={(61.23,268.67)}, rotate = 31.28] [color={rgb, 255:red, 0; green, 0; blue, 0 }  ][fill={rgb, 255:red, 0; green, 0; blue, 0 }  ][line width=0.75]      (0, 0) circle [x radius= 3.35, y radius= 3.35]   ;
    \draw [line width=0.75]    (61.23,373.88) -- (158.74,429.06) ;
    \draw [shift={(158.74,429.06)}, rotate = 29.5] [color={rgb, 255:red, 0; green, 0; blue, 0 }  ][fill={rgb, 255:red, 0; green, 0; blue, 0 }  ][line width=0.75]      (0, 0) circle [x radius= 3.35, y radius= 3.35]   ;
    \draw [color={rgb, 255:red, 255; green, 0; blue, 0 }  ,draw opacity=1 ][line width=1.5]    (613.87,354.82) -- (595.43,420.53) ;
    \draw [line width=0.75]    (594.37,420.53) -- (525.68,380.32) ;
    \draw [shift={(525.68,380.32)}, rotate = 210.34] [color={rgb, 255:red, 0; green, 0; blue, 0 }  ][fill={rgb, 255:red, 0; green, 0; blue, 0 }  ][line width=0.75]      (0, 0) circle [x radius= 3.35, y radius= 3.35]   ;
    \draw [shift={(594.37,420.53)}, rotate = 210.34] [color={rgb, 255:red, 0; green, 0; blue, 0 }  ][fill={rgb, 255:red, 0; green, 0; blue, 0 }  ][line width=0.75]      (0, 0) circle [x radius= 3.35, y radius= 3.35]   ;
    \draw [line width=0.75]    (613.87,354.82) -- (657.97,342.06) ;
    \draw [shift={(613.87,354.82)}, rotate = 343.87] [color={rgb, 255:red, 0; green, 0; blue, 0 }  ][fill={rgb, 255:red, 0; green, 0; blue, 0 }  ][line width=0.75]      (0, 0) circle [x radius= 3.35, y radius= 3.35]   ;

    \draw (154.65,158.12) node  [font=\huge]  {$\Lambda $};
    \draw (600,161.38) node  [font=\huge]  {$\Psi $};
    \draw (349.34,471.12) node  [font=\huge]  {$\Phi_{edge} =\Lambda \circ \Psi^{-1}$};
    \draw (648.24,447.95) node  [font=\huge]  {$\xi $};
    \draw (525.2,233.97) node  [font=\huge]  {$\zeta $};
    \draw (673.47,333.99) node  [font=\huge]  {$\eta $};
    \draw (614.28,285.95) node  [font=\huge]  {$\Omega_{\mathit{ref}}$};
    \draw (614.19,410.72) node  [color={rgb, 255:red, 0; green, 0; blue, 0 }  ,opacity=1,font=\Large ]  {$1$};
    \draw (511.75,393.72) node  [color={rgb, 255:red, 0; green, 0; blue, 0 }  ,opacity=1,font=\Large ]  {$2$};
    \draw (614.19,333.44) node  [color={rgb, 255:red, 0; green, 0; blue, 0 }  ,opacity=1,font=\Large ]  {$3$};
    \draw (507.46,300.57) node  [color={rgb, 255:red, 0; green, 0; blue, 0 }  ,opacity=1,font=\Large ]  {$4$};
    \draw (253.15,232.42) node  [font=\huge]  {$\lambda $};
    \draw (342.55,22.16) node  [font=\huge]  {$\sigma $};
    \draw (470.57,163.7) node  [font=\huge]  {$\vartheta $};
    \draw (399.44,221.37) node  [font=\huge]  {$H$};
    \draw (151.52,265.48) node  [font=\huge]  {$\Omega^{e}$};
    \draw (184.31,422) node  [color={rgb, 255:red, 0; green, 0; blue, 0 }  ,opacity=1 ,font=\huge]  {$x_{1}$};
    \draw (200.13,333.08) node  [color={rgb, 255:red, 0; green, 0; blue, 0 }  ,opacity=1,font=\huge ]  {$x_{3}$};
    \draw (44.58,251.57) node  [color={rgb, 255:red, 0; green, 0; blue, 0 }  ,opacity=1,font=\huge ]  {$x_{4}$};
    \draw (35.64,378.54) node  [color={rgb, 255:red, 0; green, 0; blue, 0 }  ,opacity=1,font=\huge ]  {$x_{2}$};

\end{tikzpicture}

%% file: figures/space-time-slab.tikz
\tikzset{every picture/.style={line width=0.75pt}} 

\begin{tikzpicture}[x=0.75pt,y=0.75pt,yscale=-1,xscale=1]

\draw [line width=1.5]    (200.02,194.92) -- (228,165.54) ;
\draw [line width=1.5]    (228,165.54) -- (235.55,204.43) ;
\draw [line width=1.5]    (228,165.54) -- (229.53,76.75) ;
\draw [color={rgb, 255:red, 0; green, 0; blue, 0 }  ,draw opacity=1 ] [dash pattern={on 0.84pt off 2.51pt}]  (167.55,77.55) -- (158.4,166.44) ;
\draw [color={rgb, 255:red, 0; green, 0; blue, 0 }  ,draw opacity=1 ] [dash pattern={on 0.84pt off 2.51pt}]  (224.05,42.75) -- (220.72,128.06) ;
\draw [color={rgb, 255:red, 0; green, 0; blue, 0 }  ,draw opacity=1 ] [dash pattern={on 0.84pt off 2.51pt}]  (254.45,50.5) -- (255.97,136.17) ;
\draw [color={rgb, 255:red, 0; green, 0; blue, 0 }  ,draw opacity=1 ] [dash pattern={on 0.84pt off 2.51pt}]  (284.9,59.2) -- (290.15,145.9) ;
\draw [line width=2.25]    (99.23,122.39) -- (131.13,129.75) ;
\draw [shift={(136,130.88)}, rotate = 192.99] [fill={rgb, 255:red, 0; green, 0; blue, 0 }  ][line width=0.08]  [draw opacity=0] (16.07,-7.72) -- (0,0) -- (16.07,7.72) -- (10.67,0) -- cycle    ;
\draw  [fill={rgb, 255:red, 189; green, 189; blue, 189 }  ,fill opacity=0.5, rounded corners = 1][line width=1.5]  (332.33,137.29) .. controls (278.91,170.58) and (237.51,103.72) .. (187.55,137.29) .. controls (170.09,155.19) and (153.87,170.86) .. (129.25,195.59) .. controls (177.36,163.03) and (226.32,228.48) .. (274.03,195.59) .. controls (299.89,170.3) and (314.99,153.52) .. (332.33,137.29) -- cycle ;
\draw    (129.25,195.59) -- (187.55,137.29) ;
\draw    (274.03,195.59) -- (332.33,137.29) ;
\draw [color={rgb, 255:red, 0; green, 0; blue, 0 }  ,draw opacity=1 ]   (158.4,166.44) -- (167.01,185.68) ;
\draw [color={rgb, 255:red, 0; green, 0; blue, 0 }  ,draw opacity=1 ]   (167.01,185.68) -- (193.87,156.87) ;
\draw [color={rgb, 255:red, 0; green, 0; blue, 0 }  ,draw opacity=1 ]   (200.02,194.92) -- (193.87,156.87) ;
\draw [color={rgb, 255:red, 0; green, 0; blue, 0 }  ,draw opacity=1 ]   (200.02,194.92) -- (228,165.54) ;
\draw [color={rgb, 255:red, 0; green, 0; blue, 0 }  ,draw opacity=1 ]   (235.55,204.43) -- (228,165.54) ;
\draw [color={rgb, 255:red, 0; green, 0; blue, 0 }  ,draw opacity=1 ]   (235.55,204.43) -- (261,175.05) ;
\draw [color={rgb, 255:red, 0; green, 0; blue, 0 }  ,draw opacity=1 ]   (274.03,195.59) -- (261,175.05) ;
\draw [color={rgb, 255:red, 0; green, 0; blue, 0 }  ,draw opacity=1 ]   (193.87,156.87) -- (158.4,166.44) ;
\draw [color={rgb, 255:red, 0; green, 0; blue, 0 }  ,draw opacity=1 ]   (228,165.54) -- (193.87,156.87) ;
\draw [color={rgb, 255:red, 0; green, 0; blue, 0 }  ,draw opacity=1 ]   (261,175.05) -- (228,165.54) ;
\draw [color={rgb, 255:red, 0; green, 0; blue, 0 }  ,draw opacity=1 ]   (303.18,166.44) -- (261,175.05) ;
\draw [color={rgb, 255:red, 0; green, 0; blue, 0 }  ,draw opacity=1 ]   (187.55,137.29) -- (193.87,156.87) ;
\draw [color={rgb, 255:red, 0; green, 0; blue, 0 }  ,draw opacity=1 ]   (193.87,156.87) -- (220.72,128.06) ;
\draw [color={rgb, 255:red, 0; green, 0; blue, 0 }  ,draw opacity=1 ]   (228,165.54) -- (221.84,127.5) ;
\draw [color={rgb, 255:red, 0; green, 0; blue, 0 }  ,draw opacity=1 ]   (228,165.54) -- (255.97,136.17) ;
\draw [color={rgb, 255:red, 0; green, 0; blue, 0 }  ,draw opacity=1 ]   (261,175.05) -- (255.97,136.17) ;
\draw [color={rgb, 255:red, 0; green, 0; blue, 0 }  ,draw opacity=1 ]   (261,175.05) -- (290.15,145.9) ;
\draw [color={rgb, 255:red, 0; green, 0; blue, 0 }  ,draw opacity=1 ]   (303.18,166.44) -- (290.15,145.9) ;
\draw [line width=1.5]    (141.5,103.61) -- (129.25,195.59) ;
\draw [color={rgb, 255:red, 0; green, 0; blue, 0 }  ,draw opacity=1 ][fill={rgb, 255:red, 138; green, 138; blue, 138 }  ,fill opacity=1 ][line width=0.75]    (193.61,51.5) -- (187.55,137.29) ;
\draw [line width=1.5]    (270.89,103.61) -- (274.03,195.59) ;
\draw  [fill={rgb, 255:red, 189; green, 189; blue, 189 }  ,fill opacity=0.69, rounded corners = 2 ][line width=1.5]  (323,51.5) .. controls (275.25,81.25) and (238.25,21.5) .. (193.61,51.5) .. controls (178,67.5) and (163.5,81.5) .. (141.5,103.61) .. controls (184.5,74.5) and (228.25,133) .. (270.89,103.61) .. controls (294,81) and (307.5,66) .. (323,51.5) -- cycle ;
\draw  [fill={rgb, 255:red, 189; green, 189; blue, 189 }  ,fill opacity=0.89 ][line width=1.5]  (270.54,103.61) .. controls (224.2,132.6) and (185.6,73.8) .. (141.14,103.61) .. controls (137.39,132) and (133.64,166.75) .. (128.89,195.59) .. controls (177.01,163.03) and (225.96,228.48) .. (273.68,195.59) .. controls (272.64,165.5) and (271.39,137) .. (270.54,103.61) -- cycle ;
\draw [color={rgb, 255:red, 0; green, 0; blue, 0 }  ,draw opacity=1 ] [dash pattern={on 0.84pt off 2.51pt}]  (175.25,94.75) -- (167.01,185.68) ;
\draw [color={rgb, 255:red, 0; green, 0; blue, 0 }  ,draw opacity=1 ] [dash pattern={on 0.84pt off 2.51pt}]  (204.75,103) -- (200.02,194.92) ;
\draw [color={rgb, 255:red, 0; green, 0; blue, 0 }  ,draw opacity=1 ] [dash pattern={on 0.84pt off 2.51pt}]  (236.5,111.5) -- (235.55,204.43) ;
\draw [color={rgb, 255:red, 0; green, 0; blue, 0 }  ,draw opacity=1 ][fill={rgb, 255:red, 0; green, 0; blue, 0 }  ,fill opacity=1 ]   (167.55,77.55) -- (175.22,94.75) ;
\draw [color={rgb, 255:red, 0; green, 0; blue, 0 }  ,draw opacity=1 ][fill={rgb, 255:red, 0; green, 0; blue, 0 }  ,fill opacity=1 ]   (175.22,94.75) -- (199.14,69) ;
\draw [color={rgb, 255:red, 0; green, 0; blue, 0 }  ,draw opacity=1 ][fill={rgb, 255:red, 0; green, 0; blue, 0 }  ,fill opacity=1 ]   (204.62,103) -- (199.14,69) ;
\draw [color={rgb, 255:red, 0; green, 0; blue, 0 }  ,draw opacity=1 ][fill={rgb, 255:red, 0; green, 0; blue, 0 }  ,fill opacity=1 ]   (204.62,103) -- (229.53,76.75) ;
\draw [color={rgb, 255:red, 0; green, 0; blue, 0 }  ,draw opacity=1 ][fill={rgb, 255:red, 0; green, 0; blue, 0 }  ,fill opacity=1 ]   (236.26,111.5) -- (229.53,76.75) ;
\draw [color={rgb, 255:red, 0; green, 0; blue, 0 }  ,draw opacity=1 ][fill={rgb, 255:red, 0; green, 0; blue, 0 }  ,fill opacity=1 ]   (236.26,111.5) -- (258.93,85.25) ;
\draw [color={rgb, 255:red, 0; green, 0; blue, 0 }  ,draw opacity=1 ][fill={rgb, 255:red, 0; green, 0; blue, 0 }  ,fill opacity=1 ]   (270.54,103.61) -- (258.93,85.25) ;
\draw [color={rgb, 255:red, 0; green, 0; blue, 0 }  ,draw opacity=1 ][fill={rgb, 255:red, 0; green, 0; blue, 0 }  ,fill opacity=1 ]   (199.14,69) -- (167.55,77.55) ;
\draw [color={rgb, 255:red, 0; green, 0; blue, 0 }  ,draw opacity=1 ][fill={rgb, 255:red, 0; green, 0; blue, 0 }  ,fill opacity=1 ]   (229.53,76.75) -- (199.14,69) ;
\draw [color={rgb, 255:red, 0; green, 0; blue, 0 }  ,draw opacity=1 ][fill={rgb, 255:red, 0; green, 0; blue, 0 }  ,fill opacity=1 ]   (258.93,85.25) -- (229.53,76.75) ;
\draw [color={rgb, 255:red, 0; green, 0; blue, 0 }  ,draw opacity=1 ][fill={rgb, 255:red, 0; green, 0; blue, 0 }  ,fill opacity=1 ]   (295.2,78.8) -- (258.93,85.25) ;
\draw [color={rgb, 255:red, 0; green, 0; blue, 0 }  ,draw opacity=1 ][fill={rgb, 255:red, 0; green, 0; blue, 0 }  ,fill opacity=1 ]   (193.52,51.5) -- (199.14,69) ;
\draw [color={rgb, 255:red, 0; green, 0; blue, 0 }  ,draw opacity=1 ][fill={rgb, 255:red, 0; green, 0; blue, 0 }  ,fill opacity=1 ]   (199.14,69) -- (223.06,43.25) ;
\draw [color={rgb, 255:red, 0; green, 0; blue, 0 }  ,draw opacity=1 ][fill={rgb, 255:red, 0; green, 0; blue, 0 }  ,fill opacity=1 ]   (229.53,76.75) -- (224.05,42.75) ;
\draw [color={rgb, 255:red, 0; green, 0; blue, 0 }  ,draw opacity=1 ][fill={rgb, 255:red, 0; green, 0; blue, 0 }  ,fill opacity=1 ]   (229.53,76.75) -- (254.45,50.5) ;
\draw [color={rgb, 255:red, 0; green, 0; blue, 0 }  ,draw opacity=1 ][fill={rgb, 255:red, 0; green, 0; blue, 0 }  ,fill opacity=1 ]   (258.93,85.25) -- (254.45,50.5) ;
\draw [color={rgb, 255:red, 0; green, 0; blue, 0 }  ,draw opacity=1 ][fill={rgb, 255:red, 0; green, 0; blue, 0 }  ,fill opacity=1 ]   (258.93,85.25) -- (284.9,59.2) ;
\draw [color={rgb, 255:red, 0; green, 0; blue, 0 }  ,draw opacity=1 ][fill={rgb, 255:red, 0; green, 0; blue, 0 }  ,fill opacity=1 ]   (295.2,78.8) -- (284.9,59.2) ;
\draw  [fill={rgb, 255:red, 189; green, 189; blue, 189 }  ,fill opacity=0.89 ][line width=1.5]  (322.64,51.5) -- (331.98,137.29) -- (273.68,195.59) -- (270.54,103.61) -- cycle ;
\draw [color={rgb, 255:red, 0; green, 0; blue, 0 }  ,draw opacity=1 ] [dash pattern={on 0.84pt off 2.51pt}]  (295.2,78.8) -- (303.18,166.44) ;
\draw [line width=1.5]    (204.75,103) -- (229.53,76.75) ;
\draw [line width=1.5]    (236.26,111.5) -- (229.53,76.75) ;
\draw [line width=1.5]    (200.02,194.92) -- (204.75,103) ;
\draw [line width=1.5]    (235.55,204.43) -- (236.5,111.5) ;
\draw  [draw opacity=0][fill={rgb, 255:red, 95; green, 95; blue, 95 }  ,fill opacity=0.58 ] (229.53,76.75) -- (236.26,111.5) -- (235.55,204.43) -- (200.02,194.92) -- (204.62,103) -- cycle ;
\draw  [draw opacity=0][fill={rgb, 255:red, 5; green, 5; blue, 5 }  ,fill opacity=0.07 ] (228,165.54) -- (235.55,204.43) -- (215.67,200.67) -- (200.02,194.92) -- cycle ;
\draw [line width=2.25]    (280.02,45.44) .. controls (293.67,28.51) and (305.59,18.93) .. (340,18) ;
\draw [shift={(276.89,49.39)}, rotate = 307.92] [fill={rgb, 255:red, 0; green, 0; blue, 0 }  ][line width=0.08]  [draw opacity=0] (16.07,-7.72) -- (0,0) -- (16.07,7.72) -- (10.67,0) -- cycle    ;
\draw [line width=2.25]    (336.67,209.35) .. controls (318.57,209.03) and (307.79,201.73) .. (296.85,186.87) ;
\draw [shift={(293.99,182.82)}, rotate = 415.73] [fill={rgb, 255:red, 0; green, 0; blue, 0 }  ][line width=0.08]  [draw opacity=0] (16.07,-7.72) -- (0,0) -- (16.07,7.72) -- (10.67,0) -- cycle    ;
\draw [line width=2.25]    (218.46,181.75) .. controls (206.28,206.33) and (192.5,224.06) .. (169,225) ;
\draw [shift={(220.78,176.96)}, rotate = 115.3] [fill={rgb, 255:red, 0; green, 0; blue, 0 }  ][line width=0.08]  [draw opacity=0] (16.07,-7.72) -- (0,0) -- (16.07,7.72) -- (10.67,0) -- cycle    ;
\draw [line width=1.5]    (31.91,211.72) -- (60.17,183.46) ;
\draw [shift={(63,180.64)}, rotate = 495] [fill={rgb, 255:red, 0; green, 0; blue, 0 }  ][line width=0.08]  [draw opacity=0] (13.4,-6.43) -- (0,0) -- (13.4,6.44) -- (8.9,0) -- cycle    ;
\draw [line width=1.5]    (31.91,211.72) -- (72,211.72) ;
\draw [shift={(76,211.72)}, rotate = 180] [fill={rgb, 255:red, 0; green, 0; blue, 0 }  ][line width=0.08]  [draw opacity=0] (13.4,-6.43) -- (0,0) -- (13.4,6.44) -- (8.9,0) -- cycle    ;
\draw [line width=1.5]    (31.91,172) -- (31.91,211.72) ;
\draw [shift={(31.91,168)}, rotate = 90] [fill={rgb, 255:red, 0; green, 0; blue, 0 }  ][line width=0.08]  [draw opacity=0] (13.4,-6.43) -- (0,0) -- (13.4,6.44) -- (8.9,0) -- cycle    ;

\draw (79.16,117.74) node  [font=\Large]  {$P_{n}$};
\draw (361.57,208.56) node  [font=\Large]  {$\Omega _{n}$};
\draw (372.96,18.99) node  [font=\Large]  {$\Omega _{n+1}$};
\draw (142.42,223.68) node  [font=\Large]  {$Q^{e}_{n}$};
\draw (89.89,209) node  [font=\large]  {$x$};
\draw (71.89,165.37) node  [font=\large]  {$y$};
\draw (17.95,159.69) node  [font=\large]  {$t$};

\end{tikzpicture}

%% file: figures/space-time-nefem-element.tikz

\tikzset {_n93zw4t5p/.code = {\pgfsetadditionalshadetransform{ \pgftransformshift{\pgfpoint{3.5 bp } { -29.5 bp }  }  \pgftransformrotate{-354 }  \pgftransformscale{2 }  }}}
\pgfdeclarehorizontalshading{_zayis4hkp}{150bp}{rgb(0bp)=(0,0,0);
rgb(37.5bp)=(0,0,0);
rgb(49.55357142857143bp)=(1,1,1);
rgb(62.5bp)=(0,0,0);
rgb(100bp)=(0,0,0)}
\tikzset{_yqtter8rx/.code = {\pgfsetadditionalshadetransform{\pgftransformshift{\pgfpoint{3.5 bp } { -29.5 bp }  }  \pgftransformrotate{-354 }  \pgftransformscale{2 } }}}
\pgfdeclarehorizontalshading{_mqk2p35pf} {150bp} {color(0bp)=(transparent!50);
color(37.5bp)=(transparent!50);
color(49.55357142857143bp)=(transparent!83);
color(62.5bp)=(transparent!50);
color(100bp)=(transparent!50) }
\pgfdeclarefading{_zoamf0ymh}{\tikz \fill[shading=_mqk2p35pf,_yqtter8rx] (0,0) rectangle (50bp,50bp); }
\tikzset{every picture/.style={line width=0.75pt}} 

\begin{tikzpicture}[x=0.75pt,y=0.75pt,yscale=-1,xscale=1]

\draw [line width=2.25]    (555.03,381.61) -- (480.01,356.6) ;

\draw [shift={(558.82,382.87)}, rotate = 198.43] [fill={rgb, 255:red, 0; green, 0; blue, 0 }  ][line width=2.25]  [draw opacity=0] (14.29,-6.86) -- (0,0) -- (14.29,6.86) -- cycle    ;
\draw  [fill={rgb, 255:red, 88; green, 88; blue, 88 }  ,fill opacity=0.2 ][line width=0.75]  (211.47,194.6) .. controls (316.48,209.18) and (381.43,192.4) .. (427.47,131.84) .. controls (333.99,109.21) and (303.35,100.45) .. (257.04,89.51) .. controls (232.56,144.96) and (232.56,145.69) .. (211.47,194.6) -- cycle ;
\draw [color={rgb, 255:red, 0; green, 0; blue, 0 }  ,draw opacity=1 ][line width=1.5]  [dash pattern={on 1.69pt off 2.76pt}]  (257.04,282.51) -- (480.01,356.6) ;

\draw [line width=1.5]  [dash pattern={on 1.69pt off 2.76pt}]  (257.04,89.51) -- (257.04,282.51) ;
\draw [shift={(257.04,282.51)}, rotate = 90] [color={rgb, 255:red, 0; green, 0; blue, 0 }  ][fill={rgb, 255:red, 0; green, 0; blue, 0 }  ][line width=1.5]      (0, 0) circle [x radius= 4.36, y radius= 4.36]   ;
\draw [shift={(257.04,89.51)}, rotate = 90] [color={rgb, 255:red, 0; green, 0; blue, 0 }  ][fill={rgb, 255:red, 0; green, 0; blue, 0 }  ][line width=1.5]      (0, 0) circle [x radius= 4.36, y radius= 4.36]   ;
\path  [shading=_zayis4hkp,_n93zw4t5p,path fading= _zoamf0ymh ,fading transform={xshift=2}] (211.47,194.6) .. controls (313.63,212.11) and (388.06,185.84) .. (427.47,131.84) .. controls (470.54,316.76) and (443.1,197.08) .. (480.01,356.6) .. controls (423.25,426.51) and (352.03,470.29) .. (214.39,439.79) .. controls (213.08,339.52) and (213.08,277.64) .. (211.47,194.6) -- cycle ; 
 \draw  [line width=1.5]  (211.47,194.6) .. controls (313.63,212.11) and (388.06,185.84) .. (427.47,131.84) .. controls (470.54,316.76) and (443.1,197.08) .. (480.01,356.6) .. controls (423.25,426.51) and (352.03,470.29) .. (214.39,439.79) .. controls (213.08,339.52) and (213.08,277.64) .. (211.47,194.6) -- cycle ; 

\draw [line width=1.5]  [dash pattern={on 1.69pt off 2.76pt}]  (257.04,282.51) -- (214.39,439.79) ;
\draw [shift={(214.39,439.79)}, rotate = 105.17] [color={rgb, 255:red, 0; green, 0; blue, 0 }  ][fill={rgb, 255:red, 0; green, 0; blue, 0 }  ][line width=1.5]      (0, 0) circle [x radius= 4.36, y radius= 4.36]   ;

\draw    (427.47,131.84) -- (480.01,356.6) ;
\draw [shift={(480.01,356.6)}, rotate = 76.84] [color={rgb, 255:red, 0; green, 0; blue, 0 }  ][fill={rgb, 255:red, 0; green, 0; blue, 0 }  ][line width=0.75]      (0, 0) circle [x radius= 3.35, y radius= 3.35]   ;
\draw [shift={(427.47,131.84)}, rotate = 76.84] [color={rgb, 255:red, 0; green, 0; blue, 0 }  ][fill={rgb, 255:red, 0; green, 0; blue, 0 }  ][line width=0.75]      (0, 0) circle [x radius= 3.35, y radius= 3.35]   ;
\draw    (211.47,194.6) -- (214.39,439.79) ;
\draw [shift={(214.39,439.79)}, rotate = 89.32] [color={rgb, 255:red, 0; green, 0; blue, 0 }  ][fill={rgb, 255:red, 0; green, 0; blue, 0 }  ][line width=0.75]      (0, 0) circle [x radius= 3.35, y radius= 3.35]   ;
\draw [shift={(211.47,194.6)}, rotate = 89.32] [color={rgb, 255:red, 0; green, 0; blue, 0 }  ][fill={rgb, 255:red, 0; green, 0; blue, 0 }  ][line width=0.75]      (0, 0) circle [x radius= 3.35, y radius= 3.35]   ;
\draw [line width=2.25]    (257.04,29.3) -- (257.04,89.51) ;

\draw [shift={(257.04,25.3)}, rotate = 90] [fill={rgb, 255:red, 0; green, 0; blue, 0 }  ][line width=2.25]  [draw opacity=0] (14.29,-6.86) -- (0,0) -- (14.29,6.86) -- cycle    ;
\draw [line width=0.75]    (257.04,89.51) -- (427.47,131.84) ;
\draw [shift={(427.47,131.84)}, rotate = 13.95] [color={rgb, 255:red, 0; green, 0; blue, 0 }  ][fill={rgb, 255:red, 0; green, 0; blue, 0 }  ][line width=0.75]      (0, 0) circle [x radius= 3.35, y radius= 3.35]   ;

\draw [line width=0.75]    (257.04,89.51) -- (211.47,194.6) ;
\draw [shift={(211.47,194.6)}, rotate = 113.45] [color={rgb, 255:red, 0; green, 0; blue, 0 }  ][fill={rgb, 255:red, 0; green, 0; blue, 0 }  ][line width=0.75]      (0, 0) circle [x radius= 3.35, y radius= 3.35]   ;

\draw [color={rgb, 255:red, 0; green, 0; blue, 0 }  ,draw opacity=1 ][line width=2.25]    (194.11,514.57) -- (214.39,439.79) ;

\draw [shift={(193.06,518.43)}, rotate = 285.17] [fill={rgb, 255:red, 0; green, 0; blue, 0 }  ,fill opacity=1 ][line width=2.25]  [draw opacity=0] (14.29,-6.86) -- (0,0) -- (14.29,6.86) -- cycle    ;
\draw [color={rgb, 255:red, 208; green, 2; blue, 27 }  ,draw opacity=1 ][line width=2.25]    (211.47,194.6) .. controls (299.04,213.57) and (390.98,187.3) .. (427.47,131.84) ;
\draw [shift={(427.47,131.84)}, rotate = 303.34] [color={rgb, 255:red, 208; green, 2; blue, 27 }  ,draw opacity=1 ][fill={rgb, 255:red, 208; green, 2; blue, 27 }  ,fill opacity=1 ][line width=2.25]      (0, 0) circle [x radius= 5.36, y radius= 5.36]   ;
\draw [shift={(211.47,194.6)}, rotate = 12.23] [color={rgb, 255:red, 208; green, 2; blue, 27 }  ,draw opacity=1 ][fill={rgb, 255:red, 208; green, 2; blue, 27 }  ,fill opacity=1 ][line width=2.25]      (0, 0) circle [x radius= 5.36, y radius= 5.36]   ;
\draw [color={rgb, 255:red, 208; green, 2; blue, 27 }  ,draw opacity=1 ][line width=2.25]    (214.39,439.79) .. controls (353.04,470.44) and (428.93,423.74) .. (480.01,356.6) ;
\draw [shift={(480.01,356.6)}, rotate = 307.27] [color={rgb, 255:red, 208; green, 2; blue, 27 }  ,draw opacity=1 ][fill={rgb, 255:red, 208; green, 2; blue, 27 }  ,fill opacity=1 ][line width=2.25]      (0, 0) circle [x radius= 5.36, y radius= 5.36]   ;
\draw [shift={(214.39,439.79)}, rotate = 12.46] [color={rgb, 255:red, 208; green, 2; blue, 27 }  ,draw opacity=1 ][fill={rgb, 255:red, 208; green, 2; blue, 27 }  ,fill opacity=1 ][line width=2.25]      (0, 0) circle [x radius= 5.36, y radius= 5.36]   ;

\draw (223.35,514.34) node [scale=1.7280000000000002]  {\huge$x$};
\draw (574.08,407.26) node [scale=1.7280000000000002]  {\huge$y$};
\draw (232.19,15.63) node [scale=1.7280000000000002]  {\huge$t$};
\draw (495.38,131.16) node [scale=1.44,color={rgb, 255:red, 208; green, 2; blue, 27 }  ,opacity=1 ]  {\LARGE$\boldsymbol{x}_{u}\left(\theta _{4}\right)$};
\draw (296.48,68.73) node [scale=1.44,color={rgb, 255:red, 0; green, 0; blue, 0 }  ,opacity=1 ]  {\LARGE$t^{n+1}$};
\draw (533.32,328) node [scale=1.44,color={rgb, 255:red, 208; green, 2; blue, 27 }  ,opacity=1 ]  {\LARGE$\boldsymbol{x}_{l}\left(\theta _{2}\right)$};
\draw (281.62,258.92) node [scale=1.44,color={rgb, 255:red, 0; green, 0; blue, 0 }  ,opacity=1 ]  {\LARGE$t^{n}$};
\draw (150.38,195.16) node [scale=1.44,color={rgb, 255:red, 208; green, 2; blue, 27 }  ,opacity=1 ]  {\LARGE$\boldsymbol{x}_{u}\left(\theta _{3}\right)$};
\draw (154.38,440.16) node [scale=1.44,color={rgb, 255:red, 208; green, 2; blue, 27 }  ,opacity=1 ]  {\LARGE$\boldsymbol{x}_{l}\left(\theta _{1}\right)$};
\draw (368.38,473.16) node [scale=1.44,color={rgb, 255:red, 208; green, 2; blue, 27 }  ,opacity=1 ]  {\LARGE$\boldsymbol{C}_{l}\left(\theta _{l}\right)$};
\draw (365.38,218.16) node [scale=1.44,color={rgb, 255:red, 208; green, 2; blue, 27 }  ,opacity=1 ]  {\LARGE$\boldsymbol{C}_{u}\left(\theta _{u}\right)$};

\end{tikzpicture}

%% file: figures/strong-coupling-scheme-d-n.tikz.tex
\tikzset{every picture/.style={line width=0.75pt}} 

\begin{tikzpicture}[x=0.75pt,y=0.75pt,yscale=-0.6,xscale=0.6, every node/.style={scale=0.6}]

    \draw   (96,403) -- (158,435.19) -- (96,467.38) -- (34,435.19) -- cycle ;
    \draw   (8,285) -- (184,285) -- (184,380) -- (8,380) -- cycle ;
    \draw    (96,183.5) -- (96,282) ;
    \draw [shift={(96,285)}, rotate = 270] [fill={rgb, 255:red, 0; green, 0; blue, 0 }  ][line width=0.08]  [draw opacity=0] (12.5,-6.01) -- (0,0) -- (12.5,6.01) -- cycle    ;
    \draw    (96,380) -- (96,403) ;
    \draw    (158,435.19) -- (277,435.19) ;
    \draw    (96,49) -- (96,85.5) ;
    \draw [shift={(96,88.5)}, rotate = 270] [fill={rgb, 255:red, 0; green, 0; blue, 0 }  ][line width=0.08]  [draw opacity=0] (12.5,-6.01) -- (0,0) -- (12.5,6.01) -- cycle    ;
    \draw    (96,467.38) -- (96,495) ;
    \draw [shift={(96,498)}, rotate = 270] [fill={rgb, 255:red, 0; green, 0; blue, 0 }  ][line width=0.08]  [draw opacity=0] (12.5,-6.01) -- (0,0) -- (12.5,6.01) -- cycle    ;
    \draw   (40,29) .. controls (40,17.95) and (48.95,9) .. (60,9) -- (132,9) .. controls (143.05,9) and (152,17.95) .. (152,29) -- (152,29) .. controls (152,40.05) and (143.05,49) .. (132,49) -- (60,49) .. controls (48.95,49) and (40,40.05) .. (40,29) -- cycle ;
    \draw   (33,518) .. controls (33,506.95) and (41.95,498) .. (53,498) -- (139,498) .. controls (150.05,498) and (159,506.95) .. (159,518) -- (159,518) .. controls (159,529.05) and (150.05,538) .. (139,538) -- (53,538) .. controls (41.95,538) and (33,529.05) .. (33,518) -- cycle ;
    \draw    (277,136) -- (277,238.1) ;
    \draw    (187,136) -- (277,136) ;
    \draw [shift={(184,136)}, rotate = 0] [fill={rgb, 255:red, 0; green, 0; blue, 0 }  ][line width=0.08]  [draw opacity=0] (12.5,-6.01) -- (0,0) -- (12.5,6.01) -- cycle    ;
    \draw    (277,333.1) -- (277,435.19) ;
    \draw   (215.5,238.1) -- (338.5,238.1) -- (338.5,333.1) -- (215.5,333.1) -- cycle ;
    \draw   (8,88.5) -- (184,88.5) -- (184,183.5) -- (8,183.5) -- cycle ;

    \draw (96,435.19) node   [align=left] {Converged?};
    \draw (96,29) node   [align=left] {\begin{minipage}[lt]{63.45250000000001pt}\setlength\topsep{0pt}
            \begin{center}
                Solution at $\displaystyle t_{n}$
            \end{center}

    \end{minipage}};
    \draw (96,518) node   [align=left] {\begin{minipage}[lt]{72.89804000000001pt}\setlength\topsep{0pt}
            \begin{center}
                Solution at $\displaystyle t_{n+1}$
            \end{center}

    \end{minipage}};
    \draw (96,136) node   [align=left] {\begin{minipage}[lt]{93.43608pt}\setlength\topsep{0pt}
            \begin{center}
                Solve fluid on $\displaystyle \Omega ^{f}_{k}$:\\$\displaystyle \uvec^f_{k} ,p^f_{k}$ $\displaystyle =f\left( \Omega^f_{k}\right)$
            \end{center}

    \end{minipage}};
    \draw (109,61.4) node [anchor=north west][inner sep=0.75pt]  [font=\small]  {$k=1$};
    \draw (212,115.4) node [anchor=north west][inner sep=0.75pt]  [font=\small]  {$k=k+1$};
    \draw (96,332.5) node   [align=left] {\begin{minipage}[lt]{100.76pt}\setlength\topsep{0pt}
            \begin{center}
                Solve structure:\\$\displaystyle \dvec^s_{k+1} ,\uvec^{s}_{k+1} =f\left( \tvec^f_{k}\right)$
            \end{center}

    \end{minipage}};
    \draw (103,234.25) node [anchor=west] [inner sep=0.75pt]   [align=left] {Tractions $\displaystyle \tvec^f_{k}$};
    \draw (174,421.19) node   [align=left] {no};
    \draw (121,476.19) node   [align=left] {yes};
    \draw (277,285.6) node   [align=left] {\begin{minipage}[lt]{75.26784pt}\setlength\topsep{0pt}
            Update mesh
            \begin{center}
                $\displaystyle \Omega ^{f}_{k+1} =f( \dvec^s_{k+1})$
            \end{center}

    \end{minipage}};
    \draw (277,443.19) node [anchor=north] [inner sep=0.75pt]   [align=left] {Interface deformation\\and velocity $\displaystyle \dvec^s_{k+1} ,\uvec^{s}_{k+1}$};

\end{tikzpicture}

%% file: figures/strong-coupling-scheme-r-n.tikz.tex
\tikzset{every picture/.style={line width=0.75pt}} 

\definecolor{red}{rgb}{0.0, 0.0, 0.0}

\begin{tikzpicture}[x=0.75pt,y=0.75pt,yscale=-0.6,xscale=0.6, every node/.style={scale=0.6}]

    \draw   (97,407) -- (159,439.19) -- (97,471.38) -- (35,439.19) -- cycle ;
    \draw   (9,289) -- (185,289) -- (185,384) -- (9,384) -- cycle ;
    \draw    (97,187.5) -- (97,286) ;
    \draw [shift={(97,289)}, rotate = 270] [fill={rgb, 255:red, 0; green, 0; blue, 0 }  ][line width=0.08]  [draw opacity=0] (12.5,-6.01) -- (0,0) -- (12.5,6.01) -- cycle    ;
    \draw    (97,384) -- (97,407) ;
    \draw    (160,439.19) -- (279,439.19) ;
    \draw    (97,53) -- (97,89.5) ;
    \draw [shift={(97,92.5)}, rotate = 270] [fill={rgb, 255:red, 0; green, 0; blue, 0 }  ][line width=0.08]  [draw opacity=0] (12.5,-6.01) -- (0,0) -- (12.5,6.01) -- cycle    ;
    \draw    (97,471.38) -- (97,499) ;
    \draw [shift={(97,502)}, rotate = 270] [fill={rgb, 255:red, 0; green, 0; blue, 0 }  ][line width=0.08]  [draw opacity=0] (12.5,-6.01) -- (0,0) -- (12.5,6.01) -- cycle    ;
    \draw   (41,33) .. controls (41,21.95) and (49.95,13) .. (61,13) -- (133,13) .. controls (144.05,13) and (153,21.95) .. (153,33) -- (153,33) .. controls (153,44.05) and (144.05,53) .. (133,53) -- (61,53) .. controls (49.95,53) and (41,44.05) .. (41,33) -- cycle ;
    \draw   (34,522) .. controls (34,510.95) and (42.95,502) .. (54,502) -- (140,502) .. controls (151.05,502) and (160,510.95) .. (160,522) -- (160,522) .. controls (160,533.05) and (151.05,542) .. (140,542) -- (54,542) .. controls (42.95,542) and (34,533.05) .. (34,522) -- cycle ;
    \draw    (278,140) -- (278,242.1) ;
    \draw    (188,140) -- (278,140) ;
    \draw [shift={(185,140)}, rotate = 0] [fill={rgb, 255:red, 0; green, 0; blue, 0 }  ][line width=0.08]  [draw opacity=0] (12.5,-6.01) -- (0,0) -- (12.5,6.01) -- cycle    ;
    \draw    (279,337.1) -- (279,439.19) ;
    \draw   (216.5,242.1) -- (339.5,242.1) -- (339.5,337.1) -- (216.5,337.1) -- cycle ;
    \draw   (9,92.5) -- (185,92.5) -- (185,187.5) -- (9,187.5) -- cycle ;

    \draw (97,439.19) node   [align=left] {Converged?};
    \draw (97,33) node   [align=left] {\begin{minipage}[lt]{63.45250000000001pt}\setlength\topsep{0pt}
            \begin{center}
                Solution at $\displaystyle t_{n}$
            \end{center}

    \end{minipage}};
    \draw (97,522) node   [align=left] {\begin{minipage}[lt]{72.89804000000001pt}\setlength\topsep{0pt}
            \begin{center}
                Solution at $\displaystyle t_{n+1}$
            \end{center}

    \end{minipage}};
    \draw (97,140) node   [align=left] {\begin{minipage}[lt]{113.36892pt}\setlength\topsep{0pt}
            \begin{center}
                Solve fluid on $\displaystyle \Omega ^{f}_{k}$:\\$\displaystyle \uvec^f_{k} ,p^f_{k}$ $\displaystyle =f\left(\color{red} \alpha ^{f}, \color{red}\bT^{s}_{k}\color{black}, \color{black}\Omega^{f}_{k} \right)$
            \end{center}

    \end{minipage}};
    \draw (110,65.4) node [anchor=north west][inner sep=0.75pt]  [font=\small]  {$k=1$};
    \draw (213,119.4) node [anchor=north west][inner sep=0.75pt]  [font=\small]  {$k=k+1$};
    \draw (97,336.5) node   [align=left] {\begin{minipage}[lt]{140.42304000000001pt}\setlength\topsep{0pt}
            \begin{center}
                Solve structure:\\$\displaystyle \dvec^s_{k+1} ,\uvec^{s}_{k+1,} \color{red}\bT ^{s}_{k+1} \color{black} =f\left( \tvec^f_{k}\right)$
            \end{center}

    \end{minipage}};
    \draw (104,238.25) node [anchor=west] [inner sep=0.75pt]   [align=left] {Tractions $\displaystyle \tvec^f_{k}$};
    \draw (175,425.19) node   [align=left] {no};
    \draw (122,480.19) node   [align=left] {yes};
    \draw (278,289.6) node   [align=left] {\begin{minipage}[lt]{75.26784pt}\setlength\topsep{0pt}
            \begin{center}
                Update mesh\\$\displaystyle \Omega ^{f}_{k+1} =f( \dvec^s_{k+1})$
            \end{center}

    \end{minipage}};
    \draw (279,450.19) node [anchor=north] [inner sep=0.75pt]   [align=left] {\begin{minipage}[lt]{100.81pt}\setlength\topsep{0pt}
            \begin{center}
                Interface deformation,\\velocity, \color{red} and tractions \color{black}\\ $\displaystyle \dvec^s_{k+1} ,\uvec^{s}_{k+1} , \color{red}\bT ^{s}_{k+1}$
            \end{center}

    \end{minipage}};

\end{tikzpicture}

%% file: figures/3d-bump-domain.tikz
\tikzset{every picture/.style={line width=0.75pt}} 

\begin{tikzpicture}[x=0.75pt,y=0.75pt,yscale=-1,xscale=1]

\draw  [fill={rgb, 255:red, 0; green, 0; blue, 0 }  ,fill opacity=0.15 ][line width=0.75]  (155.76,60.57) .. controls (195.43,59.91) and (454.69,59.91) .. (491.51,59.99) .. controls (493.47,109.68) and (491.15,163.5) .. (491.51,189.21) .. controls (459.32,188.97) and (403.19,188.97) .. (379.01,188.75) .. controls (329.11,144.98) and (290.34,169.29) .. (265.72,188.75) .. controls (231.31,190.7) and (186.75,188.97) .. (155.76,189.79) .. controls (157.24,146.14) and (154.92,98.11) .. (155.76,60.57) -- cycle ;
\draw  [fill={rgb, 255:red, 0; green, 0; blue, 0 }  ,fill opacity=0.22 ][line width=0.75]  (323.91,157.14) .. controls (371.94,156.56) and (384.67,157.14) .. (420.3,157.25) .. controls (446.16,168.71) and (460.92,175.95) .. (491.51,189.21) .. controls (447.03,188.97) and (421.85,188.97) .. (378.61,189.21) .. controls (366.15,172.76) and (334.32,164.66) .. (323.91,157.14) -- cycle ;
\draw  [fill={rgb, 255:red, 0; green, 0; blue, 0 }  ,fill opacity=0.22 ][line width=0.75]  (194.51,157.25) -- (265.72,189.21) -- (155.76,189.79) -- (84.55,157.83) -- cycle ;
\draw [color={rgb, 255:red, 177; green, 177; blue, 177 }  ,draw opacity=1 ][line width=0.75]    (194.51,157.25) .. controls (215.35,132.94) and (254.43,126.34) .. (278.41,137.65) ;

\draw [color={rgb, 255:red, 177; green, 177; blue, 177 }  ,draw opacity=1 ][line width=0.75]    (278.41,137.65) -- (352.65,170.8) ;

\draw [color={rgb, 255:red, 177; green, 177; blue, 177 }  ,draw opacity=1 ][line width=0.75]    (307.4,157.25) -- (323.07,164.4) ;

\draw [color={rgb, 255:red, 177; green, 177; blue, 177 }  ,draw opacity=1 ][line width=0.75]    (278.41,137.65) .. controls (291.87,143.85) and (296.74,146.91) .. (307.4,157.25) ;

\draw [line width=0.75]    (420.3,28.03) -- (491.51,59.99) ;

\draw [line width=0.75]    (491.51,59.99) -- (491.51,189.21) ;

\draw [line width=0.75]    (323.07,164.4) -- (378.61,189.21) ;

\draw [line width=0.75]    (194.51,157.25) -- (265.72,189.21) ;

\draw [line width=0.75]    (265.72,189.21) .. controls (286.56,164.89) and (335.94,147.62) .. (378.61,189.21) ;

\draw [color={rgb, 255:red, 0; green, 0; blue, 0 }  ,draw opacity=1 ][line width=0.75]  [dash pattern={on 4.5pt off 4.5pt}]  (307.8,156.79) -- (251.15,114.46) ;
\draw [shift={(251.15,114.46)}, rotate = 216.77] [color={rgb, 255:red, 0; green, 0; blue, 0 }  ,draw opacity=1 ][fill={rgb, 255:red, 0; green, 0; blue, 0 }  ,fill opacity=1 ][line width=0.75]      (0, 0) circle [x radius= 3.35, y radius= 3.35]   ;
\draw [shift={(307.8,156.79)}, rotate = 216.77] [color={rgb, 255:red, 0; green, 0; blue, 0 }  ,draw opacity=1 ][fill={rgb, 255:red, 0; green, 0; blue, 0 }  ,fill opacity=1 ][line width=0.75];

\draw [color={rgb, 255:red, 0; green, 0; blue, 0 }  ,draw opacity=1 ][line width=0.75]  [dash pattern={on 4.5pt off 4.5pt}]   (379.01,188.75) -- (307.8,156.79);
\draw  [fill={rgb, 255:red, 255; green, 0; blue, 0 }  ,fill opacity=1 ][line width=0.75]  (304.61,156.79) .. controls (304.61,155.03) and (306.04,153.61) .. (307.8,153.61) .. controls (309.55,153.61) and (310.98,155.03) .. (310.98,156.79) .. controls (310.98,158.55) and (309.55,159.97) .. (307.8,159.97) .. controls (306.04,159.97) and (304.61,158.55) .. (304.61,156.79) -- cycle ;

\draw  [fill={rgb, 255:red, 0; green, 0; blue, 0 }  ,fill opacity=0.22 ][line width=0.75]  (155.76,60.57) -- (155.76,189.79) -- (84.55,157.83) -- (84.55,28.61) -- cycle ;
\draw  [color={rgb, 255:red, 74; green, 144; blue, 226 }  ,draw opacity=1 ][fill={rgb, 255:red, 74; green, 144; blue, 226 }  ,fill opacity=0.41 ][line width=0.75]  (194.51,157.25) .. controls (227.16,124.6) and (265.37,125.29) .. (307.05,150.65) .. controls (338.32,164.89) and (320.6,156.21) .. (348.39,169.06) .. controls (355.68,172.19) and (365.06,174.27) .. (378.61,189.21) .. controls (369.23,184.87) and (334.5,169.41) .. (323.07,164.4) .. controls (296.28,161.77) and (273.88,181.22) .. (265.72,189.21) .. controls (256.08,184.95) and (208.4,163.5) .. (194.51,157.25) -- cycle ;
\draw [line width=0.75]    (77.92,171.33) -- (149.13,203.28) ;
\draw [shift={(149.13,203.28)}, rotate = 204.17000000000002] [color={rgb, 255:red, 0; green, 0; blue, 0 }  ][line width=0.75]    (0,5.59) -- (0,-5.59)(12.93,-4.9) .. controls (8.95,-2.3) and (5.31,-0.67) .. (2,0) .. controls (5.31,0.67) and (8.95,2.3) .. (12.93,4.9)   ;
\draw [shift={(77.92,171.33)}, rotate = 24.17] [color={rgb, 255:red, 0; green, 0; blue, 0 }  ][line width=0.75]    (0,5.59) -- (0,-5.59)(12.93,-4.9) .. controls (8.95,-2.3) and (5.31,-0.67) .. (2,0) .. controls (5.31,0.67) and (8.95,2.3) .. (12.93,4.9)   ;
\draw [line width=0.75]    (67.77,29.77) -- (67.77,158.99) ;
\draw [shift={(67.77,158.99)}, rotate = 270] [color={rgb, 255:red, 0; green, 0; blue, 0 }  ][line width=0.75]    (0,5.59) -- (0,-5.59)(12.93,-4.9) .. controls (8.95,-2.3) and (5.31,-0.67) .. (2,0) .. controls (5.31,0.67) and (8.95,2.3) .. (12.93,4.9)   ;
\draw [shift={(67.77,29.77)}, rotate = 90] [color={rgb, 255:red, 0; green, 0; blue, 0 }  ][line width=0.75]    (0,5.59) -- (0,-5.59)(12.93,-4.9) .. controls (8.95,-2.3) and (5.31,-0.67) .. (2,0) .. controls (5.31,0.67) and (8.95,2.3) .. (12.93,4.9)   ;
\draw [line width=0.75]    (267.21,212.64) -- (380.5,212.63) ;
\draw [shift={(380.5,212.63)}, rotate = 540] [color={rgb, 255:red, 0; green, 0; blue, 0 }  ][line width=0.75]    (0,5.59) -- (0,-5.59)(12.93,-4.9) .. controls (8.95,-2.3) and (5.31,-0.67) .. (2,0) .. controls (5.31,0.67) and (8.95,2.3) .. (12.93,4.9)   ;
\draw [shift={(267.21,212.64)}, rotate = 360] [color={rgb, 255:red, 0; green, 0; blue, 0 }  ][line width=0.75]    (0,5.59) -- (0,-5.59)(12.93,-4.9) .. controls (8.95,-2.3) and (5.31,-0.67) .. (2,0) .. controls (5.31,0.67) and (8.95,2.3) .. (12.93,4.9)   ;
\draw [line width=0.75]    (378.61,189.21) -- (491.51,189.21) ;

\draw [line width=0.75]    (420.3,28.03) -- (420.3,157.25) ;

\draw [color={rgb, 255:red, 0; green, 0; blue, 0 }  ,draw opacity=0.24 ][line width=0.75]    (307.4,157.25) -- (420.3,157.25) ;

\draw [line width=0.75]    (420.3,157.25) -- (491.51,189.21) ;

\draw [line width=0.75]    (84.55,157.83) -- (194.51,157.25) ;

\draw [line width=0.75]    (152.82,189.21) -- (265.72,189.21) ;

\draw  [fill={rgb, 255:red, 0; green, 0; blue, 0 }  ,fill opacity=0.22 ][line width=0.75]  (420.3,28.03) -- (491.51,59.99) -- (155.76,60.57) -- (84.55,28.61) -- cycle ;
\draw  [fill={rgb, 255:red, 0; green, 0; blue, 0 }  ,fill opacity=0.14 ][line width=0.75]  (420.3,28.03) -- (491.51,59.99) -- (491.51,189.21) -- (420.3,157.25) -- cycle ;
\draw [color={rgb, 255:red, 0; green, 0; blue, 0 }  ,draw opacity=1 ][line width=0.75]  [dash pattern={on 4.5pt off 4.5pt}]  (322.36,146.42) -- (251.15,114.46) ;
\draw [shift={(251.15,114.46)}, rotate = 204.17] [color={rgb, 255:red, 0; green, 0; blue, 0 }  ,draw opacity=1 ][fill={rgb, 255:red, 0; green, 0; blue, 0 }  ,fill opacity=1 ][line width=0.75]      (0, 0) circle [x radius= 3.35, y radius= 3.35]   ;
\draw [shift={(322.36,146.42)}, rotate = 204.17] [color={rgb, 255:red, 0; green, 0; blue, 0 }  ,draw opacity=1 ][fill={rgb, 255:red, 0; green, 0; blue, 0 }  ,fill opacity=1 ][line width=0.75]      (0, 0) circle [x radius= 3.35, y radius= 3.35]   ;
\draw [color={rgb, 255:red, 0; green, 0; blue, 0 }  ,draw opacity=1 ][line width=0.75]  [dash pattern={on 4.5pt off 4.5pt}]  (230.11,172.77) -- (286.76,130.44) ;
\draw [shift={(286.76,130.44)}, rotate = 323.23] [color={rgb, 255:red, 0; green, 0; blue, 0 }  ,draw opacity=1 ][fill={rgb, 255:red, 0; green, 0; blue, 0 }  ,fill opacity=1 ][line width=0.75]      (0, 0) circle [x radius= 3.35, y radius= 3.35]   ;
\draw [shift={(230.11,172.77)}, rotate = 323.23] [color={rgb, 255:red, 0; green, 0; blue, 0 }  ,draw opacity=1 ][fill={rgb, 255:red, 0; green, 0; blue, 0 }  ,fill opacity=1 ][line width=0.75]      (0, 0) circle [x radius= 3.35, y radius= 3.35]   ;
\draw [color={rgb, 255:red, 0; green, 0; blue, 0 }  ,draw opacity=1 ][line width=0.75]  [dash pattern={on 4.5pt off 4.5pt}]  (379.01,188.75) -- (322.36,146.42) ;
\draw [shift={(322.36,146.42)}, rotate = 216.77] [color={rgb, 255:red, 0; green, 0; blue, 0 }  ,draw opacity=1 ][fill={rgb, 255:red, 0; green, 0; blue, 0 }  ,fill opacity=1 ][line width=0.75]      (0, 0) circle [x radius= 3.35, y radius= 3.35]   ;
\draw [shift={(379.01,188.75)}, rotate = 216.77] [color={rgb, 255:red, 0; green, 0; blue, 0 }  ,draw opacity=1 ][fill={rgb, 255:red, 0; green, 0; blue, 0 }  ,fill opacity=1 ][line width=0.75]      (0, 0) circle [x radius= 3.35, y radius= 3.35]   ;
\draw [color={rgb, 255:red, 0; green, 0; blue, 0 }  ,draw opacity=1 ][line width=0.75]  [dash pattern={on 4.5pt off 4.5pt}]  (194.51,156.79) -- (251.15,114.46) ;
\draw [shift={(251.15,114.46)}, rotate = 323.23] [color={rgb, 255:red, 0; green, 0; blue, 0 }  ,draw opacity=1 ][fill={rgb, 255:red, 0; green, 0; blue, 0 }  ,fill opacity=1 ][line width=0.75]      (0, 0) circle [x radius= 3.35, y radius= 3.35]   ;
\draw [shift={(194.51,156.79)}, rotate = 323.23] [color={rgb, 255:red, 0; green, 0; blue, 0 }  ,draw opacity=1 ][fill={rgb, 255:red, 0; green, 0; blue, 0 }  ,fill opacity=1 ][line width=0.75]      (0, 0) circle [x radius= 3.35, y radius= 3.35]   ;

\draw [color={rgb, 255:red, 0; green, 0; blue, 0 }  ,draw opacity=1 ][line width=0.75]  [dash pattern={on 4.5pt off 4.5pt}]  (265.72,188.75) -- (194.51,156.79) ;

\draw [color={rgb, 255:red, 0; green, 0; blue, 0 }  ,draw opacity=1 ][line width=0.75]  [dash pattern={on 4.5pt off 4.5pt}]  (265.72,188.75) -- (322.36,146.42) ;
\draw [shift={(322.36,146.42)}, rotate = 323.23] [color={rgb, 255:red, 0; green, 0; blue, 0 }  ,draw opacity=1 ][fill={rgb, 255:red, 0; green, 0; blue, 0 }  ,fill opacity=1 ][line width=0.75]      (0, 0) circle [x radius= 3.35, y radius= 3.35]   ;
\draw [shift={(265.72,188.75)}, rotate = 323.23] [color={rgb, 255:red, 0; green, 0; blue, 0 }  ,draw opacity=1 ][fill={rgb, 255:red, 0; green, 0; blue, 0 }  ,fill opacity=1 ][line width=0.75]      (0, 0) circle [x radius= 3.35, y radius= 3.35]   ;
\draw [color={rgb, 255:red, 0; green, 0; blue, 0 }  ,draw opacity=1 ][line width=0.75]  [dash pattern={on 4.5pt off 4.5pt}]  (343.4,172.77) -- (286.76,130.44) ;
\draw [shift={(286.76,130.44)}, rotate = 216.77] [color={rgb, 255:red, 0; green, 0; blue, 0 }  ,draw opacity=1 ][fill={rgb, 255:red, 0; green, 0; blue, 0 }  ,fill opacity=1 ][line width=0.75]      (0, 0) circle [x radius= 3.35, y radius= 3.35]   ;
\draw [shift={(343.4,172.77)}, rotate = 216.77] [color={rgb, 255:red, 0; green, 0; blue, 0 }  ,draw opacity=1 ][fill={rgb, 255:red, 0; green, 0; blue, 0 }  ,fill opacity=1 ][line width=0.75]      (0, 0) circle [x radius= 3.35, y radius= 3.35]   ;

\draw  [fill={rgb, 255:red, 255; green, 0; blue, 0 }  ,fill opacity=1 ][line width=0.75]  (340.22,172.77) .. controls (340.22,171.01) and (341.64,169.59) .. (343.4,169.59) .. controls (345.16,169.59) and (346.58,171.01) .. (346.58,172.77) .. controls (346.58,174.53) and (345.16,175.95) .. (343.4,175.95) .. controls (341.64,175.95) and (340.22,174.53) .. (340.22,172.77) -- cycle ;
\draw  [fill={rgb, 255:red, 255; green, 0; blue, 0 }  ,fill opacity=1 ][line width=0.75]  (375.82,188.75) .. controls (375.82,186.99) and (377.25,185.56) .. (379.01,185.56) .. controls (380.76,185.56) and (382.19,186.99) .. (382.19,188.75) .. controls (382.19,190.5) and (380.76,191.93) .. (379.01,191.93) .. controls (377.25,191.93) and (375.82,190.5) .. (375.82,188.75) -- cycle ;
\draw  [fill={rgb, 255:red, 255; green, 0; blue, 0 }  ,fill opacity=1 ][line width=0.75]  (319.18,146.42) .. controls (319.18,144.66) and (320.6,143.24) .. (322.36,143.24) .. controls (324.12,143.24) and (325.54,144.66) .. (325.54,146.42) .. controls (325.54,148.18) and (324.12,149.6) .. (322.36,149.6) .. controls (320.6,149.6) and (319.18,148.18) .. (319.18,146.42) -- cycle ;
\draw  [fill={rgb, 255:red, 255; green, 0; blue, 0 }  ,fill opacity=1 ][line width=0.75]  (283.57,130.44) .. controls (283.57,128.69) and (285,127.26) .. (286.76,127.26) .. controls (288.51,127.26) and (289.94,128.69) .. (289.94,130.44) .. controls (289.94,132.2) and (288.51,133.63) .. (286.76,133.63) .. controls (285,133.63) and (283.57,132.2) .. (283.57,130.44) -- cycle ;
\draw  [fill={rgb, 255:red, 255; green, 0; blue, 0 }  ,fill opacity=1 ][line width=0.75]  (247.97,114.46) .. controls (247.97,112.71) and (249.39,111.28) .. (251.15,111.28) .. controls (252.91,111.28) and (254.33,112.71) .. (254.33,114.46) .. controls (254.33,116.22) and (252.91,117.65) .. (251.15,117.65) .. controls (249.39,117.65) and (247.97,116.22) .. (247.97,114.46) -- cycle ;
\draw  [fill={rgb, 255:red, 255; green, 0; blue, 0 }  ,fill opacity=1 ][line width=0.75]  (191.32,156.79) .. controls (191.32,155.04) and (192.75,153.61) .. (194.51,153.61) .. controls (196.26,153.61) and (197.69,155.04) .. (197.69,156.79) .. controls (197.69,158.55) and (196.26,159.98) .. (194.51,159.98) .. controls (192.75,159.98) and (191.32,158.55) .. (191.32,156.79) -- cycle ;
\draw  [fill={rgb, 255:red, 255; green, 0; blue, 0 }  ,fill opacity=1 ][line width=0.75]  (226.93,172.77) .. controls (226.93,171.01) and (228.35,169.59) .. (230.11,169.59) .. controls (231.87,169.59) and (233.29,171.01) .. (233.29,172.77) .. controls (233.29,174.53) and (231.87,175.95) .. (230.11,175.95) .. controls (228.35,175.95) and (226.93,174.53) .. (226.93,172.77) -- cycle ;
\draw  [fill={rgb, 255:red, 255; green, 0; blue, 0 }  ,fill opacity=1 ][line width=0.75]  (262.53,188.75) .. controls (262.53,186.99) and (263.96,185.57) .. (265.72,185.57) .. controls (267.47,185.57) and (268.9,186.99) .. (268.9,188.75) .. controls (268.9,190.51) and (267.47,191.93) .. (265.72,191.93) .. controls (263.96,191.93) and (262.53,190.51) .. (262.53,188.75) -- cycle ;

\draw [line width=0.75]    (157.26,213.67) -- (267.21,212.64) ;
\draw [shift={(267.21,212.64)}, rotate = 539.46] [color={rgb, 255:red, 0; green, 0; blue, 0 }  ][line width=0.75]    (0,5.59) -- (0,-5.59)(12.93,-4.9) .. controls (8.95,-2.3) and (5.31,-0.67) .. (2,0) .. controls (5.31,0.67) and (8.95,2.3) .. (12.93,4.9)   ;
\draw [shift={(157.26,213.67)}, rotate = 359.46] [color={rgb, 255:red, 0; green, 0; blue, 0 }  ][line width=0.75]    (0,5.59) -- (0,-5.59)(12.93,-4.9) .. controls (8.95,-2.3) and (5.31,-0.67) .. (2,0) .. controls (5.31,0.67) and (8.95,2.3) .. (12.93,4.9)   ;
\draw [line width=0.75]    (380.5,212.63) -- (493,213.1) ;
\draw [shift={(493,213.1)}, rotate = 180.24] [color={rgb, 255:red, 0; green, 0; blue, 0 }  ][line width=0.75]    (0,5.59) -- (0,-5.59)(12.93,-4.9) .. controls (8.95,-2.3) and (5.31,-0.67) .. (2,0) .. controls (5.31,0.67) and (8.95,2.3) .. (12.93,4.9)   ;
\draw [shift={(380.5,212.63)}, rotate = 0.24] [color={rgb, 255:red, 0; green, 0; blue, 0 }  ][line width=0.75]    (0,5.59) -- (0,-5.59)(12.93,-4.9) .. controls (8.95,-2.3) and (5.31,-0.67) .. (2,0) .. controls (5.31,0.67) and (8.95,2.3) .. (12.93,4.9)   ;
\draw [line width=0.75]    (31.29,225.4) -- (62.88,225.4) ;
\draw [shift={(64.88,225.4)}, rotate = 180] [fill={rgb, 255:red, 0; green, 0; blue, 0 }  ][line width=0.75]  [draw opacity=0] (8.93,-4.29) -- (0,0) -- (8.93,4.29) -- cycle    ;

\draw [line width=0.75]    (64.24,241.65) -- (31.29,225.4) ;

\draw [shift={(66.03,242.53)}, rotate = 206.26] [fill={rgb, 255:red, 0; green, 0; blue, 0 }  ][line width=0.75]  [draw opacity=0] (8.93,-4.29) -- (0,0) -- (8.93,4.29) -- cycle    ;
\draw [line width=0.75]    (31.29,225.4) -- (31.29,198.24) ;
\draw [shift={(31.29,196.24)}, rotate = 450] [fill={rgb, 255:red, 0; green, 0; blue, 0 }  ][line width=0.75]  [draw opacity=0] (8.93,-4.29) -- (0,0) -- (8.93,4.29) -- cycle    ;

\draw (326.09,225.94) node [scale=1] [align=left] {$L$};
\draw (55.23,95.43) node [scale=1] [align=left] {$L$};
\draw (213.35,227.52) node [scale=1] [align=left] {$L$};
\draw (435.78,227.68) node [scale=1] [align=left] {$L$};
\draw (106.62,200.97) node [scale=1] [align=left] {$L$};
\draw (70.67,213.44) node [scale=1.2]  {$x$};
\draw (19.05,195.55) node [scale=1.2]  {$y$};
\draw (53.14,252.06) node [scale=1.2]  {$z$};

\end{tikzpicture}

%% file: figures/3d-cylinder-domain.tikz
\tikzset{every picture/.style={line width=0.75pt}} 

\begin{tikzpicture}[x=0.75pt,y=0.75pt,yscale=-1,xscale=1]

\draw    (196.66,171.68) -- (196.66,192.2) ;

\draw    (245.06,171.68) -- (245.06,192.2) ;

\draw [color={rgb, 255:red, 0; green, 0; blue, 0 }  ,draw opacity=0.38 ][fill={rgb, 255:red, 0; green, 0; blue, 0 }  ,fill opacity=0.38 ][line width=0.75]  [dash pattern={on 4.5pt off 4.5pt}]  (192.8,124.4) -- (215.48,134.12) ;

\draw  [fill={rgb, 255:red, 0; green, 0; blue, 0 }  ,fill opacity=0.22 ][line width=0.75]  (455.27,212.26) -- (509.3,236.5) -- (124.33,236.94) -- (70.31,212.7) -- cycle ;
\draw  [fill={rgb, 255:red, 0; green, 0; blue, 0 }  ,fill opacity=0.22 ][line width=0.75]  (124.33,58.24) -- (124.33,236.94) -- (70.31,212.7) -- (70.31,34) -- cycle ;
\draw [line width=0.75]    (462.12,27.43) -- (513.58,51.25) ;
\draw [shift={(513.58,51.25)}, rotate = 204.84] [color={rgb, 255:red, 0; green, 0; blue, 0 }  ][line width=0.75]    (0,5.59) -- (0,-5.59)(10.93,-4.9) .. controls (6.95,-2.3) and (3.31,-0.67) .. (2,0) .. controls (3.31,0.67) and (6.95,2.3) .. (10.93,4.9)   ;
\draw [shift={(462.12,27.43)}, rotate = 24.84] [color={rgb, 255:red, 0; green, 0; blue, 0 }  ][line width=0.75]    (0,5.59) -- (0,-5.59)(10.93,-3.29) .. controls (6.95,-1.4) and (3.31,-0.3) .. (2,0) .. controls (3.31,0.3) and (6.95,1.4) .. (10.93,3.29)   ;
\draw  [fill={rgb, 255:red, 0; green, 0; blue, 0 }  ,fill opacity=0.22 ][line width=0.75]  (455.27,33.56) -- (509.3,57.8) -- (124.33,58.24) -- (70.31,34) -- cycle ;
\draw  [fill={rgb, 255:red, 0; green, 0; blue, 0 }  ,fill opacity=0.14 ][line width=0.75]  (455.27,33.56) -- (509.3,57.8) -- (509.3,236.5) -- (455.27,212.26) -- cycle ;
\draw [line width=0.75]    (125.55,248.91) -- (221.23,248.91) ;
\draw [shift={(221.23,248.91)}, rotate = 180] [color={rgb, 255:red, 0; green, 0; blue, 0 }  ][line width=0.75]    (0,5.59) -- (0,-5.59)(10.93,-4.9) .. controls (6.95,-2.3) and (3.31,-0.67) .. (2,0) .. controls (3.31,0.67) and (6.95,2.3) .. (10.93,4.9)   ;
\draw [shift={(125.55,248.91)}, rotate = 0] [color={rgb, 255:red, 0; green, 0; blue, 0 }  ][line width=0.75]    (0,5.59) -- (0,-5.59)(10.93,-3.29) .. controls (6.95,-1.4) and (3.31,-0.3) .. (2,0) .. controls (3.31,0.3) and (6.95,1.4) .. (10.93,3.29)   ;
\draw [line width=0.75]    (35.71,260.61) -- (58.85,260.61) ;
\draw [shift={(60.85,260.61)}, rotate = 180] [fill={rgb, 255:red, 0; green, 0; blue, 0 }  ][line width=0.75]  [draw opacity=0] (8.93,-4.29) -- (0,0) -- (8.93,4.29) -- cycle    ;

\draw [line width=0.75]    (59.92,272.55) -- (35.71,260.61) ;

\draw [shift={(61.71,273.44)}, rotate = 206.26] [fill={rgb, 255:red, 0; green, 0; blue, 0 }  ][line width=0.75]  [draw opacity=0] (8.93,-4.29) -- (0,0) -- (8.93,4.29) -- cycle    ;
\draw [line width=0.75]    (35.71,260.61) -- (35.71,240.78) ;
\draw [shift={(35.71,238.78)}, rotate = 450] [fill={rgb, 255:red, 0; green, 0; blue, 0 }  ][line width=0.75]  [draw opacity=0] (8.93,-4.29) -- (0,0) -- (8.93,4.29) -- cycle    ;

\draw [color={rgb, 255:red, 138; green, 138; blue, 138 }  ,draw opacity=1 ]   (156.13,145.12) -- (211.71,170.05) ;

\draw   (196.37,147.48) .. controls (196.37,134.12) and (207.21,123.28) .. (220.57,123.28) .. controls (233.94,123.28) and (244.77,134.12) .. (244.77,147.48) .. controls (244.77,160.85) and (233.94,171.68) .. (220.57,171.68) .. controls (207.21,171.68) and (196.37,160.85) .. (196.37,147.48) -- cycle ;
\draw [color={rgb, 255:red, 138; green, 138; blue, 138 }  ,draw opacity=1 ]   (175.19,100.65) -- (230.77,125.58) ;

\draw  [draw opacity=0][fill={rgb, 255:red, 74; green, 144; blue, 226 }  ,fill opacity=0.65 ] (146.8,109.11) .. controls (152.04,102.6) and (161.1,95.93) .. (175.19,100.65) .. controls (190.99,107.24) and (209.22,116.26) .. (230.77,125.58) .. controls (220.02,120.55) and (205.89,123.72) .. (199.22,136.11) .. controls (193.82,145.01) and (194.93,162.48) .. (211.71,170.05) .. controls (189.69,159.94) and (172.72,153.14) .. (156.13,145.12) .. controls (139.66,137.07) and (139.82,118.12) .. (146.8,109.11) -- cycle ;
\draw  [draw opacity=0][fill={rgb, 255:red, 74; green, 144; blue, 226 }  ,fill opacity=0.33 ] (230.77,125.58) .. controls (219.97,121.01) and (207.11,122.82) .. (199.01,136.73) .. controls (192.24,147.88) and (197.77,163.89) .. (211.71,170.05) .. controls (225.59,175.23) and (235.41,168.75) .. (241.3,159.62) .. controls (246.7,151.84) and (248.27,135.39) .. (230.77,125.58) -- cycle ;
\draw  [draw opacity=0] (158.09,145.96) .. controls (148.83,142.58) and (142.21,133.68) .. (142.21,123.24) .. controls (142.21,109.88) and (153.02,99.06) .. (166.36,99.06) .. controls (169.28,99.06) and (172.09,99.58) .. (174.68,100.53) -- (166.36,123.24) -- cycle ; \draw  [color={rgb, 255:red, 138; green, 138; blue, 138 }  ,draw opacity=1 ] (158.09,145.96) .. controls (148.83,142.58) and (142.21,133.68) .. (142.21,123.24) .. controls (142.21,109.88) and (153.02,99.06) .. (166.36,99.06) .. controls (169.28,99.06) and (172.09,99.58) .. (174.68,100.53) ;
\draw [color={rgb, 255:red, 0; green, 0; blue, 0 }  ,draw opacity=1 ][line width=0.75]  [dash pattern={on 4.5pt off 4.5pt}]  (142.45,147.44) -- (196.66,171.68) ;

\draw [color={rgb, 255:red, 0; green, 0; blue, 0 }  ,draw opacity=1 ][line width=0.75]  [dash pattern={on 4.5pt off 4.5pt}]  (142.45,147.44) -- (142.45,123.24) ;

\draw [color={rgb, 255:red, 0; green, 0; blue, 0 }  ,draw opacity=1 ][line width=0.75]  [dash pattern={on 4.5pt off 4.5pt}]  (142.45,99.04) -- (196.66,123.28) ;

\draw [color={rgb, 255:red, 0; green, 0; blue, 0 }  ,draw opacity=0.38 ][line width=0.75]  [dash pattern={on 4.5pt off 4.5pt}]  (234.92,167.09) -- (220.05,160.61) ;

\draw [color={rgb, 255:red, 0; green, 0; blue, 0 }  ,draw opacity=0.38 ][line width=0.75]  [dash pattern={on 4.5pt off 4.5pt}]  (166.65,147.44) -- (188.43,147.44) ;

\draw [color={rgb, 255:red, 0; green, 0; blue, 0 }  ,draw opacity=1 ][line width=0.75]  [dash pattern={on 4.5pt off 4.5pt}]  (142.45,99.04) -- (142.45,123.24) ;

\draw [color={rgb, 255:red, 0; green, 0; blue, 0 }  ,draw opacity=1 ][line width=0.75]  [dash pattern={on 4.5pt off 4.5pt}]  (190.85,99.04) -- (245.06,123.28) ;

\draw [color={rgb, 255:red, 0; green, 0; blue, 0 }  ,draw opacity=1 ][line width=0.75]  [dash pattern={on 4.5pt off 4.5pt}]  (142.45,147.44) -- (161.35,147.46) ;

\draw [color={rgb, 255:red, 0; green, 0; blue, 0 }  ,draw opacity=1 ][line width=0.75]  [dash pattern={on 4.5pt off 4.5pt}]  (166.65,99.04) -- (190.85,99.04) ;

\draw [color={rgb, 255:red, 0; green, 0; blue, 0 }  ,draw opacity=1 ][line width=0.75]  [dash pattern={on 4.5pt off 4.5pt}]  (142.45,99.04) -- (166.65,99.04) ;

\draw [color={rgb, 255:red, 0; green, 0; blue, 0 }  ,draw opacity=0.38 ][line width=0.75]  [dash pattern={on 4.5pt off 4.5pt}]  (190.85,145.02) -- (190.56,125.65) ;

\draw [color={rgb, 255:red, 0; green, 0; blue, 0 }  ,draw opacity=0.38 ][line width=0.75]  [dash pattern={on 4.5pt off 4.5pt}]  (190.75,107.29) -- (190.56,120.82) ;

\draw [color={rgb, 255:red, 0; green, 0; blue, 0 }  ,draw opacity=0.38 ][line width=0.75]  [dash pattern={on 4.5pt off 4.5pt}]  (196.17,159.56) -- (215.54,159.56) ;

\draw [color={rgb, 255:red, 0; green, 0; blue, 0 }  ,draw opacity=1 ][line width=0.75]  [dash pattern={on 4.5pt off 4.5pt}]  (169.55,111.16) -- (169.55,135.36) ;

\draw [color={rgb, 255:red, 0; green, 0; blue, 0 }  ,draw opacity=1 ][line width=0.75]  [dash pattern={on 4.5pt off 4.5pt}]  (169.55,159.56) -- (188.22,159.56) ;

\draw [color={rgb, 255:red, 0; green, 0; blue, 0 }  ,draw opacity=1 ][line width=0.75]  [dash pattern={on 4.5pt off 4.5pt}]  (193.75,111.16) -- (217.95,111.16) ;

\draw [color={rgb, 255:red, 0; green, 0; blue, 0 }  ,draw opacity=1 ][line width=0.75]  [dash pattern={on 4.5pt off 4.5pt}]  (169.55,111.16) -- (193.75,111.16) ;

\draw [color={rgb, 255:red, 0; green, 0; blue, 0 }  ,draw opacity=1 ][line width=0.75]  [dash pattern={on 4.5pt off 4.5pt}]  (217.96,119.45) -- (217.95,111.16) ;

\draw [color={rgb, 255:red, 0; green, 0; blue, 0 }  ,draw opacity=0.38 ][line width=0.75]  [dash pattern={on 4.5pt off 4.5pt}]  (217.96,119.45) -- (217.95,159.56) ;

\draw [color={rgb, 255:red, 0; green, 0; blue, 0 }  ,draw opacity=1 ][line width=0.75]  [dash pattern={on 4.5pt off 4.5pt}]  (220.86,171.68) -- (245.06,171.68) ;

\draw [color={rgb, 255:red, 0; green, 0; blue, 0 }  ,draw opacity=1 ][line width=0.75]  [dash pattern={on 4.5pt off 4.5pt}]  (196.66,123.28) -- (196.66,147.48) ;

\draw [color={rgb, 255:red, 0; green, 0; blue, 0 }  ,draw opacity=1 ][line width=0.75]  [dash pattern={on 4.5pt off 4.5pt}]  (196.66,171.68) -- (220.86,171.68) ;

\draw [color={rgb, 255:red, 0; green, 0; blue, 0 }  ,draw opacity=1 ][line width=0.75]  [dash pattern={on 4.5pt off 4.5pt}]  (220.86,123.28) -- (245.06,123.28) ;

\draw [color={rgb, 255:red, 0; green, 0; blue, 0 }  ,draw opacity=1 ][line width=0.75]  [dash pattern={on 4.5pt off 4.5pt}]  (196.66,123.28) -- (220.86,123.28) ;

\draw [color={rgb, 255:red, 0; green, 0; blue, 0 }  ,draw opacity=1 ][line width=0.75]  [dash pattern={on 4.5pt off 4.5pt}]  (245.06,171.68) -- (245.06,147.48) ;

\draw [color={rgb, 255:red, 0; green, 0; blue, 0 }  ,draw opacity=1 ][line width=0.75]  [dash pattern={on 4.5pt off 4.5pt}]  (245.06,123.28) -- (245.06,147.48) ;

\draw [color={rgb, 255:red, 0; green, 0; blue, 0 }  ,draw opacity=0.38 ][line width=0.75]  [dash pattern={on 4.5pt off 4.5pt}]  (167.14,147.46) -- (220.86,171.68) ;

\draw [color={rgb, 255:red, 0; green, 0; blue, 0 }  ,draw opacity=1 ][line width=0.75]  [dash pattern={on 4.5pt off 4.5pt}]  (166.65,99.04) -- (220.86,123.28) ;

\draw  [fill={rgb, 255:red, 255; green, 0; blue, 0 }  ,fill opacity=1 ][line width=0.75]  (140.03,99.04) .. controls (140.03,97.7) and (141.11,96.62) .. (142.45,96.62) .. controls (143.78,96.62) and (144.86,97.7) .. (144.86,99.04) .. controls (144.86,100.37) and (143.78,101.45) .. (142.45,101.45) .. controls (141.11,101.45) and (140.03,100.37) .. (140.03,99.04) -- cycle ;
\draw  [fill={rgb, 255:red, 255; green, 0; blue, 0 }  ,fill opacity=1 ][line width=0.75]  (164.23,99.04) .. controls (164.23,97.7) and (165.31,96.62) .. (166.65,96.62) .. controls (167.98,96.62) and (169.06,97.7) .. (169.06,99.04) .. controls (169.06,100.37) and (167.98,101.45) .. (166.65,101.45) .. controls (165.31,101.45) and (164.23,100.37) .. (164.23,99.04) -- cycle ;
\draw  [fill={rgb, 255:red, 255; green, 0; blue, 0 }  ,fill opacity=1 ][line width=0.75]  (188.43,99.04) .. controls (188.43,97.7) and (189.51,96.62) .. (190.85,96.62) .. controls (192.18,96.62) and (193.26,97.7) .. (193.26,99.04) .. controls (193.26,100.37) and (192.18,101.45) .. (190.85,101.45) .. controls (189.51,101.45) and (188.43,100.37) .. (188.43,99.04) -- cycle ;
\draw  [fill={rgb, 255:red, 255; green, 0; blue, 0 }  ,fill opacity=1 ][line width=0.75]  (140.03,147.44) .. controls (140.03,146.1) and (141.11,145.02) .. (142.45,145.02) .. controls (143.78,145.02) and (144.86,146.1) .. (144.86,147.44) .. controls (144.86,148.77) and (143.78,149.85) .. (142.45,149.85) .. controls (141.11,149.85) and (140.03,148.77) .. (140.03,147.44) -- cycle ;
\draw  [fill={rgb, 255:red, 255; green, 0; blue, 0 }  ,fill opacity=1 ][line width=0.75]  (167.14,111.16) .. controls (167.14,109.83) and (168.22,108.75) .. (169.55,108.75) .. controls (170.88,108.75) and (171.97,109.83) .. (171.97,111.16) .. controls (171.97,112.49) and (170.88,113.57) .. (169.55,113.57) .. controls (168.22,113.57) and (167.14,112.49) .. (167.14,111.16) -- cycle ;
\draw  [fill={rgb, 255:red, 255; green, 0; blue, 0 }  ,fill opacity=1 ][line width=0.75]  (190.85,111.14) .. controls (190.85,109.8) and (191.93,108.72) .. (193.26,108.72) .. controls (194.59,108.72) and (195.68,109.8) .. (195.68,111.14) .. controls (195.68,112.47) and (194.59,113.55) .. (193.26,113.55) .. controls (191.93,113.55) and (190.85,112.47) .. (190.85,111.14) -- cycle ;
\draw  [fill={rgb, 255:red, 255; green, 0; blue, 0 }  ,fill opacity=1 ][line width=0.75]  (215.54,111.16) .. controls (215.54,109.83) and (216.62,108.75) .. (217.95,108.75) .. controls (219.28,108.75) and (220.37,109.83) .. (220.37,111.16) .. controls (220.37,112.49) and (219.28,113.57) .. (217.95,113.57) .. controls (216.62,113.57) and (215.54,112.49) .. (215.54,111.16) -- cycle ;
\draw  [color={rgb, 255:red, 0; green, 0; blue, 0 }  ,draw opacity=0.22 ][fill={rgb, 255:red, 255; green, 0; blue, 0 }  ,fill opacity=0.22 ][line width=0.75]  (215.54,135.36) .. controls (215.54,134.03) and (216.62,132.95) .. (217.95,132.95) .. controls (219.28,132.95) and (220.37,134.03) .. (220.37,135.36) .. controls (220.37,136.69) and (219.28,137.77) .. (217.95,137.77) .. controls (216.62,137.77) and (215.54,136.69) .. (215.54,135.36) -- cycle ;
\draw [color={rgb, 255:red, 0; green, 0; blue, 0 }  ,draw opacity=1 ][line width=0.75]  [dash pattern={on 4.5pt off 4.5pt}]  (142.45,123.24) -- (196.66,147.48) ;

\draw [color={rgb, 255:red, 0; green, 0; blue, 0 }  ,draw opacity=1 ][line width=0.75]  [dash pattern={on 4.5pt off 4.5pt}]  (169.55,159.56) -- (169.55,135.36) ;

\draw [color={rgb, 255:red, 0; green, 0; blue, 0 }  ,draw opacity=1 ][line width=0.75]  [dash pattern={on 4.5pt off 4.5pt}]  (196.66,171.68) -- (196.66,147.48) ;

\draw  [fill={rgb, 255:red, 255; green, 0; blue, 0 }  ,fill opacity=1 ][line width=0.75]  (140.03,123.24) .. controls (140.03,121.9) and (141.11,120.82) .. (142.45,120.82) .. controls (143.78,120.82) and (144.86,121.9) .. (144.86,123.24) .. controls (144.86,124.57) and (143.78,125.65) .. (142.45,125.65) .. controls (141.11,125.65) and (140.03,124.57) .. (140.03,123.24) -- cycle ;
\draw  [fill={rgb, 255:red, 255; green, 0; blue, 0 }  ,fill opacity=1 ][line width=0.75]  (167.14,135.36) .. controls (167.14,134.03) and (168.22,132.95) .. (169.55,132.95) .. controls (170.88,132.95) and (171.97,134.03) .. (171.97,135.36) .. controls (171.97,136.69) and (170.88,137.77) .. (169.55,137.77) .. controls (168.22,137.77) and (167.14,136.69) .. (167.14,135.36) -- cycle ;
\draw  [fill={rgb, 255:red, 255; green, 0; blue, 0 }  ,fill opacity=1 ][line width=0.75]  (167.14,159.56) .. controls (167.14,158.23) and (168.22,157.15) .. (169.55,157.15) .. controls (170.88,157.15) and (171.97,158.23) .. (171.97,159.56) .. controls (171.97,160.89) and (170.88,161.97) .. (169.55,161.97) .. controls (168.22,161.97) and (167.14,160.89) .. (167.14,159.56) -- cycle ;
\draw [color={rgb, 255:red, 0; green, 0; blue, 0 }  ,draw opacity=1 ][line width=0.75]  [dash pattern={on 4.5pt off 4.5pt}]  (234.92,167.09) -- (245.06,171.68) ;

\draw  [color={rgb, 255:red, 0; green, 0; blue, 0 }  ,draw opacity=0.22 ][fill={rgb, 255:red, 255; green, 0; blue, 0 }  ,fill opacity=0.22 ][line width=0.75]  (215.54,159.56) .. controls (215.54,158.23) and (216.62,157.15) .. (217.95,157.15) .. controls (219.28,157.15) and (220.37,158.23) .. (220.37,159.56) .. controls (220.37,160.89) and (219.28,161.97) .. (217.95,161.97) .. controls (216.62,161.97) and (215.54,160.89) .. (215.54,159.56) -- cycle ;
\draw [color={rgb, 255:red, 0; green, 0; blue, 0 }  ,draw opacity=0.38 ][line width=0.75]  [dash pattern={on 4.5pt off 4.5pt}]  (215.86,158.33) -- (193.18,148.42) ;

\draw  [color={rgb, 255:red, 0; green, 0; blue, 0 }  ,draw opacity=0.22 ][fill={rgb, 255:red, 255; green, 0; blue, 0 }  ,fill opacity=0.22 ][line width=0.75]  (188.43,147.44) .. controls (188.43,146.1) and (189.51,145.02) .. (190.85,145.02) .. controls (192.18,145.02) and (193.26,146.1) .. (193.26,147.44) .. controls (193.26,148.77) and (192.18,149.85) .. (190.85,149.85) .. controls (189.51,149.85) and (188.43,148.77) .. (188.43,147.44) -- cycle ;
\draw  [color={rgb, 255:red, 0; green, 0; blue, 0 }  ,draw opacity=0.22 ][fill={rgb, 255:red, 255; green, 0; blue, 0 }  ,fill opacity=0.22 ][line width=0.75]  (188.15,123.24) .. controls (188.15,121.9) and (189.23,120.82) .. (190.56,120.82) .. controls (191.89,120.82) and (192.97,121.9) .. (192.97,123.24) .. controls (192.97,124.57) and (191.89,125.65) .. (190.56,125.65) .. controls (189.23,125.65) and (188.15,124.57) .. (188.15,123.24) -- cycle ;
\draw [color={rgb, 255:red, 0; green, 0; blue, 0 }  ,draw opacity=0.38 ][fill={rgb, 255:red, 0; green, 0; blue, 0 }  ,fill opacity=0.38 ][line width=0.75]  [dash pattern={on 4.5pt off 4.5pt}]  (220.01,136.39) -- (244.77,147.48) ;

\draw [color={rgb, 255:red, 0; green, 0; blue, 0 }  ,draw opacity=1 ][line width=0.75]  [dash pattern={on 4.5pt off 4.5pt}]  (190.85,101.45) -- (190.75,107.29) ;

\draw  [draw opacity=0][dash pattern={on 4.5pt off 4.5pt}] (190.43,125.19) .. controls (189.53,136.49) and (180.87,145.6) .. (169.78,147.18) -- (166.36,123.24) -- cycle ; \draw  [color={rgb, 255:red, 0; green, 0; blue, 0 }  ,draw opacity=0.36 ][dash pattern={on 4.5pt off 4.5pt}] (190.43,125.19) .. controls (189.53,136.49) and (180.87,145.6) .. (169.78,147.18) ;
\draw  [draw opacity=0][dash pattern={on 4.5pt off 4.5pt}] (169.49,99.26) .. controls (180.56,100.7) and (189.27,109.63) .. (190.39,120.8) -- (166.36,123.24) -- cycle ; \draw  [color={rgb, 255:red, 0; green, 0; blue, 0 }  ,draw opacity=0.45 ][dash pattern={on 4.5pt off 4.5pt}] (169.49,99.26) .. controls (180.56,100.7) and (189.27,109.63) .. (190.39,120.8) ;
\draw  [color={rgb, 255:red, 0; green, 0; blue, 0 }  ,draw opacity=0.22 ][fill={rgb, 255:red, 255; green, 0; blue, 0 }  ,fill opacity=0.22 ][line width=0.75]  (167.36,147.18) .. controls (167.36,145.84) and (168.45,144.76) .. (169.78,144.76) .. controls (171.11,144.76) and (172.19,145.84) .. (172.19,147.18) .. controls (172.19,148.51) and (171.11,149.59) .. (169.78,149.59) .. controls (168.45,149.59) and (167.36,148.51) .. (167.36,147.18) -- cycle ;
\draw [color={rgb, 255:red, 0; green, 0; blue, 0 }  ,draw opacity=0.38 ][line width=0.75]  [dash pattern={on 4.5pt off 4.5pt}]  (161.35,147.46) -- (167.36,147.18) ;

\draw  [color={rgb, 255:red, 0; green, 0; blue, 0 }  ,draw opacity=0.22 ][fill={rgb, 255:red, 255; green, 0; blue, 0 }  ,fill opacity=0.22 ][line width=0.75]  (191.34,159.56) .. controls (191.34,158.23) and (192.42,157.15) .. (193.75,157.15) .. controls (195.08,157.15) and (196.17,158.23) .. (196.17,159.56) .. controls (196.17,160.89) and (195.08,161.97) .. (193.75,161.97) .. controls (192.42,161.97) and (191.34,160.89) .. (191.34,159.56) -- cycle ;
\draw [color={rgb, 255:red, 0; green, 0; blue, 0 }  ,draw opacity=0.38 ][line width=0.75]  [dash pattern={on 4.5pt off 4.5pt}]  (188.22,159.56) -- (191.34,159.56) ;

\draw  [draw opacity=0][fill={rgb, 255:red, 0; green, 0; blue, 0 }  ,fill opacity=0.19 ] (124.62,58.24) .. controls (141.18,58.21) and (421.59,57.37) .. (509.58,57.8) .. controls (509.64,77.95) and (509.26,128.27) .. (509.58,147.15) .. controls (357.8,143.77) and (358.12,144.09) .. (245.06,147.48) .. controls (244.26,130.82) and (232.59,123.68) .. (220.86,123.28) .. controls (209.8,123.85) and (198.55,129) .. (196.66,147.48) .. controls (197.41,155.49) and (201.22,170.35) .. (220.86,171.68) .. controls (233.05,171.69) and (244.67,163.3) .. (245.06,147.48) .. controls (263.96,146.13) and (416.89,130.74) .. (509.58,147.15) .. controls (509.96,182.84) and (509.32,220.64) .. (509.58,236.5) .. controls (446.74,236.84) and (128.76,238.95) .. (124.62,236.94) .. controls (124.79,185.14) and (124.79,83.18) .. (124.62,58.24) -- cycle ;
\draw  [fill={rgb, 255:red, 255; green, 0; blue, 0 }  ,fill opacity=1 ][line width=0.75]  (194.24,123.28) .. controls (194.24,121.95) and (195.32,120.87) .. (196.66,120.87) .. controls (197.99,120.87) and (199.07,121.95) .. (199.07,123.28) .. controls (199.07,124.62) and (197.99,125.7) .. (196.66,125.7) .. controls (195.32,125.7) and (194.24,124.62) .. (194.24,123.28) -- cycle ;
\draw  [fill={rgb, 255:red, 255; green, 0; blue, 0 }  ,fill opacity=1 ][line width=0.75]  (218.44,123.28) .. controls (218.44,121.95) and (219.52,120.87) .. (220.86,120.87) .. controls (222.19,120.87) and (223.27,121.95) .. (223.27,123.28) .. controls (223.27,124.62) and (222.19,125.7) .. (220.86,125.7) .. controls (219.52,125.7) and (218.44,124.62) .. (218.44,123.28) -- cycle ;
\draw  [fill={rgb, 255:red, 255; green, 0; blue, 0 }  ,fill opacity=1 ][line width=0.75]  (242.64,123.28) .. controls (242.64,121.95) and (243.72,120.87) .. (245.06,120.87) .. controls (246.39,120.87) and (247.47,121.95) .. (247.47,123.28) .. controls (247.47,124.62) and (246.39,125.7) .. (245.06,125.7) .. controls (243.72,125.7) and (242.64,124.62) .. (242.64,123.28) -- cycle ;
\draw  [fill={rgb, 255:red, 255; green, 0; blue, 0 }  ,fill opacity=1 ][line width=0.75]  (242.64,147.48) .. controls (242.64,146.15) and (243.72,145.07) .. (245.06,145.07) .. controls (246.39,145.07) and (247.47,146.15) .. (247.47,147.48) .. controls (247.47,148.82) and (246.39,149.9) .. (245.06,149.9) .. controls (243.72,149.9) and (242.64,148.82) .. (242.64,147.48) -- cycle ;
\draw  [fill={rgb, 255:red, 255; green, 0; blue, 0 }  ,fill opacity=1 ][line width=0.75]  (218.44,171.68) .. controls (218.44,170.35) and (219.52,169.27) .. (220.86,169.27) .. controls (222.19,169.27) and (223.27,170.35) .. (223.27,171.68) .. controls (223.27,173.02) and (222.19,174.1) .. (220.86,174.1) .. controls (219.52,174.1) and (218.44,173.02) .. (218.44,171.68) -- cycle ;
\draw  [fill={rgb, 255:red, 255; green, 0; blue, 0 }  ,fill opacity=1 ][line width=0.75]  (242.64,171.68) .. controls (242.64,170.35) and (243.72,169.27) .. (245.06,169.27) .. controls (246.39,169.27) and (247.47,170.35) .. (247.47,171.68) .. controls (247.47,173.02) and (246.39,174.1) .. (245.06,174.1) .. controls (243.72,174.1) and (242.64,173.02) .. (242.64,171.68) -- cycle ;
\draw  [fill={rgb, 255:red, 255; green, 0; blue, 0 }  ,fill opacity=1 ][line width=0.75]  (194.24,147.48) .. controls (194.24,146.15) and (195.32,145.07) .. (196.66,145.07) .. controls (197.99,145.07) and (199.07,146.15) .. (199.07,147.48) .. controls (199.07,148.82) and (197.99,149.9) .. (196.66,149.9) .. controls (195.32,149.9) and (194.24,148.82) .. (194.24,147.48) -- cycle ;
\draw  [fill={rgb, 255:red, 255; green, 0; blue, 0 }  ,fill opacity=1 ][line width=0.75]  (194.24,171.68) .. controls (194.24,170.35) and (195.32,169.27) .. (196.66,169.27) .. controls (197.99,169.27) and (199.07,170.35) .. (199.07,171.68) .. controls (199.07,173.02) and (197.99,174.1) .. (196.66,174.1) .. controls (195.32,174.1) and (194.24,173.02) .. (194.24,171.68) -- cycle ;
\draw [line width=0.75]    (221.23,248.91) -- (509.11,248.91) ;
\draw [shift={(509.11,248.91)}, rotate = 180] [color={rgb, 255:red, 0; green, 0; blue, 0 }  ][line width=0.75]    (0,5.59) -- (0,-5.59)(10.93,-4.9) .. controls (6.95,-2.3) and (3.31,-0.67) .. (0,0) .. controls (3.31,0.67) and (6.95,2.3) .. (10.93,4.9)   ;
\draw [shift={(221.23,248.91)}, rotate = 0] [color={rgb, 255:red, 0; green, 0; blue, 0 }  ][line width=0.75]    (0,5.59) -- (0,-5.59)(10.93,-3.29) .. controls (6.95,-1.4) and (3.31,-0.3) .. (0,0) .. controls (3.31,0.3) and (6.95,1.4) .. (10.93,3.29)   ;
\draw [line width=0.75]    (524.15,148.05) -- (524.15,237.4) ;
\draw [shift={(524.15,237.4)}, rotate = 270] [color={rgb, 255:red, 0; green, 0; blue, 0 }  ][line width=0.75]    (0,5.59) -- (0,-5.59)(10.93,-4.9) .. controls (6.95,-2.3) and (3.31,-0.67) .. (0,0) .. controls (3.31,0.67) and (6.95,2.3) .. (10.93,4.9)   ;
\draw [shift={(524.15,148.05)}, rotate = 90] [color={rgb, 255:red, 0; green, 0; blue, 0 }  ][line width=0.75]    (0,5.59) -- (0,-5.59)(10.93,-3.29) .. controls (6.95,-1.4) and (3.31,-0.3) .. (0,0) .. controls (3.31,0.3) and (6.95,1.4) .. (10.93,3.29)   ;
\draw [line width=0.75]    (524.15,58.7) -- (524.15,148.05) ;
\draw [shift={(524.15,148.05)}, rotate = 270] [color={rgb, 255:red, 0; green, 0; blue, 0 }  ][line width=0.75]    (0,5.59) -- (0,-5.59)(10.93,-4.9) .. controls (6.95,-2.3) and (3.31,-0.67) .. (0,0) .. controls (3.31,0.67) and (6.95,2.3) .. (10.93,4.9)   ;
\draw [shift={(524.15,58.7)}, rotate = 90] [color={rgb, 255:red, 0; green, 0; blue, 0 }  ][line width=0.75]    (0,5.59) -- (0,-5.59)(10.93,-3.29) .. controls (6.95,-1.4) and (3.31,-0.3) .. (0,0) .. controls (3.31,0.3) and (6.95,1.4) .. (10.93,3.29)   ;
\draw  [dash pattern={on 0.84pt off 2.51pt}]  (96,224.57) .. controls (83.37,190.6) and (27.3,166.99) .. (27.77,135.4) .. controls (28.24,103.81) and (87.37,75.8) .. (96.75,45.87) ;

\draw  [dash pattern={on 0.84pt off 2.51pt}]  (124.2,147.17) .. controls (95.37,147) and (38.77,138.4) .. (27.77,135.4) .. controls (16.77,132.4) and (62.17,130.6) .. (69.43,122.93) ;

\draw [line width=0.75]    (94.82,135.06) -- (27.77,135.4) ;

\draw [shift={(96.82,135.05)}, rotate = 179.71] [fill={rgb, 255:red, 0; green, 0; blue, 0 }  ][line width=0.75]  [draw opacity=0] (8.93,-4.29) -- (0,0) -- (8.93,4.29) -- cycle    ;
\draw [line width=0.75]    (94.69,112.8) -- (38.97,112.8) ;

\draw [shift={(96.69,112.8)}, rotate = 180] [fill={rgb, 255:red, 0; green, 0; blue, 0 }  ][line width=0.75]  [draw opacity=0] (8.93,-4.29) -- (0,0) -- (8.93,4.29) -- cycle    ;
\draw [line width=0.75]    (94.56,90.54) -- (61.37,90.54) ;

\draw [shift={(96.56,90.54)}, rotate = 180] [fill={rgb, 255:red, 0; green, 0; blue, 0 }  ][line width=0.75]  [draw opacity=0] (8.93,-4.29) -- (0,0) -- (8.93,4.29) -- cycle    ;
\draw [line width=0.75]    (94.65,68.21) -- (81.37,68.21) ;

\draw [shift={(96.65,68.21)}, rotate = 180] [fill={rgb, 255:red, 0; green, 0; blue, 0 }  ][line width=0.75]  [draw opacity=0] (8.93,-4.29) -- (0,0) -- (8.93,4.29) -- cycle    ;
\draw [line width=0.75]    (94.28,157.56) -- (36.57,157.56) ;

\draw [shift={(96.28,157.56)}, rotate = 180] [fill={rgb, 255:red, 0; green, 0; blue, 0 }  ][line width=0.75]  [draw opacity=0] (8.93,-4.29) -- (0,0) -- (8.93,4.29) -- cycle    ;
\draw [line width=0.75]    (94.19,179.89) -- (57.5,179.89) ;

\draw [shift={(96.19,179.89)}, rotate = 180] [fill={rgb, 255:red, 0; green, 0; blue, 0 }  ][line width=0.75]  [draw opacity=0] (8.93,-4.29) -- (0,0) -- (8.93,4.29) -- cycle    ;
\draw [line width=0.75]    (94.31,202.15) -- (81.77,202.15) ;

\draw [shift={(96.31,202.15)}, rotate = 180] [fill={rgb, 255:red, 0; green, 0; blue, 0 }  ][line width=0.75]  [draw opacity=0] (8.93,-4.29) -- (0,0) -- (8.93,4.29) -- cycle    ;
\draw [line width=0.75]    (108.29,141.2) -- (60.17,141.2) ;

\draw [shift={(110.29,141.2)}, rotate = 180] [fill={rgb, 255:red, 0; green, 0; blue, 0 }  ][line width=0.75]  [draw opacity=0] (8.93,-4.29) -- (0,0) -- (8.93,4.29) -- cycle    ;
\draw [line width=0.75]    (80.46,129.24) -- (52.17,129.24) ;

\draw [shift={(82.46,129.24)}, rotate = 180] [fill={rgb, 255:red, 0; green, 0; blue, 0 }  ][line width=0.75]  [draw opacity=0] (8.93,-4.29) -- (0,0) -- (8.93,4.29) -- cycle    ;
\draw    (196.66,192.2) -- (245.06,192.2) ;
\draw [shift={(245.06,192.2)}, rotate = 180] [color={rgb, 255:red, 0; green, 0; blue, 0 }  ][line width=0.75]    (0,5.59) -- (0,-5.59)(10.93,-3.29) .. controls (6.95,-1.4) and (3.31,-0.3) .. (0,0) .. controls (3.31,0.3) and (6.95,1.4) .. (10.93,3.29)   ;
\draw [shift={(196.66,192.2)}, rotate = 0] [color={rgb, 255:red, 0; green, 0; blue, 0 }  ][line width=0.75]    (0,5.59) -- (0,-5.59)(10.93,-3.29) .. controls (6.95,-1.4) and (3.31,-0.3) .. (0,0) .. controls (3.31,0.3) and (6.95,1.4) .. (10.93,3.29)   ;
\draw    (273.5,118.86) -- (242.37,133.77) ;
\draw [shift={(240.57,134.63)}, rotate = 334.4] [fill={rgb, 255:red, 0; green, 0; blue, 0 }  ][line width=0.75]  [draw opacity=0] (8.93,-4.29) -- (0,0) -- (8.93,4.29) -- cycle    ;

\draw  [dash pattern={on 0.84pt off 2.51pt}]  (96.75,45.87) -- (96,224.57) ;

\draw  [dash pattern={on 0.84pt off 2.51pt}]  (69.43,122.93) -- (124.33,147.59) ;

\draw    (253.5,23.86) -- (234.39,43.56) ;
\draw [shift={(233,45)}, rotate = 314.11] [fill={rgb, 255:red, 0; green, 0; blue, 0 }  ][line width=0.75]  [draw opacity=0] (8.93,-4.29) -- (0,0) -- (8.93,4.29) -- cycle    ;

\draw (172.27,264.97) node [scale=1.2] [align=left] {$\displaystyle 5D$};
\draw (366.58,263.88) node [scale=1.2] [align=left] {$\displaystyle 20D$};
\draw (541.84,191.49) node [scale=1.2] [align=left] {$\displaystyle 2D$};
\draw (552.45,103.09) node [scale=1.2] [align=left] {$\displaystyle 2.1D$};
\draw (513.3,26.63) node [scale=1.2] [align=left] {$\displaystyle 4.1D$};
\draw (65.18,251.66) node [scale=1.2]  {$x$};
\draw (26.55,238.27) node [scale=1.2]  {$y$};
\draw (52.07,280.56) node [scale=1.2]  {$z$};
\draw (221.65,202.15) node [scale=1.2] [align=left] {$\displaystyle D$};
\draw (301.15,101.73) node [scale=1.2] [align=left] {$\displaystyle \uvec_{\mathbf{cyinder}} = \ovec$};
\draw (278.11,13.99) node [scale=1.2] [align=left] {$\displaystyle \uvec_{\mathbf{wall}} = \ovec$};
\draw (23.83,91.42) node [scale=1.2] [align=left] {$\displaystyle \uvec_{\mathbf{in}}$};

\end{tikzpicture}

%% file: figures/2d-inflating-balloon-domain.tikz
\tikzset{every picture/.style={line width=0.75pt}} 

\begin{tikzpicture}[x=0.75pt,y=0.75pt,yscale=-0.65,xscale=0.65, every node/.style={scale=0.65}]

\draw    (5.43,278) -- (44.5,278) ;
\draw [shift={(47.5,278)}, rotate = 180] [fill={rgb, 255:red, 0; green, 0; blue, 0 }  ][line width=0.08]  [draw opacity=0] (10.72,-5.15) -- (0,0) -- (10.72,5.15) -- (7.12,0) -- cycle    ;
\draw    (5.43,278) -- (5.43,237.82) ;
\draw [shift={(5.43,234.82)}, rotate = 450] [fill={rgb, 255:red, 0; green, 0; blue, 0 }  ][line width=0.08]  [draw opacity=0] (10.72,-5.15) -- (0,0) -- (10.72,5.15) -- (7.12,0) -- cycle    ;

\draw [line width=1.5]    (205.5,282) .. controls (173.03,282.96) and (151.4,280.23) .. (138.65,256.88) ;
\draw [shift={(136.91,253.44)}, rotate = 424.81] [fill={rgb, 255:red, 0; green, 0; blue, 0 }  ][line width=0.08]  [draw opacity=0] (13.4,-6.43) -- (0,0) -- (13.4,6.44) -- (8.9,0) -- cycle    ;
\draw    (276.5,51.64) -- (276.5,239.45) ;
\draw [shift={(276.5,242.45)}, rotate = 270] [fill={rgb, 255:red, 0; green, 0; blue, 0 }  ][line width=0.08]  [draw opacity=0] (8.93,-4.29) -- (0,0) -- (8.93,4.29) -- cycle    ;
\draw [shift={(276.5,48.64)}, rotate = 90] [fill={rgb, 255:red, 0; green, 0; blue, 0 }  ][line width=0.08]  [draw opacity=0] (8.93,-4.29) -- (0,0) -- (8.93,4.29) -- cycle    ;
\draw  [color={rgb, 255:red, 0; green, 0; blue, 0 }  ,draw opacity=1 ][fill={rgb, 255:red, 108; green, 108; blue, 108 }  ,fill opacity=0.7 ,even odd rule] (46.44,145.54) .. controls (46.44,92.02) and (89.83,48.64) .. (143.35,48.64) .. controls (196.87,48.64) and (240.26,92.02) .. (240.26,145.54) .. controls (240.26,199.06) and (196.87,242.45) .. (143.35,242.45) .. controls (89.83,242.45) and (46.44,199.06) .. (46.44,145.54)(31.5,145.54) .. controls (31.5,83.77) and (81.58,33.69) .. (143.35,33.69) .. controls (205.13,33.69) and (255.2,83.77) .. (255.2,145.54) .. controls (255.2,207.32) and (205.13,257.4) .. (143.35,257.4) .. controls (81.58,257.4) and (31.5,207.32) .. (31.5,145.54) ;
\draw    (42.78,35.88) -- (65.42,61.11) ;
\draw [shift={(67.43,63.34)}, rotate = 228.09] [fill={rgb, 255:red, 0; green, 0; blue, 0 }  ][line width=0.08]  [draw opacity=0] (8.93,-4.29) -- (0,0) -- (8.93,4.29) -- cycle    ;
\draw  [fill={rgb, 255:red, 97; green, 95; blue, 95 }  ,fill opacity=0.39 ,even odd rule] (118.65,145.54) .. controls (118.65,131.9) and (129.71,120.85) .. (143.35,120.85) .. controls (156.99,120.85) and (168.05,131.9) .. (168.05,145.54) .. controls (168.05,159.18) and (156.99,170.24) .. (143.35,170.24) .. controls (129.71,170.24) and (118.65,159.18) .. (118.65,145.54)(46.44,145.54) .. controls (46.44,92.02) and (89.83,48.64) .. (143.35,48.64) .. controls (196.87,48.64) and (240.26,92.02) .. (240.26,145.54) .. controls (240.26,199.06) and (196.87,242.45) .. (143.35,242.45) .. controls (89.83,242.45) and (46.44,199.06) .. (46.44,145.54) ;
\draw    (110.51,123.85) -- (110.51,166.06) ;
\draw [shift={(110.51,169.06)}, rotate = 270] [fill={rgb, 255:red, 0; green, 0; blue, 0 }  ][line width=0.08]  [draw opacity=0] (8.93,-4.29) -- (0,0) -- (8.93,4.29) -- cycle    ;
\draw [shift={(110.51,120.85)}, rotate = 90] [fill={rgb, 255:red, 0; green, 0; blue, 0 }  ][line width=0.08]  [draw opacity=0] (8.93,-4.29) -- (0,0) -- (8.93,4.29) -- cycle    ;
\draw  [dash pattern={on 0.84pt off 2.51pt}]  (143.35,120.85) -- (110.51,120.85) ;
\draw  [dash pattern={on 0.84pt off 2.51pt}]  (143.35,170.24) -- (110.51,170.24) ;
\draw    (121.03,145.17) -- (135.35,145.17) ;
\draw [shift={(118.03,145.17)}, rotate = 0] [fill={rgb, 255:red, 0; green, 0; blue, 0 }  ][line width=0.08]  [draw opacity=0] (8.93,-4.29) -- (0,0) -- (8.93,4.29) -- cycle    ;
\draw    (143.01,124.18) -- (143.01,138.49) ;
\draw [shift={(143.01,121.18)}, rotate = 90] [fill={rgb, 255:red, 0; green, 0; blue, 0 }  ][line width=0.08]  [draw opacity=0] (8.93,-4.29) -- (0,0) -- (8.93,4.29) -- cycle    ;
\draw    (164,145.17) -- (149.69,145.17) ;
\draw [shift={(167,145.17)}, rotate = 180] [fill={rgb, 255:red, 0; green, 0; blue, 0 }  ][line width=0.08]  [draw opacity=0] (8.93,-4.29) -- (0,0) -- (8.93,4.29) -- cycle    ;
\draw    (142.03,167.14) -- (142.03,152.83) ;
\draw [shift={(142.03,170.14)}, rotate = 270] [fill={rgb, 255:red, 0; green, 0; blue, 0 }  ][line width=0.08]  [draw opacity=0] (8.93,-4.29) -- (0,0) -- (8.93,4.29) -- cycle    ;

\draw    (127.33,129.98) -- (137.45,140.1) ;
\draw [shift={(125.21,127.86)}, rotate = 45] [fill={rgb, 255:red, 0; green, 0; blue, 0 }  ][line width=0.08]  [draw opacity=0] (8.93,-4.29) -- (0,0) -- (8.93,4.29) -- cycle    ;
\draw    (157.71,129.98) -- (147.59,140.1) ;
\draw [shift={(159.83,127.86)}, rotate = 135] [fill={rgb, 255:red, 0; green, 0; blue, 0 }  ][line width=0.08]  [draw opacity=0] (8.93,-4.29) -- (0,0) -- (8.93,4.29) -- cycle    ;
\draw    (157.71,160.36) -- (147.59,150.24) ;
\draw [shift={(159.83,162.48)}, rotate = 225] [fill={rgb, 255:red, 0; green, 0; blue, 0 }  ][line width=0.08]  [draw opacity=0] (8.93,-4.29) -- (0,0) -- (8.93,4.29) -- cycle    ;
\draw    (127.33,160.36) -- (137.45,150.24) ;
\draw [shift={(125.21,162.48)}, rotate = 315] [fill={rgb, 255:red, 0; green, 0; blue, 0 }  ][line width=0.08]  [draw opacity=0] (8.93,-4.29) -- (0,0) -- (8.93,4.29) -- cycle    ;

\draw    (143.35,48.64) -- (276.5,48.64) ;
\draw    (79.22,77.03) -- (98.12,95.93) ;
\draw [shift={(77.1,74.91)}, rotate = 45] [fill={rgb, 255:red, 0; green, 0; blue, 0 }  ][line width=0.08]  [draw opacity=0] (8.93,-4.29) -- (0,0) -- (8.93,4.29) -- cycle    ;
\draw    (143.35,242.45) -- (276.5,242.45) ;

\draw (212.5,282) node [anchor=west] [inner sep=0.75pt]  [font=\Large]  {$\Omega _{s}$};
\draw (141.38,198.15) node  [font=\Large]  {$\Omega _{f}$};
\draw (284.5,145.54) node [anchor=west] [inner sep=0.75pt]  [font=\Large]  {$6D$};
\draw (56.45,19.86) node  [font=\Large]  {$a$};
\draw (92.26,146.99) node  [font=\Large]  {$D$};
\draw (7.43,231.42) node [anchor=south west] [inner sep=0.75pt]  [font=\Large]  {$y$};
\draw (49.5,274.6) node [anchor=south west] [inner sep=0.75pt]  [font=\Large]  {$x$};

\end{tikzpicture}

%% file: figures/inflating-balloon-domain.tikz
\tikzset{every picture/.style={line width=0.75pt}} 

\begin{tikzpicture}[x=0.75pt,y=0.75pt,yscale=-0.55,xscale=0.55, every node/.style={scale=0.55}]

\draw    (24.43,84) -- (86.34,66.33) ;
\draw [shift={(89.23,65.51)}, rotate = 524.0699999999999] [fill={rgb, 255:red, 0; green, 0; blue, 0 }  ][line width=0.08]  [draw opacity=0] (10.72,-5.15) -- (0,0) -- (10.72,5.15) -- (7.12,0) -- cycle    ;
\draw    (24.43,84) -- (88.56,84) ;
\draw [shift={(91.56,84)}, rotate = 180] [fill={rgb, 255:red, 0; green, 0; blue, 0 }  ][line width=0.08]  [draw opacity=0] (10.72,-5.15) -- (0,0) -- (10.72,5.15) -- (7.12,0) -- cycle    ;
\draw    (24.43,84) -- (24.43,35.82) ;
\draw [shift={(24.43,32.82)}, rotate = 450] [fill={rgb, 255:red, 0; green, 0; blue, 0 }  ][line width=0.08]  [draw opacity=0] (10.72,-5.15) -- (0,0) -- (10.72,5.15) -- (7.12,0) -- cycle    ;

\draw [line width=1.5]    (324.94,373.08) .. controls (289.21,369.91) and (271.05,364.48) .. (258.63,340.9) ;
\draw [shift={(256.91,337.44)}, rotate = 424.81] [fill={rgb, 255:red, 0; green, 0; blue, 0 }  ][line width=0.08]  [draw opacity=0] (13.4,-6.43) -- (0,0) -- (13.4,6.44) -- (8.9,0) -- cycle    ;
\draw    (373.45,92.24) -- (424.62,273.6) ;
\draw [shift={(425.44,276.49)}, rotate = 254.24] [fill={rgb, 255:red, 0; green, 0; blue, 0 }  ][line width=0.08]  [draw opacity=0] (8.93,-4.29) -- (0,0) -- (8.93,4.29) -- cycle    ;
\draw [shift={(372.63,89.36)}, rotate = 74.24] [fill={rgb, 255:red, 0; green, 0; blue, 0 }  ][line width=0.08]  [draw opacity=0] (8.93,-4.29) -- (0,0) -- (8.93,4.29) -- cycle    ;
\draw  [dash pattern={on 0.84pt off 2.51pt}]  (120.35,114.6) -- (120.35,146.69) ;
\draw  [dash pattern={on 0.84pt off 2.51pt}] (166.76,213.24) .. controls (166.76,151.46) and (216.84,101.39) .. (278.61,101.39) .. controls (340.38,101.39) and (390.46,151.46) .. (390.46,213.24) .. controls (390.46,275.01) and (340.38,325.09) .. (278.61,325.09) .. controls (216.84,325.09) and (166.76,275.01) .. (166.76,213.24) -- cycle ;
\draw  [dash pattern={on 0.84pt off 2.51pt}] (181.39,213.24) .. controls (181.39,159.54) and (224.92,116.02) .. (278.61,116.02) .. controls (332.3,116.02) and (375.83,159.54) .. (375.83,213.24) .. controls (375.83,266.93) and (332.3,310.46) .. (278.61,310.46) .. controls (224.92,310.46) and (181.39,266.93) .. (181.39,213.24) -- cycle ;
\draw  [dash pattern={on 0.84pt off 2.51pt}]  (92.24,165.76) -- (255.2,119.03) ;
\draw  [dash pattern={on 0.84pt off 2.51pt}]  (147.41,351.75) -- (308,306.17) ;
\draw  [color={rgb, 255:red, 0; green, 0; blue, 0 }  ,draw opacity=1 ][fill={rgb, 255:red, 108; green, 108; blue, 108 }  ,fill opacity=0.7 ,even odd rule] (23.44,258.54) .. controls (23.44,205.02) and (66.83,161.64) .. (120.35,161.64) .. controls (173.87,161.64) and (217.26,205.02) .. (217.26,258.54) .. controls (217.26,312.06) and (173.87,355.45) .. (120.35,355.45) .. controls (66.83,355.45) and (23.44,312.06) .. (23.44,258.54)(8.5,258.54) .. controls (8.5,196.77) and (58.58,146.69) .. (120.35,146.69) .. controls (182.13,146.69) and (232.2,196.77) .. (232.2,258.54) .. controls (232.2,320.32) and (182.13,370.4) .. (120.35,370.4) .. controls (58.58,370.4) and (8.5,320.32) .. (8.5,258.54) ;
\draw  [fill={rgb, 255:red, 97; green, 95; blue, 95 }  ,fill opacity=0.73 ] (93.29,150) .. controls (157.79,131.75) and (162.2,128.6) .. (251.24,104.69) .. controls (344.17,84) and (393.61,165.89) .. (390.46,213.24) .. controls (390.23,276.56) and (343.34,312.89) .. (305.98,321.79) .. controls (264.13,333.96) and (280.42,329.68) .. (238.67,341.5) .. controls (174.64,359.62) and (211.83,348.43) .. (148.04,367.09) .. controls (215.06,348.21) and (234.56,289.06) .. (232.2,258.54) .. controls (232.67,190.9) and (169.75,133.01) .. (93.29,150) -- cycle ;
\draw  [dash pattern={on 0.84pt off 2.51pt}]  (278.61,66.3) -- (278.61,101.39) ;
\draw    (123.22,113.72) -- (275.74,67.18) ;
\draw [shift={(278.61,66.3)}, rotate = 523.03] [fill={rgb, 255:red, 0; green, 0; blue, 0 }  ][line width=0.08]  [draw opacity=0] (8.93,-4.29) -- (0,0) -- (8.93,4.29) -- cycle    ;
\draw [shift={(120.35,114.6)}, rotate = 343.03] [fill={rgb, 255:red, 0; green, 0; blue, 0 }  ][line width=0.08]  [draw opacity=0] (8.93,-4.29) -- (0,0) -- (8.93,4.29) -- cycle    ;
\draw  [dash pattern={on 0.84pt off 2.51pt}]  (255.2,119.03) -- (372.63,89.36) ;
\draw    (19.78,148.88) -- (42.42,174.11) ;
\draw [shift={(44.43,176.34)}, rotate = 228.09] [fill={rgb, 255:red, 0; green, 0; blue, 0 }  ][line width=0.08]  [draw opacity=0] (8.93,-4.29) -- (0,0) -- (8.93,4.29) -- cycle    ;
\draw  [fill={rgb, 255:red, 97; green, 95; blue, 95 }  ,fill opacity=0.39 ,even odd rule] (95.65,258.54) .. controls (95.65,244.9) and (106.71,233.85) .. (120.35,233.85) .. controls (133.99,233.85) and (145.05,244.9) .. (145.05,258.54) .. controls (145.05,272.18) and (133.99,283.24) .. (120.35,283.24) .. controls (106.71,283.24) and (95.65,272.18) .. (95.65,258.54)(23.44,258.54) .. controls (23.44,205.02) and (66.83,161.64) .. (120.35,161.64) .. controls (173.87,161.64) and (217.26,205.02) .. (217.26,258.54) .. controls (217.26,312.06) and (173.87,355.45) .. (120.35,355.45) .. controls (66.83,355.45) and (23.44,312.06) .. (23.44,258.54) ;
\draw  [color={rgb, 255:red, 0; green, 0; blue, 0 }  ,draw opacity=1 ][fill={rgb, 255:red, 97; green, 95; blue, 95 }  ,fill opacity=0.38 ][dash pattern={on 0.84pt off 2.51pt}] (113.78,234.35) .. controls (130.52,229.61) and (217.6,204.65) .. (271.5,189.33) .. controls (291.34,183.58) and (304.47,200.03) .. (303.61,212.86) .. controls (302.63,228.9) and (292.43,234) .. (284.83,237) .. controls (257.31,245) and (155.35,274.73) .. (127.99,282.74) .. controls (103.91,285.93) and (96.34,270.79) .. (95.65,258.54) .. controls (95.78,240.99) and (110.11,236.04) .. (113.78,234.35) -- cycle ;
\draw  [draw opacity=0][dash pattern={on 0.84pt off 2.51pt}] (278.67,237.47) .. controls (278.65,237.47) and (278.63,237.47) .. (278.61,237.47) .. controls (264.97,237.47) and (253.91,226.62) .. (253.91,213.24) .. controls (253.91,203.12) and (260.23,194.44) .. (269.21,190.82) -- (278.61,213.24) -- cycle ; \draw  [dash pattern={on 0.84pt off 2.51pt}] (278.67,237.47) .. controls (278.65,237.47) and (278.63,237.47) .. (278.61,237.47) .. controls (264.97,237.47) and (253.91,226.62) .. (253.91,213.24) .. controls (253.91,203.12) and (260.23,194.44) .. (269.21,190.82) ;
\draw  [dash pattern={on 0.84pt off 2.51pt}]  (308,306.17) -- (425.44,276.49) ;
\draw    (87.51,236.85) -- (87.51,279.06) ;
\draw [shift={(87.51,282.06)}, rotate = 270] [fill={rgb, 255:red, 0; green, 0; blue, 0 }  ][line width=0.08]  [draw opacity=0] (8.93,-4.29) -- (0,0) -- (8.93,4.29) -- cycle    ;
\draw [shift={(87.51,233.85)}, rotate = 90] [fill={rgb, 255:red, 0; green, 0; blue, 0 }  ][line width=0.08]  [draw opacity=0] (8.93,-4.29) -- (0,0) -- (8.93,4.29) -- cycle    ;
\draw  [dash pattern={on 0.84pt off 2.51pt}]  (120.35,233.85) -- (87.51,233.85) ;
\draw  [dash pattern={on 0.84pt off 2.51pt}]  (120.35,283.24) -- (87.51,283.24) ;
\draw    (98.03,258.17) -- (112.35,258.17) ;
\draw [shift={(95.03,258.17)}, rotate = 0] [fill={rgb, 255:red, 0; green, 0; blue, 0 }  ][line width=0.08]  [draw opacity=0] (8.93,-4.29) -- (0,0) -- (8.93,4.29) -- cycle    ;
\draw    (120.01,237.18) -- (120.01,251.49) ;
\draw [shift={(120.01,234.18)}, rotate = 90] [fill={rgb, 255:red, 0; green, 0; blue, 0 }  ][line width=0.08]  [draw opacity=0] (8.93,-4.29) -- (0,0) -- (8.93,4.29) -- cycle    ;
\draw    (141,258.17) -- (126.69,258.17) ;
\draw [shift={(144,258.17)}, rotate = 180] [fill={rgb, 255:red, 0; green, 0; blue, 0 }  ][line width=0.08]  [draw opacity=0] (8.93,-4.29) -- (0,0) -- (8.93,4.29) -- cycle    ;
\draw    (119.03,280.14) -- (119.03,265.83) ;
\draw [shift={(119.03,283.14)}, rotate = 270] [fill={rgb, 255:red, 0; green, 0; blue, 0 }  ][line width=0.08]  [draw opacity=0] (8.93,-4.29) -- (0,0) -- (8.93,4.29) -- cycle    ;

\draw    (104.33,242.98) -- (114.45,253.1) ;
\draw [shift={(102.21,240.86)}, rotate = 45] [fill={rgb, 255:red, 0; green, 0; blue, 0 }  ][line width=0.08]  [draw opacity=0] (8.93,-4.29) -- (0,0) -- (8.93,4.29) -- cycle    ;
\draw    (134.71,242.98) -- (124.59,253.1) ;
\draw [shift={(136.83,240.86)}, rotate = 135] [fill={rgb, 255:red, 0; green, 0; blue, 0 }  ][line width=0.08]  [draw opacity=0] (8.93,-4.29) -- (0,0) -- (8.93,4.29) -- cycle    ;
\draw    (134.71,273.36) -- (124.59,263.24) ;
\draw [shift={(136.83,275.48)}, rotate = 225] [fill={rgb, 255:red, 0; green, 0; blue, 0 }  ][line width=0.08]  [draw opacity=0] (8.93,-4.29) -- (0,0) -- (8.93,4.29) -- cycle    ;
\draw    (104.33,273.36) -- (114.45,263.24) ;
\draw [shift={(102.21,275.48)}, rotate = 315] [fill={rgb, 255:red, 0; green, 0; blue, 0 }  ][line width=0.08]  [draw opacity=0] (8.93,-4.29) -- (0,0) -- (8.93,4.29) -- cycle    ;

\draw    (92.66,150) -- (251.24,104.69) ;
\draw    (56.22,190.03) -- (75.12,208.93) ;
\draw [shift={(54.1,187.91)}, rotate = 45] [fill={rgb, 255:red, 0; green, 0; blue, 0 }  ][line width=0.08]  [draw opacity=0] (8.93,-4.29) -- (0,0) -- (8.93,4.29) -- cycle    ;

\draw (349.25,373.4) node  [font=\Large]  {$\Omega _{s}$};
\draw (118.38,311.15) node  [font=\Large]  {$\Omega _{f}$};
\draw (189.25,69.17) node  [font=\Large]  {$3D$};
\draw (421.16,175.29) node  [font=\Large]  {$6D$};
\draw (33.45,132.86) node  [font=\Large]  {$a$};
\draw (69.26,259.99) node  [font=\Large]  {$D$};
\draw (102.65,52.86) node  [font=\Large]  {$z$};
\draw (39.52,20.02) node  [font=\Large]  {$x$};
\draw (105.8,83.36) node  [font=\Large]  {$y$};

\end{tikzpicture}

%% file: figures/balloon-grid-2.pdf_tex
\begingroup%
  \makeatletter%
  \providecommand\color[2][]{%
    \errmessage{(Inkscape) Color is used for the text in Inkscape, but the package 'color.sty' is not loaded}%
    \renewcommand\color[2][]{}%
  }%
  \providecommand\transparent[1]{%
    \errmessage{(Inkscape) Transparency is used (non-zero) for the text in Inkscape, but the package 'transparent.sty' is not loaded}%
    \renewcommand\transparent[1]{}%
  }%
  \providecommand\rotatebox[2]{#2}%
  \newcommand*\fsize{\dimexpr\f@size pt\relax}%
  \newcommand*\lineheight[1]{\fontsize{\fsize}{#1\fsize}\selectfont}%
  \ifx\svgwidth\undefined%
    \setlength{\unitlength}{850bp}%
    \ifx\svgscale\undefined%
      \relax%
    \else%
      \setlength{\unitlength}{\unitlength * \real{\svgscale}}%
    \fi%
  \else%
    \setlength{\unitlength}{\svgwidth}%
  \fi%
  \global\let\svgwidth\undefined%
  \global\let\svgscale\undefined%
  \makeatother%
  \begin{picture}(1,0.86352941)%
    \lineheight{1}%
    \setlength\tabcolsep{0pt}%
    \put(0,0){\includegraphics[width=\unitlength,page=1]{figures/balloon-grid-2.pdf}}%
    \put(0.81029412,0.20617647){\makebox(0,0)[lt]{\lineheight{1.25}\smash{\begin{tabular}[t]{l}$x$\end{tabular}}}}%
    \put(0.86548713,0.12422564){\makebox(0,0)[lt]{\lineheight{1.25}\smash{\begin{tabular}[t]{l}$z$\end{tabular}}}}%
    \put(0.92284007,0.07569623){\makebox(0,0)[lt]{\lineheight{1.25}\smash{\begin{tabular}[t]{l}$y$\end{tabular}}}}%
  \end{picture}%
\endgroup%

%% file: manuscript.bbl
\begin{thebibliography}{40}
\expandafter\ifx\csname natexlab\endcsname\relax\def\natexlab#1{#1}\fi
\providecommand{\bibinfo}[2]{#2}
\ifx\xfnm\relax \def\xfnm[#1]{\unskip,\space#1}\fi
\bibitem[{Hughes et~al.(2005)Hughes, Cottrell, and Bazilevs}]{hughes2005}
\bibinfo{author}{T.~Hughes}, \bibinfo{author}{J.~Cottrell},
  \bibinfo{author}{Y.~Bazilevs},
\newblock \bibinfo{title}{Isogeometric analysis: {CAD}, finite elements,
  {NURBS}, exact geometry and mesh refinement},
\newblock \bibinfo{journal}{Computer Methods in Applied Mechanics and
  Engineering} \bibinfo{volume}{194} (\bibinfo{year}{2005})
  \bibinfo{pages}{4135--4195}.
\bibitem[{Zhang et~al.(2012)Zhang, Wang, and Hughes}]{zhang2012solid}
\bibinfo{author}{Y.~Zhang}, \bibinfo{author}{W.~Wang}, \bibinfo{author}{T.~J.
  Hughes},
\newblock \bibinfo{title}{Solid {T}-spline construction from boundary
  representations for genus-zero geometry},
\newblock \bibinfo{journal}{Computer Methods in Applied Mechanics and
  Engineering} \bibinfo{volume}{249-252} (\bibinfo{year}{2012})
  \bibinfo{pages}{185--197}.
\bibitem[{Schillinger et~al.(2012)Schillinger, Ded\`{e}, Scott, Evans, Borden,
  Rank, and Hughes}]{schillinger2012isogeometric}
\bibinfo{author}{D.~Schillinger}, \bibinfo{author}{L.~Ded\`{e}},
  \bibinfo{author}{M.~A. Scott}, \bibinfo{author}{J.~A. Evans},
  \bibinfo{author}{M.~J. Borden}, \bibinfo{author}{E.~Rank},
  \bibinfo{author}{T.~J. Hughes},
\newblock \bibinfo{title}{An isogeometric design-through-analysis methodology
  based on adaptive hierarchical refinement of {NURBS}, immersed boundary
  methods, and {T}-spline {CAD} surfaces},
\newblock \bibinfo{journal}{Computer Methods in Applied Mechanics and
  Engineering} \bibinfo{volume}{249-252} (\bibinfo{year}{2012})
  \bibinfo{pages}{116--150}.
\bibitem[{Sevilla et~al.(2011)Sevilla, Fern\'{a}ndez-M\'{e}ndez, and
  Huerta}]{sevilla2011}
\bibinfo{author}{R.~Sevilla}, \bibinfo{author}{S.~Fern\'{a}ndez-M\'{e}ndez},
  \bibinfo{author}{A.~Huerta},
\newblock \bibinfo{title}{{NURBS}-enhanced finite element method ({NEFEM})},
\newblock \bibinfo{journal}{Archives of Computational Methods in Engineering}
  \bibinfo{volume}{18} (\bibinfo{year}{2011}) \bibinfo{pages}{441}.
\bibitem[{Sevilla et~al.(2008)Sevilla, Fern\'{a}ndez-M\'{e}ndez, and
  Huerta}]{sevilla2008NEFEM}
\bibinfo{author}{R.~Sevilla}, \bibinfo{author}{S.~Fern\'{a}ndez-M\'{e}ndez},
  \bibinfo{author}{A.~Huerta},
\newblock \bibinfo{title}{{NURBS}-enhanced finite element method ({NEFEM})},
\newblock \bibinfo{journal}{International Journal for Numerical Methods in
  Engineering} \bibinfo{volume}{76} (\bibinfo{year}{2008})
  \bibinfo{pages}{56--83}.
\bibitem[{Sevilla et~al.(2011)Sevilla, Fern\'{a}ndez-M\'{e}ndez, and
  Huerta}]{sevilla20113dNEFEM}
\bibinfo{author}{R.~Sevilla}, \bibinfo{author}{S.~Fern\'{a}ndez-M\'{e}ndez},
  \bibinfo{author}{A.~Huerta},
\newblock \bibinfo{title}{{3D} {NURBS}-enhanced finite element method
  ({NEFEM})},
\newblock \bibinfo{journal}{International Journal for Numerical Methods in
  Engineering} \bibinfo{volume}{88} (\bibinfo{year}{2011})
  \bibinfo{pages}{103--125}.
\bibitem[{Stavrev et~al.(2016)Stavrev, Knechtges, Elgeti, and
  Huerta}]{stavrev2016STNEFEM}
\bibinfo{author}{A.~Stavrev}, \bibinfo{author}{P.~Knechtges},
  \bibinfo{author}{S.~Elgeti}, \bibinfo{author}{A.~Huerta},
\newblock \bibinfo{title}{Space-time {NURBS}-enhanced finite elements for
  free-surface flows in {2D}},
\newblock \bibinfo{journal}{International Journal for Numerical Methods in
  Fluids} \bibinfo{volume}{81} (\bibinfo{year}{2016})
  \bibinfo{pages}{426--450}.
\bibitem[{Cottrell et~al.(2009)Cottrell, Hughes, and
  Bazilevs}]{cottrell2009isogeometric}
\bibinfo{author}{J.~A. Cottrell}, \bibinfo{author}{T.~J. Hughes},
  \bibinfo{author}{Y.~Bazilevs}, \bibinfo{title}{Isogeometric analysis:
  {T}owards integration of {CAD} and {FEA}}, \bibinfo{publisher}{John Wiley \&
  Sons}, \bibinfo{year}{2009}.
\bibitem[{Sevilla et~al.(2016)Sevilla, Rees, and Hassan}]{Sevilla2016}
\bibinfo{author}{R.~Sevilla}, \bibinfo{author}{L.~Rees},
  \bibinfo{author}{O.~Hassan},
\newblock \bibinfo{title}{The generation of triangular meshes for
  nurbs-enhanced fem},
\newblock \bibinfo{journal}{International Journal for Numerical Methods in
  Engineering} \bibinfo{volume}{108} (\bibinfo{year}{2016})
  \bibinfo{pages}{941--968}.
\bibitem[{Hosters et~al.(2018)Hosters, Helmig, Stavrev, Behr, and
  Elgeti}]{hosters2018}
\bibinfo{author}{N.~Hosters}, \bibinfo{author}{J.~Helmig},
  \bibinfo{author}{A.~Stavrev}, \bibinfo{author}{M.~Behr},
  \bibinfo{author}{S.~Elgeti},
\newblock \bibinfo{title}{{Fluid}-structure interaction with {NURBS}-based
  coupling},
\newblock \bibinfo{journal}{Computer Methods in Applied Mechanics and
  Engineering} \bibinfo{volume}{332} (\bibinfo{year}{2018})
  \bibinfo{pages}{520--539}.
\bibitem[{Szab\'{o} et~al.(2004)Szab\'{o}, D{\"u}ster, and
  Rank}]{szabo2004pfem}
\bibinfo{author}{B.~Szab\'{o}}, \bibinfo{author}{A.~D{\"u}ster},
  \bibinfo{author}{E.~Rank},
\newblock \bibinfo{title}{The p-version of the finite element method},
\newblock in: \bibinfo{booktitle}{Encyclopedia of {C}omputational {M}echanics},
  \bibinfo{publisher}{Wiley \& Sons}, \bibinfo{year}{2004}, pp.
  \bibinfo{pages}{119--139}.
\bibitem[{Causin et~al.(2005)Causin, Gerbeau, and Nobile}]{causin2005}
\bibinfo{author}{P.~Causin}, \bibinfo{author}{J.~Gerbeau},
  \bibinfo{author}{F.~Nobile},
\newblock \bibinfo{title}{{Added-mass effect in the design of partitioned
  algorithms for fluid-structure problems}},
\newblock \bibinfo{journal}{Computer Methods in Applied Mechanics and
  Engineering} \bibinfo{volume}{194} (\bibinfo{year}{2005})
  \bibinfo{pages}{4506--4527}.
\bibitem[{Degroote et~al.(2010)Degroote, Swillens, Bruggeman, Haelterman,
  Segers, and Vierendeels}]{degroote2010}
\bibinfo{author}{J.~Degroote}, \bibinfo{author}{A.~Swillens},
  \bibinfo{author}{P.~Bruggeman}, \bibinfo{author}{R.~Haelterman},
  \bibinfo{author}{P.~Segers}, \bibinfo{author}{J.~Vierendeels},
\newblock \bibinfo{title}{Simulation of fluid–structure interaction with the
  interface artificial compressibility method},
\newblock \bibinfo{journal}{International Journal for Numerical Methods in
  Biomedical Engineering} \bibinfo{volume}{26} (\bibinfo{year}{2010})
  \bibinfo{pages}{276--289}.
\bibitem[{Spenke et~al.(2020)Spenke, Hosters, and Behr}]{spenke2020}
\bibinfo{author}{T.~Spenke}, \bibinfo{author}{N.~Hosters},
  \bibinfo{author}{M.~Behr},
\newblock \bibinfo{title}{A multi-vector interface quasi-{N}ewton method with
  linear complexity for partitioned fluid–structure interaction},
\newblock \bibinfo{journal}{Computer Methods in Applied Mechanics and
  Engineering} \bibinfo{volume}{361} (\bibinfo{year}{2020})
  \bibinfo{pages}{112810}.
\bibitem[{K{\"u}ttler et~al.(2006)K{\"u}ttler, F{\"o}rster, and
  Wall}]{kuttler2006}
\bibinfo{author}{U.~K{\"u}ttler}, \bibinfo{author}{C.~F{\"o}rster},
  \bibinfo{author}{W.~A. Wall},
\newblock \bibinfo{title}{A solution for the incompressibility dilemma in
  partitioned fluid-structure interaction with pure {Dirichlet} fluid domains},
\newblock \bibinfo{journal}{Comput Mech} \bibinfo{volume}{38}
  (\bibinfo{year}{2006}) \bibinfo{pages}{417--429}.
\bibitem[{Badia et~al.(2008)Badia, Nobile, and Vergara}]{badia2008}
\bibinfo{author}{S.~Badia}, \bibinfo{author}{F.~Nobile},
  \bibinfo{author}{C.~Vergara},
\newblock \bibinfo{title}{Fluid-structure partitioned procedures based on
  {Robin} transmission conditions},
\newblock \bibinfo{journal}{Journal of Computational Physics}
  \bibinfo{volume}{227} (\bibinfo{year}{2008}) \bibinfo{pages}{7027--7051}.
\bibitem[{Gerardo-Giorda et~al.(2010)Gerardo-Giorda, Nobile, and
  Vergara}]{gerardo-giorda2010}
\bibinfo{author}{L.~Gerardo-Giorda}, \bibinfo{author}{F.~Nobile},
  \bibinfo{author}{C.~Vergara},
\newblock \bibinfo{title}{Analysis and optimization of {Robin}-{Robin}
  partitioned procedures in fluid-structure interaction problems},
\newblock \bibinfo{journal}{SIAM J. Numer. Anal.} \bibinfo{volume}{48}
  (\bibinfo{year}{2010}) \bibinfo{pages}{2091–2116}.
\bibitem[{Nobile and Vergara(2008)}]{nobile2008}
\bibinfo{author}{F.~Nobile}, \bibinfo{author}{C.~Vergara},
\newblock \bibinfo{title}{An effective fluid-structure interaction formulation
  for vascular dynamics by generalized {Robin} conditions},
\newblock \bibinfo{journal}{SIAM J. Sci. Comput.} \bibinfo{volume}{30}
  (\bibinfo{year}{2008}) \bibinfo{pages}{731–763}.
\bibitem[{Hosters(2018)}]{hosters2018thesis}
\bibinfo{author}{N.~Hosters}, \bibinfo{title}{{S}pline-based methods for
  fluid-structure interaction}, \bibinfo{type}{{PhD} thesis}, RWTH Aachen
  University, \bibinfo{address}{Aachen}, \bibinfo{year}{2018}.
\bibitem[{Tezduyar et~al.(1992)Tezduyar, Behr, and Liou}]{TEZDUYAR1992339}
\bibinfo{author}{T.~Tezduyar}, \bibinfo{author}{M.~Behr},
  \bibinfo{author}{J.~Liou},
\newblock \bibinfo{title}{A new strategy for finite element computations
  involving moving boundaries and interfaces--{T}he
  deforming-spatial-domain/space-time procedure: {I}. {T}he concept and the
  preliminary numerical tests},
\newblock \bibinfo{journal}{Computer Methods in Applied Mechanics and
  Engineering} \bibinfo{volume}{94} (\bibinfo{year}{1992})
  \bibinfo{pages}{{339--351}}.
\bibitem[{Piegl and Tiller(1997)}]{piegl1997nurbs}
\bibinfo{author}{L.~Piegl}, \bibinfo{author}{W.~Tiller}, \bibinfo{title}{The
  {NURBS} book}, \bibinfo{publisher}{Springer-Verlag}, \bibinfo{year}{1997}.
\bibitem[{Bazilevs et~al.(2010)Bazilevs, Calo, Cottrell, Evans, Hughes, Lipton,
  Scott, and Sederberg}]{BAZILEVS2010229}
\bibinfo{author}{Y.~Bazilevs}, \bibinfo{author}{V.~Calo},
  \bibinfo{author}{J.~Cottrell}, \bibinfo{author}{J.~Evans},
  \bibinfo{author}{T.~Hughes}, \bibinfo{author}{S.~Lipton},
  \bibinfo{author}{M.~Scott}, \bibinfo{author}{T.~Sederberg},
\newblock \bibinfo{title}{Isogeometric analysis using {T}-splines},
\newblock \bibinfo{journal}{Computer Methods in Applied Mechanics and
  Engineering} \bibinfo{volume}{199} (\bibinfo{year}{2010})
  \bibinfo{pages}{229--263}.
\bibitem[{Sevilla(2009)}]{sevilla2009thesis}
\bibinfo{author}{R.~Sevilla}, \bibinfo{title}{NURBS-enhanced finite element
  method (NEFEM)}, Ph.D. thesis, Universitat Polit{\`e}cnica de Catalunya
  (UPC), \bibinfo{year}{2009}.
\bibitem[{Sevilla et~al.(2011)Sevilla, Fern\'{a}ndez-M\'{e}ndez, and
  Huerta}]{sevilla2011Comparison}
\bibinfo{author}{R.~Sevilla}, \bibinfo{author}{S.~Fern\'{a}ndez-M\'{e}ndez},
  \bibinfo{author}{A.~Huerta},
\newblock \bibinfo{title}{Comparison of high-order curved finite elements},
\newblock \bibinfo{journal}{International Journal for Numerical Methods in
  Engineering} \bibinfo{volume}{87} (\bibinfo{year}{2011})
  \bibinfo{pages}{719--734}.
\bibitem[{Zienkiewicz et~al.(2005)Zienkiewicz, Taylor, and
  Zhu}]{zienkiewicz2005finite}
\bibinfo{author}{O.~C. Zienkiewicz}, \bibinfo{author}{R.~L. Taylor},
  \bibinfo{author}{J.~Z. Zhu}, \bibinfo{title}{The finite element method: its
  basis and fundamentals}, \bibinfo{publisher}{Elsevier}, \bibinfo{year}{2005}.
\bibitem[{Williams et~al.(2014)Williams, Shunn, and
  Jameson}]{williams2014symmetric}
\bibinfo{author}{D.~Williams}, \bibinfo{author}{L.~Shunn},
  \bibinfo{author}{A.~Jameson},
\newblock \bibinfo{title}{Symmetric quadrature rules for simplexes based on
  sphere close packed lattice arrangements},
\newblock \bibinfo{journal}{Journal of Computational and Applied Mathematics}
  \bibinfo{volume}{266} (\bibinfo{year}{2014}) \bibinfo{pages}{18--38}.
\bibitem[{Shunn and Ham(2012)}]{shunn2012}
\bibinfo{author}{L.~Shunn}, \bibinfo{author}{F.~Ham},
\newblock \bibinfo{title}{Symmetric quadrature rules for tetrahedra based on a
  cubic close-packed lattice arrangement},
\newblock \bibinfo{journal}{Journal of Computational and Applied Mathematics}
  \bibinfo{volume}{236} (\bibinfo{year}{2012}) \bibinfo{pages}{4348--4364}.
\bibitem[{Fu and Ogden(2001)}]{Fu2001NonlinearE}
\bibinfo{author}{Y.~Fu}, \bibinfo{author}{R.~W. Ogden},
  \bibinfo{title}{{N}onlinear elasticity: {T}heory and {A}pplications},
  \bibinfo{publisher}{Cambridge University Press}, \bibinfo{year}{2001}.
\bibitem[{Bathe(1996)}]{bathe1996finite}
\bibinfo{author}{K.-J. Bathe}, \bibinfo{title}{Finite element procedures},
  \bibinfo{publisher}{Prentice Hall}, \bibinfo{year}{1996}.
\bibitem[{Karyofylli et~al.(2019)Karyofylli, Wendling, Make, Hosters, and
  Behr}]{karyofylli2019simplex-space-time}
\bibinfo{author}{V.~Karyofylli}, \bibinfo{author}{L.~Wendling},
  \bibinfo{author}{M.~Make}, \bibinfo{author}{N.~Hosters},
  \bibinfo{author}{M.~Behr},
\newblock \bibinfo{title}{Simplex space-time meshes in thermally coupled
  two-phase flow simulations of mold filling},
\newblock \bibinfo{journal}{Computers \& Fluids} \bibinfo{volume}{192}
  (\bibinfo{year}{2019}) \bibinfo{pages}{104261}.
\bibitem[{Pauli and Behr(2017)}]{pauli2017}
\bibinfo{author}{L.~Pauli}, \bibinfo{author}{M.~Behr},
\newblock \bibinfo{title}{On stabilized space-time {FEM} for anisotropic
  meshes: Incompressible {N}avier-{S}tokes equations and applications to blood
  flow in medical devices},
\newblock \bibinfo{journal}{International Journal for Numerical Methods in
  Fluids} \bibinfo{volume}{85} (\bibinfo{year}{2017})
  \bibinfo{pages}{189--209}.
\bibitem[{Jansen et~al.(1999)Jansen, Collis, Whiting, and
  Shakib}]{JANSEN1999153}
\bibinfo{author}{K.~E. Jansen}, \bibinfo{author}{S.~S. Collis},
  \bibinfo{author}{C.~Whiting}, \bibinfo{author}{F.~Shakib},
\newblock \bibinfo{title}{A better consistency for low-order stabilized finite
  element methods},
\newblock \bibinfo{journal}{Computer Methods in Applied Mechanics and
  Engineering} \bibinfo{volume}{174} (\bibinfo{year}{1999})
  \bibinfo{pages}{153--170}.
\bibitem[{Chung and Hulbert(1993)}]{chung1993}
\bibinfo{author}{J.~Chung}, \bibinfo{author}{G.~M. Hulbert},
\newblock \bibinfo{title}{A time integration algorithm for structural dynamics
  with improved numerical dissipation: {T}he generalized-\ensuremath{\alpha}
  method},
\newblock \bibinfo{journal}{Journal of Applied Mechanics} \bibinfo{volume}{60}
  (\bibinfo{year}{1993}) \bibinfo{pages}{371--375}.
\bibitem[{Kuhl and Crisfield(1999)}]{kuhl1999}
\bibinfo{author}{D.~Kuhl}, \bibinfo{author}{A.~Crisfield},
\newblock \bibinfo{title}{Energy-conserving and decaying algorithms in
  non-linear structural dynamics},
\newblock \bibinfo{journal}{International Journal for Numerical Methods in
  Engineering} \bibinfo{volume}{45} (\bibinfo{year}{1999})
  \bibinfo{pages}{569--599}.
\bibitem[{Ramm and Wall(2004)}]{ramm2004}
\bibinfo{author}{E.~Ramm}, \bibinfo{author}{W.~A. Wall},
\newblock \bibinfo{title}{Shell structures -- a sensitive interrelation between
  physics and numerics},
\newblock \bibinfo{journal}{International Journal for Numerical Methods in
  Engineering} \bibinfo{volume}{60} (\bibinfo{year}{2004})
  \bibinfo{pages}{381--427}.
\bibitem[{Benson et~al.(2010)Benson, Bazilevs, Hsu, and Hughes}]{benson2010}
\bibinfo{author}{D.~Benson}, \bibinfo{author}{Y.~Bazilevs},
  \bibinfo{author}{M.~Hsu}, \bibinfo{author}{T.~Hughes},
\newblock \bibinfo{title}{{Isogeometric shell analysis: The Reissner--Mindlin
  shell}},
\newblock \bibinfo{journal}{Computer Methods in Applied Mechanics and
  Engineering} \bibinfo{volume}{199} (\bibinfo{year}{2010})
  \bibinfo{pages}{276--289}.
\bibitem[{Sch{\"a}fer et~al.(1996)Sch{\"a}fer, Turek, Durst, Krause, and
  Rannacher}]{Schaefer1996}
\bibinfo{author}{M.~Sch{\"a}fer}, \bibinfo{author}{S.~Turek},
  \bibinfo{author}{F.~Durst}, \bibinfo{author}{E.~Krause},
  \bibinfo{author}{R.~Rannacher},
\newblock \bibinfo{title}{Benchmark computations of laminar flow around a
  cylinder},
\newblock in: \bibinfo{editor}{E.~H. Hirschel} (Ed.), \bibinfo{booktitle}{Flow
  Simulation with High-Performance Computers II. Notes on Numerical Fluid
  Mechanics (NNFM)}, volume~\bibinfo{volume}{48},
  \bibinfo{publisher}{Vieweg+Teubner Verlag}, \bibinfo{address}{Wiesbaden},
  \bibinfo{year}{1996}, pp. \bibinfo{pages}{547--566}.
\bibitem[{Braack and Richter(2006)}]{BRAACK2006372}
\bibinfo{author}{M.~Braack}, \bibinfo{author}{T.~Richter},
\newblock \bibinfo{title}{Solutions of {3D} {N}avier-{S}tokes benchmark
  problems with adaptive finite elements},
\newblock \bibinfo{journal}{Computers \& Fluids} \bibinfo{volume}{35}
  (\bibinfo{year}{2006}) \bibinfo{pages}{372--392}.
\bibitem[{John(2002)}]{john2002}
\bibinfo{author}{V.~John},
\newblock \bibinfo{title}{Higher order finite element methods and multigrid
  solvers in a benchmark problem for the {3D} {N}avier-{S}tokes equations},
\newblock \bibinfo{journal}{International Journal for Numerical Methods in
  Fluids} \bibinfo{volume}{40} (\bibinfo{year}{2002})
  \bibinfo{pages}{775--798}.
\bibitem[{John(2006)}]{john2006efficiency}
\bibinfo{author}{V.~John},
\newblock \bibinfo{title}{On the efficiency of linearization schemes and
  coupled multigrid methods in the simulation of a {3D} flow around a
  cylinder},
\newblock \bibinfo{journal}{International Journal for Numerical Methods in
  Fluids} \bibinfo{volume}{50} (\bibinfo{year}{2006})
  \bibinfo{pages}{845--862}.

\end{thebibliography}
